\documentclass[twoside,a4paper,10pt]{book}

\usepackage{amsmath}
\usepackage{amssymb}
\usepackage{graphicx}
\usepackage{fancyhdr}              	

\let\==\equiv

\def\qed{\raise1pt\hbox{\vrule height5pt width5pt depth0pt}}

\def\to{\rightarrow}

\mathchardef\aa   = "050B
\mathchardef\bb   = "050C
\mathchardef\ggg  = "050D
\mathchardef\xxx  = "0518
\mathchardef\zzzzz= "0510
\mathchardef\oo   = "0521
\mathchardef\lll  = "0515
\mathchardef\mm   = "0516
\mathchardef\Dp   = "0540
\mathchardef\H    = "0548
\mathchardef\FFF  = "0546
\mathchardef\ppp  = "0570
\mathchardef\nn   = "0517
\mathchardef\ff   = "0527
\mathchardef\pps  = "0520
\mathchardef\FFF  = "0508
\mathchardef\nnnnn= "056E

\newcommand{\beq}{\begin{equation}}
\newcommand{\eeq}{\end{equation}}
\newcommand{\bea}{\begin{eqnarray}}
\newcommand{\eea}{\end{eqnarray}}

\begin{document}

\frontmatter

\begin{titlepage}

\thispagestyle{empty}
\begin{center}
{\large \textsc{UNIVERSIT\`E DE PARIS 11 - U.F.R. DES SCIENCES D'ORSAY}} \\

\vspace{1cm}

{\large \textsc{Laboratoire de Physique Th\'eorique d'Orsay}} \\

\vspace{0.5cm}

\includegraphics[width=3cm]{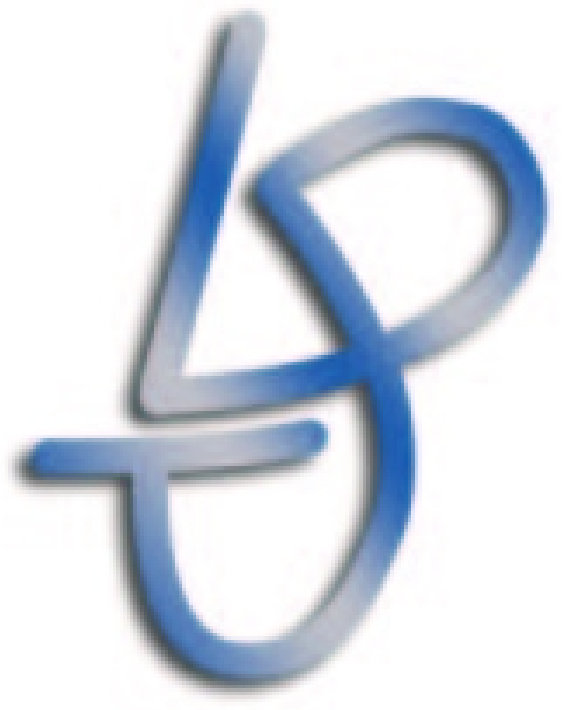}

\vspace{1cm}

{\normalsize \textbf{TH\`ESE DE DOCTORAT DE L'UNIVERSIT\'E PARIS 11}} \\[.5cm]
Sp\'ecialit\'e: {\large \textbf{PHYSIQUE TH\'EORIQUE}}\\[.25cm]
pr\'esent\'e par\\ [.5cm]
{\large \textbf{Luca DALL'ASTA}}\\[.25cm]
pour obtenir le grade de \\[.5cm]
{\large DOCTEUR DE L'UNIVERSIT\'E PARIS 11}
\vspace{1cm}

Sujet:

\vspace{0.5cm}

{\huge

\textbf{Ph\'enom\`enes dynamiques sur des r\'eseaux complexes} \\
} 

\end{center}

\vspace{1cm}
{\large

\noindent Jury compos\'e de:
\vspace{.2cm}

\noindent M.~\textsl{Alain Barrat} \hspace{1cm} (Directeur de th\`ese) 

\noindent M.~\textsl{Olivier Martin}
 
\noindent M.~\textsl{R\'emi Monasson} 

\noindent M.~\textsl{Romualdo Pastor-Satorras}  \hspace{1cm} (Rapporteur) 

\noindent M.~\textsl{Cl\'ement Sire}  \hspace{1cm} (Rapporteur) 

\noindent M.~\textsl{Alessandro Vespignani}
}

\vspace{1.cm}

\begin{center}
{\large {Juin, 2006}}
\end{center}

\end{titlepage}

\thispagestyle{empty} 


\pagestyle{empty}
\pagenumbering{roman}
\mbox{}
\newpage

\setcounter{page}{1}
\tableofcontents
\newpage
\pagestyle{fancy}

\addcontentsline{toc}{chapter}{List of Publications}

\chapter*{List of Publications}

The results exposed in this thesis have been published in a series of papers and preprints that we report here divided by argument:
\begin{itemize}

\item Exploration of networks (Chapter~\ref{CHAP3}): 
\begin{itemize}
\item[-]
Dall'Asta, L., Alvarez-Hamelin, J.~I., Barrat, A., V\'azquez, A., and Vespignani, A.\\
`Traceroute-like exploration of unknown networks: A statistical analysis,'\\
{\em Lect. Notes Comp. Sci.} {\bf 3405}, 140-153 (2005).

\item[-]
Dall'Asta, L., Alvarez-Hamelin, J.~I., Barrat, A.,  V\'azquez, A., and Vespignani, A.,\\
`Exploring networks with traceroute-like probes: theory and
simulations, \\
{\em Theor. Comp. Sci.} {\bf 355}, 6-24 (2006).

\item[-]
Dall'Asta, L.,  Alvarez-Hamelin, J.~I., Barrat, A.,  V\'azquez, A., and Vespignani, A.,\\
`Statistical theory of Internet exploration,'\\
{\em Phys. Rev. E} {\bf 71} 036135 (2005).

\item[-]
Dall'Asta, L.,  Alvarez-Hamelin, J.~I., Barrat, A., V\'azquez, A., and Vespignani, A.,\\
`How accurate are \texttt{traceroute}-like Internet mappings?,'\\
{\em Comference Proc. AlgoTel '05}, INRIA, 31-34 (2005).

\item[-]
Viger, F., Barrat, A., Dall'Asta, L.,  Zhang, C.-H., and Kolaczyk, E.,\\ 
`Network Inference from Traceroute Measurements: Internet Topology `Species','\\
preprint arxiv:cs/0510007 (2005).

\end{itemize}

\item $k$-core analysis of networks (Chapter~\ref{CHAP3}): 
\begin{itemize}
\item[-] Alvarez-Hamelin, J.~I., Dall'Asta, L., Barrat, A., and Vespignani, A.,\\
`$k$-core decomposition: a tool for the visualization of large scale networks,'\\
preprint arxiv:cs/0504107 (2005).

\item[-] Alvarez-Hamelin, J.~I., Dall'Asta, L., Barrat, A., and Vespignani, A.,\\
`k-core decomposition: a tool for the analysis of large scale Internet graphs,'\\
preprint arxiv:cs/0511007 (2005).

\item[-] Alvarez-Hamelin, J.~I., Dall'Asta, L., Barrat, A., and Vespignani, A.,\\
`Large scale networks fingerprinting and visualization using the k-core decomposition,' \\
in {\em Advances in Neural Information Processing Systems, NIPS '05} {\bf 18} (2005). 
\end{itemize}

\item Functional properties of weighted networks (Chapter~\ref{CHAP4}):
\begin{itemize}
\item[-]
Dall'Asta, L.,\\
`Inhomogeneous percolation models for spreading phenomena in random  graphs,' \\
{\em J. Stat. Mech.} P08011 (2005).

\item[-]
Dall'Asta, L., Barrat, A., Barth\'elemy, M., and Vespignani, A., \\
`Vulnerability of weighted networks,' \\
{\em J. Stat. Mech. in press},  
preprint arxiv:physics/0603163, (2006).
\end{itemize}

\item Naming Game Model (Chapter~\ref{CHAP5}):
\begin{itemize}
\item[-]
Baronchelli, A.,  Dall'Asta, L., Barrat, A., and Loreto, V.,\\ 
`Topology Induced Coarsening in Language Games,' \\
{\em Phys. Rev. E} {\bf 73}, 015102(R) (2006). 

\item[-]
Baronchelli, A.,  Dall'Asta, L.,  Barrat, A., and Loreto, V., \\  
`Strategies for fast convergence in semiotic dynamics,' \\
{\em ALIFE X}, Bloomington Indiana (2006), (preprint arxiv:physics/0511201).

\item[-] 
Dall'Asta, L., Baronchelli, A., Barrat, A., and Loreto, V.,  \\ 
`Agreement dynamics on small-world networks,' \\
{\em Europhys. Lett.} {\bf 73(6)}, 969-975 (2006). 

\item[-]
Dall'Asta, L., Baronchelli, A., Barrat, A., and Loreto, V., \\
`Non-equilibrium dynamics of language games on complex networks,' \\
{\em submitted to Phys. Rev. E} (2006).
\end{itemize}

\item Other works published during the PhD:

\item[-]
B\"orner, K., Dall'Asta, L.,  Ke, W., and Vespignani, A.,  \\
`Studying The Emerging Global Brain: Analyzing And Visualizing The Impact Of Co-Authorship Teams,' \\
{\em Complexity} {\bf 10(4)}, 57-67 (2005).

\item[-]
L. Dall'Asta,\\
`Exact Solution of the One-Dimensional Deterministic Fixed-Energy Sandpile,' \\
{\em Phys. Rev. Lett.} {\bf 96}, 058003 (2006). 

\item[-]
M. Casartelli, L. Dall'Asta, A. Vezzani, and P. Vivo, \\
`Dynamical Invariants in the Deterministic Fixed-Energy Samdpile,'  \\
{\em submitted to Eur. Phys. J. B}, preprint arxiv:cond-mat/0502208 (2006).

\end{itemize}

\mainmatter

\pagenumbering{arabic}
\setcounter{page}{1}

\chapter{Introduction}
\label{CHAP1}

\section{A networked description of Nature and Society}
The recent interest of a wide interdisciplinary scientific community for the study of complex networks is justified primary by the fact that a network description of complex systems allows to get relevant information by means of purely statistical coarse-grained analyses, without taking into account the detailed characterization of the system.
Moreover, using an abstract networked representation, it is possible to compare, in the same framework, systems that are originally very different, so that the identification of some universal properties becomes much easier. 
{\em Simplicity} and {\em universality} are two fundamental principles of the physical research, in particular of statistical physics, that is traditionally interested in the study of the emergence of collective phenomena in many interacting particles systems, even outside its classical fields of research such as condensed matter theory.\\
Along the last century, statistical physicists have developed a suite of analytical and numerical techniques by means of which it has been possible to understand the origin of phase transitions and critical phenomena in many particle systems, and that have been successfully applied also in other fields, from informatics (e.g. optimization problems) to biology (e.g. protein folding) and social sciences (e.g. opinion formation models).
The presence of disorder, randomness and heterogeneity is the other important ingredient that justifies the use of statistical physics approaches in so many different fields and in particular in the study of complex networks, that present non-trivial irregular topological structures. \\
From a mathematical point of view, complex networks are sets of many interacting components, the nodes, whose collective behavior is complex in the sense that it cannot be directly predicted and characterized in terms of the behavior of the individual components. The links connecting pairs of components correspond to the interactions that are responsible of the global behavior of the system.
It is clear that a large number of systems can be described in this manner, thus it is not difficult to find disparate examples of networks both in nature and society. 
The most evident application of networks theory is the study of the Internet~\cite{vespi_book}, whose detailed characterization is not possible, but that can be investigated using the statistical analysis of its topological and functional properties.  
In general, all infrastructures fit very well the framework of networks theory, so that most of the real networks studied
are communication or transportation networks (such as the Internet, the Web, the air-transportation network, power-grids, telephone and roads networks, etc)~\cite{mendes_book}.
A second class is represented by those networks related to social interactions~\cite{wasserman}, such as sexual-contact networks, networks of acquaintances or collaboration networks. Finally, another large class concerns biological networks~\cite{kauffman93,mendes_book} (e.g. protein interaction networks, cellular networks, neural networks, food-webs, etc.).\\ 
The massive use of statistical techniques in the characterization of complex networks is, however, closely related with the recent improvement of computers, that allow to easily retrieve, collect and handle large amounts of data. 
In the last decade, indeed, the analysis of the topological structure of large real networks such as the Internet and the Web pointed out that many real networks have unexpected topological properties, characterized by heterogeneous connectivity patterns~\cite{faloutsos}.  
These surprising results were in contrast with the common belief that real networks could be modeled using either regular networks (e.g. grids or fully connected networks) or random graphs (i.e. networks in which nodes are randomly connected in such a way that all them have approximately the same number of connections~\cite{clark_book}).  
These models have been studied for a long time without the necessity of a relevant statistical analysis, since in such  networks all nodes are approximately equivalent, and the overall behavior of the network is well represented by monitoring that of a single node. On the contrary, the recent discoveries immediately revealed a completely different scenario.\\
A large number of data about various networks have been gradually collected, ranging from social sciences to biology, all of them presenting the same type of heterogeneity in the connectivity patterns.
The necessity of a more mathematical analysis of networks excited a large number of physicists, who recognized the possibility to apply the powerful methods of statistical physics. 
Without going into a detailed description of the statistical framework that physicists have built introducing statistical physics methods into the ideas inherited from graph theory (see Chapter~\ref{CHAP2} for an introduction), 
the  main achievement in the characterization of networks topology is the identification of few universal features that are common to many networks and allow to divide them into different classes. \\
A first relevant property regards the degree of a node (i.e. the number of connections to other nodes). In real networks, the probability of finding a node with a given degree (i.e. the degree distribution) significantly deviates from the peaked distributions expected for random graphs and, in many cases, exhibits a broadly skewed shape, with power law tails with an exponent between $2$ and $3$. In this range of values for the exponent, the distribution presents diverging second moments, meaning that we can find very large fluctuations in the values of nodes connectivity ({\em scale-free property}).\\
Moreover, real networks are characterized by relatively short paths between any two nodes ({\em small-world property}~\cite{milgram,watts98}), a very important property in determining networks behavior at both structural and functional levels. 
The small world property, while intriguing, was already present in random graphs models, in which the average intervertex distance scales as the logarithm of the number of nodes. However, the novelty is due to the fact that real networks present this property together with a high density of triangles and other small cycles or motifs, that are completely absent in traditional random graphs, whose local structure is tree-like. \\
These unexpected results have initiated a revival of network modeling, resulting in the introduction and study of new classes of modeling paradigms~\cite{mendes_book,newman_review,vespi_book,lazlo_review}. 
Many efforts have been spent to conceive models that are able to reproduce and predict the statistical properties of real networks, but researchers have soon realized that the characterization of real networks is not exhausted by its topological properties and that {\em in real networks topology and dynamics are intrinsically related}.
 
\section[Relation between Topology and Dynamics]{Relation between Topology and Dynamics: a question of timescales}
The dynamical phenomena related to complex networks can be summarized in three different categories: the dynamical evolution of networks, the dynamics on networks, and the dynamical interplay between the networks topology and processes evolving on them.\\
The topology of real networks is indeed far from being fixed, the number of nodes and links changing together with local and global properties of the system. In particular, evolutionary principles are often necessary ingredients in order to explain some peculiar topological properties of networks (e.g. the preferential attachment principle is necessary to understand the emergence of degree heterogeneity in networks such as the Internet or the Web). \\
On the other hand, networks are structures on which dynamical processes take place, thus it is interesting to study the behavior of dynamical systems models evolving on networks.
Many of them, such as routing algorithms, oscillators, epidemic spreading, or searching processes, have direct applications in the study of the dynamical phenomena observed on real networks, others such as random walks, statistical mechanics models, opinion formation, percolation, and strategic games provide more general information that can be used to build a common theoretical framework by means of which the different properties of dynamical processes on networks can be analyzed and explained.\\
The third situation, characterized by the interplay of the dynamics ``of networks'' with the dynamics ``on networks'' is more complicated and such kind of problems has been only recently considered by the complex networks community. 
Moreover, this case has been usually neglected because of the different temporal scales of the two types of dynamics.
We can indeed assume that the structural properties of a network evolve with a time-scale $\tau_{T}$, while a particular dynamical process taking place on the network evolves with time-scale $\tau_{D}$, the above mentioned situation corresponding to the case $\tau_{T} \simeq \tau_{D}$.
When $\tau_{T} \gg \tau_{D}$, we can study the evolution of dynamical processes on networks with quenched topological structure; while the case $\tau_{D} \gg \tau_{T}$ means that the temporal evolution of processes on the network is neglected compared to that of the network structure itself. 
The latter case holds not only when these dynamical processes are slow compared to the changes in the topology, but also in the situation in which these processes are fast but they do not influence the structure of the network. \\
While the evolution of networks topology has been largely investigated in the past, the present thesis is devoted to study some aspects of dynamical processes on networks, i.e. the case in which $\tau_{T} \gg \tau_{D}$. 
Actually, the scenario is much more complicated since real networks are usually characterized by a large number of dynamical processes evolving at the same time, so that in addition to the mentioned topological and dynamical temporal scales, we have to distinguish the temporal scale governing the evolution of a single process $\tau_{D}^{s}$ from that governing the evolution of the overall average properties of that class of processes $\tau_{D}^{o}$. 
This is simple to understand if we think at the functioning of the Internet. Billions of data-packets are continuously transferred between the routers, each one performing a sort of (random) walk on the network from a source to a destination. But looking at the global average properties of the traffic, we observe sufficiently stable quantities, so that we can encode the average traffic between two neighboring routers (in terms of transferred bytes) using a single value, a {\em weight}, by means of which we can label the corresponding link.\\
Therefore, a first way to take into account of the dynamics is that of endowing the links with weights, representing the flow of information or the traffic among the constituent units of the system. More generally, a weighted network representation allows to take into account the functional properties of networks.\\
On the other hand, it is also important to focus on the dynamical behavior of single dynamical processes, such as the spreading of information or viruses on social and infrastructure networks, or the processes of networks exploration from a given source node. One of the striking results of this scales separation is that it is also possible to study single dynamical processes, such as spreading and percolation, in a weighted network, with quenched structural and functional properties (i.e. $\tau_{D}^{s} \ll \tau_{T}$ and $\tau_{D}^{s} \ll \tau_{D}^{o}$). \\
As a final remark, we note that also the general motivation with which we have studied dynamical phenomena on networks is twofold.
On the one hand, we wanted to study the effects of inhomogeneous topological and functional properties on the behavior of 
some classes of dynamical models; on the other hand, we have exploited some of these dynamical phenomena in order to investigate unknown topological properties of real networks.
This twofold role held by dynamical processes is maybe the most important idea that statistical physicists should learn from the new interdisciplinary field of complex networks. Physicists have been used for many year to study a variety of interactions on very well defined topologies, now we have to face a more complex scenario, in which the role of topology and dynamical rules may be even inverted, i.e. well-defined dynamical phenomena can be used to uncover topological properties of the system.

\section{Summary of the thesis}  
The work developed in this thesis concerns the study of various aspects of dynamical processes on networks: each chapter is devoted to a particular issue, but apparently different problems are related by the general scenario that we have mentioned in the previous paragraph. 

Chapter~\ref{CHAP2} provides an introduction to the science of complex networks: in the first part we recall the main statistical measures used to analyze networks; then we give some examples of real complex networks, focusing on the Internet and the World-wide Air-transportation Network; the final section is devoted to review the most important theoretical models of complex networks. This is not an exhaustive introduction, but it is conceived to give the most relevant notions that are used or mentioned in the rest of the work. 

Chapter~\ref{CHAP3} concerns the theoretical characterization of the processes of exploration of complex networks.
The relevance of this topic resides in the fact that the topology of real networks is often only partially known, and the methods used to acquire information on such topological properties may present biases affecting the reliability of the  phenomenological observations. We consider several different types of networks sampling methods, discussing their  advantages and limits with respect to their natural fields of application. 
We focus in particular on a tree-like exploration method used in real mapping processes of the Internet and referred as \texttt{traceroute} exploration. 
In order to verify the reliability of the experimental data, and consequently of the main properties, such as the existence of a broad degree distribution, that have been derived from their analysis, we propose a theoretical model of \texttt{traceroute} exploration of networks.
This model allows an analytical study by means of a mean-field approximation, providing a deeper understanding of the relation between the topological properties of the original network and those of the sampled network. 
Moreover, massive numerical simulations on computer-generated networks with various topologies allows to have also a clearer quantitative description of the mapping processes.
The general picture acquired from this study is finally exploited to introduce a statistical technique by means of which  some of the biased quantities can be opportunely corrected.  \\  
This topic is also a clear example of the possible use of dynamical processes for the characterization of unknown topological properties of real networks. 

In Chapter~\ref{CHAP4}, we take into account the weighted network representation and its relation with the functional properties of the network. In the first part of the chapter, we investigate the role of weights in determining the functional robustness of the system, and we compare the results with those based on purely topological measures. 
We use the case study of the airports network. The main idea is that of measuring the vulnerability of the network using global observables based on both topological and traffic centrality measures: we remove the most central nodes according to different centrality measures, monitoring the effects on the structural and functional integrity of the system.\\
This study gives an example of the different roles played by topology and weights, at the same time pointing out the validity of a static representation of networks functionality (i.e. encoding average flows and traffic into weights on the edges). 
The second part of the chapter is instead devoted to study weighted networks from a purely dynamical point of view. Exploiting some remarkable properties of percolation theory, we build a general theoretical framework in which spreading processes on weighted networks can be analyzed. \\ 
Using an analogy with the scenario proposed in the previous paragraph, passing from the first to the second part of the chapter, we pass from a situation in which we are interested only in the structural and functional properties determined by the average dynamical behavior of the system, to the study of the effects of such (structural and functional) properties on the evolution of a particular dynamical process on the network.

Chapter~\ref{CHAP5} is completely devoted to the analysis of the recently proposed Naming Game model, that was conceived as a model for the emergence of a communication system or a shared vocabulary in a population of agents. The rules governing the pairwise interactions between the individuals are simple but present several new features such as negotiation, feedback and memory, that are typical properties of human social dynamics. For this reason the model can be usefully applied also in different contexts, such as problems of opinion formation. \\
The dynamical evolution of the model is studied considering populations with different topologies, from regular lattices to complex networks, showing that the dynamical phenomena generated by the model depend strongly on the topological properties of the system. In the last section of the chapter, the attention is focused on the activity patterns of single agents, that display rather unexpected properties due to the non-trivial relation between memory and degree heterogeneity. 

General conclusions on the work done and possible future developments of the ideas exposed in the thesis are reported in Chapter~\ref{CHAP6}.

\chapter[Structure of complex networks: an Overview]{Structure of complex networks: an Overview}\label{CHAP2}

\section{Introduction}\label{CHAP2_1}
The first step toward a complete characterization of complex networks consists in a reliable description of their topological properties. As we will see in the following chapters, topological quantities play a
relevant role in determining the functionality of real networks as well as the dynamical patterns of processes taking place on them.  
Consequently, we devote Section~\ref{CHAP2_2} to introduce a set of mathematical tools, some of them borrowed from Graph Theory, that will be useful in the statistical investigation of complex networks.
In Section~\ref{CHAP2_3}, several examples of real complex networks are reported, together with the analysis of their most important topological properties. Special care is reserved to the Internet and the World-wide Air-transportation Network, whose topological and dynamical properties will be further investigated in Chapters~\ref{CHAP3}-\ref{CHAP4}. 
Finally, in Section~\ref{CHAP2_4}, we present a brief overview of the main models of complex networks, that are commonly used in order to reproduce topological and dynamical properties observed in real networks. 
The present chapter is not supposed to be an exhaustive review of all recent developments in the Science of Complex Networks, for which we refer to some very good books \cite{vespi_book,mendes_book}, and review articles \cite{dorogoR,lazlo_review,newman_review}. Similarly, for a simple introduction to Graph Theory we refer to Ref.~\cite{clark_book}, while a more rigorous approach is provided by the book of Bollob\'as \cite{bollobas_book}. \\
Our purpose is more properly that of providing a brief description of the measures used in networks analysis, focusing only on those concepts that are useful for a better comprehension of the work developed in the thesis.

\newpage
\section{Statistical Measures of Networks Topology}\label{CHAP2_2}

Graph theory is a fundamental field of mathematics whose modern formulation 
can be ascribed to P. Erd\"os and A. R\'enyi, for a series of papers appeared in 
the early '60s in which they laid the groundwork for the study of random graphs \cite{erdos1,erdos2}.
In the following, we go through the basic notions of graph theory, enriching them with
the definition of other more recently introduced quantities, that are commoly used for 
the statistical characterization of networks structure. 
 
\subsection{Basic notions of Graph Theory}\label{CHAP2_2_1} 

\begin{figure}[t]
\begin{center}
\includegraphics[width=12.0cm]{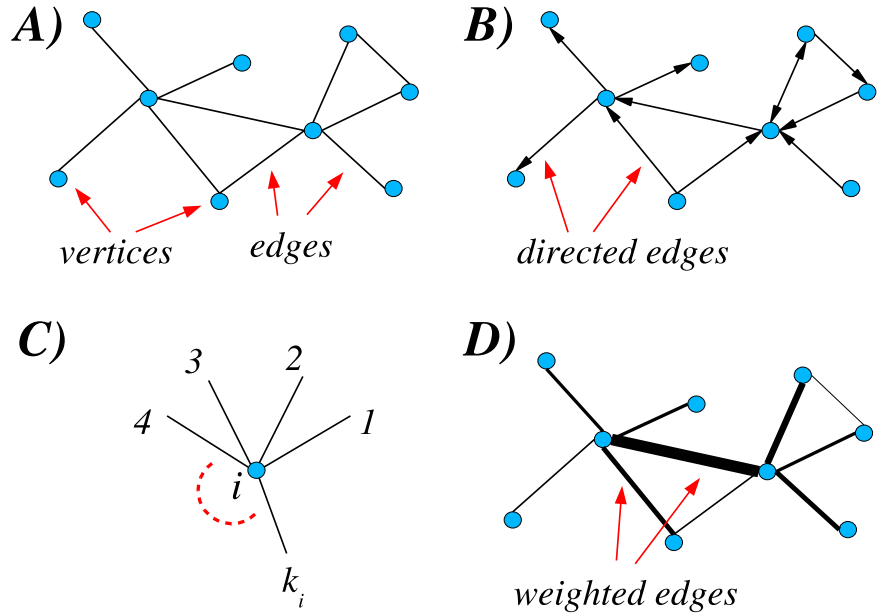} 
\end{center}
\caption{Basic elements of graph theory and their typical graphic representations: (A) vertices and edges in an undirected graph, (B) directed edges indicated by arrows in a directed graph, (C) the degree of a node, and (D) weighted edges.}
\label{ex_graphs}
\end{figure}

An {\em undirected graph} $G$ is a mathematical structure defined as the pair $G=(\mathcal{V},\mathcal{E})$, in which $\mathcal{V}$ is a non-empty set of elements, called {\em vertices}, and $\mathcal{E}$ is the set of {\em edges}, i.e. unordered pairs of vertices. More generally, each system whose elementary units are connected in pairs can be represented as a graph. 
In the interdisciplinary context the nomenclature used is not equally clear. Vertices are usually called {\em nodes} by computer scientists, {\em sites} by physicists and {\em actors} by sociologists. Edges are also addressed as {\em links}, {\em bonds}, or {\em ties}. We will use indifferently these terms, without any reference to a particular field of research.
The cardinality of the sets $\mathcal{V}$ and $\mathcal{E}$ are denoted by $N$ and $E$. The number of vertices is also referred to as the {\em size} of the graph.  
The simplest generalization of the definition of graph is that of {\em directed graph}, obtained considering oriented edges ({\em arcs}), i.e. ordered pairs of vertices.
A graphical representation of a graph consists in drawing a dot for every vertex, and a line between two vertices if they are connected by an edge (see Fig.~\ref{ex_graphs}). If the graph is directed, the direction is indicated by drawing an arrow.\\
A convenient mathematical notation to define a graph is the adjacency matrix $\mathbf{A}=\{a_{ij}\}$, a $N\times N$ matrix such that 
\begin{equation}
a_{ij} = \left\{ \begin{array}{cc} 1 & \quad \text{if} \quad (i,j) \in \mathcal{E} \\  0 & \quad \text{otherwise}~. \end{array}\right.  
\end{equation}
The adjacency matrix of undirected graphs is symmetric. Two vertices joined by an edge are called adjacent or {\em neighbors}; the {\em neighborhood} of a node $i$ is the set $\mathcal{V}(i)$ of all neighbors of the node $i$. 
The number of neighbors of a node $i$ is called the {\em degree} $k_{i}=\sum_{j} a_{ij}$ of $i$. 
In case of directed edges, we have to distinguish between incoming and outcoming edges, thus we define an in-degree ($k^{in}_{i}=\sum_{j} a_{ji}$) and an out-degree ($k^{out}_{i}=\sum_{j} a_{ij}$). We do not go deeper into the definition of properties for directed graphs, since in this thesis only undirected networks will be explicitly studied.   \\
Moreover, we consider graphs in which vertices do not present self-links (i.e. edges from a vertex to itself), or multi-links (i.e. more than one edge connecting two vertices). Such objects, whose properties are rather unusual in real networks, are known in graph theory as {\em multigraphs}.
If we exclude self-links and multiple links, the maximum possible number of edges is $N(N-1)/2$.
Those graphs, whose number of edges is close to such a value, are called {\em dense} graphs, while the graphs in which the number of edges is bounded by a linear function of $N$ are {\em sparse} graphs. \\
A generalization of the notion of graph, that will be repeatedly taken into account in the following chapters, is that of  {\em weighted graph}. In weighted (directed or undirected) graphs, each edge $(i,j)$ carries a weight, that is a variable $w_{ij}$ assuming real (or integer) values. However, also the nodes can be differentiated, introducing classes of nodes with the same set of internal variables. Graphs with distinct classes of nodes will be denoted as {\em multi-type graphs}. Graphs in which there are two or more distinct sets of nodes with no edges connecting vertices in the same set are commonly referred as {\em multipartite graphs}.
{\em Real networks are actually weighted and multi-type, though in many situations it is more convenient to study their properties by means of single-type, unipartite and/or unweighted representations}.  

\subsection{Degree distribution}\label{CHAP2_2_2}
A natural way to collect nodes in classes is that of considering nodes with the same degree $k$.
This is a convenient strategy to analyze large graphs, since the connectivity properties of the nodes are statistically represented by the histogram $P(k) = N_{k}/N$, in which $N_{k}$ is the number of vertices of degree equal to $k$.
In the infinite size limit ($N\to \infty$), $P(k)$ is called {\em degree distribution}, since it represents the probability distribution that a node has degree $k$. The degree distribution $P(k)$ satisfies a normalization condition
$\sum_{k=0}^{\infty} P(k) = 1$.  
The {\em average degree} of an undirected graph is defined as the average value of $k$ over all the vertices in the graph,
\begin{equation}
\langle k \rangle = \sum_{k} k P(k) \equiv \frac{2E}{N}~.
\end{equation}
The condition of sparseness for a graph can be translated into $\langle k\rangle \simeq \mathcal{O}(1)$. \\
However, in order to study topological properties of networks, the knowledge of higher moments of the degree distribution is also important.
For instance, the second moment $\langle k^2 \rangle$ measures the fluctuations of the degree distribution, and governs the percolation properties \cite{cohen2}; while higher moments determine conditions for the mean-field behavior of the Ising model on general networks \cite{dorogovtsev1}.\\ 
For a long time, Graph Theory has been interested in random graphs with homogeneous connectivity, i.e. with a degree distribution that is very peaked around a characteristic average degree and decays exponentially fast for $k \gg \langle k \rangle$. On the contrary, recent phenomenological findings have shown that a large number of real networks present heavy-tailed distributions, some of them close to a power-law behavior.
In these networks, there is a non-negligible probability of finding ``hubs'', i.e. nodes of degree $k \gg \langle k \rangle$.    

\subsection{Two and three points degree correlations}\label{CHAP2_2_3}
The degree distribution does not exhaust the topological characterization of a network, since it has been shown that many real networks present degree correlations between nodes, i.e. the probability that a node of degree $k$ is connected to another node of degree $k'$ depends on $k$ and $k'$ themselves. 
More rigorously, we can introduce a conditional probability $P(k'|k)$ that a vertex of degree $k$ is connected to a vertex of degree $k'$. This quantity satisfies a normalization $\sum_{k'} P(k'|k) = 1$ and a detailed balance condition \cite{boguna2} 
\begin{equation}\label{balance}
kP(k'|k)P(k) = k' P(k|k') P(k')~, 
\end{equation}
corresponding to the absence of dangling bonds.
In uncorrelated graphs, $P(k' |k)$ does not depend on $k$ and it can be easily obtained from the normalization condition and Eq.~\ref{balance}, 
\begin{equation}
P(k' | k) = \frac{k' P(k')}{\langle k \rangle}~.
\label{fact}
\end{equation}
Similarly, it is possible to define a three-points correlation function $P(k',k''|k)$, i.e. the probability that a vertex of degree $k$ is simultaneously connected to vertices of degree $k'$ and $k''$.\\
In general, the direct measurement of these two conditional probabilities is quite cumbersome and gives very noisy results on any kind of network. For this reason one usually prefers more practical estimates by means of indirect quantities, that are averaged over the neighborhood of a node. \\
In a given network with adjacency matrix $\left\{a_{ij}\right\}$, a good estimation of the degree correlations of a vertex $i$ is provided by the {\em average degree of the nearest neighbors} of $i$ 
\begin{equation}
k_{nn,i} = \frac{1}{k_{i}} \sum_{j=1}^{N} a_{ij} k_{j}~.
\end{equation}
Defining the network using its degree distribution, the average degree of the nearest neighbors of a vertex of degree $k$ is \begin{equation}
k_{nn}(k)= \sum_{k'} k' P(k'|k)~.  
\end{equation}
If the network is uncorrelated, the degree of the neighbors can assume any possible value, and the average turns out to be approximately independent of $k$, i.e. $k_{nn}(k)\simeq const$.
On the contrary, correlated networks can be schematically divided in two large classes. 
The first class is that of those presenting {\em assortative mixing}, i.e. nodes of high (small) degree are more likely to be connected with nodes of high (small) degree ($k_{nn}(k)$ grows with $k$). This seems to be a general property of social networks.
When vertices of high degree are preferentially linked with vertices of smaller degree (and viceversa), i.e. $k_{nn}(k)$ is a decreasing function of $k$, the network has {\em disassortative mixing}. Many critical infrastructures such as transportation and communication networks present a clearly disassortative behavior.\\
\begin{figure}[t]
\begin{center}
\includegraphics[width=12.0cm]{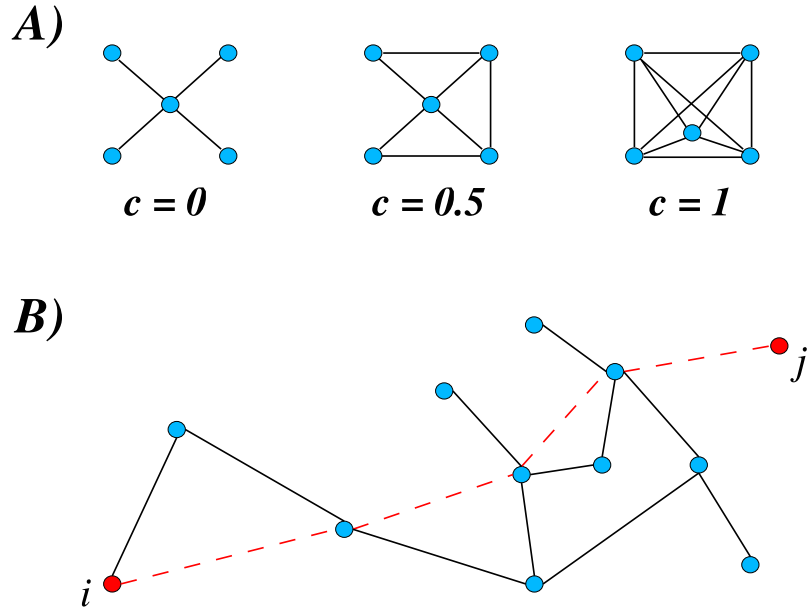} 
\end{center}
\caption{(A) The clustering coefficient gives a measure of the local cohesiveness in the neighborhood of a vertex: the central node in the example has clustering coefficient $c=1$ if all pairs of neighbors of the vertex are connected, $c=0.5$ if only half of the possible pairs are connected, and $c=0$ if no triangles are formed.
(B) The dashed path represents the shortest path (of length $\ell_{ij}= 4$) between nodes $i$ and $j$.}
\label{ex_cluster}
\end{figure}
Analogously, for three-points correlations, we can define a quantity called {\em clustering coefficient} that measures the tendency of a graph to form cliques in the neighborhood of a given node.
As depicted in Fig.~\ref{ex_cluster}-A, the clustering coefficient of a node $i$ is defined as the ratio of the actual number of edges $n_{i}$ between the neighbors of $i$, and the maximum possible number of such edges $k_{i} (k_{i}-1) / 2$, i.e.
\begin{equation}
c(i) = \frac{2n_{i}}{k_{i} (k_{i}-1)} = \frac{\sum_{jh} a_{ij}a_{jh}a_{hi}}{k_{i} (k_{i}-1)}~.
\end{equation}
The study of the clustering spectrum $c(k)$,
\begin{equation}
c(k) = \frac{1}{N_{k}} \sum_{i/ k_{i} = k} c(i)~,     
\end{equation}
provides  interesting insights on the local cohesiveness of the network.
In particular, a clustering coefficient decreasing with the degree $k$ has been put in relation with the existence of {\em hierarchical structures} (e.g. in biological networks \cite{modularity}).

\subsection{Shortest path length and distance}\label{CHAP2_2_4}
Many non-local properties of graphs are related to the reachability of a vertex starting from another. A {\em walk} from a vertex $i$ to a vertex $j$ consists in a sequence of edges and vertices joining $i$ with $j$.
The {\em length} $l$ of the walk coincides with the number of edges in the sequence.
A {\em path} is a walk in which no node is visited more than once.  A graph is {\em connected} if for any pair of vertices $i$ and $j$, there is a path from $i$ to $j$. 
The number of walks of length $l$ between two nodes $i$ and $j$ can be expressed by the $l$-th power of the adjacency matrix
\begin{equation}
\left( \mathbf{A}^{l}\right)_{ij} = \sum_{i_{1}, i_{2}, \dots, i_{l-1}} a_{i i_{1}} a_{i_{1} i_{2}}\cdots a_{i_{l -1} j}~.
\end{equation}
In particular, this definition is related to the behavior of a random walker on the graph.  
A closed walk, in which initial and final vertices coincide, is called a {\em cycle}; a $k$-cycle is a cycle of length $k$.\\
The walk of minimum length between two nodes is called {\em shortest path}, and its length corresponds to the {\em hop distance} $\ell_{ij}$ between the nodes $i$ and $j$ (see Fig.~\ref{ex_cluster}-B).
The diameter $\ell_{max}$ is the maximum distance between pairs of nodes in the graph, while the average distance $\langle \ell \rangle$ between nodes is given by
\begin{equation}
\langle \ell \rangle = \frac{1}{N(N-1)} \sum_{i\neq j} \ell_{ij}
\end{equation}
A complete characterization of the metric properties of a graph corresponds to know the full probability distribution $P_{\ell}(\ell)$ of finding two vertices separated by a distance $\ell$. In fact, many real networks present a
symmetric distribution peaked around the average value $\langle \ell \rangle$, that can be safely considered representative of the typical distance between nodes in the network.\\
From this point of view, complex networks seem to share a striking property, called {\em small-world effect} \cite{watts98,watts99}, meaning that the average intervertex distance $\langle \ell \rangle$ is very small compared to the size $N$ of the network, scaling logarithmically or slower with it.
While this property can be found also in generic random graphs, where $\langle \ell \rangle \propto \log N$,
the result is in contrast with the behavior of the distance on regular $d$-dimensional lattices, in which $\langle \ell \rangle \propto N^{1/d}$. \\
The practical implication of the small-world property is that it is possible to go from a vertex to any other in the network passing through a very small number of intermediate vertices. In this regard, the concept of small-world was firstly popularized by the sociologist S. Milgram in $1967$ by means of a famous experiment \cite{milgram}, in which he showed that a low number of acquaintances, on average only six, is actually sufficient to connect (by letter) any two individuals in the United States. 
The experiment was recently reproduced using the world-wide e-mail network and provided results consistent with the small-world hypothesis \cite{watts03}.\\
Note that the presence of the small-world property is relevant not only at a topological level, but it has also strong effects on all dynamical processes taking place on the network.\\
A plethora of different statistical measures is based on the notions of distance and shortest path, some of them are used in the topological characterization of networks, others in the study of the relation between functional properties and dynamics (see Chapter~\ref{CHAP4}); we concentrate our attention on {\em centrality measures} that will be extensively used in the following chapters. \\
 \begin{figure}[t]
\begin{center}
\includegraphics[width=12.0cm]{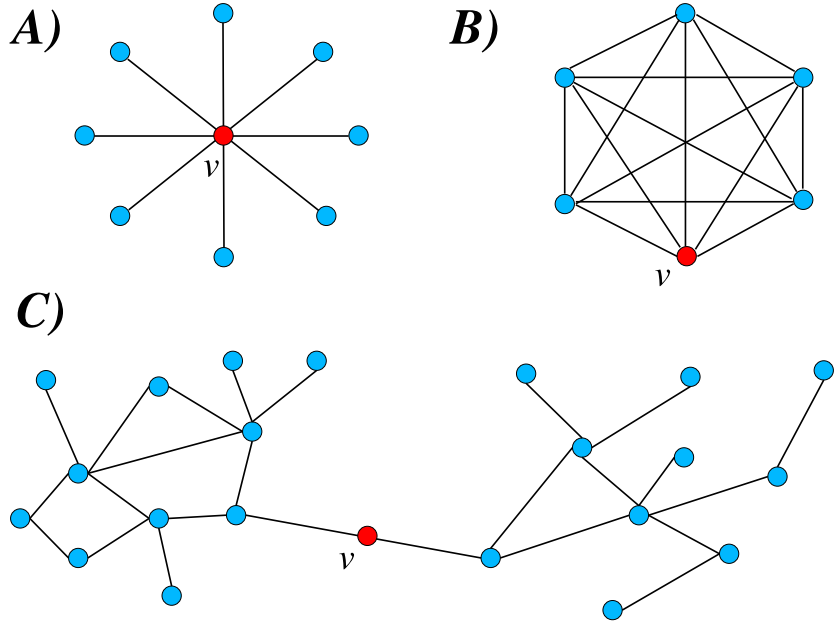} 
\end{center}
\caption{(A) Star-like network, the node $v$ (in red) has maximum betweenness because all shortest paths have to pass through the central node. (B) A fully connected clique is characterized by a zero betweenness value for all nodes. (C) The node $v$ (in red) connects two highly connected groups of nodes: its betweenness is very high even if its degree is very low.}
\label{ex_between}
\end{figure}   
The metric properties of a graph are, indeed, very appropriate to define several different measures of centrality, that are used in social sciences to estimate the importance of nodes and edges.
The most local of these measures is the {\em degree centrality}, that is proportional to the degree of a node and does not account for any metric feature of the graph.
All other centrality measures involve non-local properties in the form of the intervertex distance. For instance, the {\em closeness centrality} of a node $i$ is defined as the inverse of the sum of the distances of all nodes from $i$.\\
The most famous measure of centrality is the {\em betweenness centrality}, defined in Ref.~\cite{freeman} and recently adopted in network science as the basic definition of centrality of nodes and edges. 
The node betweenness centrality $b_{i}$ computes the relative number of all shortest paths between pairs of vertices that pass through the vertex $i$, i.e. 
\begin{equation}
\label{bet_node}
b_{i} = \sum_{s\neq i \neq t}  \frac{\sigma_{st}(i)}{\sigma_{st}}~,
\end{equation}
where $\sigma_{st}=\sigma_{ts}$ is the number of shortest paths between the vertices $s$ and $t$, and $\sigma_{st}(i)=\sigma_{ts}(i)$ is the number of them going through the node $i$.
Similarly, the edge betweenness centrality $b_{ij}$ is defined as the fraction of all shortest paths from any pair of vertices in the network that pass through the edge $(i,j)$,
\begin{equation}\label{bet_edge}
b_{i,j} = \sum_{s\neq i\neq j \neq t}  \frac{\sigma_{st}(i,j)}{\sigma_{st}}~,
\end{equation}
in which $\sigma_{st}(i,j)$ is the number of shortest paths going through the edge $(i,j)$.
It is worthy to remark that in the literature there are several slightly different definitions of 
betweenness centrality: in particular, a prefactor $1/2$ can be considered in order not to count twice the paths, while  the paths containing the interested nodes (i.e. $i$ and/or $j$) as initial or ending points can be accepted or discarded (the two cases differing just by a constant contribution).    
The computational cost of determining the betweenness centrality for all vertices (or edges) in a graph is very high, since one has to discover all existing shortest paths between pairs of vertices. An optimized algorithm, proposed by Brandes \cite{brandes}, allows to reduce the computational complexity from $\mathcal{O}(N^{2}E)$ to $\mathcal{O}(N E)$.
For sparse graphs the algorithm performs in $\mathcal{O}(N^2)$ steps, that is still a high complexity when the size of the network is very large (e.g. $N \sim \mathcal{O}(10^6)$), or when the computation has to be repeated many times, as for the measures exposed in Chapter~\ref{CHAP4}.    \\
Figuring out the meaning underlying the notion of betweenness centrality is simple by means of few examples dealing with extreme topological conditions. Let us consider a star network with a unique central vertex $v$ and $N-1$ leaves at a distance $1$ from the center (see Fig.~\ref{ex_between}-A).
The node betweenness centrality of $v$ is simple to compute because $v$ belongs to all shortest paths between pairs of leaf nodes, therefore the sum in Eq.~\ref{bet_node} becomes a sum of unit contributions and we get $b_{v} = (N-1)(N-2)$.
The opposite situation is the complete graph, in which all vertices, according to the definition in Eq.~\ref{bet_node}, have zero betweenness centrality (Fig.~\ref{ex_between}-B).  
Another interesting case is that of a node or an edge joining, as a bridge, two otherwise disconnected portions of a network (Fig.~\ref{ex_between}-C): all paths connecting pairs of nodes belonging to different regions have to pass through that particular node (edge), that turns out to have very high betweenness even if it may have very low degree. This property shows that in many networks, as we will see, betweenness centrality is non-trivially correlated with the other topological properties.

\subsection{Subgraph structures}\label{CHAP2_2_5}

\begin{figure}[t]
\begin{center}
\includegraphics[width=12.0cm]{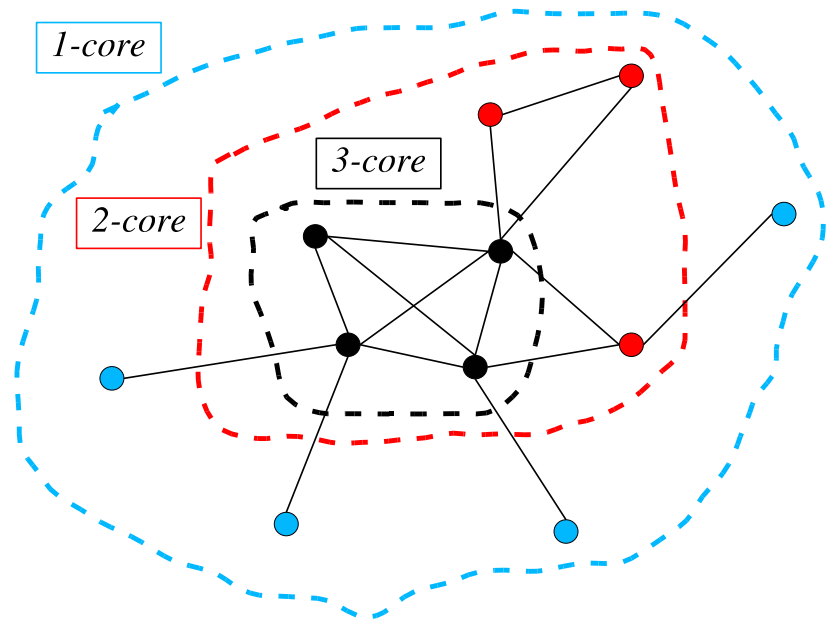}
\end{center}
\caption{Illustration of the $k$-core structure of a simple graph. The blue circle contains the $1$-core, i.e. the giant connected component of the graph. The two smaller contours draw the boundaries of the $2$-core (red) and $3$-core (black). Blue nodes belong to the $1$-shell, the red and black ones to the $2$-shell and $3$-shell respectively. 
The internal sets can be obtained from the larger ones by iterative pruning of nodes as explained in the text.}
\label{sketch_kcore}
\end{figure}

\begin{figure}[t]
\begin{center}
\includegraphics[width=12.0cm]{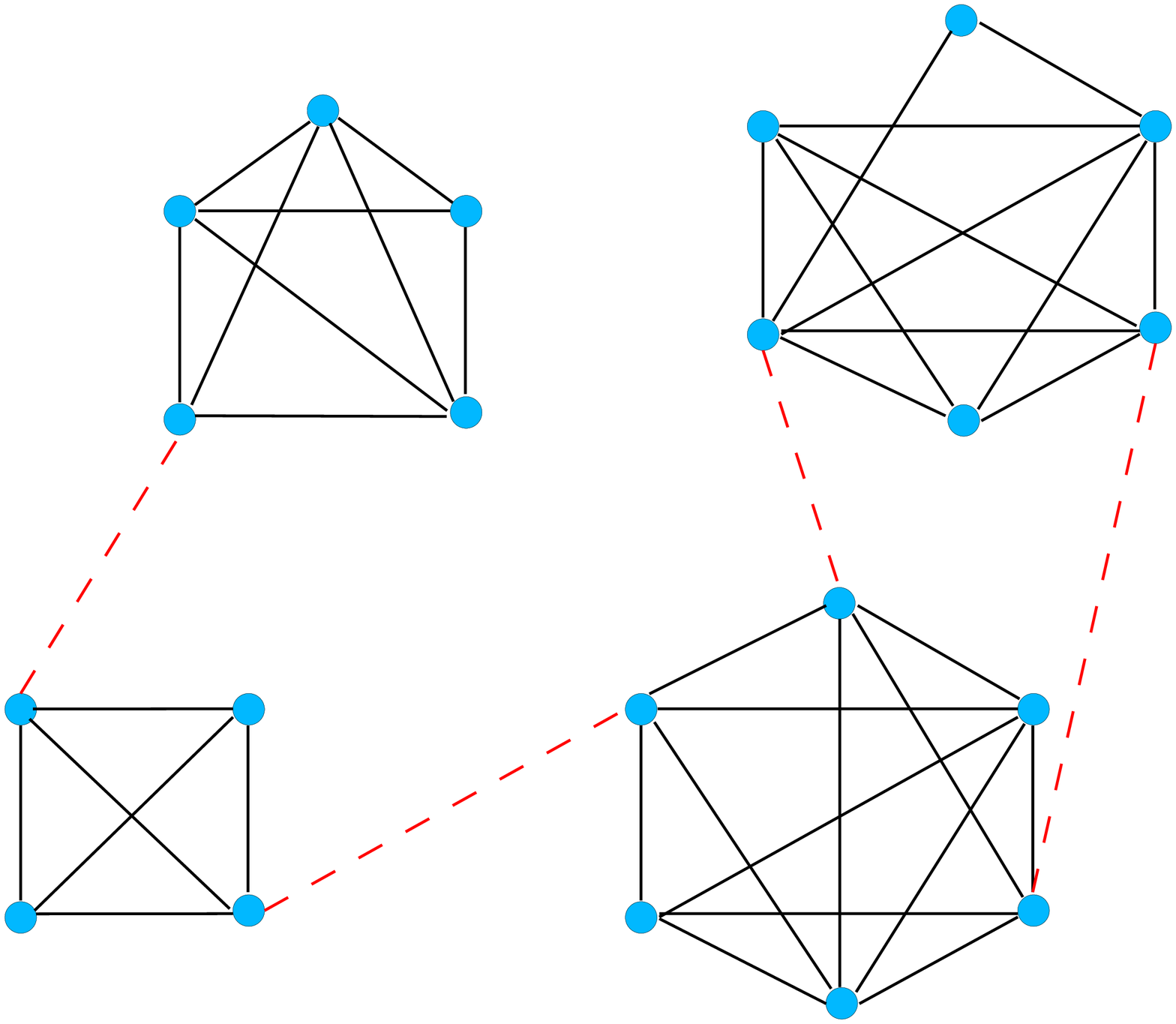} 
\end{center}
\caption{An example of network with a very strong community structure. The number of edges connecting nodes belonging to the same community (black full links) is much larger than the number of edges between nodes in different communities (red dashed links). 
}
\label{ex_comm}
\end{figure}

In this paragraph, we discuss a series of topological properties dealing with the structure of subsets of a graph. 
Firstly, a graph $G'=(\mathcal{V}', \mathcal{E}')$ is a {\em subgraph} of the graph $G=(\mathcal{V}, \mathcal{E})$ if $\mathcal{V}' \subset \mathcal{V}$ and $\mathcal{E}' \subset \mathcal{E}$. 
A {\em maximal subgraph} with respect to a given property is a subset of the graph that cannot be extended without loosing that property.
Given a subset of nodes $\mathcal{V}' \subseteq \mathcal{V}$, we call $G'=(\mathcal{V}',\mathcal{E}|\mathcal{V}')$ $\subseteq G$ the subgraph of $G$ {\em induced by} $\mathcal{V}'$. 
A {\em component} of a graph $G$ is a maximally connected subgraph of $G$; it is called {\em giant component} if its size is $\mathcal{O}(N)$.\\
We have already seen that the clustering coefficient is a measure of the cohesiveness of a graph; however, the maximal cohesiveness corresponds to sets of nodes with all-to-all connections, called {\em cliques}. Formally, a clique is a maximally complete subgraph of three or more nodes.
Though there are several other quantities involving the subgraph's definition, such as the $n$-cliques, or the $k$-plexes, for the purposes of this work, we are only interested in two of them: $k$-cores and $k$-shells.\\
The {\em $k$-core} of a graph $G$ is the maximal induced subgraph of $G$ whose vertices have the property of having degree at least $k$ \cite{bollobas_core,seidman_core}. (Note that it means that they must have degree at least $k$ inside the subgraph!). 
Such a subgraph can be obtained by recursively removing all the vertices of degree lower than $k$, using a procedure  called {\em $k$-core decomposition}.\\
Let us call $N_{k}$ the number of nodes in the graph with degree not larger than $k$ and $\mathcal{C}_{k}$ the set of nodes belonging to the $k$-core, the algorithm reads 
\begin{itemize}
\item[(1)] Set $k=0$ ($\mathcal{C}_{k}$ is empty $\forall k > 0$, $\mathcal{C}_{0} \equiv G$);
\item[(2)] $k \to k+1$;
\item[(3)] Prune all nodes with degree lower than $k$ (and the corresponding edges);
\item[(4)] Update $N_{l}$, $\forall l \geq 0$ according to the pruned network;
\item[(5)] Repeat point $(3)$ until $N_{l<k}=0$;
\item[(6)] Put all remaining nodes (and edges) in $\mathcal{C}_{k}$ and go back to point $(2)$. 
\end{itemize}
A node has shell index $k$ if it belongs to the $k$-core but not to the $k+1$-core; the {\em $k$-shell} is the set of all vertices of shell index $k$, i.e. the difference between two consecutively nested cores. 
The algorithm is very clear if we look at simple cases similar to that sketched in Fig.~\ref{sketch_kcore}, in which $k$-cores and $k$-shell are highlighted using different colors.  \\
Finally, we usually refer to {\em communities} when the graph can be reduced to a certain number of subgraphs characterized by the property that, for each of them, the number of edges connecting the subgraph with the rest of the graph is very small compared to the number of edges linking different vertices within the same subgraph.
In such a case each subgraph is a community, as depicted in Fig.~\ref{ex_comm}. 
The definition of community is not rigorous, thus the community structure of a network depends strongly on the practical method used to detect the subgraphs. 
Several algorithms have been proposed in order to find the community structure of networks. Some of them reduce iteratively the size of the different subgraphs, while others are based on the opposite principle of clustering algorithms, but all them suffer of the same incapacity of detecting the correct level at which the iterative procedure should be stopped.  This is probably an intrinsic problem due to the absence of a rigorous definition able to fix the correct resolution at which the community structure is more visible.\\
Some of the algorithms used to detect communities are based on topological properties, such as the betweenness \cite{girvan}, others exploit the properties of some dynamical systems, e.g. synchronizability of oscillators \cite{diaz-guilera}. In Chapter~\ref{CHAP5}, we will show that also non-equilibrium models of coarsening dynamics can be used to put forward alternative methods to detect communities.  

\subsection{Further metrics for weighted networks}\label{CHAP2_2_6}

\begin{figure}[t]
\begin{center}
\includegraphics[width=12.0cm]{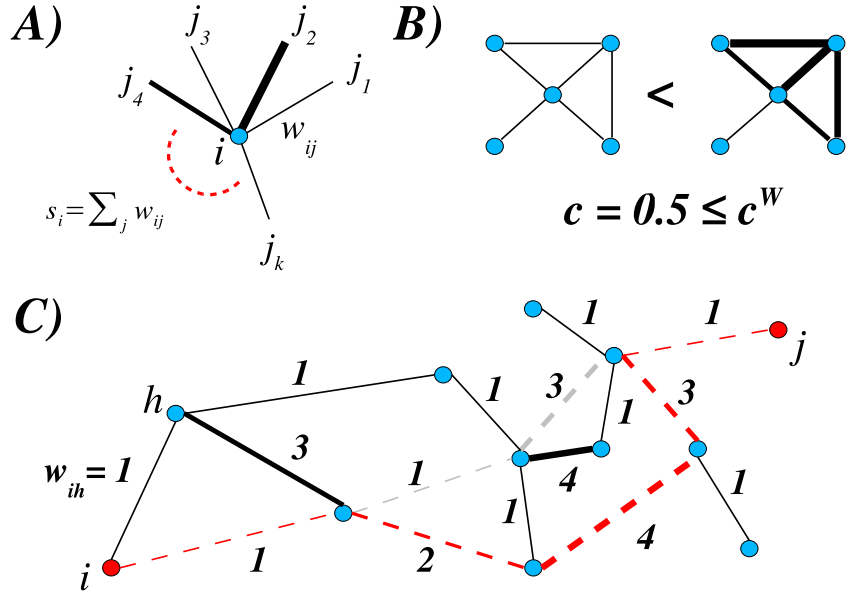} 
\end{center}
\caption{(A) Strength $s_{i}$ of the node $i$ computed as the sum of the weights along the edges connecting $i$ with its neighbors. (B) The weighted clustering coefficient $c^{W}$ gives a measure of local cohesiveness in presence of weights. Larger weights are responsible for larger weighted clustering. (C) The weighted shortest path (red dashed links) can be topologically (number of links) longer than the unweighted shortest path (grey dashed links). The weighted shortest path depends on the definition of edge length, that in this case is $\ell_{ij} = 1/w_{ij}$.}
\label{ex_clusterW}
\end{figure}

In many real networks, edges are not identical, they can have different intensities, that are related to some physical properties and are taken into account assigning them a weight. For instance, in the Internet the edges represent physical connections, cables, thus weights could be introduced to account for their bandwidth, or the traffic between routers. 
In the air-transportation network, the weights are proportional to the traffic on the airline connections. Hence, in the technological and infrastructure networks, weights usually correspond to some physical quantities (energy, information, goods, $\dots$) that are transferred between two nodes. On the other hand, in biological networks weights account for the strength of the interactions between genes or proteins; whereas in social networks they specify the intensity of interactions between the actors. \\
Many statistical quantities that have been introduced for unweighted networks can be easily generalized to weighted networks. The degree is generalized introducing the node {\em strength}; the strength $s_{i}$ of a node $i$ is
\begin{equation}
s_{i} = \sum_{j\in \mathcal{V}} w_{ij}~, 
\end{equation}
where $w_{ij}$ is the weight on the edge $(i,j)$ \cite{barrat_PNAS,yookwt} (see Fig.~\ref{ex_clusterW}-A).\\
If the weights are distributed uniformly at random, the node strength turns out to be linearly correlated with the degree, i.e. $s(k) \sim \langle w\rangle~k$. In fact, the actual degree-strength correlations observed in many real networks suggest rather a super-linear relation $s(k) \sim A k^{\beta}$, with $\beta>1$ and $A\neq \langle w\rangle$ (see Chapter~\ref{CHAP4}).\\
A standard measure in the analysis of weighted graphs is the strength distribution $P_{s}(s)$, that says which is the probability that a randomly chosen node has strength equal to $s$.
Many weighted networks with broad degree distribution, such as the World-wide Air-transportation Network, also present broad weight and strength distributions.\\
Other quantities are readily extended in order to account for weights, in particular two- and three-points correlations.
For each vertex $i$ one can define a {\em weighted average nearest neighbors degree},
\begin{equation}
k_{nn,i}^{W} = \frac{1}{s_{i}}\sum_{j\in \mathcal{V}} w_{ij} k_{j}~, 
\end{equation}
and a {\em weighted clustering coefficient}
\begin{equation}
c_{i}^{W} = \frac{1}{s_{i}(k_{i}-1)}\sum_{j,m} \frac{w_{ij}+w_{im}}{2}a_{ij}a_{jm}a_{mi}~. 
\end{equation}
The degree dependent functions, $k_{nn}^{W}(k)$ and $c^{W}(k)$ can be directly compared with the unweighted measures $k_{nn}(k)$ and $c(k)$, providing interesting information on the role of the weights.
For the clustering coefficient, the interpretation is particularly easy (Fig.~\ref{ex_clusterW}-B): when the weighted clustering coefficient is larger than the topological one, it means that triples are more likely formed by edges with larger weights. 
A similar interpretation holds for the relation between $k^{W}_{nn}(k)$ and $k_{nn}(k)$.\\
With respect to unweighted networks, when edges are weighted the neighborhood of a node is not homogeneous, namely the edges outgoing from a given vertex can in principle carry very different weights.
In order to quantify the homogeneity of the neighborhood of a node $i$, we consider the {\em disparity measure} \cite{derrida,barthelemy_Y},
\begin{equation}
Y_{i} = \sum_{j \in \mathcal{V}(i)} {\left( \frac{w_{ij}}{s_{i}}\right)}^{2}~.
\end{equation}
This quantity depends on the degree, in such a way that, when the weights are comparable, $Y(k) \sim 1/k$, while when an edge dominates on the others $Y(k) \sim 1$. \\
Many other weighted quantities have been defined, but the most important for the topics discussed in the rest of this thesis are the {\em weighted centrality measures} and, in particular, the weighted betweenness centrality.\\
Actually, it is sufficient to define the {\em weighted shortest path} and all centrality quantities can be directly constructed from it (Fig.~\ref{ex_clusterW}-C). To each edge $(i,j)$ in the network, we associate a distance that is a function of the weight, i.e. $\ell^{W}_{ij}= \ell^{W}(w_{ij})$, whose explicit form depends on the nature of the weights. 
For instance, in the airports network weights are proportional to the traffic capacity, and a larger traffic capacity leads to a better transmission along an edge, that from a ``functional'' point of view corresponds to decrease the distances. Hence, we expect that the weighted distance along the edge $(i,j)$ is inversely proportional to the weight, i.e. $\ell^{W}_{ij} \propto 1/w_{ij}$. \\ 
The definition of the node {\em weighted betweenness centrality} $b^{W}_{i}$ consists in replacing all shortest paths with their weighted versions. For any two nodes $h$ and $j$, the weighted shortest path between $h$
and $j$ is the one for which the total sum of the lengths of the edges forming the path from $h$ to $j$ is minimum, independently from the number of traversed edges.  We denote by $\sigma^{W}_{hj}$ the total
number of weighted shortest paths from $h$ to $j$ and $\sigma^{W}_{hj}(i)$ the number of them that pass 
through the vertex $i$ (with $h,j \neq i$); the weighted betweenness centrality (WBC) of the 
vertex $i$ is then defined as
\begin{equation}
b^{W}_{i}=\sum_{h,j \neq i} \frac{\sigma^{W}_{hj}(i)}{\sigma^{W}_{hj}}~,
\label{wbc}
\end{equation}
where the sum is over all the pairs with $j \neq h \neq i$. Similarly, we can define a weighted edge betweenness.
The algorithm proposed by Brandes in Ref.~\cite{brandes} can be easily extended 
to weighted graphs, with no further increase in complexity.  

\newpage
\section{Examples of real networks}\label{CHAP2_3}

In this section, we review of the phenomenological properties of two important real networks: the Internet (in Section~\ref{CHAP2_3_1}) and the World-wide Air-transportation Network (in Section~\ref{CHAP2_3_2}).
At the same time it is an opportunity to show the practical use of the statistical measures defined in the previous section.
Some of these properties are reliably considered quite universal in complex networks, such as heavy-tailed degree distributions or the small-world property, others are typical features of the particular network under study. 
Though a detailed discussion is reserved only for those networks that have been objects of direct investigation in this thesis, for the sake of completeness we provide in Section~\ref{CHAP2_3_3} a brief overview on the phenomenology of some other real networks. 

\begin{figure}[t]
\begin{center}
\includegraphics[width=7.0cm]{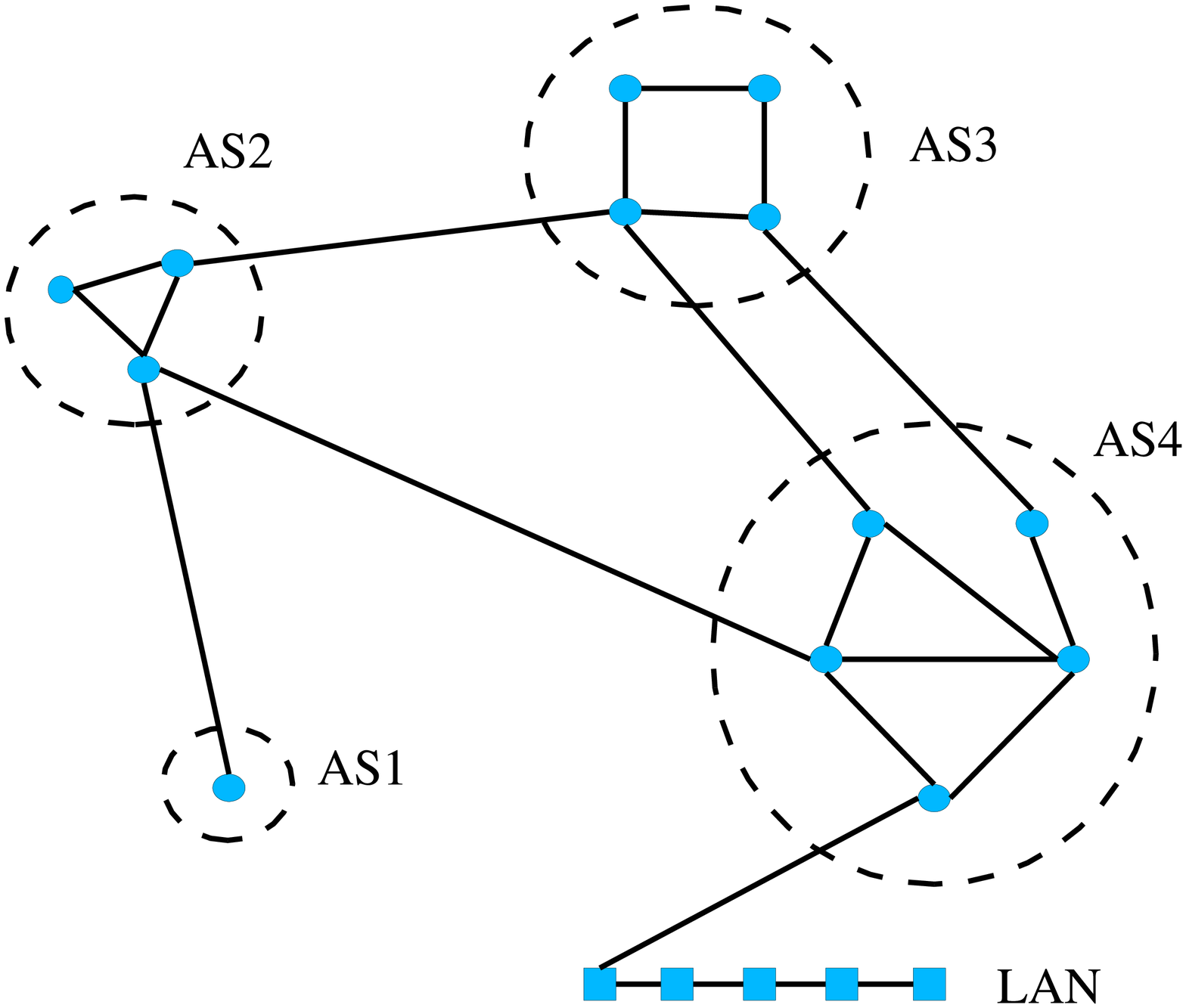} \quad
\includegraphics[width=7.0cm]{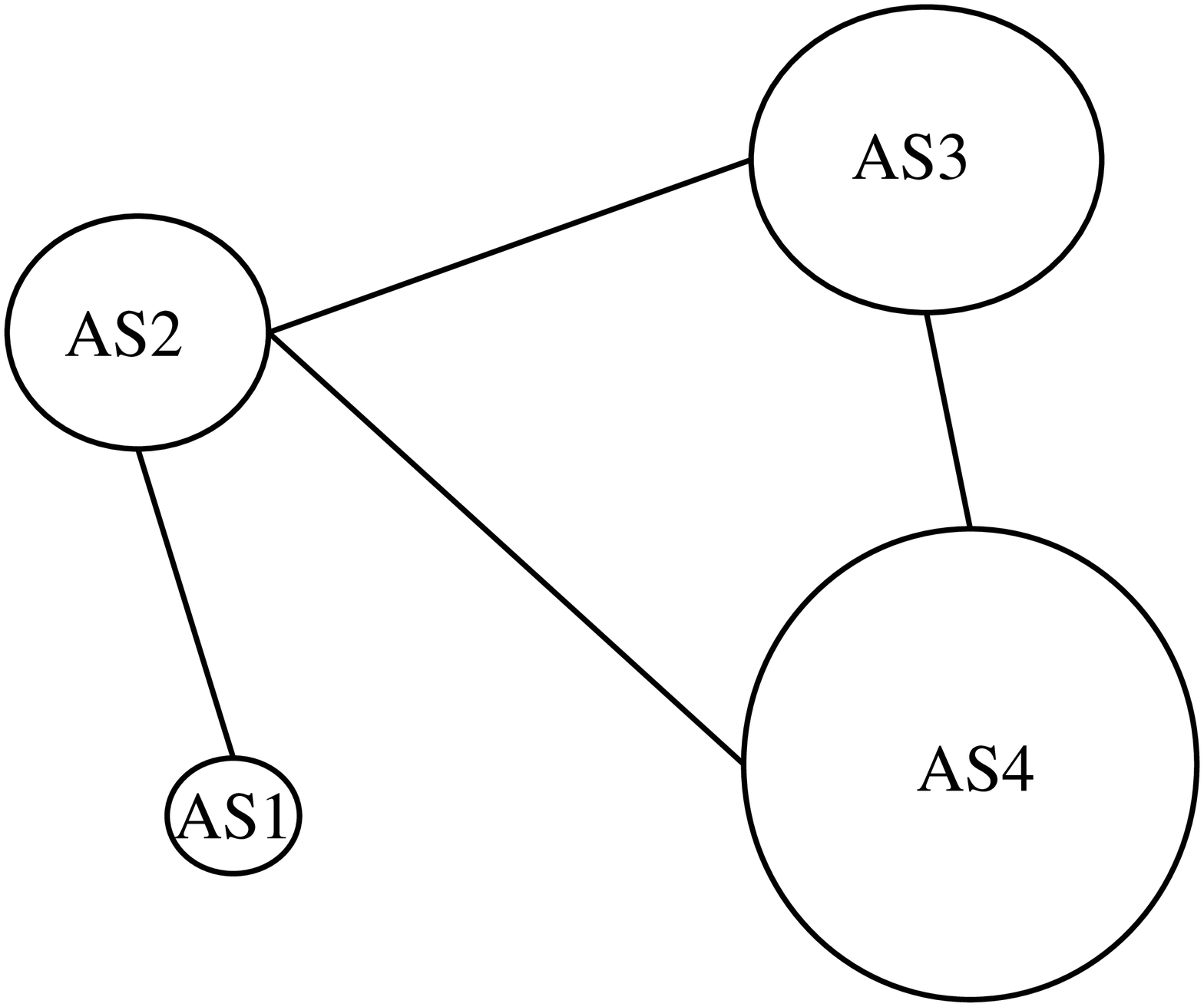}
\end{center}
\caption{Different granularity representations of the Internet: routers level (left) and autonomous systems level (right). Routers are represented by blues nodes grouped in autonomous systems (circles).}
\label{schemesInternet}
\end{figure}
 
\subsection{The Internet}\label{CHAP2_3_1}

\begin{figure}[t]
\begin{center}
\includegraphics*[width=14.0cm]{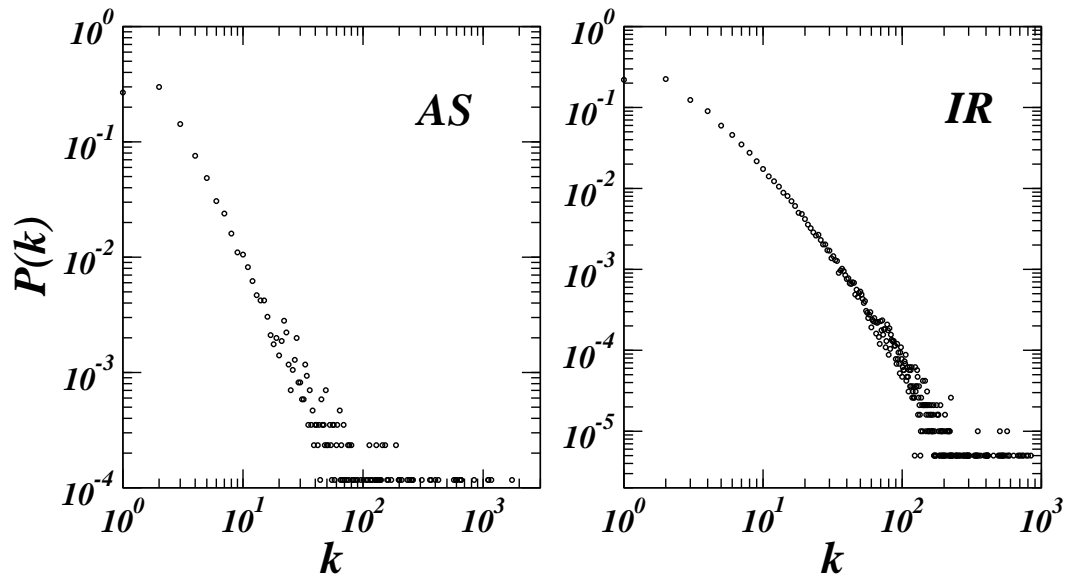}
\end{center}
\caption{Two examples of the degree distributions obtained from Internet mapping projects. $P(k)$ is clearly power-law at the AS level (left), while it is broad but slightly bended at the IR level (right). Both graphs are based on data 
collected by the Internet's mapping projects at CAIDA \cite{caida}.}
\label{degreesInternet}
\end{figure}

The Internet is a communication network in which the vertices are computers and the edges are the physical connections among them. The existence of various types of vertices reflects the high level of complexity and heterogeneity of the system: the {\em hosts} correspond to single-user's computers; the servers are computers or programs providing network services; the {\em routers} are computers devoted to arrange traffic and data exchange across the Internet.  
Similarly, hosts and routers are linked by various types of connections, that are undirected and have different traffic capacity depending on their bandwidth.\\
The structure of the Internet is the result of a complex interplay between growth and self-organization, involving processes of cooperation and competition, without central administration or external control.
{\em A good knowledge of the topological structure of the Internet is necessary to improve its functionality, and prevent the system from faults and traffic congestions}. This is the main reason for the great interest of researchers in studying the structure of the Internet. \\ 
An exhaustive introduction to the network properties of the Internet is provided by Refs.~\cite{vespi_book,baldi}; we give here only some necessary information on its structure.\\  
The first important observation is that it is impossible to keep track of all single hosts, that are hundreds of millions all around the world, organized in complex hierarchically structures, whose smaller units are Local Area Networks, connected to the main net by means of routers. Monitoring the structure of single local networks is thus too difficult, and partly useless since these networks are created just to connect hosts inside buildings, university departments, corporate networks, city areas etc, and their properties depend on very local administrative policies. 
Hence, the lowest level of granularity at which we can analyze the Internet topology is the so-called {\em router level} (IR), that is a graph with routers as vertices and the physical connections among them as edges.\\
At a higher granularity level, the Internet can be partitioned into autonomously administered domains, called {\em Autonomous Systems} (AS). Within each autonomous system, whose structure is not defined on the basis of geographical proximity but more frequently on commercial agreements and policies, the traffic is handled following proper internal control strategies and restrictions, that can vary considerably from AS to AS. In order to better understand routing problems,  it is very convenient to study the Internet topology at the level of the autonomous systems, considering each AS as a node and connections between different ASs as the links. \\
The mapping projects of the Internet deal mainly with these two scales of description: the autonomous systems level (AS), and the router level (IR).
Fig.~\ref{schemesInternet} reports a scheme of the structure of the Internet at both levels.
\begin{figure}[t]
\begin{center}
\begin{tabular}{|c|}\hline \\ \includegraphics*[width=7.0cm,height=7.0cm]{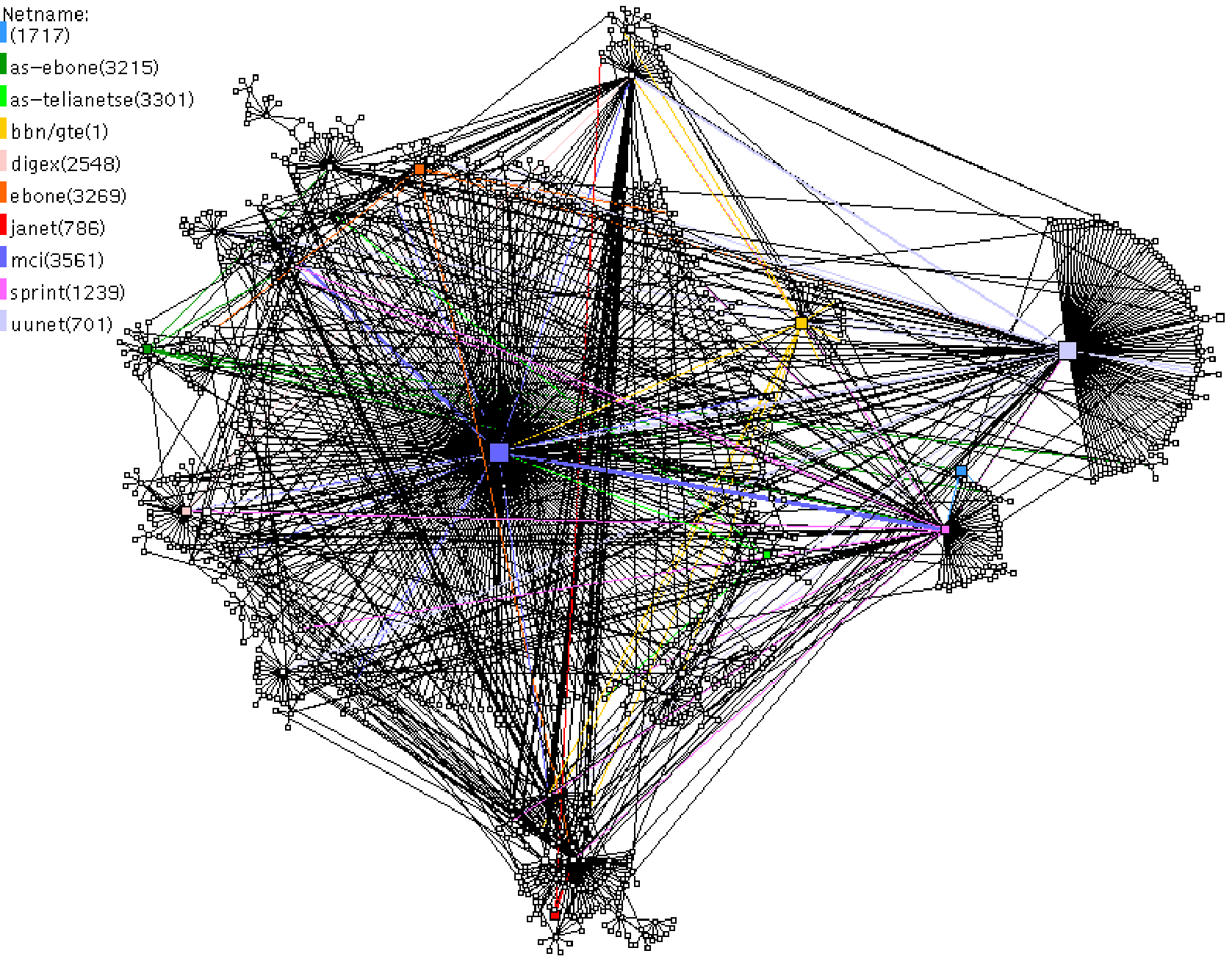}\\ \hline\end{tabular}~~
\begin{tabular}{|c|}\hline \\ \includegraphics*[width=7.0cm,height=7.0cm]{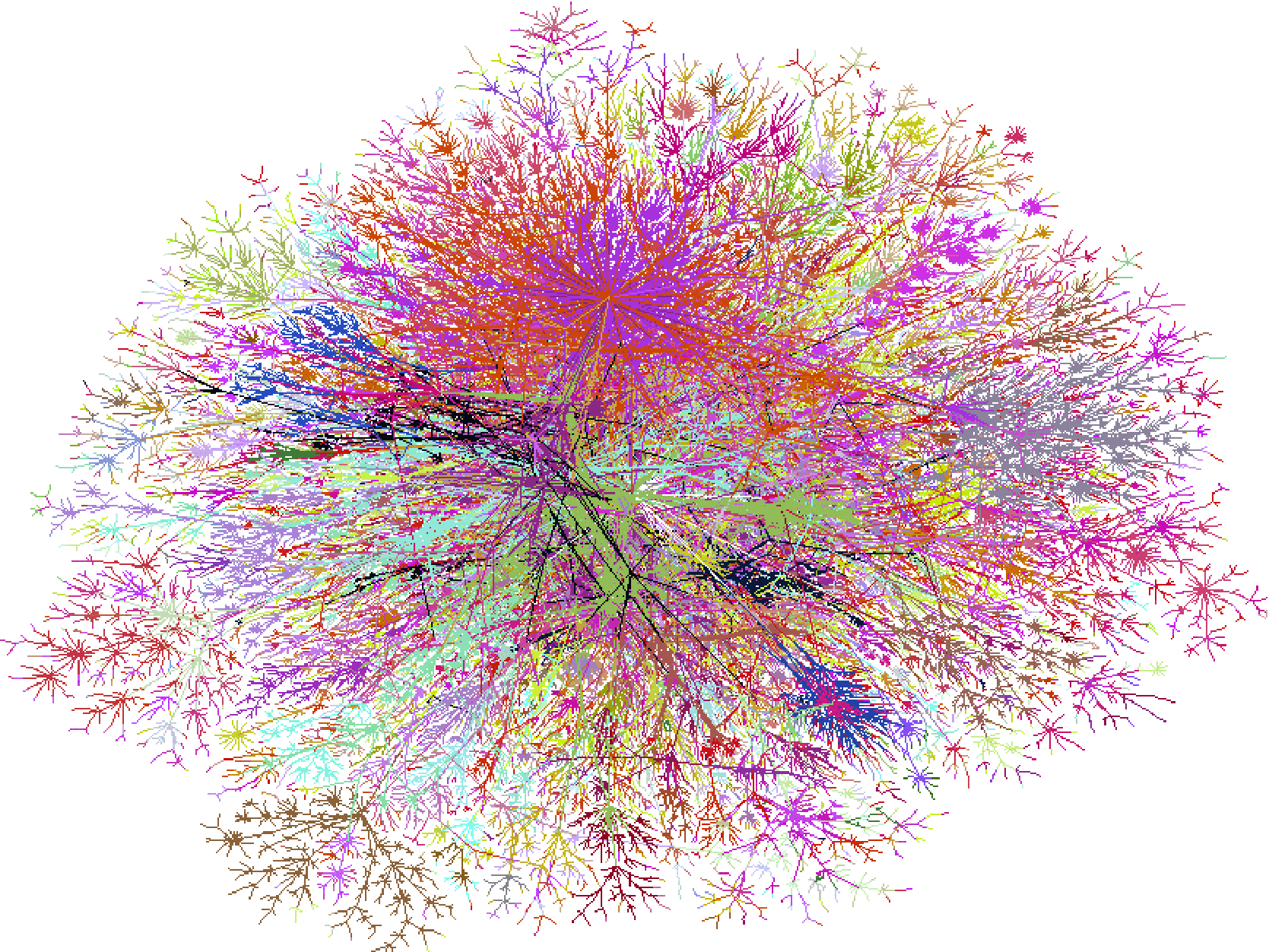}\\ \hline\end{tabular}
\end{center}
\caption{Graphs representations of the Internet at the autonomous systems level (left) and the router level (right). Both graphs are based on data collected by the Internet's mapping projects at CAIDA \cite{caida}.}
\label{mapsInternet}
\end{figure}
For many years, the structure of the Internet has been considered similar to that of a random graph, with a homogeneous degree distribution peaked around a characteristic degree value. 
In the last decade, on the contrary, massive Internet measurements provided evidences against such kind of modeling and in favor of topologies with heterogeneous degree distributions.
Historically, the first experimental evidence of a power-law degree distribution at the AS level is contained in a famous paper by Faloutsos et al. \cite{faloutsos}, who analyzed the data collected during the period $1997-1998$ by the National Laboratory for Applied Network Research (NLANR) \cite{nlanr}. In the following years many other studies, both at the AS and the router levels, always confirmed this important discovery~\cite{nlanr,caida,asdata,oregon,mercator,scan,lucent,us02,chen02}.
Though the qualitative picture coming from these two different scales is the same, the two graphs show relevant quantitative differences, that can be examined in depth using the statistical tools introduced in the previous section.
Note that, the size of the Internet is exponentially growing, but the major statistical measures do not change in time, suggesting the idea that the Internet, as a physical system, is in a sort of non-equilibrium stationary state.  
Two typical pictures of the Internet's AS and IR levels, obtained from the data of the CAIDA's mapping project \cite{caida}, are  displayed in Fig.~\ref{mapsInternet}.\\
Apart from the quantitative estimation of the exponent, that can be possibly affected by measurement biases, the existence of heavy-tails seems to be a solid feature of the Internet (see Chapter~\ref{CHAP3} for a complete statistical analysis of the Internet exploration's technique).
In Fig.~\ref{degreesInternet}, we report the degree distributions for the AS (left) and IR (right) levels, that are clearly power-laws $P(k) \sim k^{-\gamma}$ (with $\gamma \simeq 2.1$).\\
In both cases, the distribution of the shortest path length is peaked around the average distance $\langle \ell \rangle$, whose very small value is signature of the occurrence of the small-world property.\\ 
Differences in the two levels of descriptions can be found looking at the degree correlations.
The degree dependent spectrum of the average degree of nearest neighbors $k_{nn}(k)$ is roughly flat or slightly increasing with $k$ for IR maps, while it shows a very clear disassortative behavior at the AS level.
The reason for disassortative correlations can be found in the {\em strong hierarchical structure of the Internet at the autonomous system level}, that is absent at the router level. This hierarchical organization of the Internet at the AS level is reflected also in other measurements, as for instance its $k$-core structure, that is discussed in Refs.~\cite{medusa,medusa2,ignacio1,ignacio2,ignacio3,lanetvi} (see also Section~\ref{CHAP3_2_5}). \\
Both levels show similar average clustering coefficient, that is considerably higher than in standard random graphs, reinforcing the idea that the Internet is far from being locally tree-like. 
The spectrum of the clustering coefficient is almost constant at the router level and clearly decreasing at the AS level. The power-law functional dependence of $c(k)$ on the degree $k$ has been interpreted as a signature of the presence of modular structures at different scales \cite{modularity}.\\
Finally, the distribution of the node betweenness centrality $P_{b}(b)$ is clearly power-law for the AS ($P_{b}(b) \sim b^{-\gamma_{b}}$, with $\gamma_{b} \simeq 2.0$), while it shows a very broad distribution with a more bended shape 
at the IR level. The correlation between betweenness and degree is almost linear \cite{vespi_book} (but large fluctuations emerge using scatterplots instead of average values).\\
This picture of the Internet, however, is correct only at a qualitative level, whereas quantitatively, different mapping projects provide slightly different results for the average properties of the network, in relation to the unequal node and edge coverage of the measurements. An idea of such a variety of results is given by the list reported in Table \ref{tavola_Internet}. \\
The validation of Internet's large scale measurements is of primary interest to understand correctly topological and functional properties of the real Internet, and will be the subject of the next chapter.
\begin{table}[t]
\begin{center}
\begin{tabular}{l|c|c|c|c|c|c|c}
\hline
name       & date & $N$ & $\langle k \rangle$ & $k_{\max}$ & $\langle k^2 \rangle/\langle k\rangle$ & 
$\langle \ell \rangle$ & $\langle c \rangle$ \\
\hline \hline
AS RV     & 2005/04& 18119& 3.54771& 1382& 2369.82& 3.92 & 0.083\\
AS CAIDA   & 2005/04&  8542& 5.96851& 1171& 521.751& 3.18 & 0.222\\
AS Dimes   & 2005/04& 20455& 6.03862& 2800& 1556.24& 3.35 & 0.236\\
\hline
IR Mercator& 2001  & 228297& 2.79635& 1314& 36838.6 & 11.5 & 0.013\\
IR CAIDA   & 2003  & 192243& 6.33085&  841&  8884.23& 6.1 & 0.08\\
IP Dimes   & 2005  & 328011& 8.2142 & 1453& 10954   & 6.7 & 0.066\\
\hline
\end{tabular}
\caption{Main characteristics of the Internet obtained by various mapping projects:
number $N$ of vertices, average ($\langle k \rangle$)
and maximal ($k_{\max}$) degree, ratio 
$\langle k^2\rangle/\langle k\rangle$, average shortest path
$\langle \ell \rangle$ between pairs of vertices,
and average clustering coefficient $\langle c\rangle$.
For the Autonomous Systems level we consider maps of the {\em Oregon route-views} (RV) project \cite{oregon}, 
the {\em skitter} project at CAIDA \cite{caida}, and the {\em DIMES} project \cite{dimes}; at the router level, data about recent maps by CAIDA and DIMES, together with older results from the {\em Mercator} project \cite{mercator}, are reported.}
\label{tavola_Internet}
\end{center}
\end{table}
In summary, while the topological structure of the Internet at the router level is still hard to detect probably because of the unreliability of mapping processes, at higher granularity level, {\em for the ASs, the Internet structure is much  clearer and appears as a mixture of hierarchical modular structures with degree heterogeneity and small-world property}. 

\subsection{The World-wide Air-transportation Network}\label{CHAP2_3_2}

The World-wide Air-transportation Network (WAN) is the network of airplane connections all around the world, in which the vertices are the airports and the edges are non-stop direct flight connections among them.
As for the Internet, this is a physical network, in the sense that both the nodes and the edges are well-defined objects. 
We can easily as well define weights on the edges, since different flights are characterized by a different number of 
passengers and by the geographical distance they have to cover.\\
For the moment we do not consider geographical properties, that will be taken into account in Chapter~\ref{CHAP4}, and define the weight $w_{ij}$ of the link $(i,j)$ as the total number of available seats per year in flights between airports $i$ and $j$.\\
The analyzed dataset was collected by the International Air Transportation Association (IATA) for the year $2002$. It contains $N=3880$ interconnected airports (vertices) and $E=18810$ direct flight connections (edges).
This corresponds to an average degree $\langle k \rangle = 9.7$, with maximal value $k_{c}=318$. Degrees are strongly heterogeneously distributed, as confirmed by the shape of the degree distribution, that can be described by the functional form $P(k) \sim k^{-\gamma} f(k/k_{c})$, where $\gamma \simeq 2.0$ and $f(k/k_{c})$ is an exponential cut-off which finds its origin in physical constraints on the maximum number of connections that can be handled by a single airport \cite{luisair,guimera}.\\ 
Moreover, the WAN shows small-world property, since the average distance is $\langle \ell \rangle =4.4$.
It is worthy noting that weights from IATA database are symmetrical, that is probably a consequence of the traffic properties and allows to consider the network as a symmetric undirected graph.
\begin{figure}[t]
\begin{center}
\includegraphics[width=7.0cm]{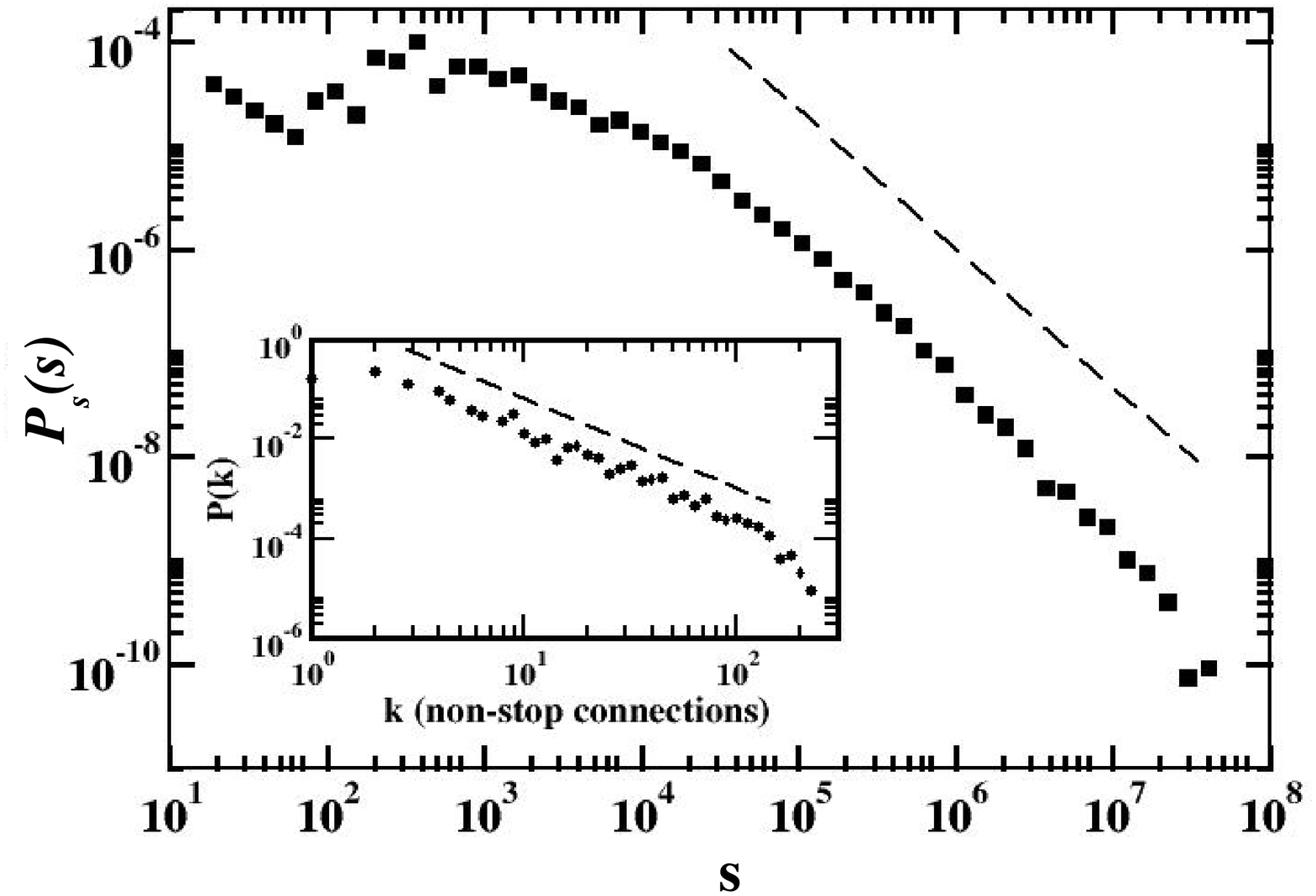}~~
\includegraphics[width=7.0cm]{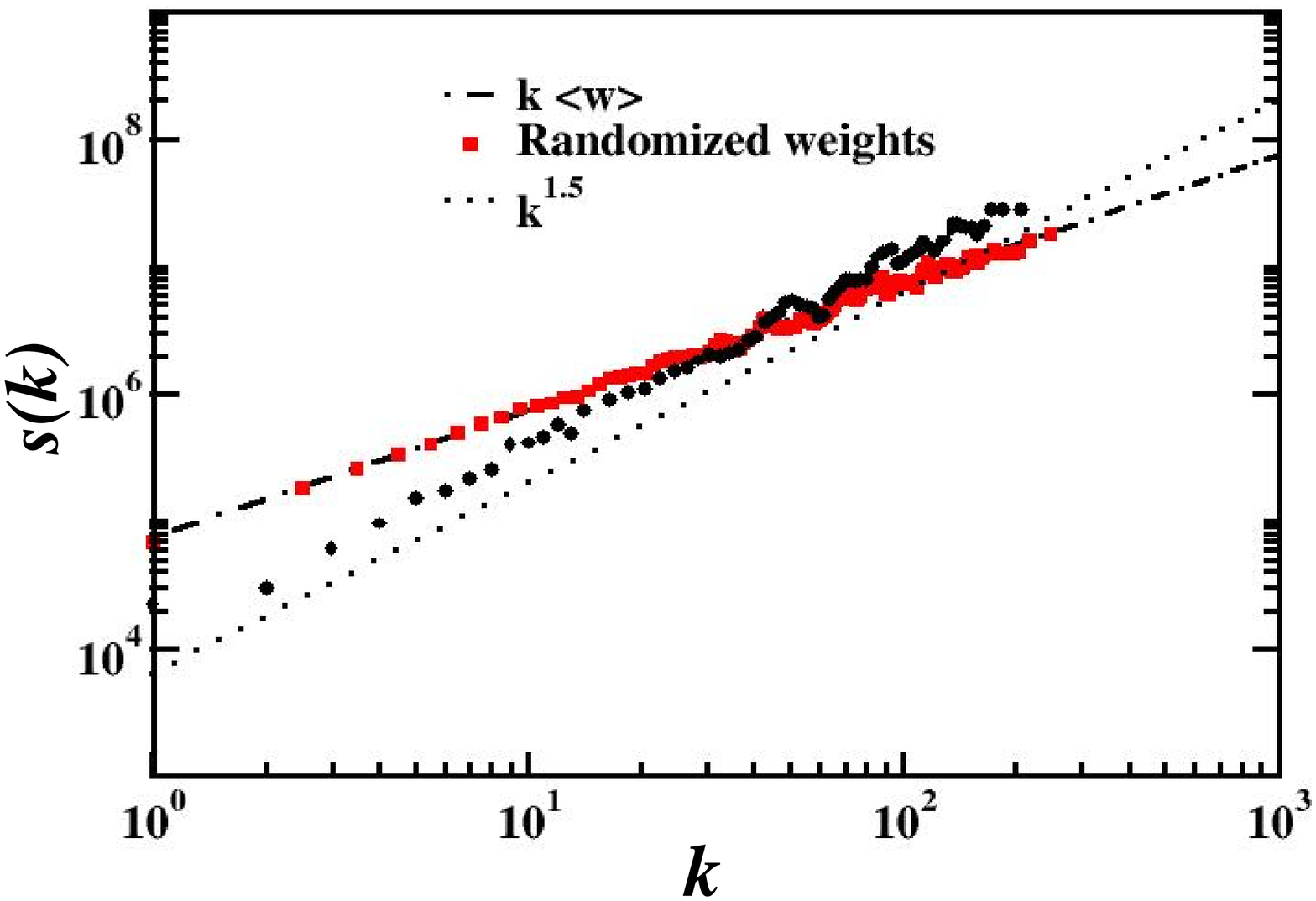}
\end{center}
\caption{The strength distribution $P_{s}(s)$ (left) and the diagram strength vs. degree (right) for the World-wide Airport Network. The diagram $s(k)$ shows the non-linear correlation between strength and degree (black circles), whereas the same quantity grows linearly after weights randomization (red squares). The inset reports the behavior of the degree distribution.}
\label{WAN_ps_sk}
\end{figure}
The analysis of weighted quantities reveals that both weights and strength are broadly distributed, 
and that they are non-trivially correlated with the degree, since the average weight $\langle w_{ij}\rangle \sim {(k_{i}k_{j})}^{\theta}$ with $\theta \simeq 0.5$ and $s(k) \propto k^{\beta}$ with $\beta \simeq 1.5$ \cite{barrat_PNAS}.
The sets of data in Figure~\ref{WAN_ps_sk}, representing the strength distribution $P(s)$ and the strength-degree diagram $s(k)$, are borrowed from Ref.~\cite{barrat_PNAS}.  
Such super-linear relation points out that highly connected airports tend to collect more and more traffic, that could in principle yield to a condensation process of the traffic on the hubs (i.e. a finite fraction of traffic handled by a small number of airports).\\
The non-trivial role of the weights is witnessed also by degree-degree correlations and clustering that show a slightly different behavior if weights are considered \cite{barrat_PNAS} (not shown).
The topological average nearest neighbors degree $k_{nn}(k)$, indeed, reaches a plateau for high degrees, while the weighted quantity $k_{nn}^{W}(k)$ still increases, showing that even if the degree of neighboring nodes is various, the hubs are preferentially connected with high traffic nodes. A similar interpretation holds also for the observed values of the weighted clustering coefficient $c^{W}(k)$ for high degree nodes, that is larger compared to the topological one. 

\subsection{Other Examples}\label{CHAP2_3_3}
 
It is possible to distinguish three main types of real complex networks: 
\begin{itemize}
\item artificial infrastructures and technological networks;
\item social networks generated by interactions between individuals; 
\item networks existing in nature, such  as food webs or biological networks.
\end{itemize}
The Internet and the WAN are the most popular examples of the first class, that contains many other communication and transportation networks \cite{eubank,gastner,watts98,aiello,oil,roads}. Not all of them show a broad degree distribution, but a considerable amount of heterogeneity can be recovered at a traffic level, as recently  shown by De Montis et al. \cite{demontis} in the case of the Sardinian transportation network. \\
The panorama of biological networks is very wide, and its analysis goes beyond the purposes of this thesis.
For instance, in {\em cellular networks}, the nodes are genes or proteins and the links are metabolic fluxes regulating cellular activity. In a seminal work \cite{kauffman69}, Kauffman showed that chemical processes can be conveniently represented by  {\em chemical reactions networks}. The vertices are substrates, connected by the chemical reactions in which they take part. The orientation of the corresponding edge says if a substrate is involved in the reaction as product or ``educt''. The average size of these networks is quite small ($N\sim \mathcal{O}(10^{3})$), but despite of the small size, the degree distribution is fairly broadly shaped. These networks are small-worlds and their structure seems to be rather robust under random defects, errors and mutations. \\
Another class of biological networks are  {\em protein-protein interaction networks} (PIN) (see for instance Refs.~\cite{kauffman93,uetz,pellegrini,jeong}), in which the edges identify the existence of interactions between two proteins.
They present heterogeneous degree distribution and low average distance, but also non-trivial pair correlations, that are related with the proteins functions. Indeed, by means of modularity analyses \cite{modularity,milo}, it is possible to uncover the relation between some small topological structures (like triangles or small cliques) corresponding to some functional modules of proteins. \\ 
Other interesting natural networks are {\em food webs} \cite{foodwebs1,foodwebs2,foodwebs3}, i.e.  networks of animals in a given ecosystem, in which directed arrows establish ``who eats who'' in the food chain. In this case also self-links must be considered, as consequence of 
the presence of cannibalism in many animal species.
However, as we will show in another context in Chapter~\ref{CHAP3}, the problem of counting all species of an ecosystem makes the definition of these networks quite difficult. Similar difficulties are faced in the definition of trophic links between species. In addition, these networks are very small  ($N\sim \mathcal{O}(10^{2})$) and their degree distributions are not clearly fat-tailed \cite{dunne}. The clustering coefficient is, instead, very large showing that they are far from being  random graphs, even if they present features that are typical of small-world models.
\begin{figure}[t]
\begin{center}
\includegraphics[width=12.0cm]{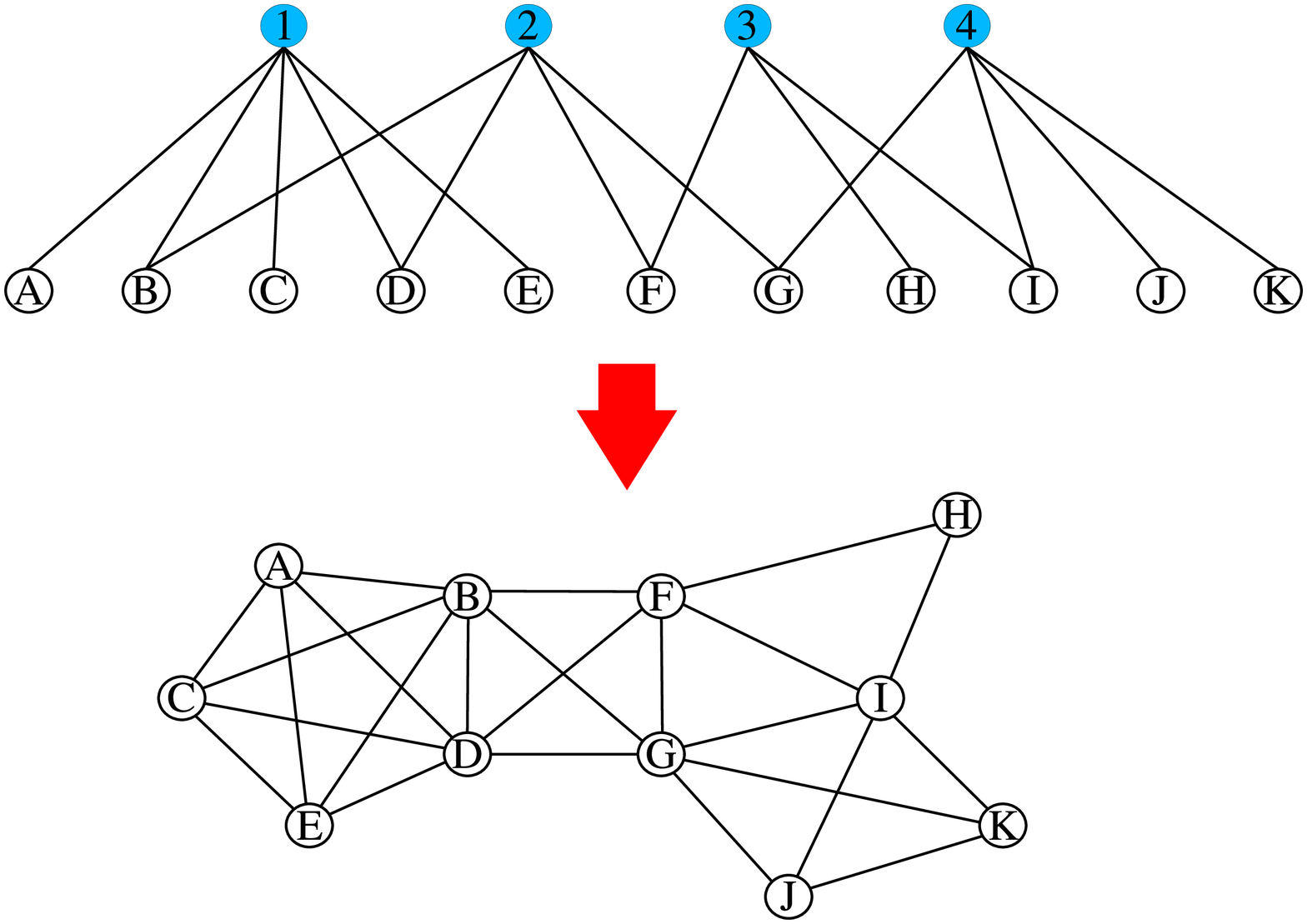}
\end{center}
\caption{(Top) Example of bipartite representation of a social network: blue nodes with numbers are the affiliations, and white nodes with letters are the actors. (Bottom) The figure represents the unipartite projection of the network, in which pairs of actors are connected by links if in the original network they have at least one common affiliation.}
\label{bipartite}
\end{figure}
Another large field of application of the network description are social sciences.
{\em Social networks} are used to represent social interactions among individuals (called {\em actors}), such as acquaintances, collaborations, sexual relations, etc. The typical structure of social networks is that of multi-partite networks, with a set of nodes representing actors and other sets of nodes representing {\em affiliations} they belong to (Fig.~\ref{bipartite}). 
Actors are indirectly linked together by means of common {\em affiliations}, i.e. the places they frequent, the office in which they work, etc. 
A standard technique to study social networks consists of projecting multi-partite networks on unipartite representations, in which nodes of a unique type (the actors) are linked by an edge if they have at least one common affiliation.   
Interesting examples of such networks are co-authorship networks, the most popular one being the network of collaborations among physicists submitting manuscripts on the well-known archive called ``{\em cond-mat}'' \cite{condmat_network1} \footnote{It is a public archive of condensed matter preprints and e-prints, 
see `{\tt http://www.arxiv.org/archive/cond-mat }'.}.
This database of condensed-matter physics contains $N=12722$ scientists who authored e-prints during the period from $1995$ to $1998$. 
According to the unipartite description, the presence of an edge between two authors means that they have co-authored at least one paper. Obviously, the link between authors who have co-authored many works together should be stronger, and we can assign weights to the edges following the definition proposed by Newman \cite{newman_coauthor},
\begin{equation}
w_{ij} = \sum_{p} \frac{\delta^{p}_{i} \delta^{p}_{j}}{n_{p}-1}~,  
\end{equation}
where $\delta_{i}^{p}$ is $1$ if the author $i$ has contributed to the paper $p$ and $0$ otherwise, and $n_{p}$ is the number of authors of the paper $p$.
Both the degree distribution and the strength distribution are broad, but the weights are linearly correlated with the degree, revealing that their distribution is independent of the topology.     
The network of ``{\em cond-mat}'' shows, moreover, an interesting community structure, with many induced subgraphs of different sizes, corresponding to different research fields \cite{palla,social_community}.\\
A further level of information that sometimes is available in many co-authorship networks is the number of citations gained by a paper. In such a case the weights are defined {\em \'a la Newman} but including the number of citations \cite{borner}. 
A very interesting issue is the study of the temporal evolution of co-authorship networks, identifying those topological and weighted structures that reinforced during the years and those which got lost. 
This kind of analysis has been carried out in Ref.~\cite{borner}, focusing on the study of the temporal evolution of the impact of co-authors groups in a particular scientific community (the database analyzed was the InfoVis Contest dataset, a network of $614$ papers by $1036$ authors, published between $1974$ and $2004$).

\newpage
\section{Networks modeling}\label{CHAP2_4}

In the previous section, we have reviewed some examples of real networks, from which we conclude that 
a networked description can be applied to a variety of systems with a large number of interacting units, independently of their functions and role in nature or society.
This makes evident the lack of a unique underlying theoretical framework in which all networks properties may be analyzed and interpreted.\\  
In order to build such a theoretical framework starting from phenomenological data, the first step consists of networks modeling.
We can roughly divide networks in two classes: {\em static networks} and {\em evolving networks}.
In the first class, the overall statistical properties are fixed and single networks are generated using static algorithms. A typical example of this class is the static random graphs ensemble defined by Erd\"os and R\'enyi \cite{erdos1,erdos2}. 
Classical random graph models are generated drawing edges uniformly at random between pairs of vertices with a fixed probability. The resulting graphs have poissonian degree distributions ({\em Erd\"os-R\'enyi Model}). Recently, this ensemble of graphs has been extended in order to include graphs with any possible form of the degree distribution ({\em Configuration Model}) \cite{molloy1,molloy2,canfield,aiello,catanzaro}. 
In this section, we introduce these two models together with another static model, that was proposed by Watts and  Strogatz  as a toy-model able to reproduce the main properties of real small-world networks ({\em Watts-Strogatz Model}) \cite{watts98}. \\
Evolving networks, on the contrary, are a very recent topic, that has attracted most attention after the discovery of broad degree distributions in real growing networks such as the Internet, and the possibility of producing power-law degree distributions by means of very simple evolution rules. In this case, the generation algorithm of the network implies a non-equilibrium process in which the statistical quantities evolve in time. 
However, all these growing models are built in such a way that the infinite size (and thus time) limit gives well-defined
statistical properties.
In the following, we will describe only two models of growing networks with power-law degree distribution, the famous {\em Barab\'asi-Albert Model} \cite{BA_model} and its clustered version proposed by Dorogovtsev, Mendes, and Samukhin ({\em DMS Model}) \cite{DMS_model}.\\
Though from the point of view of the application to the description of real networks the division in static and evolving networks is important, for the purposes of this thesis, i.e. for the study of dynamical processes taking place on networks, a better classification is that of considering the two following classes: {\em homogeneous networks}, with degree distribution peaked around the average value; and {\em heterogeneous networks}, in which the degree distribution is skewed and may present heavy-tails, or more generally, large fluctuations around the average value. \\
Many real networks evolve in time, but, in fact, the temporal scale of their topological rearrangements is usually much longer than the time-scale of the dynamical processes occurring on the network.
This property allows to study dynamical processes on networks that have been obtained with a growing mechanism as if they were static models.
    
\subsection{Homogeneous Networks}\label{CHAP2_4_1}

\begin{figure}[t]
 \vskip .5in
\begin{center}
\includegraphics[width=7.0cm]{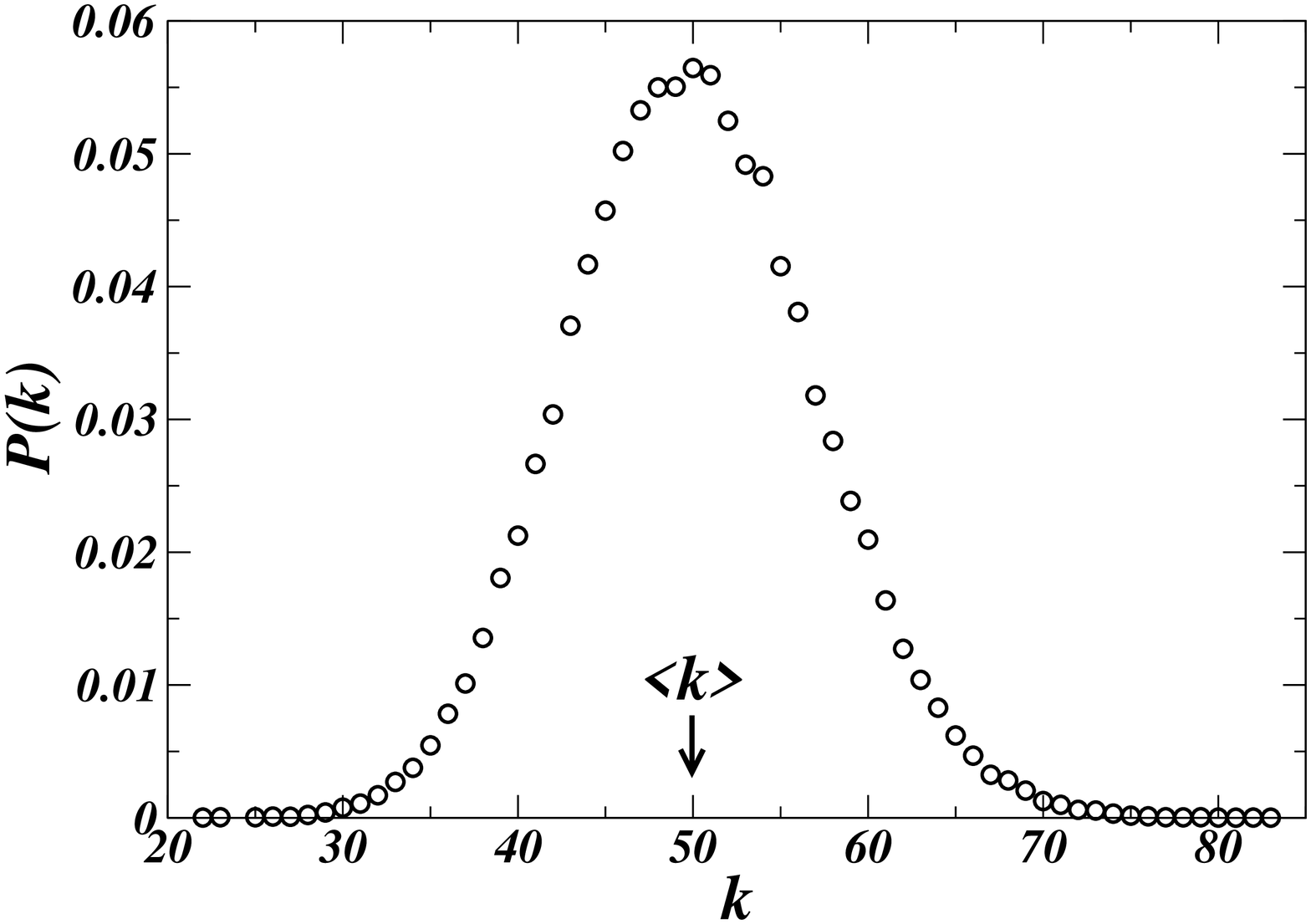}
\includegraphics[width=7.0cm]{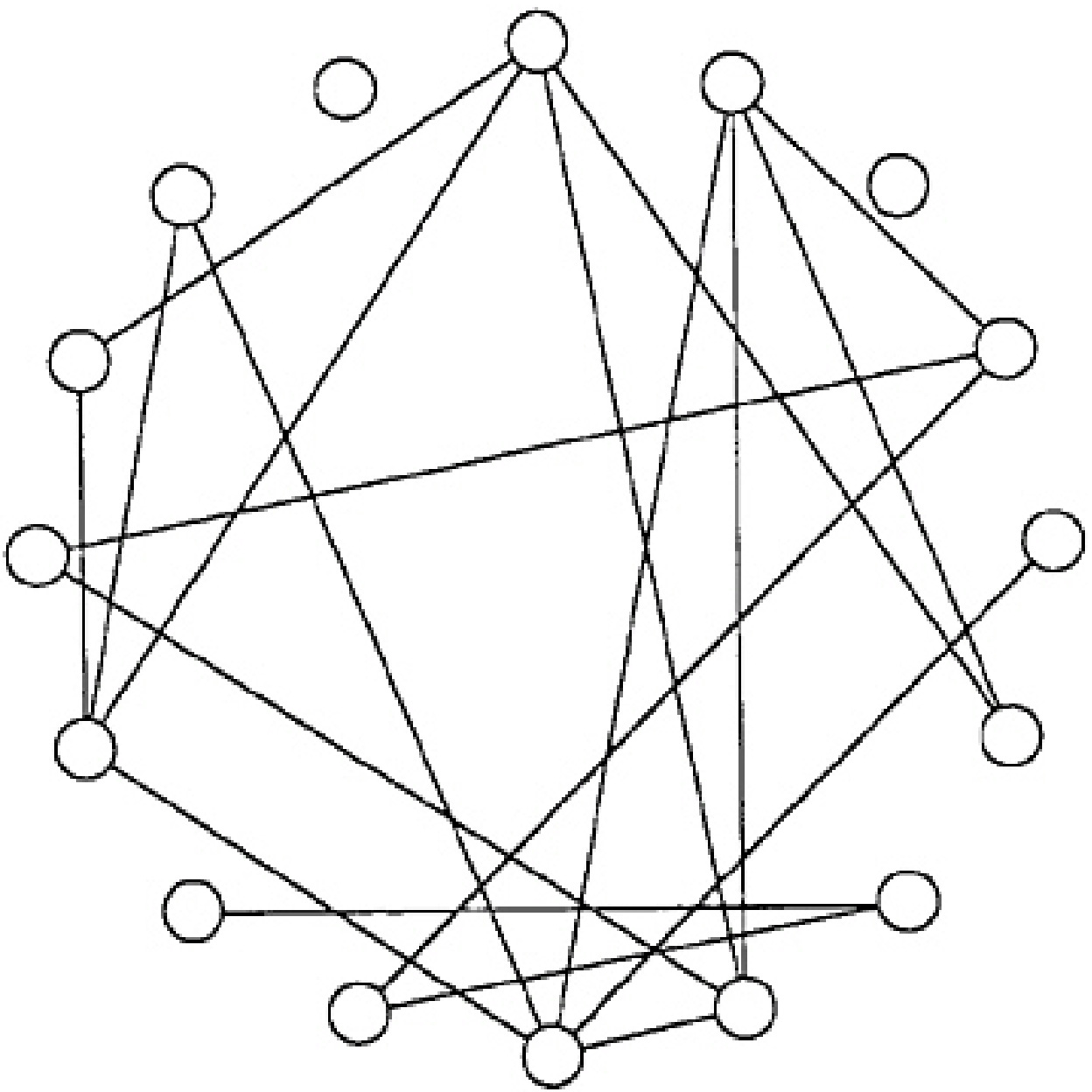}
\end{center}
\caption{Typical degree distribution (left) for a homogeneous poissonian random graph (right).}
\label{histo1}
\end{figure}

{\bf \em Erd\"os-R\'enyi Model (ER) -\quad} As already mentioned the theory of random graphs was founded by P. Erd\"os and A. R\'enyi \cite{erdos1,erdos2}, who defined two random graphs ensembles $\mathcal{G}_{N,E}$ and $\mathcal{G}_{N,p}$, in which the fixed quantities are, respectively, the number of nodes and edges in the former case and the number of nodes and the linking probability in the latter. In $\mathcal{G}_{N,E}$, the graphs are defined choosing uniformly at random pairs of nodes and connecting them with an edge, until the number of edges equals $E$. The second case is practically more convenient, and is defined as follows. 
Starting from $N$ nodes, one connects with probability $p$ each pair of nodes.
At the end of the procedure, the average number of edges is $E = pN(N -1)/2$ and the average degree is $2E/N$ $= p(N-1) \simeq pN$, if $N$ is sufficiently large.
For a finite number of nodes, the probability that a node has degree $k$ is given by a binomial law
\begin{equation}
P(k) = C^{k}_{N-1} p^{k}{(1-p)}^{N-1-k}~,
\end{equation}
where $p^{k}{(1-p)}^{N-1-k}$ is the probability of having exactly $k$ edges and $C^{k}_{N-1} = \frac{(N-1)!}{k! (N-1-k)!}$ is the number of possible ways these edges can be selected.
Taking the limit $N\to \infty$ and $p \to 0$, in such a way that $pN \to \langle k \rangle = const$, the binomial law tends to a Poissonian distribution
\begin{equation}
P(k) \simeq \frac{{\langle k \rangle}^{k}}{k!} e^{-\langle k\rangle}~.
\end{equation}
Poissonian distributions are very peaked around the average value, with bounded fluctuations; indeed, the second moment is finite, $\langle k^{2}\rangle$ = ${\langle k\rangle}^{2} +\langle k\rangle$. An example of such degree distribution is displayed in Fig.~\ref{histo1} (left), while the right panel in the same figure displays a sketch of its graphical representation.
Since the probability of finding nodes of degree much larger than $\langle k\rangle$ decreases exponentially, 
these graphs are prototypes for homogeneous networks.\\
However, ER random graphs show a giant component only for $p > p_{c}$ with $p_{c} = 1/N$, that corresponds to the critical average connectivity $\langle k\rangle =1$.  
Erd\"os and R\'enyi (\cite{erdos1,erdos2}) proved the existence of a transition between a phase, for $p < p_{c}$, in which with probability $1$ the graph has no component of size larger of $\mathcal{O}(\log N)$, and a phase, for $p > p_{c}$, in which the graph has a giant component or order $\mathcal{O}(N)$. In the marginal case $p=p_{c}$, the largest component has size $\mathcal{O}(N^{2/3})$.
This second order phase transition belongs to the same universality class of the mean-field percolation transitions \cite{bollobas_book}.\\  
The ER random graphs are completely uncorrelated, thus the average degree of the nearest neighbors is a constant independent of $k$. The clustering coefficient is simply the probability that any two neighbors of a given vertex are also connected to each others, i.e.
\begin{equation}
\langle c \rangle = p = \frac{\langle k\rangle}{N-1}~. 
\end{equation}
The fact that in random graphs the clustering coefficient vanishes in the limit of large size $N$ justifies the local tree-like approximation (Bethe Tree) used to obtain many relevant results. \\
For instance, the tree-like approximation allows to compute how the diameter of the graph scales with $N$. Indeed, since each node has typically $\langle k\rangle$ neighbors, at distance $\ell$ in a tree-like topology, the number of visited nodes scales as ${\langle k \rangle}^{\ell}$. The diameter is reached when the number of visited nodes is equal to $N$, but the shortest path distribution is very peaked, thus we can approximate the diameter with the average distance, $\langle \ell \rangle$, obtaining 
\begin{equation}
\langle \ell\rangle \sim \frac{\log N}{\log \langle k\rangle}~.  
\end{equation}

\begin{figure}[t]
\begin{center}
\includegraphics[width=5.0cm]{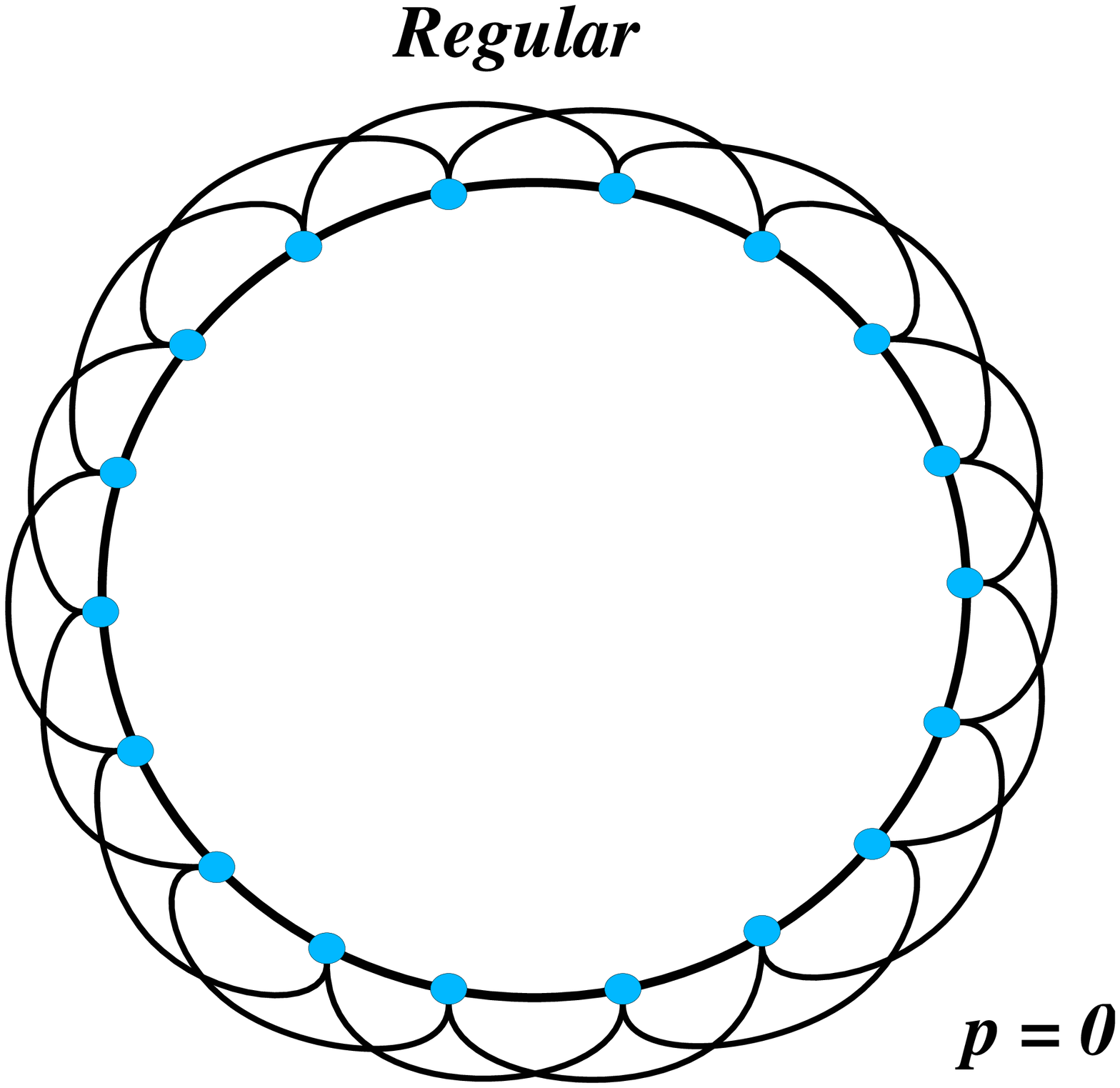}
\includegraphics[width=5.0cm]{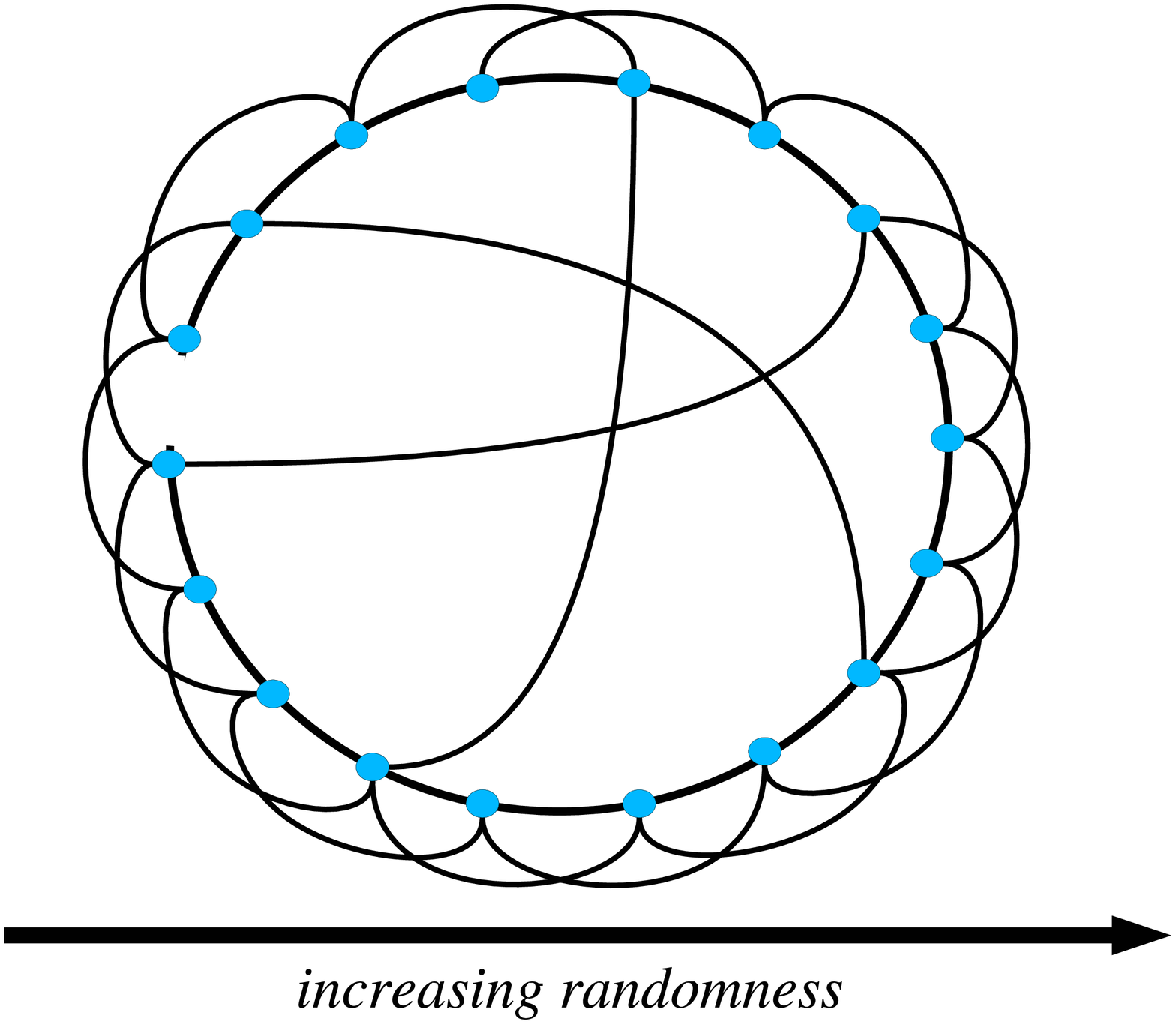}
\includegraphics[width=5.0cm]{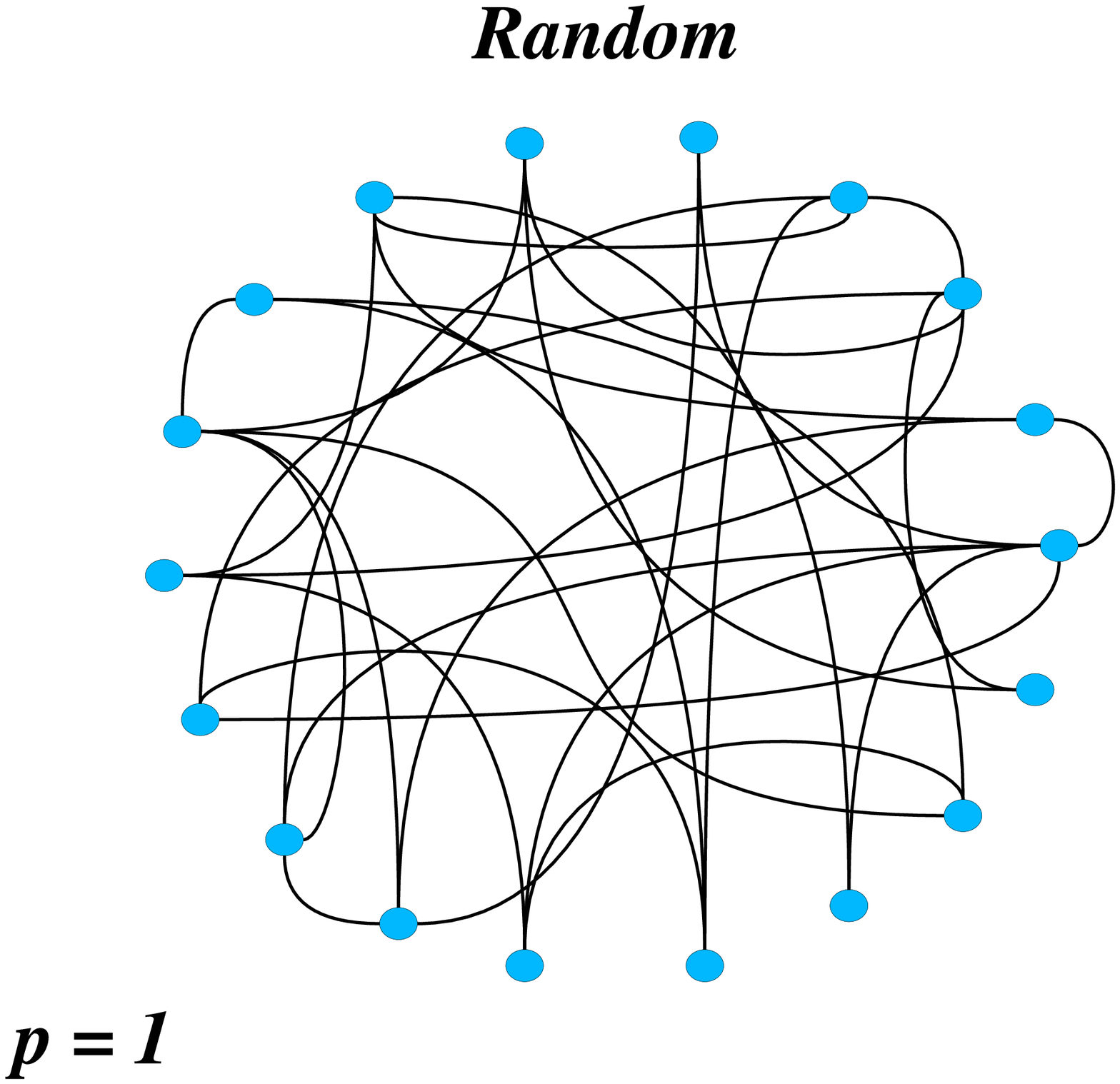}
\end{center}
\caption{Rewiring process in the Watts-Strogatz model of smallworld network. Increasing the rewiring probability $p$, the network passes from a regular $2$-banded network (left) to a smallworld network (center) and finally to a random graph (right).}
\label{rewiring}
\end{figure}

{\bf \em Watts-Strogatz Model (WS) -\quad} In Section~\ref{CHAP2_3}, we have shown experimental data from which it emerges that the average hop distance between two vertices in real complex networks is very small, and it is possible to reach every vertex in a small number of steps. Nevertheless, random graphs are not optimal models for the study of real networks, since the most of them are clustered, i.e. they contain a lot of triangles, whereas random graphs are locally tree-like. In order to overcome this problem, Watts and Strogatz proposed a simple model interpolating between a regular lattice with large average distance but strong clustering and the random graph with small diameter and small clustering \cite{watts98}.\\
The initial network is a one dimensional $m$-banded graph, i.e. a ring of $N$ sites in which each vertex is connected to its $2m$ nearest neighbors. The vertices are then visited one after the other: each link connecting a vertex to one of its $m$ nearest neighbors in the clockwise sense is left in place with probability $1-p$, and with probability $p$ is rewired to a randomly chosen other vertex. The long-range connections introduced play the role of shortcuts connecting regions that are very distant in the original network. Figure~\ref{rewiring} displays a sketch of the rewiring mechanism.\\
For $p=1$, the network is completely randomized but, since it has at least degree $2m$, it is not equivalent to a random graph. The interesting regime is for $1/N \ll p \ll 1$, in which a still rather large clustering coefficient coexists with 
a logarithmically scaling average distance. 
In the limit $p\to 0$, the small-world property disappears and the metric structure of the lattice is restored. It has been shown (\cite{barrat_weigt,amaral_sw,newman_sw,newman_sw2}) that the transition occurs precisely at $p=0$, and in the infinite size limit the average distance diverges as $\langle \ell \rangle \sim 1/p$. This cross-over phenomenon for increasing rewiring probability plays an important role in determining the behavior of dynamical processes defined on this type of network (see Chapter~\ref{CHAP5} for the case of Naming Game on Watts-Strogatz small-world networks).    
The degree distribution of this model in the regime $1/N \ll p \ll 1$ has the form \cite{barrat_weigt}
\begin{equation}
P(k) = \sum_{i=0}^{\min(k-m,m)} \left(\begin{array}{c} m \\ i \end{array}\right) {(1-p)}^{i} p^{m-i} \frac{{(pm)}^{k-m-i}}{(k-m-i)!} e^{-pm}~,
\end{equation}
for $k\geq m$, and is equal to zero for $k<m$.\\
The clustering coefficient can be easily computed recalling that two neighbors that are linked together in the original model remain connected with probability ${(1-p)}^{3}$,
\begin{equation}
\langle c\rangle (p) \sim \frac{3(m-1)}{2(2m-1)}{(1-p)}^{3}~. 
\end{equation}

\subsection{Heterogeneous Networks}\label{CHAP2_4_2}

In the last years, a huge amount of experimental data yielded undoubtful evidences that real networks present a strong degree heterogeneity, expressed by a broad degree distribution.    
In order to reproduce the main features of this new class of heterogeneous networks, a big effort has been devoted to network modeling, and a large number of models with degree heterogeneity has been put forward.
The main feature of these networks is that the average degree is not representative of the distribution, and the second moment $\langle k^2 \rangle$ is very large, possibly diverging in the infinite size limit.
A characterization of the heterogeneity level of a degree distribution is given by the parameter $\langle k^2 \rangle/\langle k\rangle$, that is strictly related with the expression of the normalized fluctuations and enters in the description of many dynamical phenomena on networks, such as percolation and epidemic spreading.\\
When the degree distribution is a power law $P(k) \sim k^{-\gamma}$ ($2 < \gamma \leq 3$), the fluctuations diverge with $N$ (the average remaining finite), and nodes with very large degree appear in the network. However, not all heterogeneous networks are power-law, many of them possess bended degree distributions that cannot be classified as power-laws. A broad distribution, that is not scale-free, but has been used to fit a number of data coming from the Internet's measurements (\cite{broido}) is the Weibull (WEI), whose form is $P(k) = (a/c){(k/c)}^{a-1}\exp(-{(k/c)}^{a})$, with $a, c$ real positive constants. Weibull distributions are good candidates as degree distributions also for networks of scientific collaborations, wordwebs and biological networks, where the existence of a neat power-law is still 
under debate. \\
Now, we introduce the most relevant models of heterogeneous networks, that will be used in the numerical simulations related to the investigations reported in the next chapters. 

\begin{figure}[t]
\begin{center}
\includegraphics[width=12.0cm]{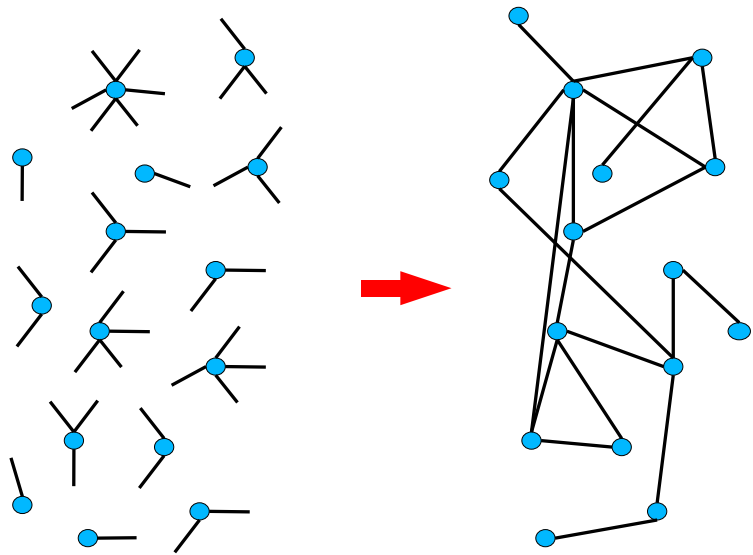}
\end{center}
\caption{Illustration of the Molloy-Reed algorithm used to generate a generalized random graph starting from a degree sequence.
The stubs (left) are linked in such a way that self-links and multi-links are avoided.}
\label{ucm}
\end{figure}

{\bf \em Configuration Model (CM) -\quad} This is a static model of scale-free graphs that generalizes the random graph ensemble of Erd\"os and R\'enyi to generic degree distributions. It is particularly useful to study dynamics on network models with a given degree distribution and controlled correlations.
A famous algorithm to generate generalized random graphs was proposed by Molloy and Reed \cite{molloy1,molloy2}. \\
A degree sequence $\{k_{i}\}$ ($i=1, \dots, N$) is drawn randomly from the desired degree distribution $P(k)$ and assigned to the $N$ nodes of the network, with the additional constraint that the sum $\sum_{i} k_{i}$ must be even. At this point, the vertices are connected by $\sum_{i} k_{i}/2$ edges, respecting the assigned degrees and avoiding self- and multiple-connections. Figure~\ref{ucm} reports a sketch of the generation procedure. 
The last condition introduces unexpected correlations, producing a slightly disassortative behavior.
In order to eliminate correlations, Catanzaro et al. \cite{catanzaro} have proposed a variation of the model characterized by a cut-off at $\sqrt{N}$ for the possible degree values. In the rest of this work, when talking of configuration model we will always refer to this particular model, that is known as {\em Uncorrelated Configuration Model (UCM)} and presents flat $k_{nn}(k)$ and $c(k)$.

\begin{figure}[t]
\begin{center}
\includegraphics[width=7.0cm]{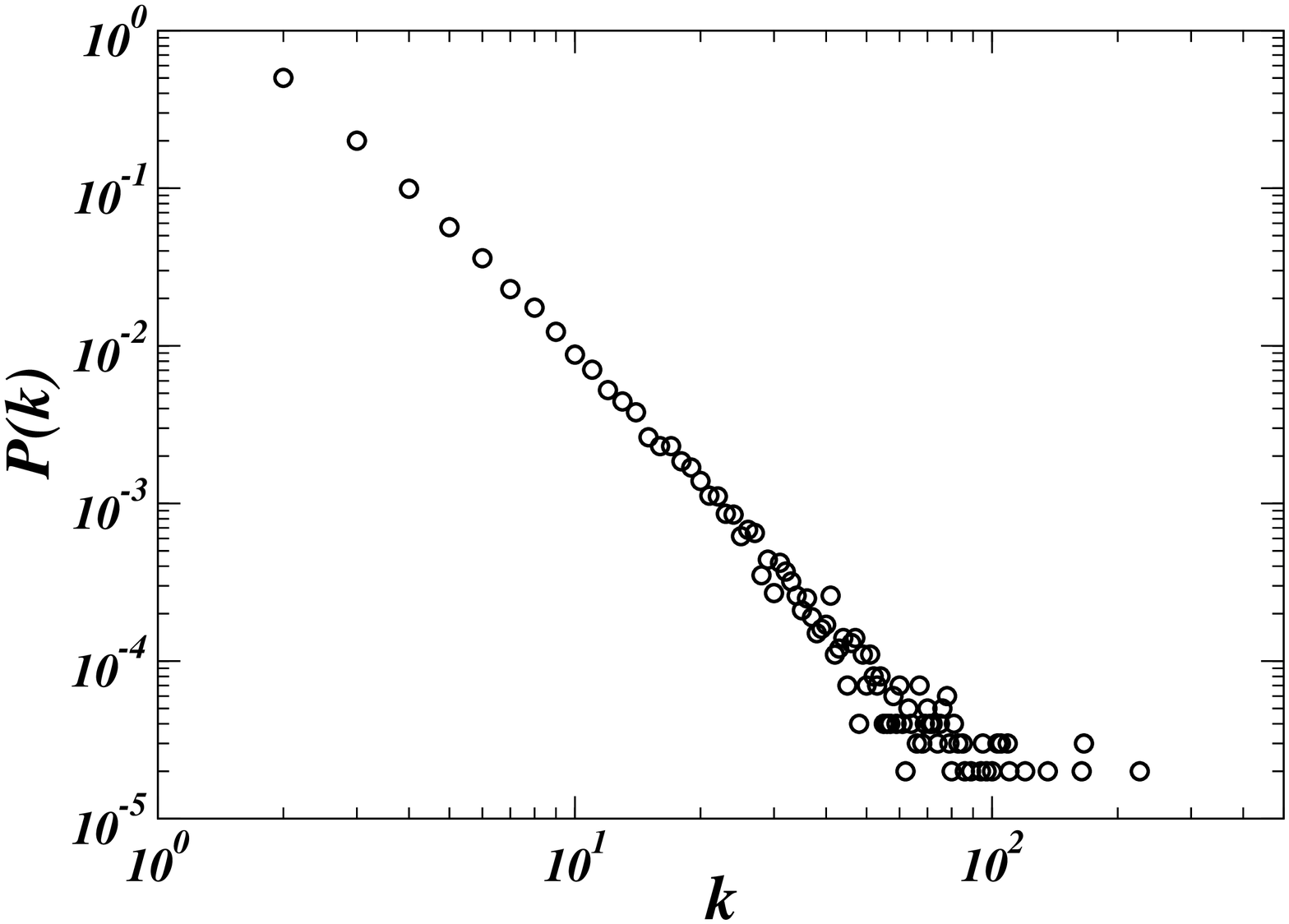}
\includegraphics[width=7.0cm]{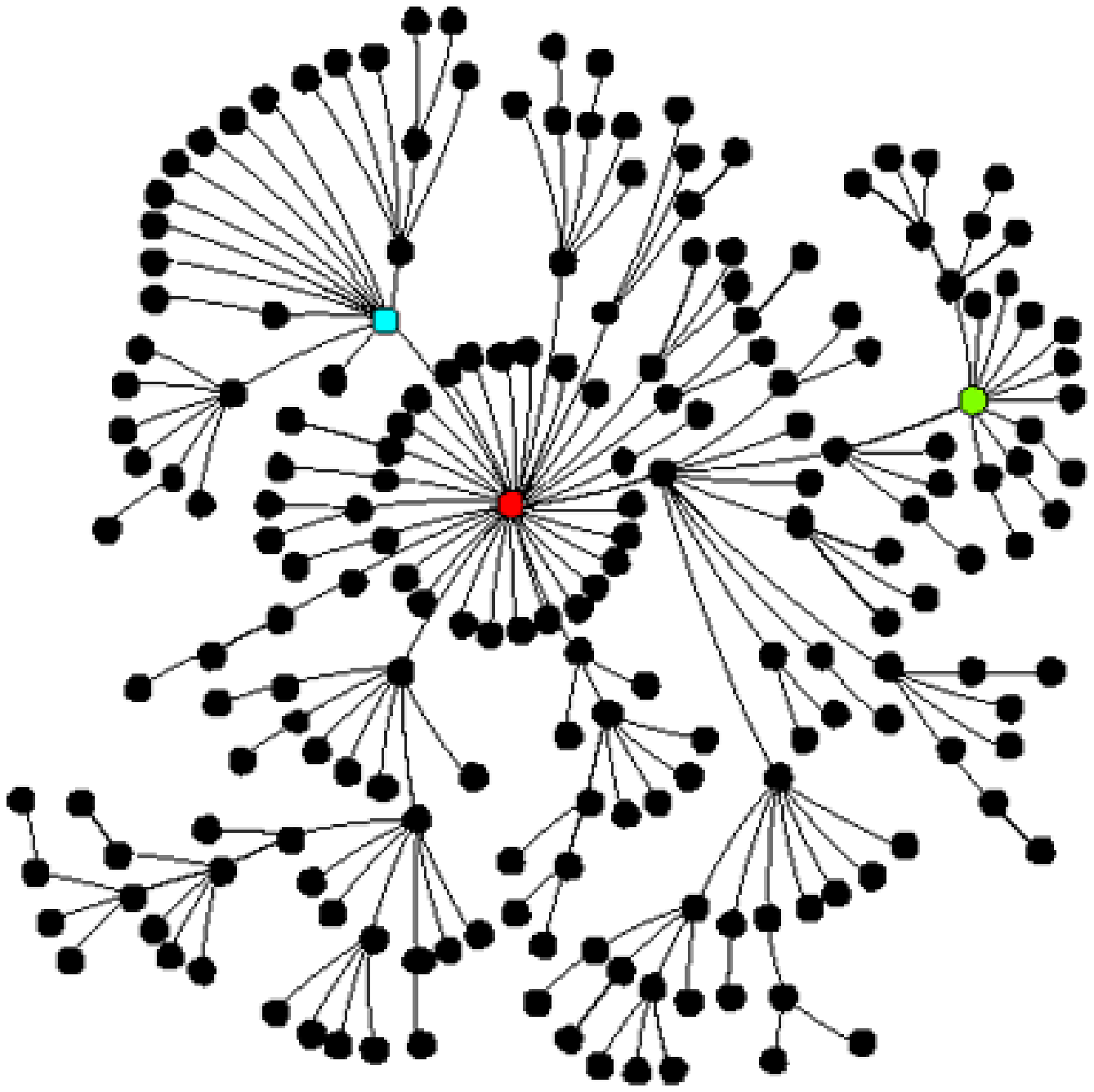}
\end{center}
\caption{Typical degree distribution (left) for a heterogeneous BA model (right).}
\label{histo2}
\end{figure}

{\bf \em Barab\'asi-Albert Model (BA) -\quad} The first attempt to model real growing networks such as the Web was provided by  Barab\'asi and Albert, who proposed the idea of {\em preferential attachment} as the central ingredient in order to get a power-law degree distribution \cite{BA_model}. The preferential attachment is based on the simple idea that, during the network's evolution, new coming nodes become preferentially connected with nodes that already have a large number of connections.
It was proposed as a ``construction recipe'' for the Web, in which new pages acquire more visibility if they link to very important webpages, but it can be assumed as a valuable principle for a large number of technological and social networks, in which nodes want to optimize their conditions connecting with very important and central nodes.
During the growth, the ``rich gets richer'' effect is produced: large degree nodes more easily increase their degree compared to low degree nodes.\\ 
The algorithm starts from a small fully connected core of $m_{0}$ nodes (their precise properties do not change the statistical properties of the model in the large size limit). 
At each time step $t =1, 2, \dots, N-m_{0}$, a new node $j$ enters the network and forms $m\leq m_{0}$ edges with distinct existing nodes: the probability that an edge is created between $j$ and the node $i$ is 
\begin{equation}\label{attaching_kernel}
\Pi_{j\to i} = \frac{k_{i}}{\sum_{l}k_{l}}~. 
\end{equation}
Every new node has $m$ links and the network size at time $t$ is $N(t)=t+m_{0}$; since the number of links is $E=mt$, in the large time limit the average degree is simply $\langle k\rangle = 2m$. 
The degree distribution of the BA model can be obtained by means of different methods (mean-field approximation \cite{BA_model,reka_MF}, rate equation \cite{BA_pkRE}, or master equation \cite{BA_pkME}), and shows, in the limit $t\to \infty$, a power law behavior $P(k) \sim k^{-\gamma}$ with $\gamma =3$. Figure~\ref{histo2} displays the degree distribution and a graphical representation of a BA network.\\
Apart from the very interesting idea of preferential attachment, the Barab\'asi-Albert model is a very peculiar network,
with flat degree correlations and almost vanishing clustering.
Many variations of the model have been proposed, including node aging~\cite{PA_aging1}, fitness~\cite{PA_fitness1,PA_fitness2}, edge rewiring~\cite{PA_rewiring}, limited information~\cite{PA_limited}, etc; in particular, the addition of a constant $A$ representing the initial attractivity ($k_{i} \to k_{i} +A$ in the kernel of Eq.~\ref{attaching_kernel}) allows to generate power law networks with desired exponent $2 \leq \gamma \leq \infty$ \cite{BA_pkME}. 
Note that the linearity (in $k$) of the attaching kernel in Eq.~\ref{attaching_kernel} is a necessary condition to get a power-law distribution. It has been indeed proved (\cite{BA_pkRE,PA_nonlinear}) that using generalized kernels of the type  $\Pi_{j\to i}$ = $\frac{k_{i}^{\beta}}{\sum_{l}k_{l}^{\beta}}$ the degrees of the emerging network are power-law distributed only for $\beta=1$. When $\beta < 1$, the degree distribution is exponentially shaped; when $\beta >1$ the evolution produces edge-condensation on few vertices.

{\bf \em Dorogovtsev-Mendes-Samukhin Model (DMS) - \quad} Another interesting variation of BA model is the growing clustered network proposed by Dorogovtsev, Mendes and Samukhin (DMS) \cite{DMS_model}, in which the evolving rule consists in attaching new vertices to the extremities of a randomly chosen edge. The probability to choose a given vertex is thus proportional to the number of edges of which it is an extremity, i.e. to its degree. 
While the DMS model generates a power law distribution ($\gamma = 3$), the growing rule does not explicitely need any a priori knowledge of node's degrees.
The only real difference between this model and the BA model is in the clustering coefficient, that in the DMS model is very high ($\langle c\rangle \simeq 0.739$ for $m=1$). The general behavior of the clustering spectrum $c(k)$ and the  average value $\langle c\rangle$ for the DMS model has been computed using the rate equation formalism in Ref.~\cite{barrat_romu}.

\subsection{Weighted Networks}\label{CHAP2_4_3}
In this last paragraph, we discuss a model of growing weighted network that is particularly appropriate for the 
description of the World-wide Air-transportation Network.\\
In real weighted networks, the weights are not fixed but evolve in time together with the topology, so that the characteristics of the networks depend on the interplay of these two types of dynamics (topological and weighted) and on the relation between their time-scales.
As for the purely topological models, also for weighted evolving networks there is a variety of slightly different models, each one pinpointing a particular aspect of the evolution.
We are interested in the situation in which {\em the evolutions of links and weights are coupled and have the same temporal  scale}, that is similar to what happens in the real network of airports.\\ 
In the case of the WAN, in fact, each new airport $j$ that enters the existing network brings new flights, and new passengers are introduced into the system. A fraction of these passengers will not stop at the destinations of the direct flights they have taken from $j$ (e.g. a new airport is connected only to New York but part of the passengers would like to go to other cities, such as Philadelphia or Chicago); hence, part of the traffic is immediately locally redirected on the network. 
Moreover, new airports try to connect themselves to the most connected hubs, in order to guarantee many correspondences to the passengers.
Consequently, a good way to naively model the growth of a network like the WAN is that of considering 1) a preferential attachment procedure based on the strength and 2) a redistribution rule for the local traffic brought by the new node.
\begin{figure}[t]
\begin{center}
\includegraphics[width=12.0cm]{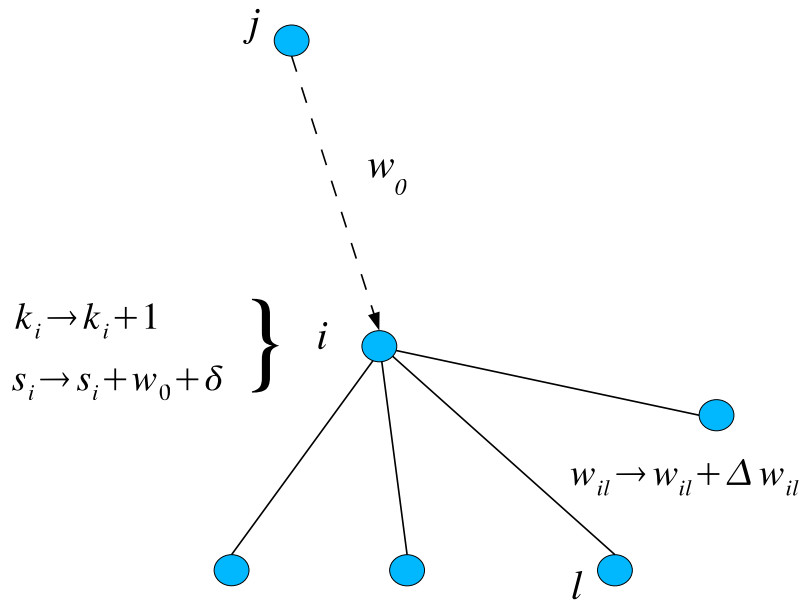}
\end{center}
\caption{Weights redistribution rule in the BBV model. When a new node $j$ enters the network attaching to a node $i$, it carries new traffic that is redistributed to the neighbors of $i$ by means of a contribution to their weights. The edge $(j,i)$ assumes a fixed weight $w_{0}$.}
\label{redistribution_BBV}
\end{figure}
More precisely, the {\em Barrat-Barth\'elemy-Vespignani (BBV) model} \cite{BBV_model,BBV_model2,BBV_model3} starts from an initial clique of $m_{0}$ nodes with a fixed weight $w_{0}$. At each time step, a new vertex $j$ is added to the network with $m$ edges (of weight $w_{0}$) that are 
randomly attached to a previously existing vertex $i$ according to the probability
\begin{equation}
\Pi_{j\to i} = \frac{s_{i}}{\sum_{l}s_{l}}~. 
\end{equation}
Immediately after the creation of the new link $(i,j)$, the weight of links $(i,l)$ connecting $i$ with each other neighbor $l$ is locally redistributed according to the rule (see Fig.~\ref{redistribution_BBV})
\begin{equation}
w_{il} \to w_{il}+ \delta \frac{w_{il}}{s_{i}} \quad \quad \forall l \in \mathcal{V}(i)~,
\end{equation}
i.e. the increase of the traffic $\delta$ is locally distributed among the neighboring connections, each link receiving a fraction of traffic that is proportional to the amount of traffic already handled by that connection.\\
This network model displays power law distributions of degree, strength and weights.
An interesting feature of the model is the presence of non-linear correlations between degree and strength. 
However, it is worthy noting that the model fails in reproducing the large fluctuations that characterize the quantity $s(k)$ in the case of the WAN (and other growing weighted networks), and that are clearly visible in the scatter-plot degree vs. strength.\\
In a recent publication \cite{spatial_BBV}, the authors of the model have shown that large fluctuations may be due to the existence of spatial constraints coming from the embedding of the nodes in a two-dimensional euclidean space.
This modified model consists in considering the nodes deployed on a two-dimensional euclidean space: each new node is situated in a randomly chosen point of the lattice, and the preferential attachment kernel is modified in order to account for the fact that a node prefers to connect to nodes that are well-connected but also spatially close to itself,
\begin{equation}
\Pi_{j\to i} \propto s_{i} \exp(-\eta d_{ij})~, 
\end{equation}
where $d_{ij}$ is the euclidean distance between nodes $i$ and $j$, and $\eta$ is related to a characteristic length scale.
When $\eta$ is small enough that spatial effects cannot be avoided, the relation between degree and strength is still non-linear but now presents very large fluctuations (around the average value expressed by $s(k)$).  \\
This model will be furtherly investigated in the analysis of the vulnerability of weighted networks (Chapter~\ref{CHAP4}), in order to explain some results obtained for the real airport network.

\chapter[Exploration of complex networks]{Exploration of complex networks}\label{CHAP3}

\section{Introduction}\label{CHAP3_1}

The present chapter is devoted to describe and study the exploration techniques of complex networks.
Motivations of this research and a general introduction to the problem, in which we highlight several 
different sampling methods, are provided in Sections~\ref{CHAP3_1_1}-\ref{CHAP3_1_2}. 
Then, in Section~\ref{CHAP3_2}, we focus on a theoretical model for the exploration of the Internet, that is analyzed
using a typical mean-field statistical physics approach. A variety of different measures are introduced to investigate the main properties of the exploration process and its biases. Finally, exploiting an interesting application of non-parametric statistics, we propose an approach to compensate the biases (Section~\ref{CHAP3_3}).

\subsection{Motivations}\label{CHAP3_1_1}
Network modeling is undoubtedly the favorite tool used by researchers to 
understand the origins of the ubiquity of complex networks in the real world. 
In particular, the presence, in biological as well as technological systems, 
of the same peculiar topological properties, such as a broad degree distribution 
and very small average inter-vertex distances, is very intriguing. 
By means of network models, some of the phenomenological results have been reproduced, 
and possible explanations for many observed properties have been put forward.
An aspect, on the contrary, that has been relatively disregarded is the validation of 
phenomenological data, and the identification of possible errors or biases occurred during the process 
of data collection and analysis.
The importance of this issue resides in the fact that ``systematic'' errors in the statistics, due 
to sampling biases, could compromise the reliability of the data and of the observed properties of 
real networks. \\
The idea that the sampling of networks may introduce biases is far from being unrealistic, 
as proved by several examples reported in the Section~\ref{CHAP3_1_2} and coming from different fields of complex networks research.\\
In social and biological networks, the limited information about the exact mechanisms generating the original network and the amount of arbitrariness in the definition of the edges (e.g. relations among individuals, interactions between proteins, etc), makes sampling processes extremely problematic, since we do not have a real control on 
the origins of possible biases. \\
Completely different is the case of the physical Internet, in which nodes and edges are well-defined,   
but the dynamical nature of its structure and the lack 
of any centralized control have favored a self-organized evolution of the system, without any information on 
its topological properties.
In practice, we do not have a complete knowledge of router's neighbors, since routing decisions depend 
on optimized traffic protocols by means of which {\em data packets should be sent along the shortest 
path available to the destination}. This means that routers only know which are the neighbors belonging 
to the shortest paths; i.e., they could ignore the existence of other neighbors. 
Actually, traffic congestions and local policies can force routers to deliver 
packets through some preferential neighbors, causing small path inflations with respect to the shortest one.\\
Internet's explorations, obtained by means of tree-like probes based on \texttt{traceroute} processes, exploit 
the routing protocols in order to trace a path between different nodes of the network. In this way, they suffer of important 
biases due to the loss of {\em lateral connectivity}, i.e. of those nodes or links which do not lay on the shortest path  between two nodes (or on its small perturbations).
We will see that \texttt{traceroute} explorations can seriously misrepresent the degree distribution 
of the original network.\\
On the other hand, a good knowledge of the Internet topology is fundamental in order to 
improve its performances, minimize traffic congestions and protect the system against malicious attacks.
For this reason, the study of Internet's sampling biases is of primary interest, not only for the scientific
community but also for practitioners and common users.\\
The investigation has to be carried on at different levels; our theoretical formulation of the problem is aimed at   
\begin{itemize}
\item understanding what is the origin of the exploration biases and at what stage they affect the observed properties,
\item identifying which kind of topologies yield the most accurate sampling,
\item providing some ``rules of thumb'' for the optimization of mapping strategies,
\item obtaining alternative approaches able to correct the biases, at least in some simple cases. 
\end{itemize}
In the next section, we introduce the issue of sampling biases in complex networks, 
highlighting which dramatic effects of distortion of the shape of the degree distribution 
can be produced by a bad sampling of the network.

\subsection{Networks sampling methods and their biases}\label{CHAP3_1_2}

There are many possible sources of sampling biases in complex networks, depending on the field of research and the type of sampling method used in the experiments. Here we provide a short survey of examples, focusing in particular on the mechanisms of homogeneous sampling and tree-like explorations and on the dramatic effects they can have in misrepresenting the degree distribution of the underlying network.

{\bf \em Sampling biological and social networks - \quad}
In the context of social and biological networks, practitioners have developed many types of experiments in order to 
gain information on the topology of the networks of interest. Apart from the complexity of the experimental set up necessary for such experiments, a deep conceptual problem emerges: ties between nodes are usually associated with relations or interactions, thus they can have different nature or intensity, that are difficult to evaluate and take into account correctly. 
In social networks, this is due to the level of arbitrariness in defining relations between actors, and in biological networks to the fact that measures are usually indirect and may be influenced by the effects of external unknown variables. 
\begin{figure}[t]
\begin{center}
\includegraphics[width=10.0cm]{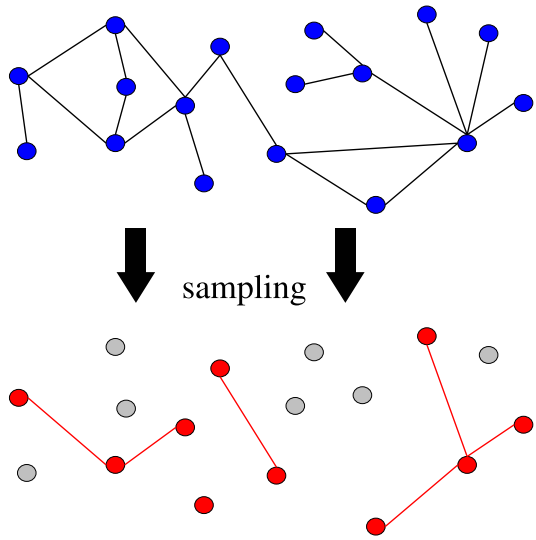}
\end{center}
\caption{Illustration of a node-picking sampling algorithm used to model sampling of biological and social networks.
The original network (top) is sampled picking up nodes at random and retaining common links (red nodes in the bottom figure).}
\label{node_picking}
\end{figure}
Nevertheless, many authors have modeled the collection of data in social as well as in
biological experiments by means of {\em node- or edge-picking sampling algorithms}.
The two algorithms give similar results, thus let us focus on the node-picking case.
Fig.~\ref{node_picking} illustrates a node-picking sampling process. 
In absence of further information on the way a node is selected, we can assume a homogeneous sampling,
i.e. a node is included in the sampled subnet with fixed probability $p$, 
and left out with probability $1-p$; only edges between sampled nodes are retained. 
Then, if $P(k)$ and $\tilde{P}(k)$ are respectively the original and sampled degree distributions,  
they are related by a ``poissonian filter'',
\begin{equation}\label{stumpf1}
\tilde{P}(k) = \sum_{i\geq k}^{\infty} P(i) \left(\begin{array}{c} i\\ k \end{array}\right) p^{k} {(1-p)}^{i-k}~.
\end{equation}
Using generating functions formalism \cite{stumpf,sampling}, it is easy to show that, when $p\simeq 1$, 
the deviation from the original distribution is negligible both for homogeneous and heterogeneous degree distributions. When the sampling probability is low ($p \ll 1$), homogeneous distributions are conserved, even if the average 
connectivity is reduced by a factor $p$, whereas in the case of power-law distributions the observed exponent may be different from the real one. Subnets have more nodes with relatively few connections, due to the sampling process, but very large degree nodes are usually well-represented, so that the original power-law behavior is recovered for $k \gg 1$. Consequently, the degree distribution appears slightly concave in the middle (in log-log scale), introducing biases in the measurement of the exponent (it is systematically reduced). Actually, a good strategy is that of looking at the tail of the distribution in which the sampling is systematically more accurate.   \\
Unfortunately, {\em sampling by means of real experiments is far from being uniform}, since external factors can be correlated to the properties of some nodes, favoring their sampling. 
For instance, social networks are usually bipartite, i.e. actors (individuals) are linked together 
via multiple interaction contexts or affiliations. Following Refs.~\cite{kossinets}, together with the random exclusion of actors or affiliations, there are other two principal mechanisms causing data missing: actors unpredictable decision of non-responding to a particular survey, and of providing fixed or preferential choices in the answer. A whole field of social network analysis is involved in studying how to predict the correlations among such data missing events (see Refs.~\cite{granovetter0,rothenberg} and in particular Ref.~\cite{wasserman} and references therein). 
We prefer to consider a simpler example coming from biology and reported in Ref.~\cite{delosrios1}, that shows how hidden variables influencing the sampling process can have dramatic effects on the results.\\
Let us consider a protein-protein interaction network (PIN), in which the nodes are proteins and the edges are the interactions between them. The standard methods to detect interactions are two-hybrid assays and mass spectrometry \cite{delosrios1}. 
Both of them are sensitive to the physical conditions in which the experiment is performed (e.g. the temperature, the solubility degree of proteins, etc).
The detectability of an interaction can be affected by these external variables, whose effect on the process is not completely known. Similarly, in neural networks, and cell regulatory networks, some nodes or edges may be ignored in the experiments only because their functions are temporarily inhibited by the activation of other functions. \\
In order to model the sampling, let us consider a network with a homogeneous degree distribution, for instance an Erd\"os-R\'enyi random graph, and for each node $i$ assign a variable $x_{i}$, taken from a probability distribution $p(x)$. Such a hidden variable is also known in literature as ``fitness'' \cite{fitness_calda,goh01,PA_fitness1,PA_fitness2}. 
Now we prune the graph leaving the edge $(i,j)$ with a probability $q(x_{i},x_{j})$.
In a biological framework, $x_{i}$ is for instance the free energy gain of a protein from being solved during the experiment. The interaction takes place only if the free energy loss $x_{c}$ in breaking a bond is compensated by the gain $x_{i}+x_{j}$ of staying together in the solution. 
On the other hand, it is reasonable to assume an exponential distribution for the free energy, i.e. $p(x) \propto \exp(-x)$.
The average degree $k(x)$ of a sampled node as a function of its free energy is readily computed as
\begin{equation}\label{rios1}
k(x) \sim pN \int_{S} q(x,x') p(x') dx'~,  
\end{equation}
where $pN$ comes from the approximation that all nodes have about $pN$ neighbors, and 
$q(x,x') = \theta(x+x'-x_{c})$. 
The integral in Eq.~\ref{rios1} yields $k(x) \propto Np \exp(x)$, that inserted into the probability relation 
$P(k) dk = p(x) dx$, provides a power-law expression for the degree distribution of the sampled network, 
$P(k) \sim k^{-2}$.
This striking result shows that, as a consequence of sampling biases, one may observe heterogeneous degree distributions even when the underlying network is homogeneous. 
In their work \cite{delosrios1}, Petermann and De Los Rios show as well that even when the original network has power-law distributed connectivity, the exponent can be considerably underestimated.  

{\bf \em Sampling technological networks - \quad}
As already stressed in this section, the case of technological networks is completely different, 
since we do not have any idea of the topology of the real graph, but the sampling methods are based on 
very well-defined probing processes, that can be modeled using {\em tree-like explorations}.\\
The first example of this class of networks is the World Wide Web, that is usually efficiently explored using so-called ``crawling processes'' \cite{crawling}.
The WWW, indeed, possesses the remarkable property that the links outgoing from a page are directly visible, thus we can apply snowball sampling methods, that are related to well-known processes like epidemic spreading and percolation (see Chapter~\ref{CHAP4}).  
A single node is firstly chosen together with its outcoming links and the nodes 
connected to them. Then, new nodes connected to those picked in the last step are selected. The process
continues recursively until the desired number of nodes are gathered. The main limitation of using this method on the Web is the huge size of the network itself, that makes difficult to reach all remote regions. 
In general, in each layer only a fraction of the nodes is sampled and this may introduce some inaccuracies \cite{snowball1}.\\
\begin{figure}[t]
\begin{center}
\includegraphics[width=10.0cm]{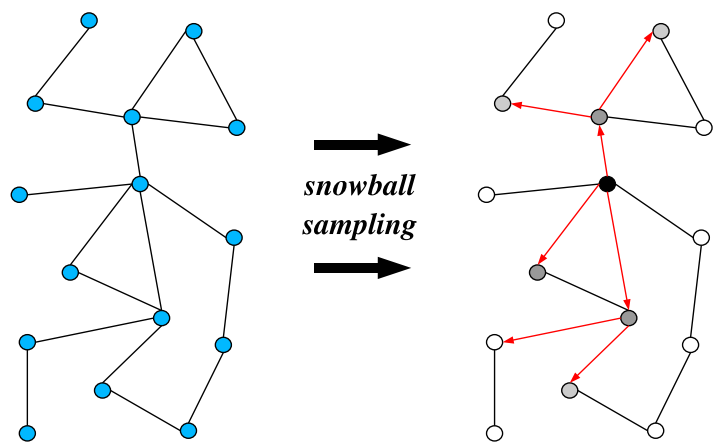}
\end{center}
\caption{
Illustration of the snowball sampling technique. The original graph (left) is sampled using a snowball algorithm starting from a root vertex (black node on the right part of the figure).  The first layer is composed of a fraction its neighboring nodes (dark grey nodes), that are used as starting nodes to explore the second layer (bright grey nodes), etc. 
}
\label{snowball}
\end{figure}
Snowball-like samplings do not work on the Internet, since routing protocols redirect probes along preferential (shortest) paths, thus preventing exploration algorithms from getting a complete knowledge of nodes neighborhood.
The common sampling strategy consists in acquiring local views of the network from several 
vantage points, merging these views in order to get a presumably accurate global map.
Local views are obtained by evaluating a certain number of paths to different destinations 
by using specific tools such as \texttt{traceroute}-like commands or by the analysis of BGP tables. 
{\em At first approximation these processes amount to the collection of
shortest paths from a source node to a set of target nodes}, obtaining
a partial spanning tree of the network. The merging of several of
these views provides the map of the Internet from which the statistical
properties of the network are evaluated. 
{\em According to this description, discovering the Internet topology is more 
than a simple network sampling problem, it consists in a real dynamical exploration process.}\\
The first contribution to the problem of sampling biases in the Internet was given by Lakhina et al. 
\cite{crovella}, who showed that \texttt{traceroute}-like explorations can seriously affect the
estimation of degree distributions. 
In particular, when the number of sources and destinations is small, 
one can observe power-law like distributions even in the sampling of  Erd\"os-R\'enyi random
graphs, whose original degree distribution is poissonian. 
Since the first data showing heavy-tailed distributions for the Internet topology have been collected 
gathering \texttt{traceroute} paths from a very limited number of sources and destinations \cite{pansiot}, 
they concluded that the Internet maps could be wrong, or at least the node degree distribution is not a sufficiently robust metric to characterize Internet's topology.
Nevertheless, the same exploration performed on a network with a power-law degree distribution behaves very differently, since a handful number of sources are sufficient to yield a sampled graph with degree distribution that looks very similar to the original one. 
The authors of Ref.~\cite{latapy} have moreover monitored numerically the observed degree of sampled nodes as a function of their real degree, finding that it is systematically underestimated.\\
The analytical foundation to the numerical work of Lakhina et al.
\cite{crovella} was provided in Refs.~\cite{clauset,achlioptas}, in which the authors
modeled \texttt{traceroute} explorations as single-source, all-destinations, shortest-path trees.
Using breadth-first search spanning trees, they rigorously proved that, for an Erd\"os-R\'enyi random graph
with average degree $\langle k\rangle$, the connectivity
distribution of the obtained spanning tree displays a power-law behavior
$k^{-1}$, with an exponential cut-off setting in at a characteristic degree
$k_c\sim \langle k \rangle$.
We give here a non-rigorous derivation of this result using differential equations.\\
A typical breadth-first search algorithm is the following. There are three types of nodes: {\em explored}, {\em untouched}, 
and {\em pending}. All edges are labeled {\em invisible}. The final observed network will be composed only of explored vertices and visible edges. 
The process starts with the root vertex labeled pending into a queue, all the others are untouched.
Vertices are chosen from the queue in the first-in-first-out order, thus at the beginning the root is popped out.
All the untouched neighbors of the vertex chosen from the queue (i.e. explored) are appended to the queue as pending vertices. The edges going from the explored vertex to these appended neighbors are made visible.
Let now be $u(t)$ and $s(t)$ the densities of untouched and pending vertices respectively, the process can be described by the following system of differential equations \cite{clauset},
\begin{equation}
\frac{d u(t)}{dt} = -\langle k \rangle u(t) \quad \quad \quad \frac{d s(t)}{dt} = \langle k \rangle u(t) -1~.
\end{equation}
Using the initial conditions $u(0)=1$ and $s(0)=0$, we get a solution of the form $u(t)=e^{-\langle k\rangle t}$ and $s(t) = 1-t-e^{-\langle k\rangle t}$.
If a node is chosen at time $t$, its observed degree is the number of previously untouched neighbors plus one given by the edge we used to reach it. Since a node can be discovered at any time from $t=0$ to $t=t_{0}$ (the smallest root of $s(t)=0$), we get the degree distribution by means of the following temporal average, i.e.
\begin{equation}
\tilde{P}(k+1) \sim \frac{1}{t_{0}}\int_{0}^{t_{0}} e^{-\langle k\rangle u(t)} \frac{{[\langle k\rangle u(t)]}^{k}}{k!} dt~, 
\end{equation}
where we used the fact that the real distribution is poissonian and the untouched nodes at time $t$ are 
homogeneously sampled from it with density $u(t)$. 
After computing the integral and making some approximations it is easy to show that the observed degree distribution
$\tilde{P}(k) \propto 1/k$ up to a degree $k \sim \langle k\rangle$ where a cut-off sets in.\\
From such analysis one could conclude that the observation of heavy-tails in Internet's degree distribution is a 
fake effect due to the use of tree-like explorations; nonetheless, this result is strictly correct only in the case of single-source probing, as will be clearer in the following.     \\
All these results stress on {\em the relevance of determining up to which extent 
the topological properties observed in sampled graphs are representative of that of the real networks}.
We have tried to answer this issue in the case of \texttt{traceroute}-like explorations using methods of statistical physics. The main results of this study, that led to the publications in 
Refs.~\cite{dallasta_traceREV,dallasta_traceTCS,dallasta_traceLNCS,dallasta_traceOTHER2}, are illustrated 
in the next sections. 

\newpage
\section{Statistical physics approach to traceroute explorations}\label{CHAP3_2}

This section gives a formal statistical description of 
 \texttt{traceroute}-like processes in terms of a simple model that provides
 a qualitative and quantitative understanding of the properties observed in real experiments.

\subsection{The model}\label{CHAP3_2_1}
\begin{figure}[t]
\begin{center}
\includegraphics[width=10.0cm]{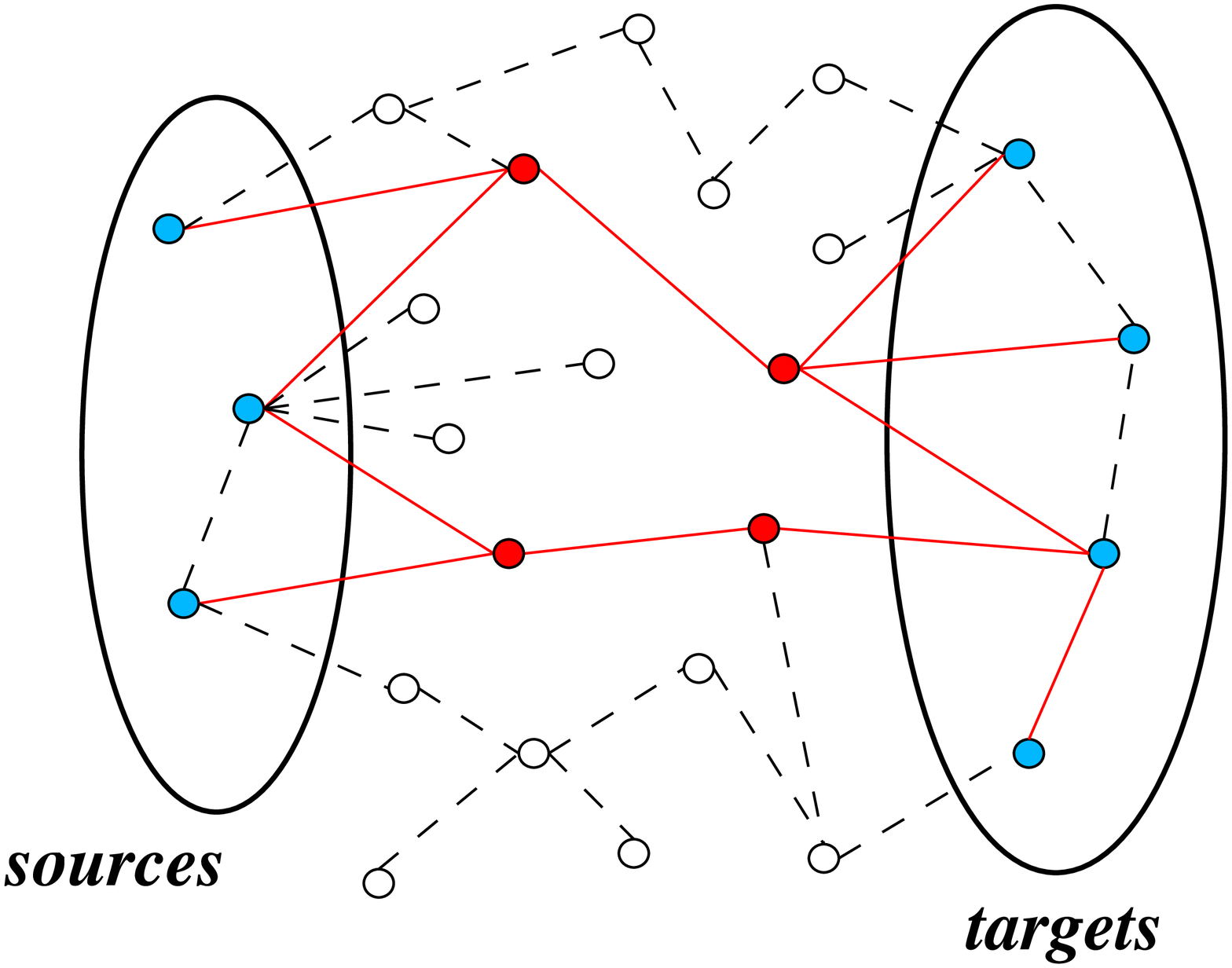}
\end{center}
\caption{Illustration of the \texttt{traceroute}-like procedure. 
Shortest paths between the set of sources and the set of destination 
targets are discovered (red full lines) while other edges are not found
(dashed black lines). Note that not all shortest paths are found since the
``Unique Shortest Path'' procedure is used. }
\label{traceroute}
\end{figure}

In a typical exploration, a set of active sources deployed in
the network sends \texttt{traceroute} probes to a set of destination nodes.
Each probe collects information on all the nodes and edges traversed along the
path connecting the source to the destination~\cite{burch99}.  
By merging the information collected on each path it
is then possible to reconstruct a partial map of the network
(Fig.~\ref{traceroute}).  More precisely, the set of edges and nodes discovered by each
probe depend on the ``path selection criterion'' (p.s.c) used to choose the path
between a pair of nodes. In the real Internet, many factors, including
commercial agreement, traffic congestion and administrative routing policies,
contribute to determine the actual path, causing it to differ even
considerably from the shortest path.  Despite these local, often unpredictable
path distortions or inflations, a reasonable first approximation of the route
traversed by \texttt{traceroute}-like probes is the shortest path between the
two nodes.  This assumption, however, is not sufficient for a proper
definition of a \texttt{traceroute} model in that equivalent shortest paths
between two nodes may exist.  In the presence of a degeneracy of shortest
paths we must therefore specify the path selection criterion by providing a
resolution algorithm for the selection of shortest paths.\\
For the sake of simplicity we can define three selection
mechanisms among equivalent paths  
that may account for some of the features encountered in the Internet
discovery:
\begin{itemize}
\item Unique Shortest Path (USP) probe. In this case the shortest
path route selected between a node $i$ and the destination target $T$ 
is always the same independently of the source $S$ (the path being
initially chosen at random among all the equivalent ones). 

\item Random Shortest Path (RSP) probe. The shortest path between
any source-destination pair is chosen randomly among the set of
equivalent shortest paths. This might mimic different peering
agreements that make independent the paths among couples of nodes.

\item All Shortest Paths (ASP) probe. The selection criterion discovers all
  the equivalent shortest paths between source-destination pairs. This might
  happen in the case of probing repeated in time (long time exploration), so
  that back-up paths and equivalent paths are discovered in different runs.
\end{itemize}
We will generically call $\mathcal{M}$-path the path found using one of these
``metrics'' or path selection criteria.  Actual \texttt{traceroute} probes
contain a mixture of the three mechanisms defined above, even if 
 many effective heuristic strategies are commonly
applied to improve the reliability and the performances of the sampling.  
An example of such heuristic tricks is the interface resolution algorithm called 
{\em iffinder}, proposed by Broido and Claffy \cite{broido}. 
In fact, a router can have more than one interface with the external world, thus different 
paths passing through different interfaces might erroneously consider two interfaces 
as two independent routers. Algorithms such as {\em iffinder} allow to avoid these type of errors.\\
As remarked by Guillaume and Latapy \cite{latapy}, the different path selection criteria may have influence 
on the general picture emerging from the theoretical model, but the   
 USP procedure clearly represents the worst scenario among the three different
methods, yielding the minimum number of discoveries. 
For this reason, we will focus only on the USP data. 
The interest of this analysis resides properly in the choice of working in the most pessimistic 
case, being aware that path inflations should actually provide a more pervasive sampling of the
real network.\\
Formally, the \texttt{traceroute} model is the following. 
Let $G=(\mathcal{V},\mathcal{E})$ be a sparse undirected graph with
vertices (nodes) $\mathcal{V}=\{1,2,\cdots,N\}$ and edges (links) $\mathcal{E}$. Then let us
define the sets of vertices $\mathcal{ S}=\{i_1,i_2,\cdots,i_{N_S}\}$ and
$\mathcal{ T}=\{j_1,j_2,\cdots,j_{N_T}\}$ specifying the random placement of
$N_S$ sources and $N_T$ destination targets.  For each ensemble of
source-target pairs $\Omega=\{ \mathcal{S}, \mathcal{T} \}$, we compute with
our p.s.c. the paths connecting each source-target pair. The sampled graph
$\mathcal{G}=(\mathcal{V}^*,\mathcal{E}^*)$ is defined as the set of vertices $\mathcal{V}^*$ (with
$N^*=|\mathcal{V}^*|$) and edges $\mathcal{E}^*$ ($E^*=|\mathcal{E}^*|$) induced by considering the union of all the
$\mathcal{M}$-paths connecting the source-target pairs.  The sampled graph should
thus be analogous to the maps obtained from real \texttt{traceroute} sampling of
the Internet.\\
In the next section, we provide a mean-field analysis of the discovery process
as function of the density $\rho_T=N_T/N$ and $\rho_S=N_S/N$ of targets and sources.
In general, \texttt{traceroute}-driven studies run from a relatively
small number of sources to a much larger set of destinations.
For this reason, in many cases it is appropriate to work with 
the density of targets $\rho_T$ while still considering 
$N_S$ instead of the corresponding density. This combination
of the parameters allows us to compare mapping 
processes on networks with different sizes.
An appropriate quantity representing the level of sampling of the networks is  
the {\em probing effort} $\epsilon= \frac{N_S N_T} {N}$,
that measures the density of probes imposed to the system. In real situations
it represents the density of \texttt{traceroute} probes in the network and
therefore a measure of the load provided to the network by the measuring
infrastructure.

\subsection{Mean-field analysis}\label{CHAP3_2_2}

Here, we provide a statistical estimate for the probability of edge and node
detection as a function of $N_S$, $N_T$ and the topology of the
underlying graph. The method is based on a simple mean-field statistical 
analysis of the simulated \texttt{traceroute} mapping.\\
For each set $\Omega=\{ \mathcal{S}, \mathcal{T} \}$ we define the
quantities 
\begin{equation}
\sum_{t=1}^{N_T} \delta_{i,j_t}= \left\{
\begin{array}{ll}
1 & \mbox{ if vertex $i$ is a target;}\\
0 & \mbox{ otherwise,}
\end{array}
\right.
\label{pik1}
\end{equation}
\begin{equation}
\sum_{s=1}^{N_S} \delta_{i,i_s}= \left\{
\begin{array}{ll}
1 & \mbox{ if vertex $i$ is a source;}\\
0 & \mbox{ otherwise,}
\end{array}
\right.
\label{pik2}
\end{equation}
where $\delta_{i,j}$ is the Kronecker symbol.
These quantities tell us if any given node $i$ belongs to the
set of sources or targets, and obey the sum rules 
$\sum_i\sum_{t=1}^{N_T} \delta_{i,j_t}=N_T$ and 
$\sum_i \sum_{s=1}^{N_S} \delta_{i,i_s}=N_S$. 
Analogously, we define the quantity $\sigma_{i,j}^{(l,m)}$ that takes 
the value $1$ if the edge $(i,j)$ belongs to the selected path  
between nodes $l$ and $m$, and $0$ otherwise. \\
For a given set of sources and targets $\Omega$, the indicator function that
a given edge $(i,j)$ will be discovered and belongs to the sampled graph is simply 
$\pi_{i,j}=1$ if the edge $(i,j)$ belongs to at least one of the $\mathcal{M}$-paths connecting 
the source-target pairs, and $0$ otherwise.
We can obtain an exact expression for $\pi_{i,j}$ by noting that $1-\pi_{i,j}$ is $1$ if and only if $(i,j)$
does not belong to any of the paths between sources and targets, i.e. if and only if  $\sigma_{i,j}^{(l,m)}=0$
for all $(l,m) \in \Omega$. 
This leads to  
\begin{equation}
\pi_{i,j}=1-
\prod_{l\neq m}\left(
1 - 
\sum_{s=1}^{N_S} \delta_{l,i_s} 
\sum_{t=1}^{N_T} \delta_{m,j_t} 
\sigma_{i,j}^{(l,m)}
\right).
\end{equation}
For a given set $\Omega=\{ \mathcal{S}, \mathcal{T} \}$, this 
function is simply $\pi_{i,j}=1$ if the edge $(i,j)$ 
belongs to at least one of the $\mathcal{M}$-paths connecting the 
source-target pairs, and $0$ otherwise.
Since we are looking at a purely statistical level, in order to get more useful expressions, 
we perform the average over all possible realizations of the set $\Omega=\{ \mathcal{S}, \mathcal{T} \}$.
By definition we have that 
\begin{equation}
\left\langle \sum_{t=1}^{N_T} \delta_{i,j_t} \right\rangle_{\Omega}
= \rho_T~~~\mbox{and}~~
\left\langle \sum_{s=1}^{N_S} \delta_{i,i_s} \right\rangle_{\Omega}
= \rho_S,
\label{aver}
\end{equation}
where $\left\langle\cdots\right\rangle_{\Omega}$ identifies the average over all
possible deployment of sources and targets $\Omega$.  These equalities simply
state that each node $i$ has, on average, a probability to be a source or a
target that is proportional to their respective densities.  
In real processes there are correlations among the paths, due to the relative position of sources and targets, 
thus in general the average is a complicate quantity, that we cannot easily compute analytically.   
A first approximation can be obtained making an uncorrelation assumption 
that yields an explicit expression for the discovery probability. {\em The assumption consists in computing the discovery probability neglecting the correlations among different paths originated by the position of sources and targets}. 
While this assumption does not
provide an exact treatment for the problem it generally conveys a qualitative
understanding of the statistical properties of the system.  In this
approximation, the average discovery probability of an edge is
\begin{eqnarray}
\nonumber&&\left\langle\pi_{i,j}\right\rangle_{\Omega}=1-
\left\langle\prod_{l\neq m}\left(
1 - 
\sum_{s=1}^{N_S} \delta_{l,i_s} 
\sum_{t=1}^{N_T} \delta_{m,j_t} 
\sigma_{i,j}^{(l,m)}
\right)\right\rangle_{\Omega}\\
&&~~~~~~~~~~\simeq 1-\prod_{l\neq m}\left(1 -
\rho_T\rho_S \left\langle\sigma_{i,j}^{(l,m)}\right\rangle_{\Omega}\right),
\label{meanedge}
\end{eqnarray}
where in the last term we take advantage of neglecting correlations by
replacing the average of the product of variables with the product of the
averages and using Eq.~\ref{aver}. 
This expression simply states that each
possible source-target pair weights in the average with the product of the
probability that the end nodes are a source and a target. The discovery
probability is thus obtained by considering the edge in an average effective
media ({\em mean-field}) of sources and targets homogeneously distributed in
the network. The realization average of
$\left\langle\sigma_{i,j}^{(l,m)}\right\rangle_{\Omega}$ is very simple in the
uncorrelated picture, depending only of the kind of the probing model. In the
case of the ASP probing, $\left\langle\sigma_{i,j}^{(l,m)} \right\rangle_{\Omega}$ is
just $1$ if $(i,j)$ belongs to one of the shortest paths between $l$ and $m$,
and $0$ otherwise.  In the case of the USP and the RSP, on the contrary, only
one path among all the equivalent ones is selected. If we denote by
$\sigma^{(l,m)}$ the number of shortest paths between vertices $l$ and $m$,
and by $x_{i,j}^{(l,m)}$ the number of these paths passing through the edge
$(i,j)$, the probability that the \texttt{traceroute} model chooses a path
going through the edge $(i,j)$ between $l$ and $m$ is
$\left\langle\sigma_{i,j}^{(l,m)}\right\rangle_{\Omega}$ = $x^{(l,m)}_{i,j} /
\sigma^{(l,m)}$.\\
The standard situation we consider is the one in which 
$\rho_T\rho_S\ll 1$ and since 
$\left\langle\sigma_{i,j}^{(l,m)}\right\rangle_{\Omega}\leq 1 $, we have  
\begin{equation}
\prod_{l\neq m}
\left(1 -\rho_T\rho_S \left\langle\sigma_{i,j}^{(l,m)}
\right\rangle_{\Omega}\right)\simeq
\prod_{l\neq m}\exp \left(-\rho_T\rho_S
\left\langle\sigma_{i,j}^{(l,m)}\right\rangle_{\Omega}\right),
\end{equation}
that inserted in Eq.~\ref{meanedge} yields 
\begin{equation}
\left\langle\pi_{i,j}\right\rangle_{\Omega} \simeq 1-
\prod_{l\neq m} \left(\exp \left(- \rho_T\rho_S
\left\langle\sigma_{i,j}^{(l,m)}\right\rangle_{\Omega}\right)\right)
= 1-\exp
\left(-\rho_T\rho_S b_{ij}
\right),
\label{edgedisc}
\end{equation}
where $b_{ij}=\sum_{l\neq m}\left\langle\sigma_{i,j}^{(l,m)}\right\rangle_{\Omega}$. In
the case of the USP and RSP probing, the quantity $b_{ij}$ is by definition
the edge betweenness centrality $\sum_{l \neq m} x^{(l,m)}_{i,j} /
\sigma^{(l,m)}$ \cite{freeman,brandes}. For the ASP probing, 
it is a closely related quantity.  Indeed, if the
shortest path is used as the metric defining the optimal path between pairs of
vertices, the betweenness gives a measure of the
amount of all-to-all traffic that goes through an edge or a vertex. 
We also recall, that the betweeness can be considered as a non-local measure of the
\textit{centrality} of an edge or vertex in the graph (see Section~\ref{CHAP2_2_4}).\\
Since the edge betweenness assumes values between $2$ and $N(N-1)$, the
discovery probability of an edge will therefore depend strongly on
its betweenness. For instance, for edges with minimum 
betweenness \mbox{$b_{ij}=2$}, we have 
$\left\langle\pi_{i,j}\right\rangle_{\Omega} \simeq 2\rho_T\rho_S$,
that recovers the probability that the two end vertices of the edge
are chosen as source and target. This implies that if the densities 
of sources and targets are small but finite in the limit of very large
$N$, all the edges of the underlying graph have a finite
probability to be discovered. On the other hand, {\em the discovery 
probability approaches one for edges with high betweenness, thus predicting a fair
sampling of the network.}\\
In most of the current realistic samplings, the situation is different.
While it is reasonable to consider $\rho_T$ a small but finite value, the
number of sources is not extensive ($N_S\sim \mathcal{O}(1)$) and their
density tends to zero as $N^{-1}$. In this case it is more convenient to
express the edge discovery probability as
\begin{equation}
\left\langle\pi_{i,j}\right\rangle_{\Omega} \simeq 
 1-\exp\left(-\epsilon \hat{b}_{ij}
\right),
\label{edgesample}
\end{equation}
where $\epsilon=\rho_T N_S$ is the density of probes imposed to the system and
the rescaled betweenness $\hat{b}_{ij}=N^{-1}b_{ij}$ is now limited in
the interval $[2N^{-1},N-1]$. In the limit of large networks ($N\to\infty$), it
is clear that edges with low betweenness have
$\left\langle\pi_{i,j}\right\rangle_{\Omega} \sim \mathcal{O}(N^{-1})$, for any finite
value of $\epsilon$. This readily implies that in real situations the
discovery process is generally not complete, a large part of low betweenness
edges not being discovered, and that the network sampling is made
progressively more accurate by increasing the density of probes $\epsilon$.\\
A similar analysis can be performed for the discovery probability
$\pi_{i}$ of a vertex $i$. For each source-target set $\Omega$ we have that 
\begin{equation}
\pi_{i}=1-\left(
1-\sum_{s=1}^{N_S} \delta_{i,i_s}- \sum_{t=1}^{N_T} \delta_{i,j_t}
\right)
\prod_{l\neq m\neq i}\left(
1 - \sum_{s=1}^{N_S} \delta_{l,i_s} 
\sum_{t=1}^{N_T} \delta_{m,j_t} \sigma_{i}^{(l,m)}
\right).
\end{equation}
where $\sigma_{i}^{(l,m)}=1$ if the vertex $i$ belongs to the
$\mathcal{M}$-path between nodes $l$ and $m$, and $0$ otherwise. Note that it
has been considered that a vertex belonging to the set of sources and targets is 
discovered with probability one.  The second term on the right hand
side therefore expresses the fact that the vertex $i$ does not belong to the
set of sources and targets and it is not discovered by any $\mathcal{M}$-path
between source-target pairs.  By using the same {\em mean-field} approximation
as previously, the average vertex discovery probability reads as
\begin{equation}
\left\langle\pi_{i}\right\rangle_{\Omega} \simeq 1 - (1-\rho_S-\rho_T)
\prod_{l\neq m\neq i}\left(1 -\rho_T\rho_S
\left\langle\sigma_{i}^{(l,m)}\right\rangle_{\Omega}\right). 
\end{equation}
As for the case of the edge discovery probability, the average considers all
possible source-target pairs weighted with probability $\rho_T\rho_S$. In the
ASP model, the average $\left\langle\sigma_{i}^{(l,m)}\right\rangle_{\Omega}$ is $1$ if
$i$ belongs to one of the shortest paths between $l$ and $m$, and $0$
otherwise.  For the USP and RSP models,
$\left\langle \sigma_{i}^{(l,m)}\right\rangle_{\Omega}$ = $x^{(l,m)}_{i} /
\sigma^{(l,m)}$ where $x^{(l,m)}_{i}$ is the number of shortest paths between
$l$ and $m$ going through $i$.  If $\rho_T\rho_S\ll 1$, by using the same
approximations used for Eq.~\ref{edgedisc} we obtain
\begin{equation}
\left\langle\pi_{i}\right\rangle_{\Omega}\simeq 1 - (1-\rho_S-\rho_T)
\exp \left(-\rho_T\rho_S b_{i}
\right),
\end{equation}
where $b_{i}=\sum_{l\neq m\neq
i}\left\langle\sigma_{i}^{(l,m)}\right\rangle_{\Omega}$. For the USP and RSP cases, 
$b_{i} = \sum_{l \neq m\neq i} x^{(l,m)}_{i} / \sigma^{(l,m)}$ is the vertex 
betweenness centrality, that is limited in the interval $[0,N(N-1)]$
\cite{freeman,brandes,goh01}.
The betweenness value $b_i=0$ holds for the leaves of the graph,
i.e. vertices with a single edge, for
which we recover
$\left\langle\pi_{i}\right\rangle_{\Omega}\simeq\rho_S+\rho_T$.
Indeed, this kind of vertices are dangling ends, that can be discovered only if
they are either sources or targets.\\
As discussed before, the most usual setup corresponds to a density 
$\rho_S\sim\mathcal{O}(N^{-1})$ and in the large $N$ limit we can
conveniently write 
\begin{equation}
\left\langle\pi_{i}\right\rangle_{\Omega}\simeq 1 - (1-\rho_T)
\exp \left(-\epsilon\hat{b}_{i}
\right),
\label{vertexsample}
\end{equation}
where we have neglected terms of order $\mathcal{O}(N^{-1})$ and the rescaled
betweenness $\hat{b}_{i}=N^{-1}b_{i}$ is now defined in the interval
$[0,N-1]$.  This expression points out that the probability of vertex
discovery is favored by the deployment of a finite density of targets that
defines its lower bound.\\
We can also provide a simple approximation for the effective
average degree $\left\langle k_i^*\right\rangle_{\Omega}$ of the node $i$ 
discovered by our sampling process. Each edge departing from the vertex 
will contribute proportionally to its discovery probability, yielding   
\begin{equation}
\left\langle k_i^*\right\rangle_{\Omega}= 
\sum_j \left( 1 - \exp \left(-\epsilon\hat{b}_{ij}\right) \right)
\simeq \epsilon\sum_j\hat{b}_{ij}. 
\end{equation}
The final expression is obtained for  edges with 
$\epsilon\hat{b}_{ij}\ll 1$. Since
the sum over all neighbors of the edge betweenness is
simply related to the vertex betweenness as 
$\sum_j{b_{ij}}= 2(b_{i}+ N-1)$, where the factor $2$ considers that
each vertex path traverses two edges and the term $N-1$ accounts for
all the edge paths for which the vertex is an endpoint, this finally
yields 
\begin{equation}
\left\langle k_i^*\right\rangle_{\Omega}\simeq
2\epsilon + 2\epsilon\hat{b}_{i}.
\label{degdisc}
\end{equation}
The present analysis shows that {\em the measured quantities and statistical
properties of the sampled graph strongly depend on the parameters of the
experimental setup and the topology of the underlying graph}. The latter
dependence is exploited by the key role played by edge and vertex betweenness
in the expressions characterizing the graph discovery. The betweenness is a
nonlocal topological quantity whose properties change considerably depending
on the kind of graph considered. This allows an intuitive understanding of the
fact that graphs with diverse topological properties deliver different answer
to sampling experiments.

\subsection{Numerical simulations on computer generated networks}\label{CHAP3_2_3}

The previous theoretical results have provided some interesting insights on the 
topological properties that are responsible of the efficiency and the accuracy of the 
sampling. In this section, we present the results of extensive numerical simulations 
in which the sampling algorithm has been reproduced on computer generated graphs with 
different topological properties. In particular, {\em we consider the two separated classes of 
homogeneous and heterogeneous networks}. 
We use degree dependent quantities {\em to monitor the efficiency of the 
sampling process as a function of the probing effort}. The results are then exploited
to understand the properties of the degree distributions of the sampled networks.\\
Our data report the various measures for three different graphs: the Erd\"os-R\'enyi (ER)
random graph as representative of the homogeneous class, and two heterogeneous random graphs,   
one with power-law distribution of the form $P(k) \sim k^{-\gamma}$ (random scale free - RSF), and 
the other with Weibull distribution (WEI) $P(k)=(a/c) (k/c)^{a-1} \exp(-(k/c)^a)$. 
Both forms have been in fact proposed as representing the topological properties of the Internet \cite{broido}. 
In both cases, we have generated random graphs using the comfiguration model (see Section~\ref{CHAP2_4_2}).  
The parameter choice is $a=0.25$ and $c=0.6$ for the Weibull distribution, and
$\gamma=2.3$ for the RSF case. Two different average degree values $\langle k \rangle=20 ,100$ have been used for the ER model.
In all cases networks are of $N=10^4$ nodes. 
The main properties of the various graphs are summarized in Table~\ref{table_trenets}.\\
\begin{table}[t]
\begin{center}
\begin{tabular}{|c|c|c|c|c|c|c|}\hline
  & ER & ER& RSF & Weibull\\
\hline
 $N$ & $10^4$ & $10^4$  &  $10^4$  &  $10^4$\\
 $E$ &  $10^5$  &  $5. 10^5$ & $22000$ & $55000$ \\
 $\langle k\rangle$  & $20$ & $100$  &  $4.4$  & $11$  \\
 $k_{c}$  &   $40$  &  $140$   &  $3500$  &  $2000$\\
\hline
\end{tabular}
\end{center}
\caption{\small 
Main characteristics of the graphs used in the numerical exploration.
}
\label{table_trenets}
\end{table}

\begin{figure}[t]
\vskip .4in
\begin{center}
\includegraphics[width=10.0cm]{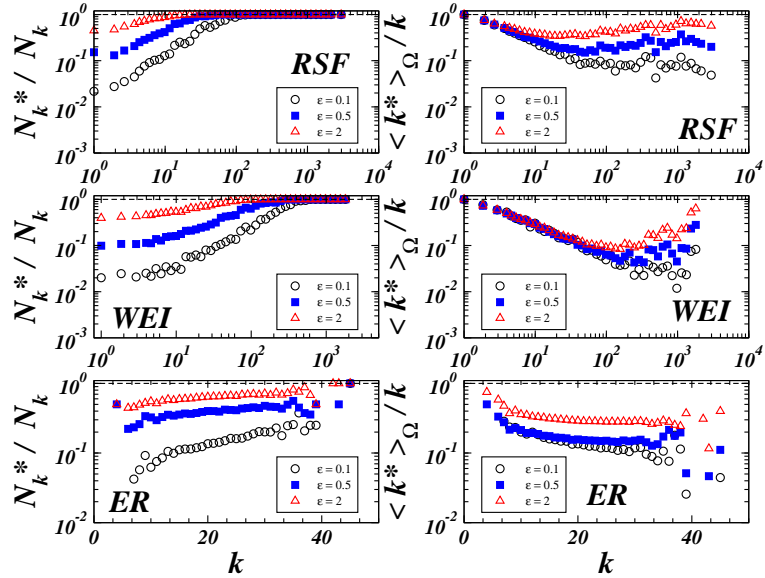}
\end{center}
\caption{Frequency $N_k^*/N_k$ of detecting a
vertex of degree $k$ (left) and proportion of discovered edges 
$\left\langle k^*\right\rangle_{\Omega}/k$ (right) as a function of the degree
in the RSF, WEI, and ER graph models. The exploration setup considers $N_S=5$
and increasing probing level $\epsilon$ obtained by progressively
higher density of targets $\rho_T$. The axis of ordinates is in log scale.}
\label{fig:3}
\end{figure}

{\bf \em Efficiency in the numerical sampling of graphs -\quad}\label{efficiency}
The first case we consider is that of homogeneous graphs (ER model).  As shown in Ref.~\cite{dallasta_traceLNCS},
vertex and edge betweenness are homogeneous quantities and their distributions are 
peaked around their average values $\langle b\rangle$ and $\langle b_{e}\rangle$, respectively, 
spanning only a small range of variation. These typical values can be inserted into 
Eqs.~\ref{edgesample} and \ref{vertexsample} to estimate the
order of magnitude of probes that allows a fair sampling of the graph. 
Both $\langle\pi_{i,j}\rangle_{\Omega}$ and $\langle\pi_{i}\rangle_{\Omega}$ tend to $1$ if
$\epsilon\gg$ max$\left[{\langle b\rangle}^{-1}, {\langle b_{e}\rangle}^{-1}\right]$. In
this limit all edges and vertices will have probability to be discovered very
close to one.
At lower values of $\epsilon$, obtained by varying $\rho_T$ and $N_S$, the
underlying graph is only partially discovered. Fig.~\ref{fig:3} shows 
the behavior
of the fraction $N_k^*/N_k$ of discovered vertices of degree $k$, where $N_k$
is the total number of vertices of degree $k$ in the underlying graph, and the
fraction of discovered edges $\left\langle k^*\right\rangle_{\Omega}/k$ in vertices of
degree $k$. $N_k^*/N_k$
naturally increases with the density of targets and sources, and it
is slightly increasing with $k$. The latter behavior can be easily
understood by noticing that vertices with larger degree have on average a
larger betweenness. On the other hand, the range of variation of
$k$ in homogeneous graphs is very narrow and only a large level of probing
may guarantee large discovery probabilities.  Similarly, the behavior of
the effective discovered degree can be understood by looking at
Eq.~\ref{degdisc}. 
Indeed the initial decrease of
$\left\langle k^*\right\rangle_{\Omega}/k$ is finally compensated by the increase of
$\langle b \rangle(k)$.
The situation is different in graphs with heavy-tailed connectivity
distributions (RSF and WEI models), for which the betweenness spans various orders
of magnitude and the fraction of vertices with very high betweenness is not negligible. 
In such a situation, even in the case of small $\epsilon$, vertices whose betweenness is
large enough ($b_i \epsilon\gg 1$) have $\left\langle
  \pi_i\right\rangle_{\Omega}\simeq 1$. Therefore all vertices with degree $k\gg
\epsilon^{-1/\beta}$ will be detected with probability one. This is clearly
visible in Fig.~\ref{fig:3} where the discovery probability $N_k^*/N_k$ of
vertices with degree $k$ saturates to one for large degree values.
Consistently, the degree value at which the curve saturates decreases with
increasing $\epsilon$.  A similar effect occurs in the measurements
concerning $\left\langle k^*\right\rangle_{\Omega}/k$. After an initial decay 
(Fig.~\ref{fig:3}) the effective discovered degree increases with the
degree of the vertices. This qualitative feature is captured by
Eq.~\ref{degdisc} that gives $\left\langle k^*\right\rangle_{\Omega}/k\simeq \epsilon
k^{-1}(1 + \langle b \rangle(k))$.  At large $k$ the term
$k^{-1}\langle b\rangle (k)\sim k^{\beta -1}$ takes over and the effective
discovered degree approaches the real degree $k$. Moreover, the broader the 
distribution of betweenness or connectivity,
the better the sampling obtained.

\begin{figure}[t]
\vskip .3in
\begin{center}
\includegraphics[width=7.5cm]{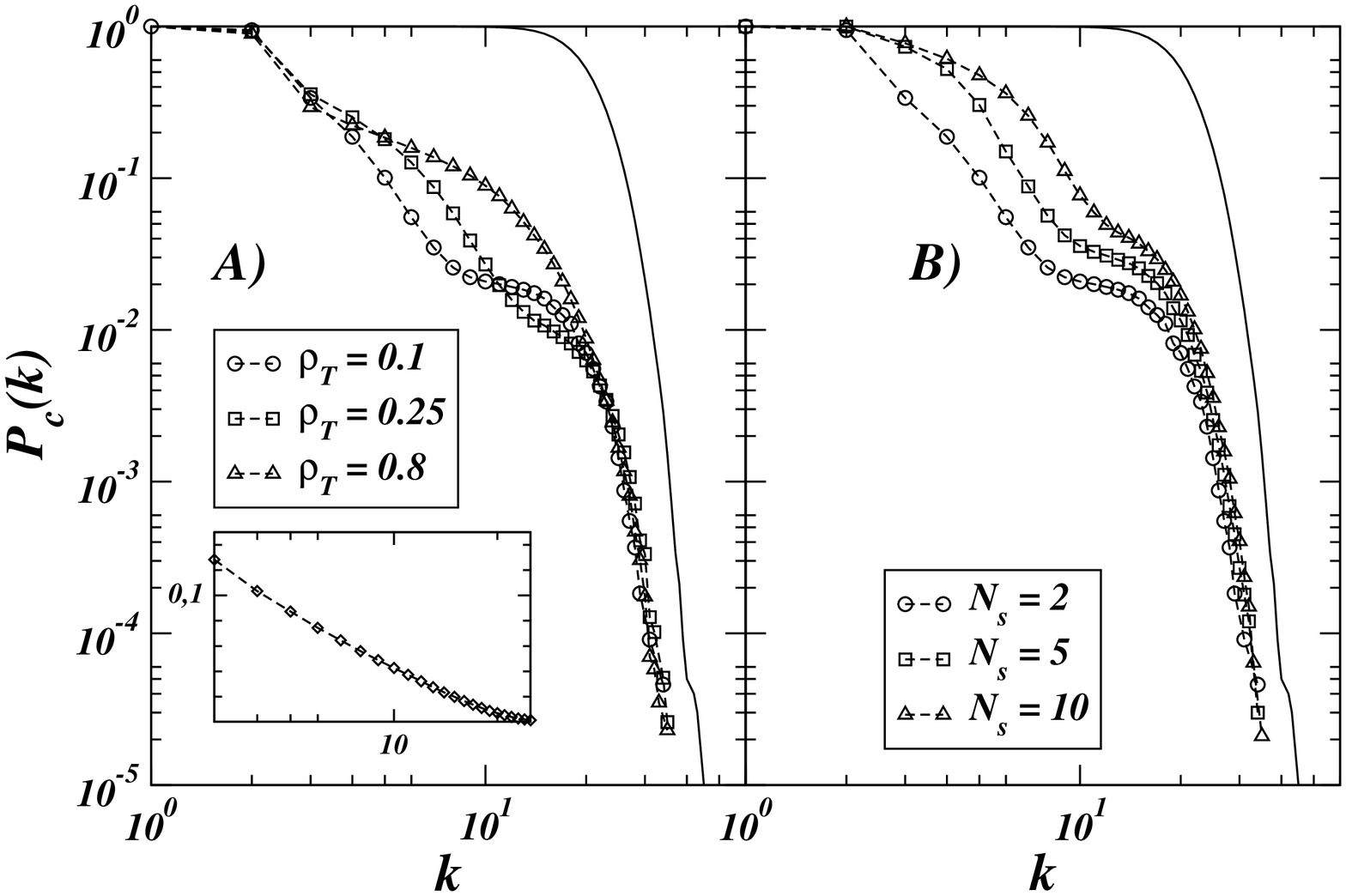}
\hskip .3in
\includegraphics[width=6.8cm]{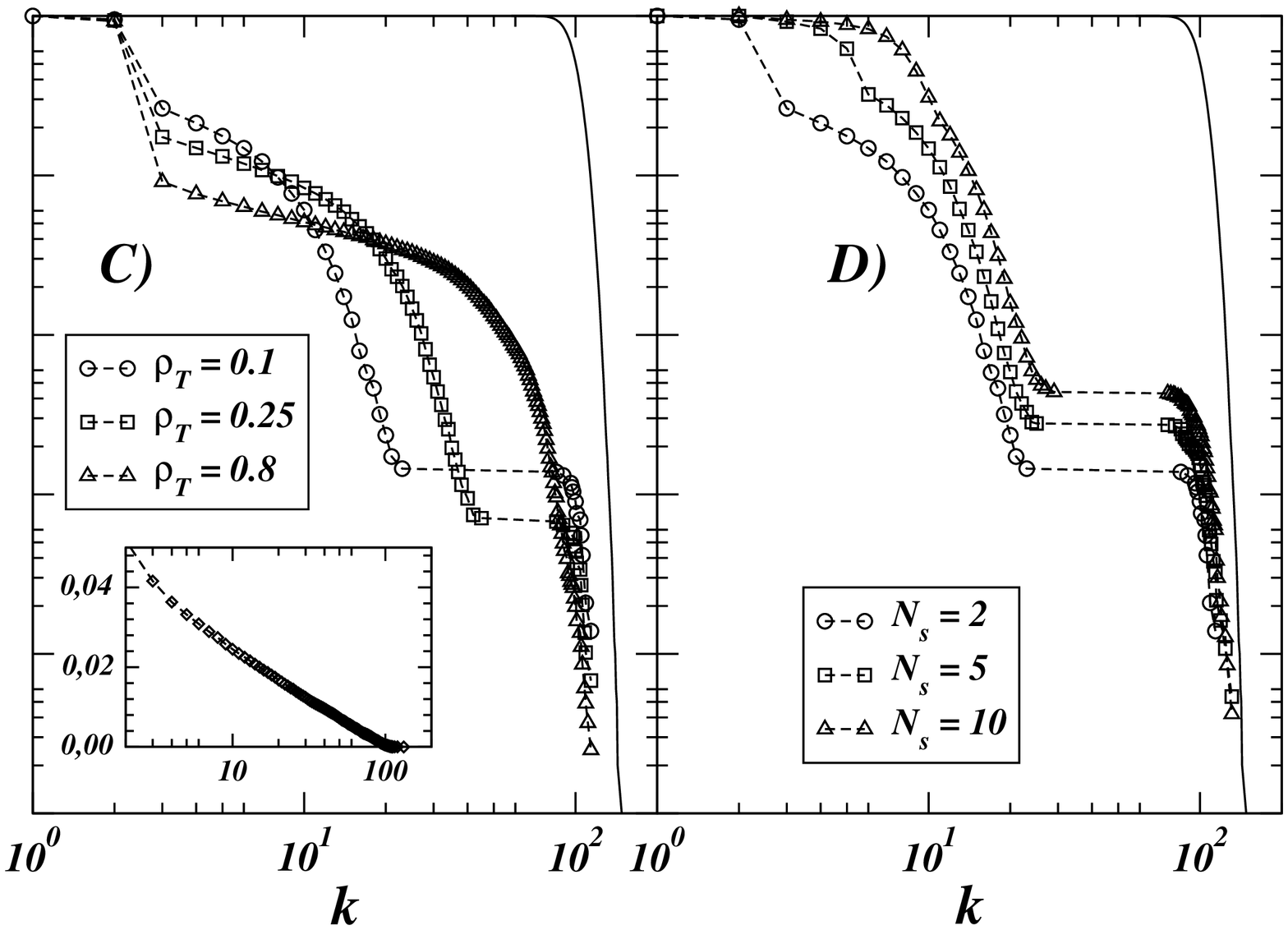}
\end{center}
\caption{
Cumulative degree distribution of the sampled ER graph for
  USP probes.  Figures (A) and (B) correspond to 
$\langle k\rangle =20$, and  (C) and (D) to $\langle k\rangle=100$.
Figures (A) and (C) show sampled distributions obtained
  with $N_S=2$ and varying density target $\rho_T$. In the insets we report the
  peculiar case $N_S=1$ that provides an apparent power-law behavior with
  exponent $-1$ at all values of $\rho_T$, with a cut-off depending
  on $\langle k\rangle$. 
  The insets are in lin-log scale to
  show the logarithmic behavior of the corresponding cumulative distribution.
  Figures (B) and (D) correspond to $\rho_T=0.1$ and
  varying number of sources $N_S$. The solid lines are
 the degree distributions of the underlying graph. For $\langle k\rangle=100$,
 the sampled cumulative
 distributions display plateaus corresponding to peaks in the degree
 distributions, induced by the sampling process.}
\label{fig:7}
\end{figure}

\begin{figure}[t]
\vskip .3in
\begin{center}
\includegraphics[width=10.0cm]{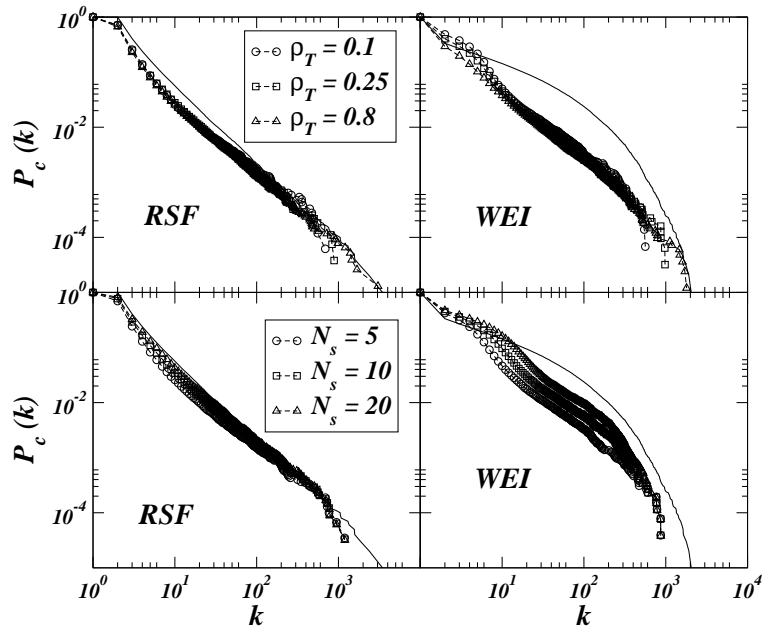}
\end{center}
\caption{Cumulative degree distributions of the sampled RSF and WEI graphs for
USP probes.  
The top figures show sampled distributions obtained with $N_S=2$ 
and varying density target $\rho_T$. 
The figures on the bottom correspond to
$\rho_T=0.1$ and varying number of sources $N_S$. The solid lines are the
degree distributions of the underlying graph.}
\label{fig:8}
\end{figure}

{\bf \em Degree distributions - \quad} 
We now get a clearer picture of the relation between the exploration process and the underlying graph, and 
we can tackle the important issue of {\em determining the origin of sampling biases in the observed 
degree distributions}. 
Fig.~\ref{fig:7} shows the cumulative degree distribution
$P_{c}(k) \equiv \sum_{k' \geq k} P(k')$ of the sampled graph defined by the ER model for increasing
density of targets and sources.  Sampled distributions look only approximately
like the genuine distribution; however, for $N_S\geq 2$ they are far from true
heavy-tail distributions at any appreciable level of probing.  
Indeed, the
distribution runs generally over a small range of degrees, with a cut-off that
sets in at the average degree $\langle k\rangle$ of the underlying graph. In order
to stretch the distribution range, homogeneous graphs with very large average
degree $\langle k\rangle$ must be considered, that emerges also from the rigorous
proof in Ref.~\cite{achlioptas} provided for single-source explorations.\\ 
However, other distinctive spurious effects appear in this case. 
In particular, since the best sampling occurs
around the high degree values, the distributions develop peaks appearing as plateaus in
the cumulative distribution (see Fig.\ref{fig:7}). 
The inset of Fig.\ref{fig:7} displays the single-source case,
in which we recover the apparent scale-free behavior with
slope $-1$. 
It is worth noting that the experimental setup with a single
source corresponds to a highly asymmetric probing process, in which the mean-field
approach, and consequently our theoretical predictions, are not valid.\\
The present analysis shows that in order to obtain a sampled graph with
apparent scale-free behavior on a degree range varying over $n$ orders of
magnitude we would need the very peculiar sampling of a homogeneous underlying
graph with an average degree $\langle k \rangle\simeq 10^n$, that is a rather unrealistic
situation in the Internet and many other information systems, where the observed cut-off sets in at least at $k \sim \mathcal{O}(10^2)$ (i.e. $n\geq 2$). Indeed, it would mean that on average Internet's autonomous systems should have at least $\mathcal{O}(10^2)$ connections, that is a completely unrealistic huge number.
On the contrary, in the case of RSP and ASP model, we observe that the obtained
distributions are closer to the real one, almost independently of the probing effort. 
On graphs with heavy-tailed distributions (see
Fig.~\ref{fig:8}), we observe completely different results, due to the fact that 
the distribution tail is fairly reproduced even at rather small
values of $\epsilon$. 
Despite both underlying graphs (WEI and RSF) have a small average
degree, the degree distribution spans more than two orders of
magnitude, and the whole range is sufficiently well sampled by the exploration process. \\
Some distortions occur for low and average degree nodes, that are 
under-sampled. This undersampling can either yield an 
apparent change in the exponent of the degree distribution (as also
noticed by Petterman and De Los Rios in Ref.~\cite{delosrios1} for single source experiments), 
or, if $N_S$ is small, yield a power-law like distribution for an underlying Weibull
distribution. As shown in Fig.~\ref{fig:8}, a small increase
in the number of sources allows to discriminate between both forms even at
small $\rho_T$.\\
The disparity in the quality of the results for homogeneous and heterogenous networks is due to
the different discovery efficiency reached by the process.
In heterogeneous graphs, vertices with high degree are efficiently sampled with
an effective measured degree that is rather close to the real one. {\em This means
that the degree distribution tail is fairly well sampled while deviations
should be expected at lower degree values}.  
In conclusion, {\em graphs with heavy-tailed degree
distribution allow a better qualitative representation of their
statistical features in sampling experiments}. Indeed, the most
important properties of these graphs are related to the heavy-tail
part of the statistical distributions that are indeed well
discriminated by the \texttt{traceroute}-like exploration. 

\subsection{Accuracy of the mapping process}\label{CHAP3_2_4}

Up to now, we have focused on the efficiency of the sampling process,
indicating which is the density of probes we have to deploy throughout the network
in order to get reasonably good maps of the network.
Another important aspect is related to the level of accuracy that the exploration is able to 
achieve in the description of the local topology. 
The most common biases affecting the mapping process concern {\em 1) the miss of
lateral connectivity, 
and 2) the multiple sampling of central nodes (and edges)}, which
may affect the efficiency of the whole process.\\
While the first problem might be solved by an optimization in the deployment of probes, actually
relying on a criterion of decentralization of sources and targets,
multiple sampling can be studied through some general concepts like
{\em redundancy} and {\em dissymmetry} of the discovery process.
A sampling is redundant when nodes (edges)
are discovered many times during the \texttt{traceroute}; it is
locally symmetric when the neighborhood of the nodes is equally sampled, i.e.
there are no preferential paths by which a node is traversed.  In the
following, we give quantitative measures of the level of redundancy and
dissymmetry of a \texttt{traceroute} mapping process, revealing their relation
with the topology of the sampled graph.
  
{\bf \em Redundancy - \quad}
On the one side, the node discovery process requires a certain level 
of redundancy, since each new passage might, in principle, contribute to a more detailed
exploration of the neighborhood.  When, however, the
discovery frequency is too large, it can seriously affect the efficiency of
the whole process. 
Let us define the edge redundancy $r_{e}(i,j)$ of an edge $(i,j)$ in a
\texttt{traceroute}-sampling 
as the number of probes passing through the edge $(i,j)$.
Using the notations of Section \ref{CHAP3_2_2}, this quantity is written
for a given set of probes and targets as 
\begin{equation}
r_{e}(i,j) = \sum_{l \neq m} 
\left( \sum_{s=1}^{N_{S}} \delta_{l,i_{s}}
 \sum_{t=1}^{N_{T}} \delta_{m,i_{t}} \sigma_{i,j}^{(l,m)} \right) .
\label{edgeredund}
\end{equation}
Averaging over all possible realizations and assuming the uncorrelation 
hypothesis, we obtain
\begin{equation}
\left\langle r_{e}(i,j) \right\rangle_{\Omega}  
\simeq \sum_{l\neq m} \rho_T\rho_S \left\langle \sigma_{i,j}^{(l,m)}
\right\rangle_{\Omega} = \rho_T\rho_S b_{ij} \ . \ \ \ \ \ \ 
\label{averedgeredun}
\end{equation}
This result implies that the average redundancy of an edge is related to the
density of sources and targets, but also to the edge betweenness. For example,
an edge of minimum betweenness $b_{ij} = 2$ can be discovered at most twice in
the extreme limit of an all-to-all probing. On the contrary, a very central
edge of betweenness $b_{ij}$ close to the maximum $N (N-1)$, would be
discovered approximately $\mathcal{O}(N)$ times by a \texttt{traceroute}-probing
from a single source to all the possible destinations.\\
Similarly, the
redundancy $r_{n}(i)$ of a node $i$, intended as the number of times the
probes cross the node $i$, can be obtained:
\begin{equation}
r_{n}(i) =  \sum_{l \neq m }
\sigma_{i}^{(l,m)} \sum_{s=1}^{N_{S}} \delta_{l,i_{s}} 
\sum_{t=1}^{N_{T}} \delta_{m,i_{t}} \ .
\end{equation}
After separating the cases $l=i$ and $m=i$ in the sum, the averaging over
the positions of sources and targets yields in the 
mean-field approximation:
\begin{equation}
\left\langle r_{n}(i)\right\rangle_{\Omega}  =  \sum_{l \neq m \neq i}  \rho_{S}\rho_{T}
 \langle \sigma_{i}^{(l,m)} \rangle_{\Omega} + 
2 \rho_{S} \rho_{T} N  \simeq  2\epsilon + \rho_{S}\rho_{T} b_{i} \ .
\label{avernoderedun}
\end{equation}
In this case, a term related to the number of \texttt{traceroute} probes $\epsilon$
appears, showing that unavoidably a part of the mapping effort goes to 
generate node redundancy. 
\footnote{By simple manipulation of formulas 
\ref{averedgeredun} and \ref{avernoderedun}, an equivalent 
of the identity $\sum_{j} b_{i,j} = 2(b_{i} + N -1)$ for redundancies is 
$\sum_{j} \left\langle r_{e}(i,j) \right\rangle_{\Omega} \simeq 2 
\left\langle r_{n}(i) \right\rangle_{\Omega} - 2\epsilon$ .}
\begin{figure}[t]
\vskip .3in
\begin{center}
\includegraphics[width=10.0cm]{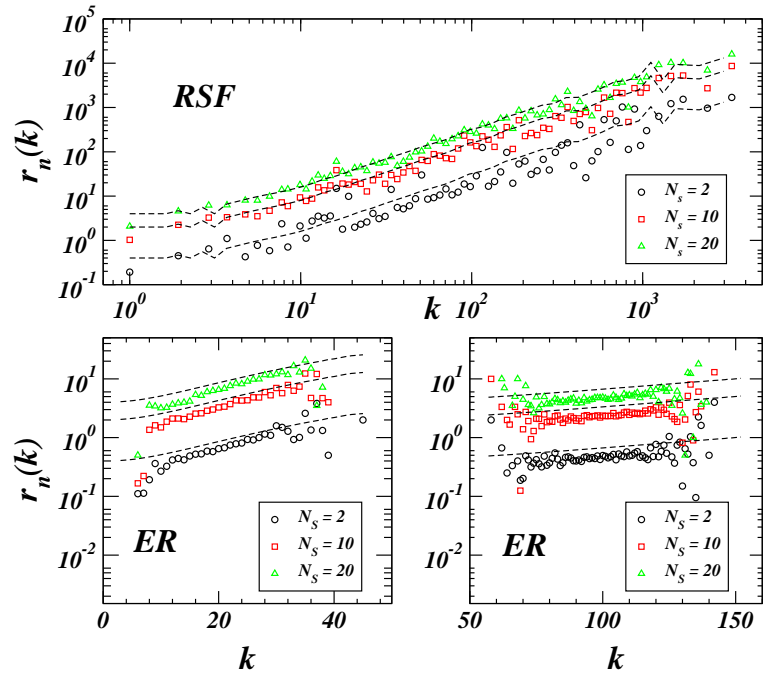}
\end{center}
\caption{
  Average node redundancy as a function of the degree $k$ for RSF (top) and ER
  (bottom) model ($N = 10^{4}$). For the ER model, two blocks of data are
  plotted, for $\langle k\rangle=20$ (left) and for $\langle k\rangle=100$ (right) The
  target density is fixed ($\rho_{T}=0.1$), and $N_S=2$ (circles), $10$
  (squares), $20$ (triangles). The dashed lines represent the analytical
  prediction $2\epsilon + \rho_{S} \rho_{T} \langle b\rangle (k)$ in perfect
  agreement with the simulations.}
\label{fig:4}
\end{figure}
In Fig.~\ref{fig:4} we report the behavior of the average node redundancy as a
function of the degree $k$ for both homogeneous and heterogeneous graphs.  For
both models, the behaviors are in good agreement with the mean-field
prediction, showing the tight relation between redundancy
and betweenness centrality.\\
In the case of heavy-tailed underlying networks, the node redundancy typically
grows as a power-law of the degree, while the values for random graphs vary on
a smaller scale. This behavior points out that {\em the intrinsic hierarchical
structure of scale-free networks plays a fundamental role even in the process
of path routing}, resulting in a huge number of probes iteratively passing
through the same set of few hubs.  On the other hand, for homogeneous graphs
the total number of node visits is quite uniformly distributed on the whole
range of connectivity, independently of the relative importance of the nodes.
This is a further element in favor of the argument that homogeneous graphs 
with large mean connectivity are pretty badly sampled. 
Indeed, the local topology of well-connected vertices is analyzed with the same level of
accuracy as for low degree nodes, yielding to generally
dissatisfying results.

{\bf \em Dissymmetry: Participation Ratio - \quad}
The high rate of redundancy found in the numerical data does not necessarily
imply that the local topology close to a node is well discovered:
preferential paths could indeed carry most of the probing effort.
Let us consider the
relative number of occurrences of a given edge $(i,j)$ during the
\texttt{traceroute}, with respect to the total occurrence for the edges in the
neighborhood of $i$.  For each discovered node $i$, we can thus define a set of
frequencies $\{f_{j}^{(i)}\}_{j \in \mathcal{V}(i)}$ for the edges $(i,j)$ of
its neighborhood; in terms of redundancy, the \textit{edge frequency}
$f_{j}^{(i)}$ is defined by
\begin{equation} 
f_{j}^{(i)} = \frac{r_{e}(i,j)}{\sum_{j \in \mathcal{V}(i)} r_{e}(i,j)}, 
\ \ \ \  0 \leq f_{j}^{(i)} \leq 1 \ \ \forall (i,j) \in \mathcal{E} .
\end{equation}         
Neglecting the correlations, we can write an approximation for the average edge 
frequency as 
\begin{eqnarray}
\nonumber\left\langle f_{j}^{(i)} \right\rangle_{\Omega}  =  \left\langle
\frac{r_{e}(i,j)}{\sum_{j \in \mathcal{V}(i)} r_{e}(i,j)}  \right\rangle_{\Omega} \simeq
\frac{\left\langle r_{e}(i,j) \right\rangle_{\Omega} }{\left\langle \sum_{j \in \mathcal{V}(i)} r_{e}(i,j) \right\rangle_{\Omega}} \\ \simeq \frac{\rho_{S}\rho_{T}b_{ij}}{2\rho_{S}\rho_{T}(b_{i}+N-1)} = \frac{b_{ij}}{2(b_{i}+N-1)},
\end{eqnarray}
where we have used the identity $\sum_{j} b_{ij} = 2(b_{i}+N-1)$.
The calculation reveals that, at a first approximation, the edge 
frequencies are topological quantities, independent of the probing effort, in agreement with the fact that 
frequencies are relative quantities.
\begin{figure}[t]
\vskip .4in
\begin{center}
\includegraphics[width=10.0cm]{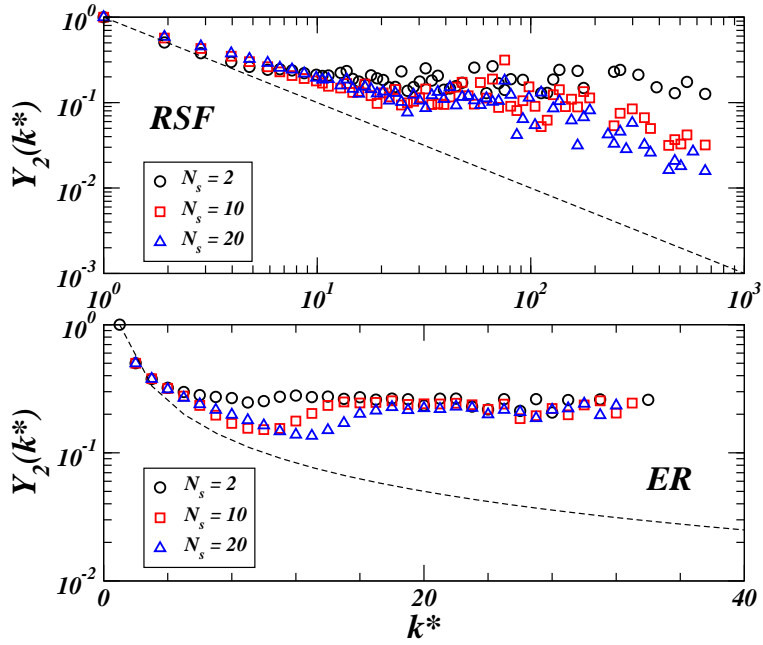}
\end{center}
\caption{Participation ratio as a function of discovered ($k^*$)
  connectivity for RSF (top) and ER (bottom) models ($N=10^{4}$). The target
  density is fixed ($\rho_{T}=0.1$) and three value of $N_S$ are presented: 2
  (circles), 10 (squares), 20 (triangles).  The dashed lines correspond to
  the $1/k^*$ bound.}
\label{fig:5}
\end{figure}
The dissymmetry of the discovery of the neighborhood of a node may be
quantified through the \textit{participation ratio} of these frequencies \cite{derrida,barthelemy_Y}:
\begin{equation}
Y_{2}(i) = \sum_{j \in \mathcal{V}(i)} {\left(f_{j}^{(i)} \right)}^2 \ .
\end{equation}   
If all the edge frequencies of $i$ are of the same
order $\sim 1/{k^{*}_{i}}$ (only discovered links give a finite contribution),
the participation ratio should decrease as $1/{k^{*}_{i}}$ with increasing
discovered connectivity ${k^{*}_{i}}$.  Hence, in the limit of an optimally
symmetric sampling, it should yield a  power law behavior
$Y_{2}(k^{*}) \sim {k^{*}}^{-1}$. When only few links are
preferred, for instance because more central in the shortest path routing, the
sum is dominated by these terms, leading to a value closer to the upper bound
$1$.  Numerical data for $Y_{2}$ as a function of the discovered
($k^{*}$) connectivity for different probing efforts, are displayed in
Fig.~\ref{fig:5}.  For heterogeneous graphs, the values of $Y_{2}$
tend towards the curve ${k^{*}}^{-1}$ for increasing $\epsilon$.
The average local topology of low degree nodes seems to be
sampled more homogeneously than the larger degree nodes. 
On the contrary, in the homogeneous case (ER), the figures show a general high level of dissymmetry persistent
at all degree values, only slightly dependent on the actual connectivity.

{\bf \em Dissymmetry: Entropy Measure - \quad}
The edge frequency is influenced by the presence of sources
and targets, thus we introduce a more refined frequency, 
$f_{kj}^{(i)}$ defined as the
number of probes passing through the {\em pair} 
$(k,i)-(i,j)$ of edges centered on
the node $i$, with respect to the total number of transits through any of the
possible couples of edges in the neighborhood of $i$. 
This frequency does not take into account single edges, but the path traversing 
each vertex and the dissymmetry of the flow. \\
A simple qualitative estimation for the average frequency is 
obtained using the usual first approximation for the edge redundancy \footnote{In this case the redundancy
appearing in the fraction defining the frequency $f_{kj}^{(i)}$ is the contribution $\rho_{S} \rho_{T} \sum_{l \neq m \neq i}\left\langle \sigma_{ki}^{(l,m)} \sigma_{ij}^{(l,m)}\right\rangle_{\Omega}$ by the edges-pair $(k,i)-(i,j)$.
Considering separately the portion of shortest path from $l$ to $i$ through $k$
and from $i$ to $m$ through $j$, we replace the sum of average products with
the approximation $(1-c_{i}) b_{ki} \left\langle \frac{\sum_{m \neq i} \sigma_{ij}^{(i,m)}}{\sum_{h \in \mathcal{V}(i)}\sum_{m \neq i} \sigma_{ih}^{(i,m)}}  \right\rangle_{\Omega}$ . Up to a factor the last averaged term is 
the redundancy $\frac{b_{ij}}{2(b_{i} + N - 1)}$ of the edge $(i,j)$. Since the sum of edge-pairs redundancies over the neighborhood gives essentially the redundancy of the central node $i$, the final expression follows easily.}    
\begin{eqnarray}
\nonumber
\left\langle f_{kj}^{(i)} \right\rangle_{\Omega} \simeq \frac{\rho_{S} \rho_{T} b_{ki}
b_{ij} (1 - c_i)}{2(b_i+N-1)} \frac{1}{\rho_{S} \rho_{T} b_i} \\ 
\nonumber \simeq  \frac{1-c_i}{2} \frac{b_{ki} b_{ij}}{b_i (b_i+N-1)}.
\end{eqnarray}
Even in this more complex situation, the approximated expression for the frequencies depends
only on topological properties of the underlying graph (such as the betweenness centrality and the clustering coefficient $c_i$).\\
By means of this frequency, we define an entropy measure providing supplementary
evidence of the tight relation between local accuracy, homogeneous sampling
and topological characterization of graphs.  Indeed, a \texttt{traceroute} that
discovers nodes crossing a larger variety of their links, and with different
paths, is expected to be more accurate (and likely efficient) than the one
always selecting the same path. \\
In the same spirit of the Shannon entropy \cite{billingsley}, which is a good indicator
of homogeneity, we
define the \textit{local traceroute entropy} of a node $i$ by
\begin{equation}
h_i = - \frac{1}{\log{k^{*}_i}} 
\sum_{k \neq j \in \mathcal{V}(i)} f_{kj}^{(i)}
\log{f_{kj}^{(i)}},
\end{equation}
where $\log{{k^{*}_i}}$ is simply a normalization factor. As usual, we define
$H(k)$ as the entropy averaged over the nodes of degree $k$.
\begin{figure}[t]
\vskip .5in
\begin{center}
\includegraphics[width=8.0cm]{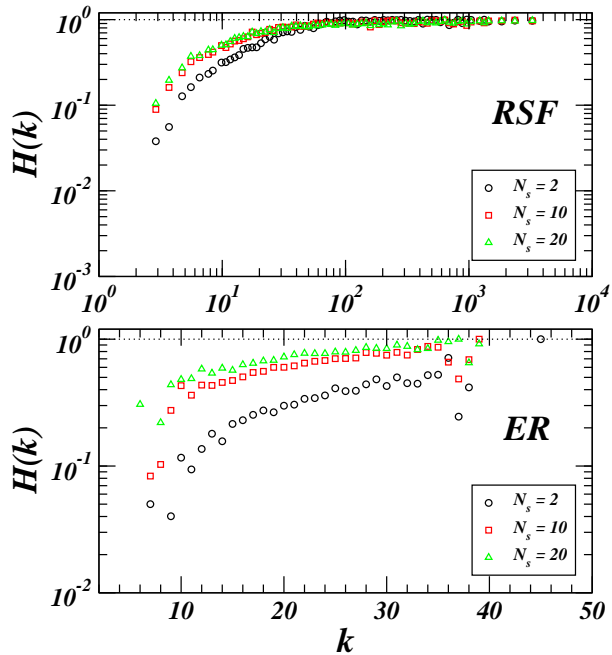}
\end{center}
\caption{
 Entropy vs.~$k$: a saturation effect is clear at 
  medium-high degree nodes for scale free topologies (RSF), instead of a more
  regular increase for homogeneous graphs (ER). In the figure there are
  different curves for $N_S$ = $2$ (circles), $10$ (squares), $20$ (triangles)
  and $\rho_{T} = 0.1$.}
\label{fig:6}
\end{figure}
The numerical data of $H(k)$ for RSF and ER models and for different levels of
probing are reported in Fig.~\ref{fig:6}. The values for ER are slightly
increasing both for increasing degree $k$ and number of sources $N_S$, with no
qualitative difference in the behavior at low or high degree regions.  On the
other hand, the case of heterogeneous networks agrees with the previous
observations. The curve for $H(k)$, indeed, shows a saturation
phenomenon to values very close to the maximum $1$ at large enough degree,
indicating a very homogeneous sampling of these nodes.\\
Summarizing, in the case of heterogeneous networks, the nodes with high degree and betweenness
are in general redundantly sampled, but present a rather symmetrical discovery of their neighborhood.
On the contrary, in homogeneous networks vertices suffer a less redundant sampling, showing a higher dissymmetry 
of the local exploration process.
This result should be taken into account in deciding source-target deployment strategies, in order
to minimize both dissymmetry and redundancy.

\subsection{Optimization}\label{CHAP3_2_5}
\label{sec:opt}
\begin{figure}[t]
\vskip .7cm
\begin{center}
\includegraphics[width=10.0cm]{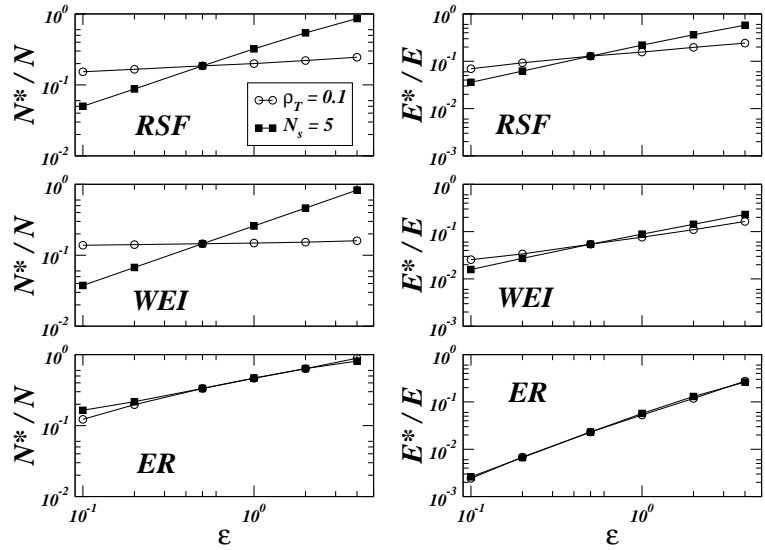}
\end{center}
\caption{Behavior of the fraction of discovered edges in explorations 
with increasing $\epsilon$. For each underlying graph studied we 
report two curves corresponding to larger $\epsilon$ achieved by increasing
the target density $\rho_T$ at constant $N_S=5$ (squares)
or the number of sources $N_S$
at constant $\rho_T=0.1$ (circles).}
\label{fig:14}
\end{figure}

In the previous sections we have provided a
general qualitative understanding of the efficiency of \texttt{traceroute}-like
exploration and the induced biases on the statistical properties. 
The quantitative analysis of the sampling strategies, however, is a
much harder task that calls for a detailed study of the discovered 
proportion of the underlying graph and the precise deployment of
sources and targets. In this perspective, Guillaume and Latapy have shown in Ref.~\cite{latapy}
that the fraction $N^*/N$ and $E^*/E$ of 
vertices and edges discovered in the sampled graph depend on the probing effort.
Unfortunately, the mean-field approximation breaks down when we aim at a quantitative representation
of the results, since the neglected correlations are necessary 
for a precise estimate of the various quantities of interest. 
For this reason we performed an exhaustive set of numerical
explorations aimed at a fine determination of the level of sampling
achieved for different experimental setups.\\
In Fig.~\ref{fig:14} we report the proportion of discovered edges in
the numerical exploration of homogeneous (ER model) and heterogeneous (RSF and WEI models) 
graphs for increasing level of probing effort $\epsilon$. The level of probing is
increased either by raising the number of sources at fixed target density
or by raising the target density at fixed number of sources.
As expected, both strategies are progressively more efficient with
increasing levels of probing. 

\begin{figure}[t]
\vskip .7cm
\begin{center}
\includegraphics[width=10.0cm]{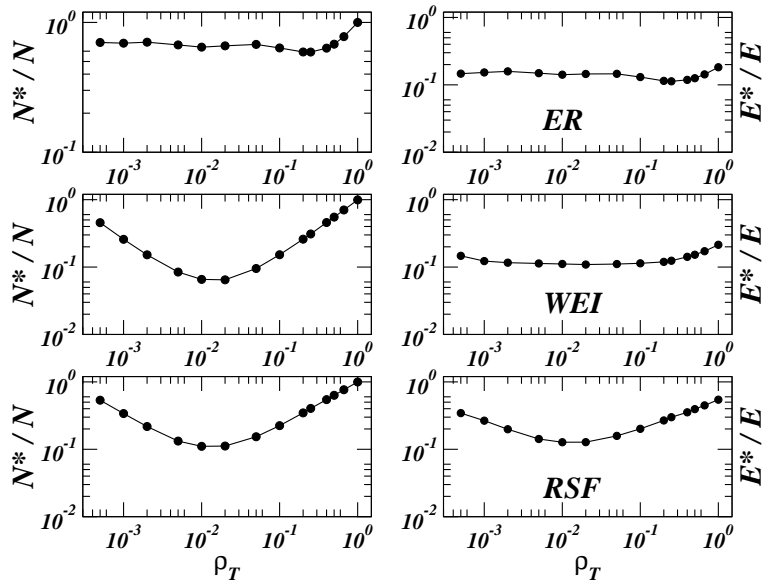}
\end{center}
\caption{Behavior as a function of $\rho_T$ 
  of the fraction of discovered edges and nodes in explorations with fixed
  $\epsilon$ (here $\epsilon=2$).  Since $\epsilon=\rho_T N_S$, the increase
  of $\rho_T$ corresponds to a lowering of the number of sources $N_S$.}
\label{fig:15}
\end{figure}
In heterogeneous graphs, it is also possible to see that when the 
number of sources is $N_S\sim \mathcal{O}(1)$ the increase of the 
number of targets achieves better
sampling than increasing the deployed sources. On the other hand,
our model of shortest path exploration is 
symmetric if we exchange sources with targets; therefore  
in the numerical experiments we can easily verify that if the number 
of sources is very large and $\rho_T\sim
\mathcal{O}(1/N)$, then the increase of 
the number of sources achieves better sampling than increasing the 
deployed targets.\\
This finding hints toward a behavior that is determined by the
number of sources and targets, $N_S$ and $N_T$ (or equivalently of $N_S$ and $\rho_T$). 
This point is clearly
illustrated in Fig.~\ref{fig:15}, where we report the behavior of
$E^*/E$ and $N^*/N$ at fixed $\epsilon$ and varying $N_S$ and $\rho_T$. 
The curves exhibit a non-trivial behavior and since we work at
fixed $\epsilon=\rho_T N_S$, any measured quantity can then be written as
$f(\rho_T,\epsilon/\rho_T)=g_\epsilon(\rho_T)$.
It is worthy noting that the curves show a structure allowing for local
minima and maxima in the discovered portion of the underlying graph.\\
This feature is a consequence of the symmetry by the exchange of sources
and targets of the model, i.e. an exploration with
$(N_T,N_S)=(N_1,N_2)$ is equivalent to one with  $(N_T,N_S)=(N_2,N_1)$.
In other words, at fixed $\epsilon=N_1 N_2/N$, a density of targets
$\rho_T=N_1/N$ is equivalent to a density
$\rho'_T= N_2/N$. Since $N_2=\epsilon/\rho_T$, we get that at
constant $\epsilon$, experiments with $\rho_T$ 
and $\rho'_T=\epsilon/(N\rho_T)$ are equivalent, obtaining by symmetry
that any measured quantity obeys the equality 
\begin{equation}
g_\epsilon(\rho_T)= g_\epsilon \left( \frac{\epsilon}{N\rho_T} \right) .
\end{equation}
This relation implies a symmetry point signaling  the presence of a
maximum or a minimum at $\rho_T =\epsilon/(N\rho_T)$. We therefore 
expect the occurrence of a symmetry in the graphs of 
Fig.~\ref{fig:15} at $\rho_T\simeq \sqrt{\epsilon/N}$. Indeed, 
the symmetry point is clearly
visible and in quantitative good agreement with the previous
estimate in the case of heterogeneous graphs. 
For homogeneous topology the curves have a smooth behavior that makes
difficult the clear identification of the symmetry point. 
Moreover, USP probes create a certain level of
correlations in the exploration that tends to hide the complete
symmetry of the curves.
\begin{figure}[t]
\vskip .3in
\begin{center}
\includegraphics[width=10.0cm]{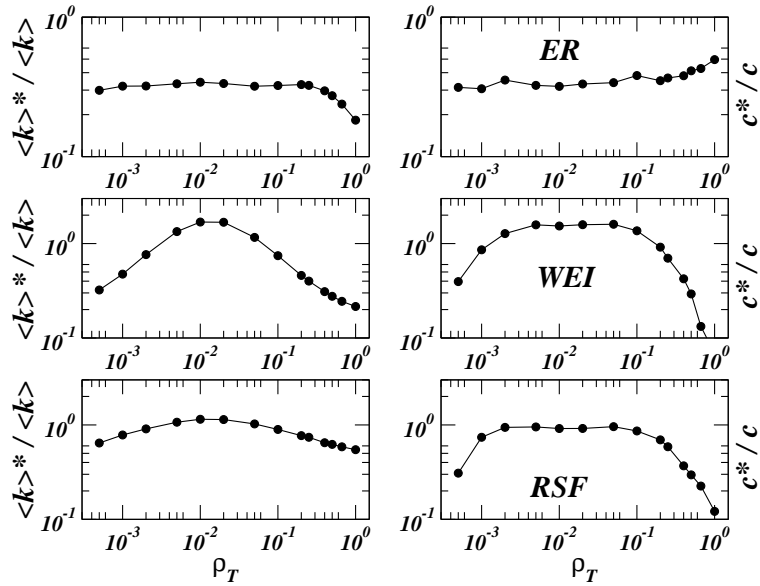}
\end{center}
\caption{Behavior as a function of $\rho_T$ 
of the fraction of the normalized average degree 
${\langle k\rangle}^*/\langle k\rangle$ 
and of the fraction of the normalized average clustering coefficient 
${\langle c\rangle}^*/\langle c\rangle$ for 
a fixed probing level $\epsilon$ (here $\epsilon=2$).}
\label{fig:18}
\end{figure}
The previous results imply that at fixed levels of probing $\epsilon$
different proportions of sources and targets may achieve different levels of
sampling. This hints to the search for optimal strategies in the relative
deployment of sources and targets.  The picture, however, is more complicate
if we look at other quantities in the sampled graph. In Fig.~\ref{fig:18} we
show the behavior at fixed $\epsilon$ of the measured average degree ${\langle k\rangle}^*$
normalized by the actual average degree
${\langle k\rangle}$ of the underlying graph as a function of $\rho_T$. The plot
shows also in this case a symmetric structure.
By comparing Fig.~\ref{fig:18} with Fig.~\ref{fig:15} we notice that the
symmetry point is of a different nature for different quantities:
the minimum in the fraction of discovered edges corresponds to the best
estimate of the average degree. 
A similar result is obtained for the behavior of the ratio ${\langle c\rangle}^*/{\langle c\rangle}$
between the clustering coefficient of the sampled and the underlying graph:
as shown in Fig.~\ref{fig:18}, the best level of
sampling is achieved at particular values of $\epsilon$ and $N_S$ that are
conflicting with the best sampling of other quantities.\\
The numerical data obtained with different source and target densities 
hint to a possible optimization of the sampling strategy. 
{\em The optimal solution, however, appears as a trade-off
strategy between the different levels of efficiency achieved in competing
ranges of the experimental setup}. In this respect, a detailed and quantitative
investigation of the various quantities of interest in different experimental
setups is needed to pinpoint the most efficient deployment of
source-target pairs depending on the underlying graph topology.

\subsection[$k$-core structure under sampling]{Non-local measures under sampling: the case of $k$-core structures}
\label{CHAP3_2_6}
Up to now, all statistical quantities studied on sampled networks are 
related to local properties and local correlations; but real networks may present 
non-locally correlated structures, whose integrity under sampling is even more questionable.  
For this reason, we have investigated the effects of \texttt{traceroute}-like sampling on the 
$k$-core organization of networks.\\
The $k$-core analysis of a network is based on a non-trivial decomposition in subgraphs, 
that has recently attracted the interest of physicists working in this field for its relation
with box counting methods in the study of self-similarity of natural systems \cite{havlin_box1,havlin_box2,salvi_box}.\\
The $k$-core of a network, defined in Section~\ref{CHAP2_2_5}, is the maximal subset induced 
by all nodes having at least $k$ neighbors in it, and the $k$-shell is the set of nodes 
belonging to the $k$-core but not to the $(k+1)$-core.\\
The $k$-core decomposition of a network, going from $k=1$ (i.e. the whole network without isolated nodes) up to the maximum available value $k_{max}$, provides a hierarchical structure in which most internal $k$-shells contain high degree nodes belonging to the very fundamental backbone of the network, whereas the external ones are formed by low-degree and more peripheral nodes. 
In Ref.~\cite{ignacio3,ignacio1,ignacio2} (see also the web-site of the visualization tool Lanet-VI \cite{lanetvi}), we have used the $k$-core decomposition as a tool for analyzing networks, discovering hierarchical structures, and studying the main statistical properties at the different scales (i.e. in the different $k$-cores).\\ 
Almost all complex networks (real and synthetic) seem to share a common ``scale-invariance'' property with respect to the $k$-core organization: indeed, after a simple rescaling the curves corresponding to quantities like the degree distribution $P(k)$, the average nearest neighbor degree $k_{nn}(k)$ and the clustering coefficient $c(k)$, computed in the different $k$-cores show a nice data collapse.       
It is thus interesting to test the behavior of the $k$-core decomposition in sampled networks, 
in order to check the robustness of these properties with respect to possible sampling biases. 
\footnote{ We will refer to cores and degrees using the same index $k$, the distinction between 
the two cases being evident from the different context. When the context is not clear, we will however specify the meaning of $k$ (if it indicates the degree or the $k$-core index).} \\
Note that a single source \texttt{traceroute}-like probing yields
essentially a tree, then the $k$-core decomposition is by definition
trivial (with maximum core $k_{max}=1$). Yet, a sampling cannot
discover paths or edges that do not exist, so that the maximal shell
index of a network, $k_{max}$, is not increased by partial sampling (as the maximal degree observed), 
and reversely, the actual $k_{max}$ is
{\em at least} equal to the one found by a sampling of the true
network.\\
Internal cores are more connected and are traversed by a larger number of paths, therefore we expect that a path-based sampling should intuitively discover and sample
better the central cores, introducing stronger biases in the structure of the peripheral shells. 
Moreover, the shell index of a node is directly related to its routing capacity, since two nodes belonging to the same shell of index $k$ have exactly $k$ ``distinct'' paths between them, where distinct means that no node and no edge are used more than once. The abundance of paths between nodes corresponds also to a higher level of structural and functional robustness of the system. Hence, nodes with high shell index are expected to perform better in routing processes.\\
We have checked such ideas performing a \texttt{traceroute}-like probing of various 
networks, and comparing their $k$-core decomposition before and after sampling. 
We have used $N_S=50$ sources, and various values of probing efforts from $\epsilon=0.1$ to $\epsilon=5$.\\
\begin{figure}[t]
\vskip .3in
\begin{center}
\includegraphics[width=10.0cm]{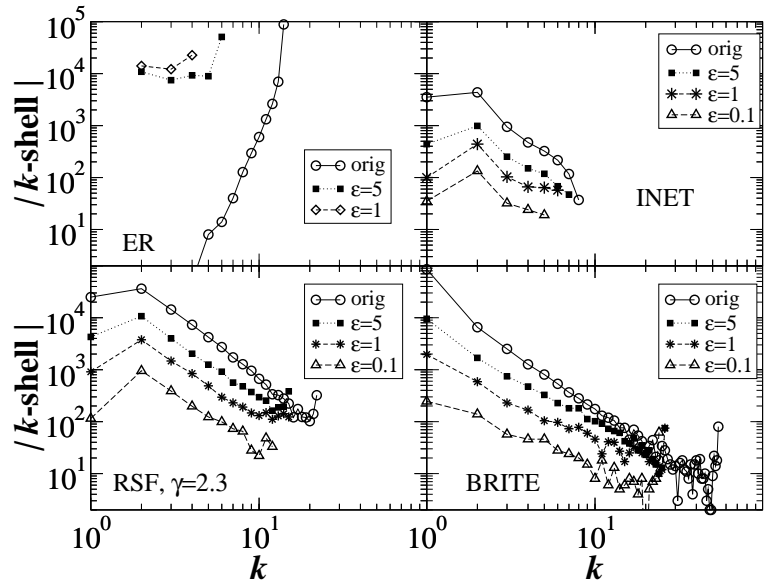}
\end{center}
\caption{Plot of the size of the $k$-shells vs. $k$ for various models,
before and after \texttt{traceroute}-like sampling, with different
probing efforts $\epsilon$. We used an Erd\"os-R\'enyi (ER) random graph with ${\langle k\rangle}=20$,
a random scale-free network (RSF) with exponent $\gamma=2.3$, and two networks obtained by the generators BRITE and INET, popularly used in the computer scientist community to reproduce some features of the Internet topology. All networks have size $N=10^5$, except for INET ($N=10^4$). Note that the $k$ index on the $x$-axis indicates the $k$-shell index, not the degree.}
\label{biases_shell}
\end{figure} 
Figure~\ref{biases_shell} reports the curves of the $k$-shell size as
a function of the index for various network models and various
sampling efforts. The numerical measures have been performed on four types of networks: the ER and RSF models, and 
two network obtained using the generators BRITE \cite{brite} and INET \cite{inet}, that are based on optimization strategies and are commonly used by computer scientist in order to reproduce some specific features of the Internet, such as power-laws and hierarchy. 
Both yield broad degree distributions and general properties similar to RSF graphs.
For ER networks, shells are almost uniformly populated 
and concentrated in a range of index values around ${\langle k\rangle}$.  
Such networks, whose $k$-core structure is very different from that
observed for AS maps (see Ref.~\cite{ignacio1}), show a rather peculiar behavior also 
after sampling.\\ 
\begin{figure}[t] 
\begin{center} 
\includegraphics[width=7.0cm]{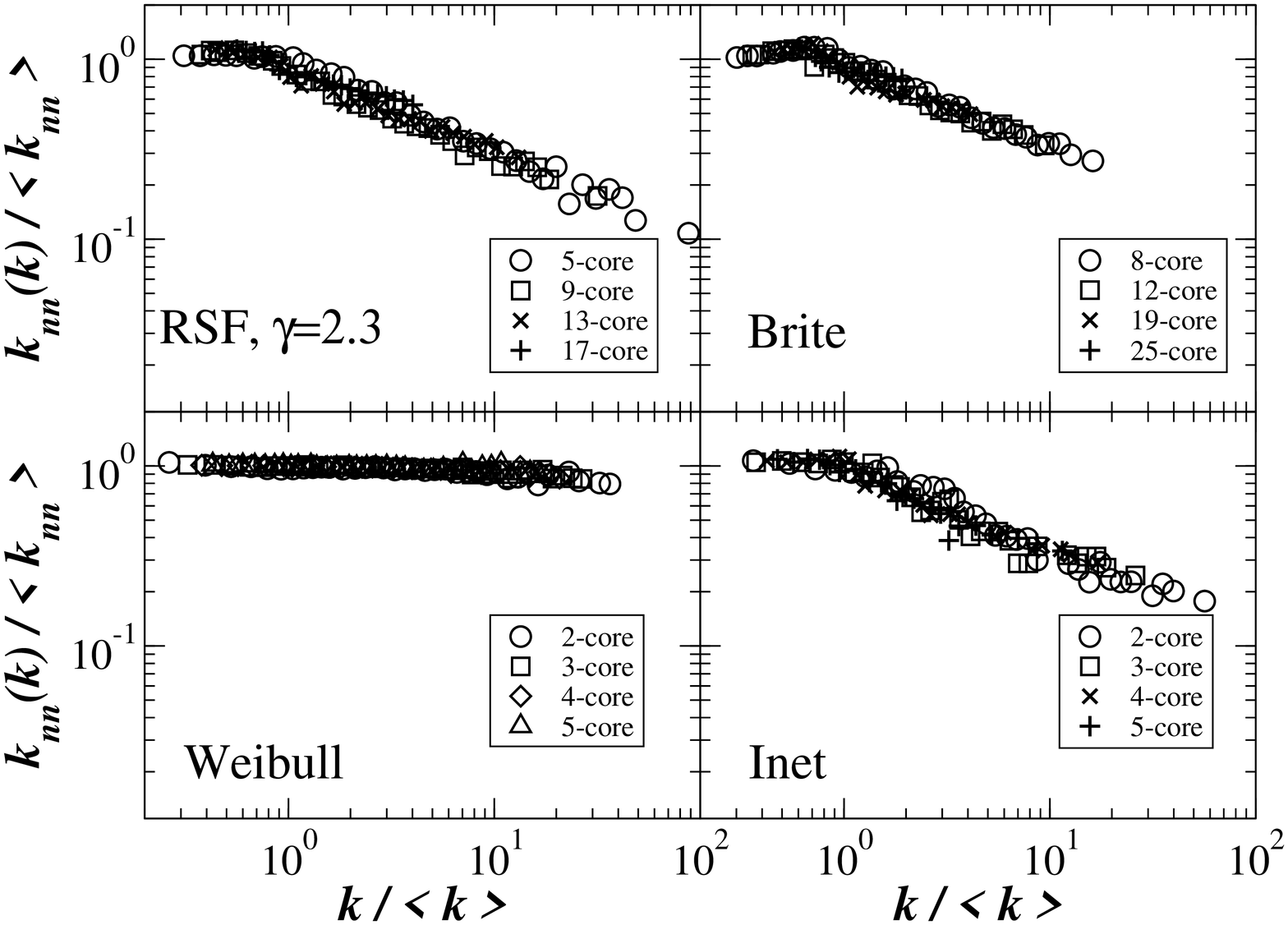}
\includegraphics[width=7.0cm]{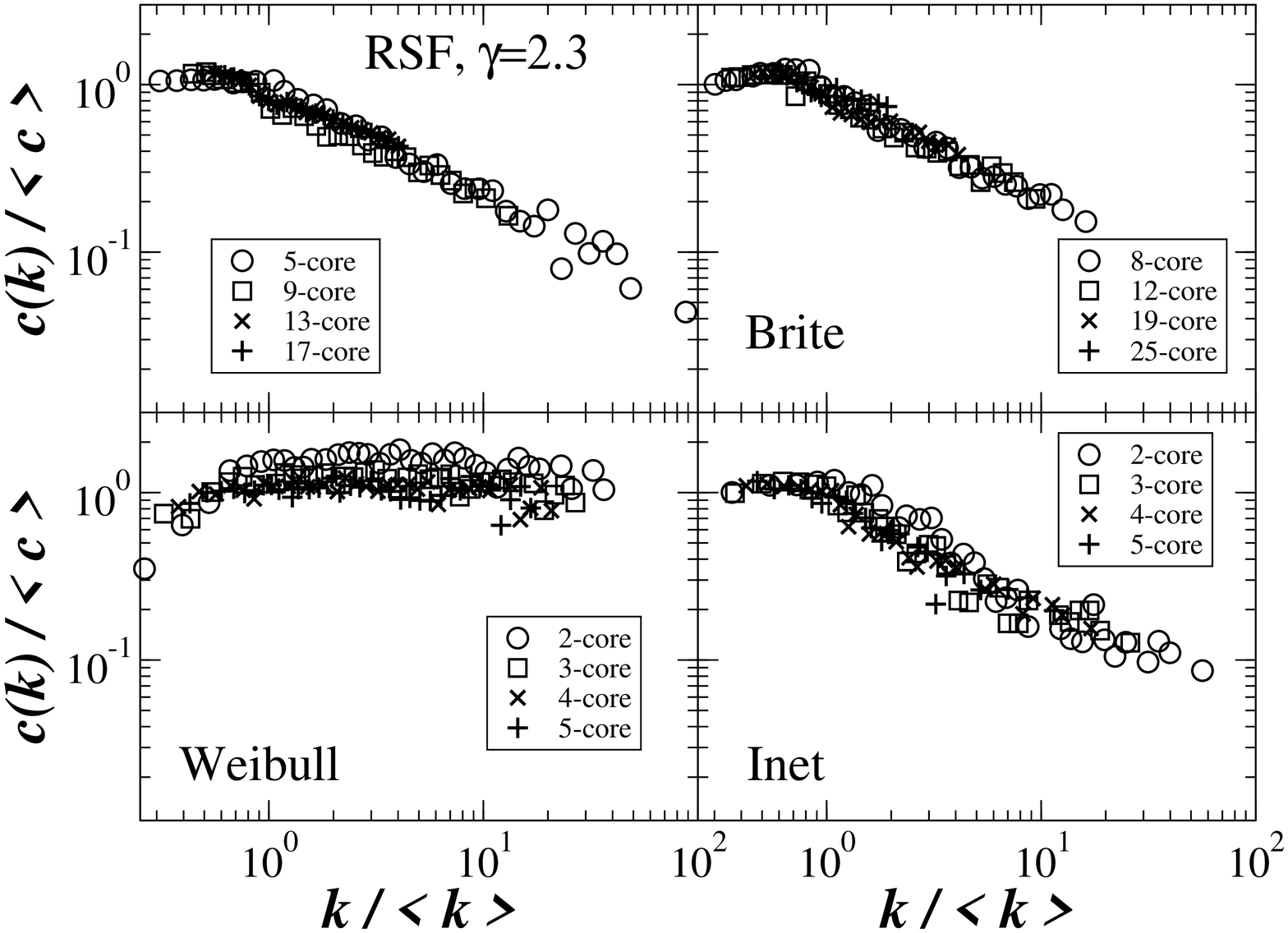}
\end{center} 
\caption{Nearest neighbors degree distribution (left) and clustering spectrum (right) of some $k$-cores,
rescaled by the corresponding average values, for some network models
after sampling through a \texttt{traceroute}-like process
with $N_S=50$ sources and target density $N_T/N=0.1$. Here, the index $k$ refers to the degree.
}
\label{KnnCnn_TR} 
\end{figure}
On the contrary, the power-law shapes obtained
for RSF or BRITE networks, are comparable to the
one observed in the AS maps and look very robust under sampling; even if the slope is
affected. Indeed, shells of smaller indices are less well sampled. 
In particular, the size of the first shell is strongly decreased by
the sampling procedure; in some cases in fact, the first shell is
larger than the second in the original network, but becomes smaller in
the sampled network.  We note that in the available AS maps, the first
shell is indeed typically smaller than the second, and that the true
AS network thus very probably exhibits a much larger shell of index
$k=1$. 
This is consistent with the idea that {\em the proportion of leaves is extremely 
underestimated in current Internet mapping data}.
A similar argument should hold for the value of the exponent in
the power-law behavior of the shell size vs. its index (see Ref.~\cite{ignacio1,ignacio2}).\\
Figure \ref{KnnCnn_TR} reports the behavior of other typical quantities 
of the network: the average degree of the nearest neighbors of a node of 
degree $k$, and the clustering coefficient of nodes of degree $k$.
These properties show self-similar features in the $k$-core decomposition that seem to be 
preserved by the sampling process. Although the precise form of the
degree distribution of the whole network is slightly altered, {\em the
basic correlation properties are conserved in the sampling}.\\
\begin{figure}[t]
\begin{center}
\includegraphics[width=5.0cm,angle=-90]{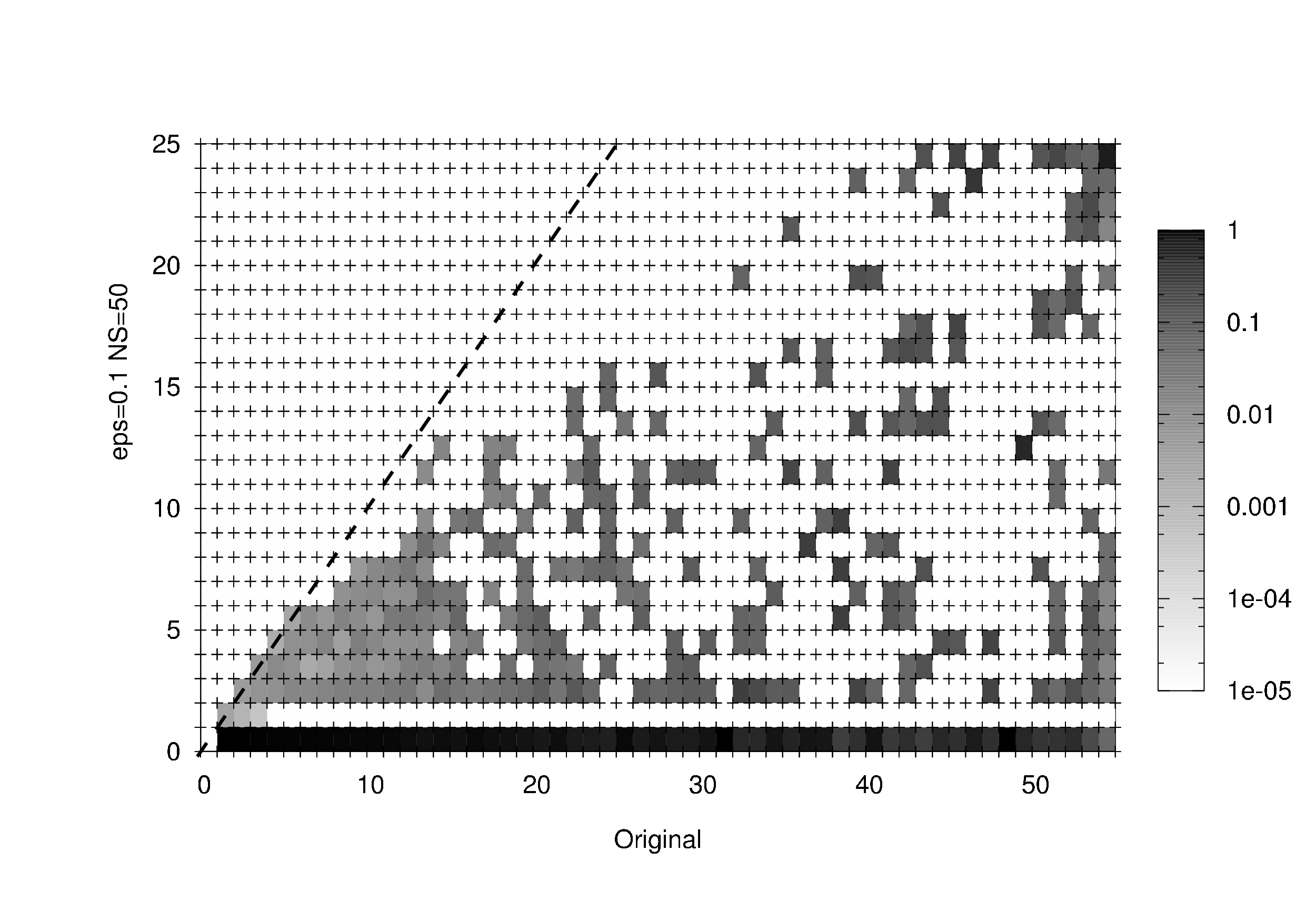}
\includegraphics[width=5.0cm,angle=-90]{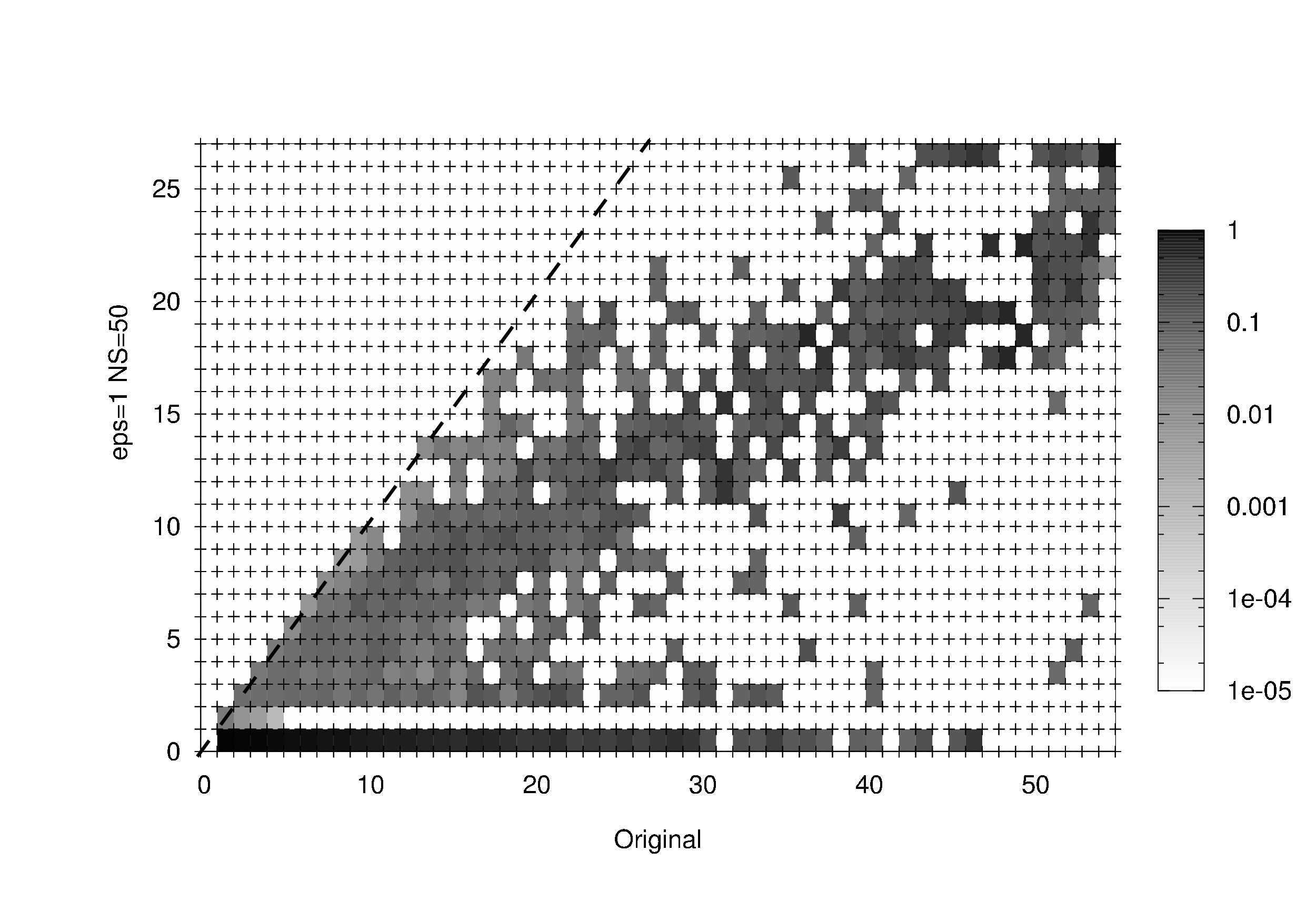}
\end{center}
\caption{The grayscale code gives the probability of a change in shell
index due to the \texttt{traceroute}-like sampling, from a certain
index before sampling ($x$ axis) to another one after sampling ($y$
axis).  The line at $y=0$ represents the probability of vertices of shell
index $x$ to be absent from the sampled graph. The initial network is
obtained by the BRITE generator. Here $N_S=50$ sources and a fraction
$N_T/N=2. 10^{-3}$ (left) and $N_T/N=2. 10^{-2}$ (right) of targets are used.}
\label{diff}
\end{figure}
While on a qualitative level, it seems possible to
distinguish between networks with different topological structures,
important quantitative biases appear, that are related with the arguments exposed in Section~\ref{CHAP3_2_3} about
the efficiency of nodes discovery. 
For a network obtained with BRITE, Fig.~\ref{diff} displays 
the probability that the original shell index $k$ of a vertex has changed in $k'$ due to the sampling process.  
At low sampling effort, many vertices remain completely undiscovered, and in general shell index properties are
strongly affected in a seemingly erratic way (see the plot for $\epsilon = 0.1$). For larger
values of sampling effort a strong correlation appears between the two shell indices, 
even if a systematic downwards trend is observed.\\
A naive explanation for such persistence of the $k$-core structure under \texttt{traceroute}-like sampling 
is given looking at the process of path merging by which we get the maps.
Let us consider three \texttt{traceroute} paths between three different pairs of vertices: 
if these paths meet two-by-two in some nodes, such three nodes result to be connected by a cycle, that is a $2$-core.
Increasing the number of paths, it is possible that strongly connected sets of nodes emerge (in the sense of a large 
number of existing paths between them).
This picture is likely to be true in scale-free networks, in which the central nodes are redundantly sampled by \texttt{traceroute} (as seen in Section~\ref{CHAP3_2_4}).
Hence, the presence of a genuine $k$-core structure in sampled graphs is due to the fact that nodes are sampled by means of paths. On the contrary, since node-picking algorithms do not guarantee that nodes are fairly connected, {\em in many biological networks the $k$-core structure might be strongly biased by sampling methods}.  \\
In summary, the results presented here indicate that the sampling biases 
do in fact affect only slightly the measure of the statistical properties 
of the $k$-core organization in heterogeneous graphs, even at
relatively low levels of sampling. This corroborates the idea that {\em the $k$-core properties 
observed for the Internet are genuine}. 
Quantitative analysis is more problematic, due to the incomplete sampling of the edges. 
In fact, the routing properties of a network are related with the multiplicity of paths between nodes, and thus with 
$k$-core properties. Hence, {\em we conclude that ``measured'' routing capacity of nodes, if limited to the analysis of 
Internet maps, are certainly rather underestimated compared to real performances}.

\newpage
\section[Network Species Problem]{Network Species Problem: a statistical method to correct biases}
\label{CHAP3_3} 
An unexpected application of the \texttt{traceroute} model of 
Internet explorations is that of providing a theoretical framework in which 
it is possible to define and test statistical estimators for important unknown properties of the
networks and study which is the best estimator for a given underlying topology.
In fact, in Ref.~\cite{viger}, we have shown that the inference from \texttt{traceroute}-like measurements
of many of the most basic topological quantities, including network's size and 
degree characteristics, is a version of the so-called `species problem' in statistics.  
This observation has important implications, since the species problem is known to be of a 
particularly challenging nature. \\ 
A basic example of a \texttt{traceroute}-based species problem is the estimate of the number
of nodes in a network (Section~\ref{CHAP3_3_1}). Using statistical subsampling principles we have derived two 
estimators for this quantity (Section~\ref{CHAP3_3_2}), the performances of which will be illustrated by means of 
numerical simulations on networks with various topological characteristics (Section~\ref{CHAP3_3_3}).\\
According to the results exposed in the previous section, one can conclude that at a 
qualitative level \texttt{traceroute}-like samplings are reliable.  
On the other hand, at a quantitative level real networks (e.g. the Internet at different levels) 
can considerably differ from sampled maps. 
The species approach seems to be valuable to estimate quantities, such as the 
 size, the number of links, the average degree, and the precise analytic
form of the heavy-tailed degree distribution, that cannot be estimated using previously 
exposed techniques.

\subsection{The Species Problem in Networks}\label{CHAP3_3_1}

Let us call $\eta\equiv \eta(G)$ a generic global quantity characterizing 
a graph $G$. In general, the real value of $\eta$ is not known, thus
 it is natural to wish to produce an estimate, say $\hat\eta$, based
on the network sampling, i.e. on the \texttt{traceroute}-sampled graph $G^*$.  
However, for quantities like the size $N$, the number of edges $E$, the average degree ${\langle k\rangle}$, 
the problem of their inference is closely related to
the {\it species problem} in statistics.  In general,
the species problem refers to the situation in which, {\em having observed
$n$ members of a (finite or infinite) population, each of whom falls
into one of $C$ distinct classes (or `species'), an estimate $\hat C$
of $C$ is desired}.  This problem arises in numerous contexts, such
as numismatics (e.g., how many of an ancient coin were 
minted~\cite{esty}), linguistics (e.g., what was the size of
an author's apparent vocabulary~\cite{mcneil,et}), 
and biology (e.g., how many species of animals inhabit a given region).\\
The species problem has received a good deal of attention in statistics 
(see, for instance, Ref.~\cite{bungefitz}), but it is, in general, a 
difficult problem, since we need to estimate the number of species {\it not} observed. 
In practice, the species expected to be missed are  those that are present in relatively low 
proportions in the population, and there could be an arbitrarily 
large number of such species in arbitrarily low proportions.
The methods proposed for its solution differ in the assumptions regarding 
the nature of the population, the type of sampling involved, 
and the statistical machinery used.
We have shown that it can also be associated with the 
inference of graph characteristics $\eta(G)$
in \texttt{traceroute}-like samplings.\\  
For example, the problem of estimating the number of vertices and edges in a 
network $G$ i.e., $N$ and $E$ may be mapped on the species problem by
considering each separate vertex $i$ (or edge $e$) as a `species'
and declaring a `member' of the species $i$ (or $e$) to have been observed
each time that $i$ (or $e$) is encountered on one of the $N_S\times N_T$
\texttt{traceroute} paths. \\ 
Again, the problem of inferring the degree $k_i$ of a vertex $i$
from \texttt{traceroute} measurements can also be mapped on the species problem,
by letting all edges incident to $i$ constitute a species and declaring
a member of that species to have been observed every time one of those
edges is encountered.  Because the values $N$, $E$, and 
$\{k_i\}_{i\in V}$ serve as basic components of many of the
other standard quantities listed above, obtaining an
accurate inference of the former could directly
impact our ability to make accurate inferences on the latter.  
In the following, we will consider only the inference of the network
size $N$, but we are now developing a similar formalism in order to extend our 
analysis to the number of edges $E$.

\subsection{Inferring $N$: Estimators of Networks Size}
\label{CHAP3_3_2}

Before proceeding to the construction of estimators for $N$,
it is useful to first better understand the relation between this quantity and known
characteristics of the Internet topology.  \\
Since the main property affecting the exploration process is the betweenness centrality,
one could argue that $N$ should be estimated using quantities derived from the betweenness.
In particular, the network's size is related to the betweenness centrality by the following simple expression \cite{goh_bet},
\begin{equation}
\sum_i b_i = N(N-1)({\langle \ell \rangle} -1) \ ,
\end{equation}
in which $\langle \ell \rangle$ is the average distance between pairs of nodes.
This may be rewritten in the form
\begin{equation}
N= 1 + \frac{\mathbb{E}[b]}{{\langle \ell \rangle} -1} \ ,
\label{eq:Nandaveb}
\end{equation}
where the expectation $\mathbb{E}[\cdot]$ is with respect to the
distribution of betweenness across nodes in the network.\\
In general, the average shortest path length ${\langle \ell \rangle}$ can be estimated quite accurately,
since \texttt{traceroute} probes lay on the shortest paths and 
the corresponding distribution is very peaked. 
Therefore, the problem of estimating $N$
is essentially equivalent to that of estimating the
average betweenness centrality.  
From the theoretical analysis in previous sections, we know that \texttt{traceroute} experiments 
give a good estimate of degree and betweenness distributions tails, thus we can assume that the 
form $P(b)\sim b^{-\beta}$ for $b>>1$ is sufficiently accurate, but 
low betweenness nodes are considerably undersampled, preventing us from having a correct 
quantitative knowledge of the full distribution.
Hence, the undersampling of low-betweenness nodes does affect the average value of the betweenness. 
Additionally, even if we neglect this problem and divide the expectation in two contributions for low and high betweenness nodes ($\mathbb{E}[b]=\mathbb{E}_{1}[b]+\mathbb{E}_{2}[b]$), in order to compute the average betweenness, we should perform an integral
of the type 
\begin{equation}
\mathbb{E}_{2}[b] \simeq \frac{1}{K}\int_{b_{min}}^{b_{max}} b^{1-\beta} db~,
\end{equation}
that has to be handled carefully since the experimental values of $\beta$ are very close to $2$ \cite{goh_bet,barthelemy_bc}, thus the integral diverges with the upper cut-off.\\
These simple arguments give an idea of the difficulty of estimating $N$
from \texttt{traceroute} measurements and suggest
{\em the futility of attempting a parametric approach with current measurement
tools and information}.  There is still the alternative of a {\em nonparametric
approach}, in which assumed parametric distributions are eschewed.
We have proposed two estimators for $N$, using
subsampling principles: one is based on the resampling of the network, the other 
is a refined estimator based on the ``leave-one-out'' principle \cite{leaveoneout1,leaveoneout2,leaveoneout3}.

{\bf \em Resampling Estimator - \quad}
Let us call {\it discovery ratio} $\theta = \mathbb{E}[N^*]/N$ the average
fraction of nodes discovered.  From the mean-field theory we have learned that 
the quantity varies smoothly as a
function of the fraction $\rho_T=N_T/N$ of targets sampled, for a given
number $N_S$ of sources.  We use this
fact, paired with the assumption of a type of scaling relation on $G$,
to construct an estimator for $N$.\\
Specifically, let $H$ be an arbitrary subgraph, of size $N(H)$,
of the network graph $G$.  
We will assume that, for roughly similar numbers $N_S(H)$ and
$N_S(G)$ of sources used, the discovery ratios for \texttt{traceroute}
sampling on $H$ and $G$ are such that 
\begin{itemize}
\item[(i)] they vary as smooth functions $\theta(H;\rho_T(H))$ and $\theta(G;\rho_T(G))$ of $\rho_T(H)$ and
$\rho_T(G)$, respectively; 
\item[(ii)] if $\rho_T(H)=\rho_T(G)$, then $\theta(H;\rho_T(H)) = \theta(G;\rho_T(G))$.  
\end{itemize}
In other words, we expect similar proportions of targets to yield similar proportions of discovered
nodes.  Our choice to use $N_S(H)=N_S(G)$ stems from the fact that
typical \texttt{traceroute}-driven studies run from a relatively
small number of sources to a much larger set of destinations.
Rewriting
the expression $\theta(H;\rho_T(H)) = \theta(G;\rho_T(G))$ yields the
equation
\begin{equation}
N(G) = N(H)\, \rho(G,H)\enskip , \hbox{ where }
\rho(G,H) = \frac{\mathbb{E}[N^*(G)]}{\mathbb{E}[N^*(H)]} \enskip .
\label{eq:scaling}
\end{equation}
Now if $H$ is a known subgraph, $N(H)$ is known as well, and
the problem of inferring $N=N(G)$ can be reduced to one of inferring
$\rho(G,H)$.  A natural candidate for such a subgraph is the choice
$H=G^*$ i.e., the graph produced by an initial \texttt{traceroute}
sampling on $G$.  In this case, $N(H)=N^*$.  \\
It remains then to estimate $\rho(G,G^*)$, which must be defined conditional
on $G^*$.  In that case, the expectation in the numerator is
simply $\mathbb{E}[N^*(G)\, |\, G^*] =N^*$.  To estimate the other expectation,
$\mathbb{E}[N^*(G^*)\, |\, G^*]$, we use a strategy
based on the resampling of paths in $G^*$.  In particular, 
for a given sampling rate $\rho^*_T= \rho_T(G^*)$, we sample $N^*_T=\rho^*_T N^*$
targets on $G^*$ and create a resampled graph, say $G^{**}$, from 
the corresponding \texttt{traceroute} paths.  Let $N^{**}=N(G^{**})$.
We do this some number of times, say $B$, forming the collection
$N^{**}_1,\ldots,N^{**}_B$. Then, we estimate $\mathbb{E}[N^*(G^*)|G^*]$ by the
average $\bar{N}^{**}_B = (1/B)\sum_r N^{**}_r$.  \\
Plugging these quantities into the expression for $N$ in Eq.~\ref{eq:scaling} we get 
\begin{equation}
\label{eq:NhatRS}
\hat{N}_{RS} = N^*\,\cdot\, \frac{N^*}{\bar{N}^{**}_B}
\end{equation}
as a resampling-based estimator for $N$. \\ 
Note, however, that its derivation is based upon the premise that $\rho^*_T=\rho_T$, 
and $\rho_T$ is unknown (i.e., since $N$ is unknown).  This issue may
be addressed by noting that the equations $\rho_T(H)=\rho_T(G)$
and $\theta(H;\rho_T(H))=\theta(G;\rho_T(G))$ together imply
the equation $N_T(G)/N_T(H) = \mathbb{E}[N^*(G)]/\mathbb{E}[N^*(H)]$.  With respect
to the calculation of $\hat{N}_{RS}$, this fact suggests the strategy
of iteratively adjusting $N^*_T=N_T(G^*)$ until the relation
$N_T/N^*_T\approx N^*/\bar{N}^{**}_B$ holds (Robbins-Monroe algorithm \cite{robbins}).  
The value of $\bar{N}^{**}_B$
for the appropriate $N^*_T$ is then substituted into 
Eq.~\ref{eq:NhatRS} to produce $\hat{N}_{RS}$.
 In practice, one may increase $B$ as the algorithm approaches the condition 
$N^*_{T}/N_{T}$ $\approx$ $\bar{N}^{**}/N^{*}$.

{\bf \em Leave-One-Out Estimator - \quad}
Various other subsampling paradigms might be used to
construct an estimator. A popular one is the `leave-one-out'
strategy, which amounts to subsampling $G^*$ with 
$N^*_T=N_T-1$.  We apply such a principle  
to the problem of estimating $N$, in a way that 
does not require the subsampling assumptions in Eq.~\ref{eq:scaling}.\\
Recall that $\mathcal{V}^*$ is the set of all vertices discovered by a
\texttt{traceroute} study, including the $N_S$ sources 
$\mathcal{S}=\{s_1,\ldots,s_{N_S}\}$ and the $N_T$ targets 
$\mathcal{T}=\{t_1,\ldots,t_{N_T}\}$.  Our approach will be to connect
$N$ to the frequency with which individual targets $t_j$ are included
in traces from the sources in $S$ to the other targets in 
$\mathcal{T}\setminus \{t_j\}$.  Accordingly, let $\mathcal{V}^*_{i,j}$ be the
set of vertices discovered on the path from source $s_i$ to target
$t_j$, inclusive of $s_i$ and $t_j$.  Then the set of vertices
discovered as a result of targets other than a given $t_j$ can be
represented as $\mathcal{V}^*_{(-j)} = \cup_i\cup_{j'\ne j} \mathcal{V}^*_{i,j'}$.
Next define $\delta_j = I\left\{t_j\notin \mathcal{V}^*_{(-j)}\right\}$ 
to be the indicator
of the event that target $t_j$ is {\em not} `discovered' by traces to
any other target.  The total number of such targets 
is $X=\sum_j \delta_j$.\\
We derive a relation between $X$ and $N$. Assuming a 
random sampling model for selection of source and target nodes from $\mathcal{V}$, we have
\begin{equation}
\Pr\left(\delta_j=1\,|\, \mathcal{V}^*_{(-j)}\right) = 
\frac{N - N^*_{(-j)}}{N-N_S-N_T+1} \enskip ,
\label{eq:dgivenV}
\end{equation}
where $N^*_{(-j)}=\left|\mathcal{V}^*_{(-j)}\right|$.  Note that, by symmetry,
the expectation $\mathbb{E}\left[N^*_{(-j)}\right]$ is the same for all $j$: we denote
this quantity by $\mathbb{E}\left[N^*_{(-)}\right]$.  As a result, 
we may write
\begin{equation}
\label{eq:E[X]}
\mathbb{E}[X] = \sum_j \frac{N - \mathbb{E}\left[N^*_{(-j)}\right]}{N-N_S-N_T+1} 
     = \frac{N_T\left(N - \mathbb{E}\left[N^*_{(-)}\right]\right)}{N-N_S-N_T+1}
\enskip ,
\end{equation}
which may be rewritten as
\begin{equation}
\label{eq:Nl1out}
N = \frac{N_T \mathbb{E}\left[N^*_{(-)}\right] - (N_S+N_T-1) \mathbb{E}[X]}{N_T - \mathbb{E}[X]}
\enskip .
\end{equation}
To obtain an estimator for $N$ from this expression it is necessary to
estimate $\mathbb{E}\left[N^*_{(-)}\right]$ and $\mathbb{E}[X]$, for which it is natural
to use the unbiased estimators $\bar{N}^*_{(-)}= (1/N_T)\sum_j
N^*_{(-j)}$ and $X$ itself, measured during the 
\texttt{traceroute} study.  However, while substitution of these
quantities in the numerator of Eq.~\ref{eq:Nl1out} is fine, substitution
of $X$ for $\mathbb{E}[X]$ in the denominator can be problematic in the event
that $X=N_T$. Indeed, the estimator of $N$ diverges when none of the targets 
$t_j$ are discovered by traces to other targets, that is possible if $\rho_T=N_T/N$ is
small. A better strategy is to estimate the quantity $1/(N_T-X)$ directly. 
We assume that the overlap between different sampled set of vertices is 
very high. Using this condition, it is possible to derive an approximately unbiased estimator of $1/(N_T-X)$.
(Note that empirical data on the Internet collected by the Skitter project at CAIDA \cite{caida} 
show that the discovery rate is rather uniform, validating our assumption.)  
The same overlapping argument implies that $N^*_{(-j)}\approx N^*$, for 
all $j$, which suggests replacement of $\bar{N}^*_{(-)}$ by $N^*$.
Putting together all these quantities in Eq.~\ref{eq:Nl1out}, and with a bit of algebra 
(see Ref.~\cite{viger} for a detailed derivation) we get 
\begin{equation}
\label{eq:NhatL1Ofinal}
\hat{N}_{L1O}\approx (N_S+N_T) \,+\, \frac{N^* - (N_S+N_T)}{1-w^*}
\enskip ,
\end{equation}
where $w^* = X/(N_T+1)$, $X$ being the number of targets not
discovered by traces to any other target.\\
In other words, $\hat{N}_{L1O}$ can be seen as
counting the $N_S+N_T$ vertices in $\mathcal{S}\cup \mathcal{T}$ separately, and then
taking the remaining $N^*-(N_S+N_T)$ nodes that were `discovered'
by traces and adjusting that number upward by a factor of $(1-w^*)^{-1}$.
This form is in fact analogous to that of a classical method in the
literature on species problems, due to Good \cite{good}, in which the
observed number of species is adjusted upwards by a similar factor
that attempts to estimate the proportion of the overall population
for which no members of species were observed. 

\subsection{Numerical Results}\label{CHAP3_3_3}

We have tested the performances of the estimators on three different types of networks,
two computer-generated networks, with homogeneous (ER model) and heterogeneous (BA model) degree distribution, and a network
based on measurements of the real Internet (Mercator mapping project \cite{mercator}).
For the synthetic networks, we have considered average degree $6$, and sizes ranging
from $10^3$ to $10^6$ nodes.
The Mercator network ($N=228263$ nodes and $E=320149$ edges) has been used to see if more realistic topologies give results in agreement with that from the models.\\  
\begin{figure}[t]
\vskip .4in
\centering
\includegraphics[width=10.0cm]{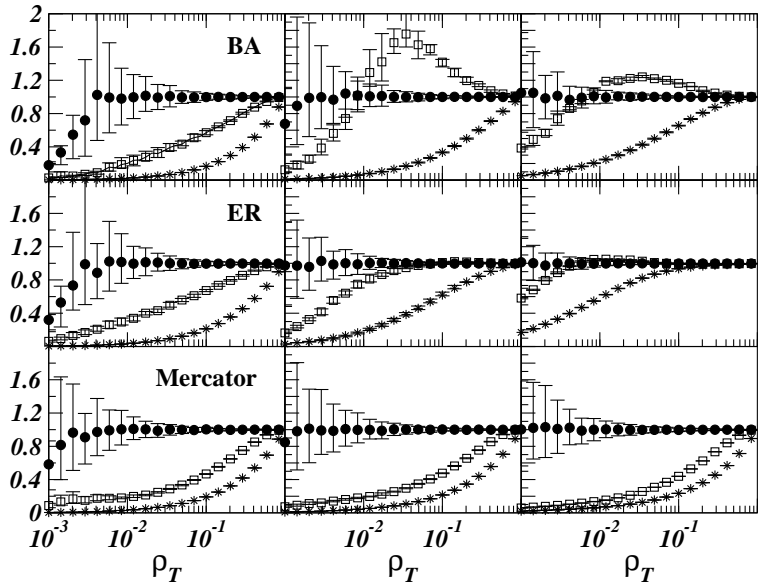}
\caption{Comparison of the various estimators for
the BA (top), ER (middle) and Mercator (bottom) networks. 
The curves show the ratios of the various estimators
to the true network size, as a function of the target density $\rho_T$.
Full circles: $\hat{N}_{L1O}/N$ ; Empty squares: 
$\hat{N}_{RS}/N$; Stars: $N^*/N$. The errorbars go from
the $10\%$ to the $90\%$ percentiles. 
Left figures: $N_S=1$ source; Middle: $N_S=10$ sources; Right:
$N_S=100$ sources} 
\label{estimators}
\end{figure}
We plot in Fig.~\ref{estimators} the ratio of the estimators to the
true size, $\hat{N}_{RS}/N$, and $\hat{N}_{L1O}/N$, together with $N^*/N$, for
the various investigated graphs, number of sources $N_S=1,10,$ and $100$, 
as a function of the target density $\rho_T$.
The improvement with respect to the ``trivial''
estimation by the size $N^*$ of the sampled graph is impressive, the optimal value 
being $1$ for all these curves. 
Increasing either the number of sources $N_S$ or density of targets $\rho_T$
yields better results, even for $N^*$, but the estimators we have introduced
converge much faster than $N^*$ towards values close to the true size
$N$.  In fact, a relatively small number of sources and targets is
sufficient for these estimators to perform very accurately, 
in particular, in the case of the ``leave-one-out'' estimator.
Note, however, that the ``leave-one-out'' estimator has a larger variability
at small values of $\rho_{T}$, while that of the resampling estimator is fairly constant
throughout.
This is because in calculating $\hat{N}_{RS}$ the uncertainty scales in the same way 
both in the sampling and in resampling, whereas for $\hat{N}_{L1O}$ the uncertainty scales with $\rho_{T}$
that changes passing from sampling to resampling.\\
\begin{figure}[t]
\vskip .4in
\centering
\includegraphics[width=10.0cm]{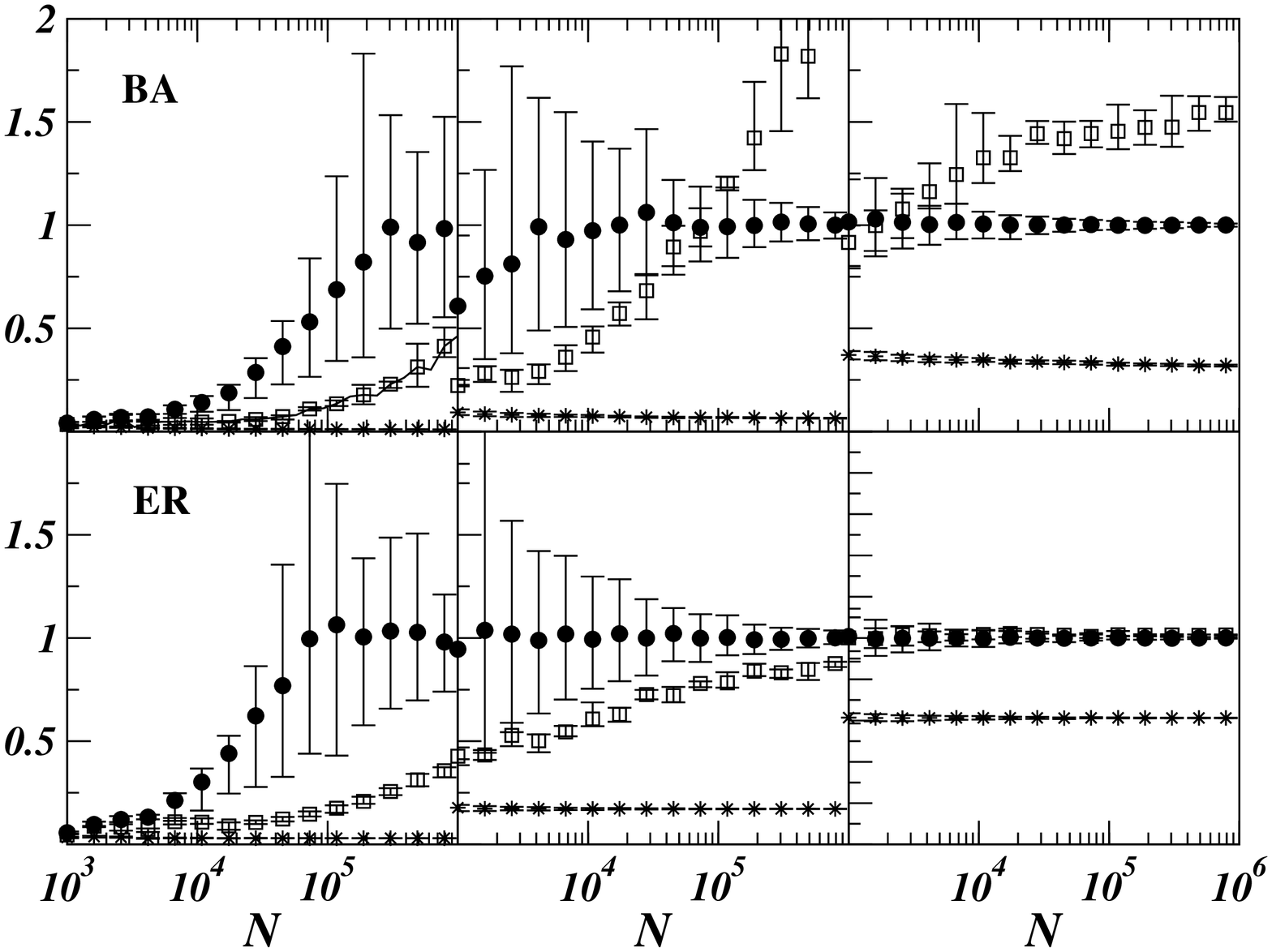}
\caption{Effect of the size $N$ of the graph $G$ for BA and ER graphs
at constant number of sources and density of targets.
The curves show the ratios of the various estimators
to the true network size, as a function of the graph size $N$.
Full circles: $\hat{N}_{L1O}/N$ ; Empty squares: 
$\hat{N}_{RS}/N$; Stars: $N^*/N$. The errorbars go from
the $10\%$ to the $90\%$ percentiles. $N_S=10$.
Left figures: $\rho_T=10^{-3}$; Middle: $\rho_T=10^{-2}$; Right:
$\rho_T=10^{-1}$.
}
\label{effectsize}
\end{figure}
In terms of topology, the estimation of $N$ appears to be easiest for the ER model.
Even $N^{*}$ is more accurate, i.e. the discovery rate is higher.
The estimation for the Mercator graph seems to be the hardest, possibly because 
it is the graph with the largest fraction of low degree (and thus low betweenness) 
nodes among the three studied.
It is worthy noting that the ``leave-one-out'' estimation seems to depend only on $N^{*}/N$ and $N_{T}$,
thus being quite stable in the three graphs.\\
Interestingly, the estimators perform better for larger sizes, as shown by 
Fig.~\ref{effectsize}, in which we investigate, at fixed $N_S$ and $\rho_T$, 
the effect of the real size of the graph $N$.  On the contrary, $N^*/N$ decreases. 
This is due to the fact that the sample graph $G^{*}$ gets bigger providing more and richer 
information, even if the discovery ratio does not grow.
The odd nature of the results for the BA graph comes from the peak associated with the resampling
estimator, see Fig.~\ref{estimators}. At fixed number of targets
however, $\hat{N}_{RS}/N$ and $\hat{N}_{L1O}/N$ decrease as $N$ increases (not shown, see Ref.~\cite{viger}).\\
The comparison of the two estimators show that the resampling estimator,
although yielding a clear improvement with respect to $N^*$, systematically
performs less well. This is probably due to the fact that the 
basic hypothesis of scaling (Eq.~\ref{eq:scaling}) is only approximately
satisfied, while for $\hat{N}_{L1O}$ the underlying hypotheses are 
well satisfied.
Nonetheless, the resampling estimator needs less information on the graph than $\hat{N}_{L1O}$, since
the ``leave-one-out'' estimator uses the knowledge that some targets
are not discovered by the paths to other targets.\\
Of course, in  order to apply this inference model to the Internet
one should take into account further issues, as the  
effects of non-random deployment of sources or the relation with more realistic models of \texttt{traceroute}
exploration.
However, the ``leave-one-out'' estimator should not suffer much in performance, since its derivation assumes 
only uniform random choice of targets, not that of sources and does not make any assumption on the routing strategy.\\
Finally, we provide a phenomenological validation of our technique.
The Internet size can also be estimated using other techniques, for instance with \texttt{ping} probes in order to test the response of some sufficient number  of randomly chosen IP addresses (the total number of possible IP addresses is $2^{32}$).
Computing the fraction $\hat{a}$ of `alive' addresses, we get the estimator $\hat{N}_{ping} = 2^{32}\hat{a}$ for the size of the Internet.\\
In order to check if the ``leave-one-out'' estimator gives results in agreement with the latter one, we performed about $3.7 \cdot 10^{6}$ \texttt{ping}'s probing of the net, obtaining $61 246$ valid responses and, in parallel, a \texttt{traceroute} study from the same source towards the same number of unique IP addresses. 
The estimated size was $\hat{N}_{ping} = 7,06 \cdot 10^{7}$ using the {\texttt ping} estimator, and $\hat{N}_{ping} = 7,23\cdot 10^{7}$ using the ``leave-one-out'' estimator, confirming a good agreement between the two.
Note that the numbers are not intended to taken too seriously (we used only one source, toward a small number of targets),
but it is important that they give consistent results.

\newpage
\section{Conclusions}\label{CHAP3_4}

In contrast with usual physical disciplines, the science of complex networks is characterized by the absence of 
first principles from which a theoretical framework can be deduced; for this reason, a very actual 
and urgent problem is that of testing the reliability of phenomenological data, on which the whole theoretical 
analysis and modeling of networks is based.
The first step in this direction is that of assessing the validity of the peculiar topological properties observed in
so many real networks, from biological to social and technological ones. 
This means that we need an accurate description of the sampling processes by which data are collected.\\
In this chapter, we have highlighted the main features of different mechanisms of  sampling that are 
involved in the experimental analysis of networks. In some cases, such as for biological and social networks, sampling methods 
may suffer of strong biases that are often unpredictable, because of the incomplete knowledge of all (hidden) processes involved in the experiments.
In the case of the Internet, instead, there exists a well-known and very useful practical method to extract information from the network, that consists in the dynamical exploration by means of \texttt{traceroute}-like probes.
Unfortunately, also this method suffers of biases that could in principle cause a misrepresentation of the topological
properties of the network, in particular of the degree distribution. \\
By means of a mean-field statistical approach, we have provided an insight into the relation between sampling biases and
topological properties of the network, {\em showing that the efficiency and accuracy of the exploration process depend mainly on the betweenness centrality distribution of the underlying network}.  
The other important parameters are the densities of sources and targets deployed in the system, or more simply, the probing effort, i.e. the number of probes used in the exploration.
In particular, we have explained how the observed distortions of the degree distribution are actually possible only when the underlying network is homogeneous and the \texttt{traceroute} process is performed using very few sources.
Moreover, in order to observe power-laws starting from homogeneous networks, the average degree of the original network must be of the same order as the power-law cut-off, a rather unlikely situation for any real technological network (in particular for the Internet at the AS level).\\
Our results provide a strong evidence that in order to observe a power-law distribution, we need at the origin a broad distribution, but show as well that {\em networks with other heterogeneous distributions, different from power-laws, can be seen as scale-free networks under sampling}. 
With a low number of sources and targets, as that involved in outdated Internet mapping projects, we cannot completely exclude the possibility that the degree distribution of the real network has the shape of a Weibull distribution rather than of a stretched exponential.
According to our analysis, in case of heterogeneous networks with a non power-law degree distribution, an increase of both source and target densities should make visible departures from it. \\
In agreement with our conclusion on the necessity of increasing the number of sources used in the Internet mapping projects in order to get more accurate data, it has been recently developed a distributed mapping project called DIMES (Distributed Internet MEasurements \& Simulations) \cite{dimes,dimes_tech}, with the aim of studying the structure and topology of the Internet using a very large number of sources. The idea is that of involving in the measurements common Internet users, creating a volunteer community that sends \texttt{traceroute}-like probes throughout the Net.  
Preliminary results from the DIMES project confirm the power-law shape for the degree distribution.
The fact that DIMES is observing power-laws seems to be the ultimate evidence of the genuine nature of scale-free topologies in the Internet.
Nonetheless, it is still very likely that the real exponent is different from (i.e. larger than) the observed one. \\
In conclusion, the main improvement upon this field due to the results obtained in this thesis are summarized in the following points:
\begin{itemize}
\item We have identified, at a mean-field level, the relations between observed properties of the sampled networks and the topological properties of the underlying network. 
\item The origin of sampling biases in the Internet measurements has been uncovered. These properties, together with a minimal set of external parameters (the density of sources and targets) govern the efficiency and accuracy of the sampling process. 
\item Tuning the external parameters we have provided some results that should serve as a guidance for more realistic optimization strategies.  
\item The effects of \texttt{traceroute} explorations on local and non-local structures have been analyzed.
\item The \texttt{traceroute}-like explorations can be translated in the framework of statistical species problems, leading to the formulation of non-parametric statistical approaches for the inference of Internet's global properties. 
As an example of the validity of this method to correct the biases introduced by the sampling process, 
we have introduced unbiased estimators of the number of nodes in the network, that can be applied to the case of the Internet, the real size of which is still unknown.
\end{itemize}

\chapter[Spreading and Vulnerability]{Spreading and Vulnerability in Complex Networks}
\label{CHAP4}

\section{Introduction}\label{CHAP4_1}
A second topic that has been object of investigation in this thesis concerns the role played by some topological and dynamical properties in determining the functionality of weighted networks. In particular, we focus on the analysis of vulnerability and spreading properties. The motivations for this type of analysis, coming from the observation of the functional properties of real infrastructure networks, are exposed in Section~\ref{CHAP4_1_1}; while in Section~\ref{CHAP4_1_2}, we introduce the main idea of the chapter, consisting in the possibility of linking two apparently different subjects as vulnerability and spreading by means of percolation-like approaches.
Then, in Section~\ref{CHAP4_2}, we focus on the vulnerability of weighted networks, taking into account the case study of the airports network. 
In Section~\ref{CHAP4_3}, we propose a theoretical framework for the study of spreading phenomena in weighted networks, using the methods of percolation theory.

\subsection{Motivations}\label{CHAP4_1_1}

From a practical point of view, an accurate knowledge of structural and dynamical properties of networks is of primary interest, particularly in the case of technological networks, that form the backbone of world-wide communication and transportation infrastructures. 
On the other hand, such networks are intrinsically weighted networks, so that neglecting weights could lead to wrong conclusions about these properties.
As mentioned in Chapter~\ref{CHAP1}, weights are usually related to dynamics: in some cases they give just a measure of the traffic on the edges, in other cases they pinpoint the different attitude of the edges in exchanging physical quantities (energy, information, goods, etc).\\
From the point of view of a single dynamical process evolving on a network (e.g. a spreading process), the introduction of weights on the edges corresponds, using a pictorial representation, to add an energy-like variable to the system and look at it as it was in an energy landscape.  
In this scenario, some paths with larger weights (or smaller, it depends on weight's definition) will be dynamically more convenient than others, and thus they will be preferentially chosen in the spreading.   
In addition, as the trajectory of a particle in a potential is deformed when we change the shape of the potential, fluxes on weighted networks adjust themselves in order to follow variations of both topology and weights. In other words, the presence of weights on the edges tunes the functionality of the system towards an optimal point.
This picture suggests us the idea of technological networks as {\em critical infrastructures}, whose topological, dynamical and functional properties are intimately related.\\
The motivations of our research in the field of weighted networks can be summarized in the following issues:
\begin{itemize}
\item Which are the levels of structural and functional stability of such critical structures if we perturb it in different ways? 
\item Which are the conditions for a macroscopic spreading? Which is the relation with the functional properties of the network?  
\end{itemize}
Both our queries are concerned with the prediction of extreme events such as that of a global collapse triggered by the spreading of a virus, a cascade failure or malicious attacks, and with the necessity to protect real networks, ensuring their functioning. 
Nevertheless, the two arguments are developed separately. \\
We will first study the vulnerability properties of weighted graphs, showing that structural robustness does not coincide with functional robustness if traffic is taken into account. 
The analysis is carried out by means of a relevant case study, where practical implications are visible: the airports network. 
The weighted representation is, furthermore, general enough to include other contributions, not strictly due to the traffic. For instance, infrastructure networks are embedded in the real space, with euclidean distances between nodes, and longer connections have higher costs.
It is thus possible to verify the role played by all these quantities in determining the functional integrity of the 
system.\\
We will then turn to consider the conditions for observing macroscopic spreading phenomena in the context of weighted networks, keeping in mind the previous results about the functional properties of infrastructure networks. 
We provide {\em a general theory for spreading phenomena in weighted networks}.
Some analytical results are presented in the case of (correlated) generalized random graphs, for which we derive 
a very general criterion for the percolation threshold, that governs the spreading properties of 
many dynamical phenomena (see Appendix~\ref{APP4_2}).\\
In order to have a unitary view of these two topics, we can actually consider the idea of the existence of 
different scales for the dynamics that we have mentioned as an introductory argument in Chapter~\ref{CHAP1}: 
a slower timescale governs the evolution of the average traffic (i.e. the weights), and a faster one characterizes single spreading processes (and thus also the perturbations to the average quantities).\\
When a cascade failure, a congestion phenomenon, or a disease pops out producing a perturbation of the normal functioning of the network, the underlying weighted structure appears as a fixed landscape, whose (weighted) properties can decisively influence the global effect of such a perturbation (e.g. the occurrence of a traffic collapse, a pandemic state, etc).  
The advantage of using quenched weighted networks is related to the possibility of describing dynamical phenomena as well as vulnerability easily using percolation-like approaches.

\subsection{Relation between percolation, vulnerability and spreading}\label{CHAP4_1_2}

Percolation theory has been largely used in statistical mechanics in order to describe topological phase transitions in lattice systems, such as the emergence of global properties when some physical parameter (temperature, concentration, etc) exceeds a critical value \cite{stauffer,essam,bunde12,kirkpatrick}. 
More precisely, a site percolation process can be sketched as follows. Each site of the lattice is occupied with a uniform probability $q$, called {\em occupation probability}, and empty with probability $1-q$. A cluster is a connected set of occupied sites. When the value $q$ of the occupation probability is larger than a critical value $q_c$, the {\em percolation threshold}, a cluster with size of the same order of the system, i.e. infinite in the thermodynamic limit, appears.
Analogously, the process can be defined on the bonds joining neighboring sites, with the only difference that the site occupation probability $q$ is replaced by a bond occupation probability. The bond percolation corresponds to the site percolation on the conjugate lattice (obtained exchanging bonds with sites and viceversa), and gives similar results.\\
\begin{figure} 
\centerline{
\includegraphics[width=12.0cm]{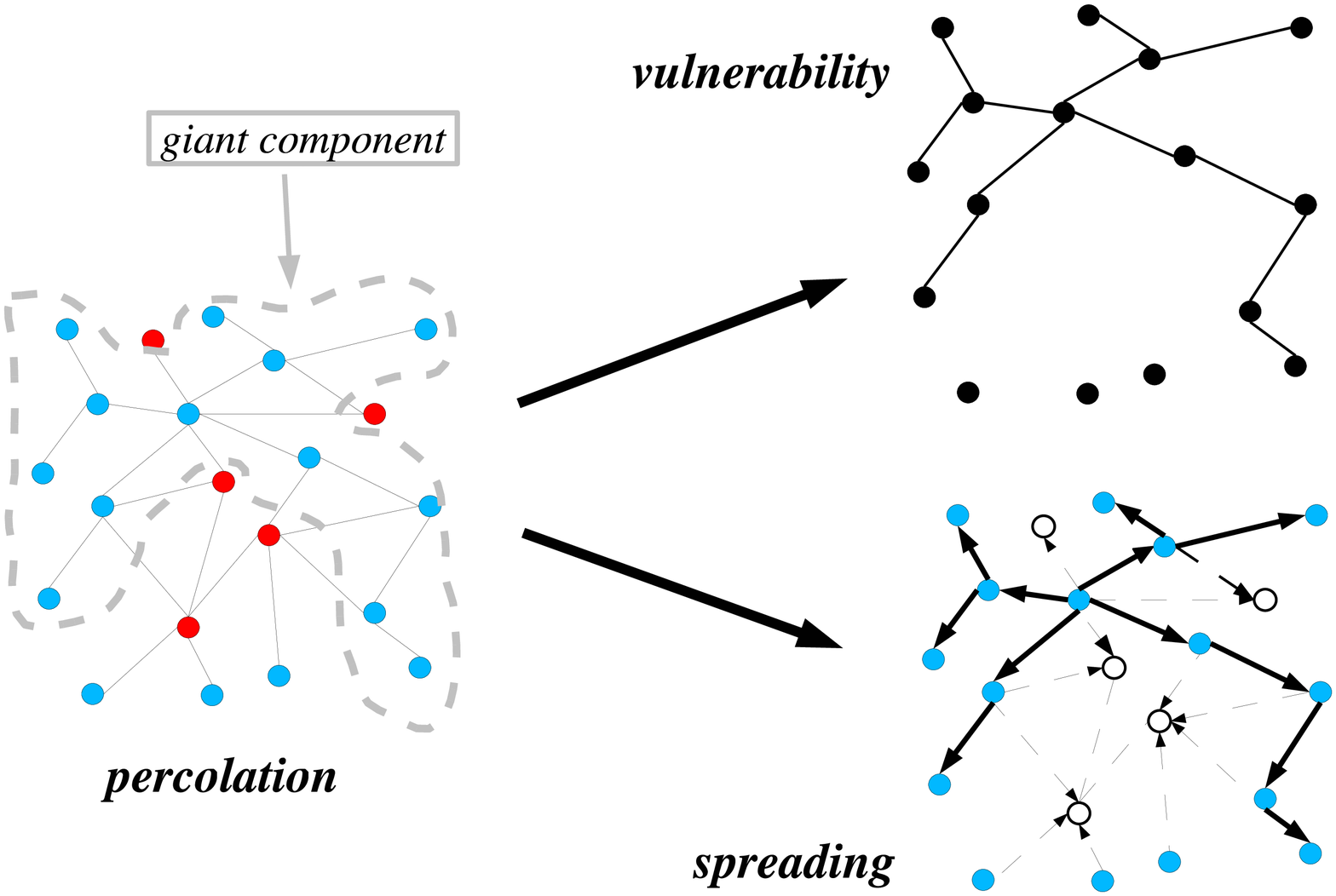}
}
\caption{Illustration of the relation between percolation, vulnerability and spreading processes. As a result of percolation (left), the nodes of the original graph are divided into occupied (blue) and unoccupied (red) nodes. On the right: the correspondence between node removal (top) or spreading (bottom). On average, the subgraph induced by occupied nodes (black nodes in the top figure) coincides with that covered by spreading (blue nodes in the bottom figure).}
\label{perco_ill}
\end{figure}
Recently, percolation theory has been successfully introduced in the field of complex networks, with some seminal works on undirected uncorrelated random graphs \cite{callaway,cohen2,cohen5}. A number of other works followed, studying
the cases of correlated \cite{newman3,newman4,vazquez,hopcroft} and directed \cite{dorogo2,schwartz,boguna} random graphs.\\
{\em Percolation is intrinsically related to vulnerability}. The process of random removal of a fraction $g$ of nodes can be seen as a percolation process in which the occupation probability of the nodes is reduced to $1-g$. Actually, we can exactly map a random removal of nodes on a site percolation problem, with $g = 1- q$ and the same order parameter (i.e. the relative size of the giant component, or topological integrity).
Percolation theory predicts the presence of a threshold value $g_c = 1- q_c$ above which the giant component $\mathcal{G}_{c}$ disappears, i.e. the network is fragmented in many disconnected components of very small sizes ($N_{g} \sim \mathcal{O}(\log N)$) with respect to the total number $N$ of vertices in the graph. \\
The notion of percolation has then {\em found applications in the context of spreading phenomena on networks, such as  epidemic models} \cite{moore1,moore2,newman1}: for some of them, the connection with the percolation is a longtime topic (see for example \cite{frisch,grassberger,warren,warren2}), while for other models, more rigorous descriptions, based on the analysis of population equations of epidemiology, are required \cite{may,pastor1,pastor2,boguna3,moreno1,moreno2,pastor1,pastor2,boguna4,boguna2}.
The models of epidemic spreading solved using percolation theory identify vertices with
individuals and edges as pairwise contacts between them \cite{newman1}. 
In this context, the existence of a giant component means that we can reach a large (infinite) number of other nodes of the same graph starting by one of them and moving along the edges. 
The spreading process takes place on the edges, that have a fixed probability to be disease-causing contacts, called {\it transmissibility}. The model's behavior is easily determined using edge percolation, that establishes the presence of a giant component as function of the value of transmissibility, and states the possibility for global epidemic outbreaks. 
Figure~\ref{perco_ill} illustrates the relation between percolation, vulnerability and spreading processes.
In Ref.~\cite{spreading_dallasta}, we have extended the analogy between spreading processes and percolation to a very general situation in which node and edge inhomogeneity is considered. Part of the analysis is reported in Appendices~\ref{APP4_1}-\ref{APP4_2}. 
As a special case of such inhomogeneous percolation, in Section~\ref{CHAP4_3}, we expose some results that are useful to study spreading processes on weighted networks. 

\newpage
\section{Vulnerability of Weighted Networks: a case study}\label{CHAP4_2}

Recent studies on phenomenological data of real weighted networks 
(communication and infrastructure networks, scientific collaboration
networks, metabolic networks, etc) have given much attention to 
the relation between weighted properties and topological quantities~\cite{barrat_PNAS,almaas:2004,Li:2003a}.
The most striking result concerns the existence of non-trivial correlation between weights and topology,
corroborating the idea that topological and weighted properties are mutually 
influenced, and both take part into the evolution process of real weighted networks.\\ 
Several groups have proposed models of evolving weighted networks  \cite{BBV_model,BBV_model3,spatial_BBV,bianconi_weight,dorogovtsev:2004,wang:2005};
we will consider only one of them, the spatial BBV model, that has been described in Section~\ref{CHAP2_4_3}.\\
In the present section, we face the issue of defining and determining the vulnerability of 
weighted networks. It is firstly a problem of definition, since in many real networks, 
together with the topological connectivity there are also weighted properties, that are usually related 
with the functionality of the network. \\
From a topological point of view, the only relevant measure of the vulnerability of a network
is the size $N_{g}$ of the giant component after the removal of a fraction $g$ of the nodes.
If we compare this size with the original size $N_{0}$ of the network's giant component, we have an estimate of the 
vulnerability of the network under node (or edge) removal.\\
Using the already mentioned analogy with percolation, it has been shown 
that, in contrast with regular lattices and homogeneous random
graphs, heavy-tailed networks can tolerate very high levels of random
failure, without falling into pieces ~\cite{cohen2,callaway}. 
On the other hand, malicious attacks to the hubs can easily break 
the entire network into small components, providing a clear identification 
of the elements which need the highest level of protection against such attacks
\cite{barabasi00,pastor1}.\\
In order to extend the vulnerability analysis to weighted complex networks, 
we have investigated how the introduction of traffic and spatial 
properties may alter or confirm the above findings. 
In particular, in Ref.~\cite{vulnera_dallasta}, we address three main questions: 
\begin{itemize}
\item[(i)] on the basis of which measures a weighted network is most efficiently protected from damage; 
\item[(ii)] how traffic and spatial constraints influence the system's robustness;
\item[(iii)] how to measure the damage (different integrity measures). 
\end{itemize}
The first step in order to face these issues consists in proposing a series of topological and 
weight-dependent centrality measures that
can be used to identify the most important vertices of a weighted
network. 
The functional integrity of the whole network depends on the
protection of these central nodes, as we will show applying these 
considerations to the World-wide Air-transportation Network.  
The main result is that {\em under malicious attacks, weighted networks turn out 
to be more vulnerable than expected, in that the traffic integrity is destroyed 
when the topological integrity of the network is still very high}. 

\subsection{Weighted centrality measures}\label{CHAP4_2_1}
We introduce a series of weighted centrality measures, that will be used to define attack 
strategies of node removal on the network. Since we consider the case of technological 
networks, such as the airport network, we focus on weighted centrality measures based on 
traffic and spatial properties.
The presence of non-trivial correlations between the different centrality measures 
reveals that {\em all of them should be considered in order to give an exhaustive description 
of weighted network's structural and functional properties}.\\  
 
\noindent
{\bf \em Measures definition - \quad} 
The centrality of nodes can be quantified by various measures, the most intuitive of which is certainly the degree. 
The natural generalization of the degree to a weighted graph is given by the strength of
vertices $s_{i}$ (see Section~\ref{CHAP2_2_6} for definitions), that in
the case of the air-transportation network quantifies the traffic
of passengers handled by the airport $i$. 
The corresponding geographical notion is the {\em distance strength} $D_i$, 
that we defined as
\begin{equation}
D_i = \sum_{j \in {\cal V}(i)} d_{ij}~,
\end{equation}
where $d_{ij}$ is the {\em Euclidean} distance between $i$ and
$j$, and quantifies how long are the connections supported by an airport.
Another interesting definition is given by mixing the two ingredients; 
the {\em outreach} 
\begin{equation}
O_i = \sum_{j \in {\cal V}(i)} w_{ij} d_{ij}~,
\end{equation}
measures the total distance traveled by passengers from the airport $i$ and 
is possibly related to the total cost of connections.  \\
In general, non-local effects are taken into account introducing the betweenness centrality
(see Section~\ref{CHAP2_2_4}), that identifies crucial nodes which may have small
degree or strength but act as bridges between different parts of the
network.  
In weighted networks, it seems natural to generalize the notion of 
betweenness centrality through a {\em weighted betweenness centrality} 
in which shortest paths are replaced with their weighted versions.  
As already anticipated in Section~\ref{CHAP2_2_6}, a straightforward way to generalize the 
hop distance in a weighted graph consists in assigning to each
edge $(i,j)$ a length $l_{ij}$ that is a function of the
characteristics of the link. In our case, both 
the weight $w_{ij}$ and the Euclidean distance
$d_{ij}$ between airports $i$ and $j$ are reasonable quantities. 
A quite natural assumption is that the effective distance between 
two linked nodes should be a decreasing function of the weight of the link. 
Indeed, larger flows (traffic) correspond to more frequent and faster exchange 
of physical quantities (e.g. information, people, goods, energy, etc). 
Moreover, when spatial embedding is considered, the edge length must be directly proportional to 
the geographical distance.
In other words, we define the length $l_{ij}$ of an edge $(i,j)$ as
\begin{equation}
l_{ij}=\frac{d_{ij}}{w_{ij}} \ .
\end{equation}
The definition of weighted betweenness centrality follows straightforwardly.\\
According to this definition, the centrality of a node provides a trade-off between 
the idea of the topological ``bridge'' connecting different
parts of a network, and the requirement of carrying a sufficient level of traffic. \\
\begin{table}[t]
\begin{center}
\begin{tabular}{cccccc}
\hline
{\footnotesize Rank} & {\footnotesize Degree} & {\footnotesize Betweenness} & {\footnotesize Strength}  & 
{\footnotesize Outreach} & {\footnotesize W. Betweenness}\\
 \hline
{\footnotesize 1} & {\footnotesize Frankfurt}   & {\footnotesize Frankfurt}      & {\footnotesize Atlanta}        &  {\footnotesize London-LHR} & {\footnotesize London-LHR} \\ 
{\footnotesize 2} & {\footnotesize Paris-CDG}   & {\footnotesize Paris-CDG}      & {\footnotesize Chicago-ORD}    & {\footnotesize L.Angeles-LAX} & {\footnotesize L.Angeles-LAX} \\
{\footnotesize 3} & {\footnotesize Munich}      & {\footnotesize Anchorage}      & {\footnotesize London-LHR}     & {\footnotesize Tokyo-HND} & {\footnotesize Singapore}   \\ 
{\footnotesize 4} & {\footnotesize Amsterdam}   & {\footnotesize Tokyo-HND}      & {\footnotesize Tokyo-HND}      & {\footnotesize Frankfurt} & {\footnotesize New-York-JFK}  \\
{\footnotesize 5} & {\footnotesize Atlanta}     & {\footnotesize Port Moresby}   & {\footnotesize L.Angeles-LAX} & {\footnotesize Paris-CDG} & {\footnotesize Miami-MIA}  \\
{\footnotesize 6} & {\footnotesize London-LHR}  & {\footnotesize London-LHR}     & {\footnotesize Dallas-DFW}     & {\footnotesize Chicago-ORD} & {\footnotesize Tokyo-HND}  \\
{\footnotesize 7} & {\footnotesize Chicago-ORD} & {\footnotesize L.Angeles-LAX} & {\footnotesize Paris-CDG}      & {\footnotesize New-York-JFK} & {\footnotesize Hong-Kong}  \\
{\footnotesize 8} &  {\footnotesize Duesseldorf} & {\footnotesize Singapore}      & {\footnotesize Frankfurt}      & {\footnotesize Singapore} & {\footnotesize Sydney}    \\ 
{\footnotesize 9} & {\footnotesize Vienna}      & {\footnotesize Vancouver }     & {\footnotesize Phoenix }       & {\footnotesize San Francisco} & {\footnotesize Chicago-ORD}  \\
{\footnotesize 10} & {\footnotesize Brussels}   & {\footnotesize Bangkok}        & {\footnotesize Denver}         & {\footnotesize Atlanta} & {\footnotesize Paris-CDG}   \\
{\footnotesize 11} & {\footnotesize Dallas-DFW}  & {\footnotesize Johannesburg}   & {\footnotesize Hong Kong}      & {\footnotesize Hong-Kong} & {\footnotesize Seattle}  \\
{\footnotesize 12} &  {\footnotesize Houston-IAH} & {\footnotesize Toronto-YYZ}    & {\footnotesize Detroit-DTW}    & {\footnotesize Bangkok} & {\footnotesize Atlanta}  \\
{\footnotesize 13} & {\footnotesize Rome-FCO}    & {\footnotesize Amsterdam}      & {\footnotesize Minneapolis-MSP}& {\footnotesize Amsterdam} & {\footnotesize Dubai} \\
{\footnotesize 14} & {\footnotesize Minneapolis} & {\footnotesize Seoul-ICN}  & {\footnotesize Madrid}         & {\footnotesize Dallas-DFW} &  {\footnotesize Frankfurt}  \\
{\footnotesize 15} & {\footnotesize Zurich}      & {\footnotesize Hong Kong}      & {\footnotesize Houston}        & {\footnotesize New-York-EWR} &  {\footnotesize Brisbane} \\
\hline
 \end{tabular}
\caption{Ranking of the fifteen world's top airports for different centrality measures:
degree, topological betweenness, strength, outreach and weighted
betweenness. Note that the lists report different single airports, not
just the total data relative to the cities; for cities with more
than one airport, the acronym of the corresponding one is indicated.}
\label{tab_top15}
\end{center}
\end{table}


\noindent {\bf \em Centrality Measures Correlation - \quad} 
On the World-wide Air-transportation Network, all the proposed measures of centrality
are broadly distributed, and they are on average non-trivially 
correlated one with the other, and in particular with the degree 
(e.g. $D(k) \sim k^{\beta_D}$ with $\beta_D \approx 1.5$~\cite{spatial_BBV}, and $O(k)
\sim k^{\beta_O}$, with $\beta_O \approx 1.8$ \cite{vulnera_dallasta}).\\ 
Vertices with large degree have also typically
large strength and betweenness, but under a detailed analysis, 
we observe that deviations are possible. 
This fact has been already noted in Refs.\cite{luisair,spatial_BBV}, where it is stressed 
that the most connected airports do not necessarily have the largest betweenness centrality. \\
Here, we observe a similar effect about the relation between topological and weighted measures.
For instance, the scatter plot of the weighted betweenness vs.
topological betweenness (not shown) shows that departures from a perfect correlation are not that rare.
Let us consider the list of Top 15 airports according to different measures of centrality (Table~\ref{tab_top15}):
strikingly, each definition provides a different ranking. In addition, some airports which
are very central according to a given definition, become peripheral
according to another criteria. For example, Anchorage has a large
betweenness centrality but ranks only $138^{th}$ and $147^{th}$ in terms
of degree and strength, respectively. Similarly, Phoenix or Detroit
have large strength but low ranks ($>40$) in terms of degree and
betweenness.\\
\begin{figure}
\centerline{
\includegraphics[width=9.0cm]{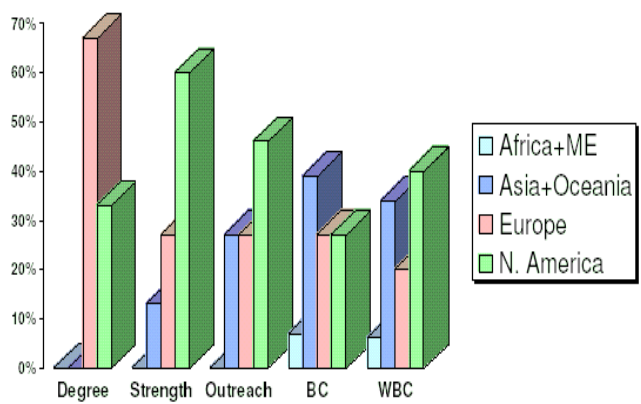}}
\caption{Geographical distribution of the world's $15$ most central
airports ranked according to different centrality measures.}
\label{fig:geo}
\end{figure}
A quantitative analysis of the correlations between two rankings of
$n$ objects can be done using rank correlations such as Kendall's
$\tau$~\cite{kendall,num_rec}
\begin{equation}
\tau=\frac{n_c-n_d}{n(n-1)/2}
\end{equation}
where $n_c$ is the number of pairs whose order does not change 
in the two different lists and $n_d$
is the number of pairs whose order is inverted. This quantity is
normalized between $-1$ and $1$:
$\tau=1$ corresponds to identical ranking while $\tau=0$
is the average for two uncorrelated rankings and $\tau=-1$ is a
perfect anticorrelation.
Table~\ref{tab2} gives the values of $\tau$ for all possible
pairs of centrality rankings. For $N=3,880$, two random rankings
yield a typical value of $\pm 10^{-2}$ so that even the smallest
observed $\tau=0.21$ is the sign of a strong correlation (All the
values in this table were already attained for a sublist of only
the first $n$ most central nodes, with $n\approx 500$). 
Remarkably enough, there is a strong correlation even between local and non-local 
quantities (degree vs. betweenness), but the weighted quantities seem 
to have a lower level of correlation with the topological ones.\\
\begin{center}
\begin{table}[b]
\begin{center}
\begin{tabular}{lccccccc}
\hline
  & $k$  & $D$ & $s$  &  $O$ &$BC$ & $WBC$ &\\
 \hline
Degree   $k$           &  1   & 0.7&   0.58  & 0.584  & 0.63 & 0.39 & \\ 
Distance strength $D$   & 0.7  & 1  &  0.56    & 0.68    & 0.48 & 0.23 & \\ 
Strength       $s$      & 0.58 & 0.56   &  1   & 0.83 &  0.404 & 0.24 & \\ 
Outreach   $O$    & 0.584 & 0.68 & 0.83 & 1  &  0.404  & 0.21 & \\ 
Betweenness     $BC$  & 0.63 &  0.48    &  0.404 & 0.404 & 1& 0.566& \\ 
Weighted $BC$  & 0.39 &  0.23 &  0.24 & 0.21 & 0.566 & 1 & \\ 
\hline
 \end{tabular}
 \end{center}
\caption{Similarity between the various rankings as measured
by Kendall's $\tau$. For random rankings of $N$ values, the typical
$\tau$ is of order $10^{-2}$.}
\label{tab2}
\end{table}
\end{center}
Another important aspect concerns the geographical 
relation between centrality measures. 
Indeed, they are non homogeneously distributed in different 
geographical regions, providing a further precious hint to understand 
the functioning of infrastructure networks such as the WAN. 
Figure~\ref{fig:geo} displays the geographical
distribution of the world's fifteen most central airports ranked
according to different centrality measures. On the one hand,
it is clear that topological measures miss the economic dimension 
of the WAN, while weighted measures reflect traffic and economic
realities. \\
European airports are very well connected but the core of the traffic 
is North America, where we find the peaks of strength and 
outreach.  
Betweenness based measures on the other hand pinpoint the
most important nodes in each geographical zone. In particular, the
weighted betweenness appears as a balanced measure which combines
traffic importance with topological centrality, leading to a more
uniform geographical distribution of the most important nodes.

\subsection{Vulnerability of the WAN}\label{CHAP4_2_2}

The analysis of complex networks robustness has been largely investigated in 
the case of unweighted networks~\cite{barabasi00,cohen2,callaway,holme02}, focusing on 
the topological integrity of the network $N_g/N_0$. 
When, increasing $g$, we reach $N_g \simeq \mathcal{O}(1)$, the entire
network has been destroyed. 
We say that the attack strategy is ``driven'' by a property of the network, if the order in which the nodes are one-by-one removed corresponds to the ranking list according to that property. 
On heterogeneous networks, a degree-driven damage strategy (i.e. nodes are removed in order of degree, starting from the maximum one) is extremely effective, 
leading to the total fragmentation of the network at very low values of
$g$~\cite{cohen2,callaway,barabasi00}, but the removal of the
nodes with largest betweenness typically leads to an even faster
destruction of the network~\cite{holme02}.\\
\begin{figure}[t]
\vskip 0.3in
\centerline{
\includegraphics[angle=-90,width=9.0cm]{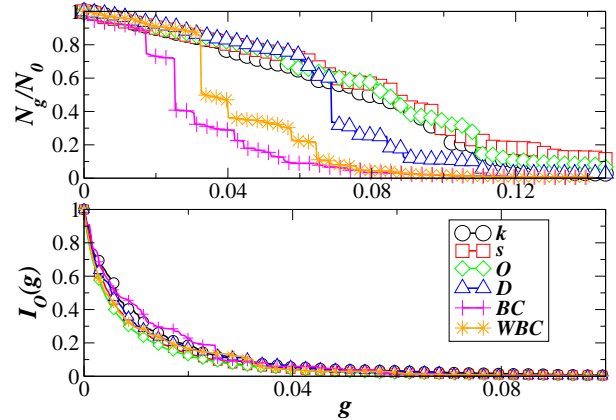}
}
\caption{Effect of different attack strategies on the size of the
connected giant component (top) and the outreach integrity (bottom). The attack strategies consist in removing nodes in order of: degree (black circles), strength (red squares), outreach (green diamonds), distance strength (blue triangles), betweenness (magenta crosses) and weighted betweenness (orange stars).}
\label{fig:damage}
\end{figure}
In the case of weighted networks, we consider additional quantities related 
to the functionality of the network, such as the
largest traffic or strength still carried by a connected component of
the network. In Ref.~\cite{vulnera_dallasta}, we have defined three new measures of network's damage
\begin{equation}
I_s(g)=\frac{{\cal S}_g}{{\cal S}_0}, \ \ \
I_O(g)=\frac{{\cal O}_g}{{\cal O}_0}~,\ \ \ 
I_D(g)=\frac{{\cal D}_g}{{\cal D}_0} ~,
\end{equation}
where ${\cal S}_{0}=\sum_i S_i$, ${\cal O}_0=\sum_i O_i$ and
${\cal D}_{0}=\sum_i D_i$ are the {\em total strength}, {\em outreach}
and {\em distance strength} in the undamaged network and
${\cal S}_g=\mbox{max}_H\sum_{i\in H} S_i$,
${\cal O}_g =\mbox{max}_H\sum_{i\in H} O_i$ 
and ${\cal D}_g=\mbox{max}_H\sum_{i\in H} D_i$
correspond to the largest strength, outreach or distance strength 
carried by any connected component $H$ in the network, after
the removal of a density $g$ of nodes. These quantities measure
the {\em integrity} of the network with respect to either
strength, outreach or distance strength, since they refer to the
relative traffic or flow that is still handled in the largest operating
component of the network.\\
Under random damage, all the integrity measures behave similarly to the simple topological case,
i.e. weighted networks are inherently resilient to random damages. 
This result is in agreement with the theoretical prediction of the absence 
of a percolation threshold in highly heterogeneous graphs
\cite{cohen2,callaway}, but it is not completely expected in weighted graphs.
The scenario corresponding to the damage of the most central nodes in the network 
is very different and depends on the strategy considered.
We have eliminated nodes (and corresponding links) according to their rank in 
terms of degree, strength, outreach, distance strength, topological betweenness, and
weighted betweenness.
\begin{figure}[t] 
\vskip .3in
\centerline{\includegraphics[width=9.0cm]{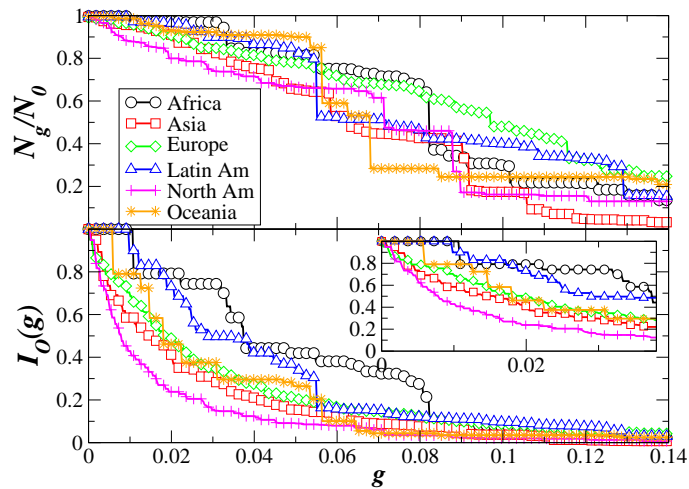}}
\caption{Geographical effect of the removal of nodes with largest
strength. The integrity decreases strongly in regions such as
North-America, while a ``delay'' is observed for the zones with
smaller initial outreach or strength.}
\label{fig:geo2}
\end{figure}
In Fig.~\ref{fig:damage} we report the behavior of $N_g/N_0$ and of
the outreach integrity $I_O(g)$ for all cases (the other integrities giving similar results). 
As expected, all strategies lead to a rapid breakdown of the network 
with a very small fraction of removed nodes. 
The structural integrity of a network is principally due to strategic points 
such as bridges and bottle-neck structures; as evidenced by the
fact that the size of the giant component decreases faster using non-local measures of
centrality (i.e. betweenness) instead of local ones (i.e.  degree, strength).
In Ref.~\cite{holme02}, it was already shown that the betweenness is the most effective 
quantity in order to pinpoint such nodes.  \\
However, the effects are reduced if we take into account weights:
 some of the important topological bridges
carry a small amount of traffic and are therefore part of more
shortest paths than weighted shortest paths. These bridges have
therefore a lower rank according to the weighted betweenness.\\ 
Note that, among the local quantities, the
 distance strength is rather effective since it targets
nodes which connect very distant parts of the network.\\
The picture changes when the attention is shifted to the weighted 
integrity measures. In this case, all strategies achieve the same level of damage, 
and their decrease is even faster than for topological quantities.
In other words, the functionality of the network 
can be temporarily jeopardized in terms of traffic even if the
physical structure is still globally well-connected.
This implies that {\em weighted networks appear more fragile than expected by   
considering only topological properties}. All targeted strategies
are very effective in dramatically damaging the network, 
reaching the complete destruction at a very
small threshold value of the fraction of removed nodes. \\
Finally, we consider the role of the spatial constraints.
As shown in Fig.~\ref{fig:geo}, various geographical zones contain
different numbers of central airports. The immediate consequence is
that the different strategies for node removal have different impacts
in different geographical areas. 
Figure~\ref{fig:geo2} displays the case of a removal of nodes
according to their strength (other removal strategies lead to similar
data), monitoring topological and outreach integrity. 
Each geographical zones is more sensitive to a particular removal strategy, 
leading to the idea that {\em large weighted networks can be composed by different 
subgraphs with very different traffic structure and thus different responses to
attacks}.

\subsection{Comparison with the spatial BBV model}\label{CHAP4_2_3}
In Section~\ref{CHAP2_4_3}, we have presented a model of growing weighted networks, proposed in Refs.~\cite{BBV_model,BBV_model2} and based on the preferential attachment. By means of a strength-driven preferential attachment (i.e. a sort of ``busy-get-busier'' principle), the model generates networks with properties that are observed in many real growing weighted networks, such as the small-world property and power-law distributed degree, strength and weights. We have already observed that the BBV model pinpoints the basic mechanisms governing the evolution of the airport network, even if it fails in the existing large fluctuations between topological and weighted quantities (such as degree vs. strength or topological betweenness vs. weighted betweenness). \\
A possible reason for the existence of these non-trivial correlations is the presence of spatial constraints, i.e. the node of the real networks are embedded in a two-dimensional space. The authors of Ref.~\cite{BBV_model} have thus put forward a modified model, in which spatial constraints are taken into account \cite{spatial_BBV}. The spatial model shows more realistic correlations, but it does not explain all the features observed in the real airport network.\\
In this section, we present some preliminary results on the study of the vulnerability properties of the model, pointing out the relation with the effects observed for the WAN. In particular, we look at the role played by spatial constraints, fixing the parameter $\delta=1$ and varying the typical length scale of the connections governed by the parameter $\eta$. The system is embedded in a square of linear size $L=1$, endowed with euclidean metric structure, so that small values of $\eta$ ($\eta \ll 1$) favor the creation of "regional" structures with hubs of smaller connectivity than the global hubs obtained in the original BBV model or in the present model when $\eta$ assumes larger values. We have considered the two cases $\eta = 0.5$ and $\eta=0.005$.\\
\begin{figure}[t] 
\vskip 0.3in
\centerline{
\includegraphics[width=7cm,clip=true]{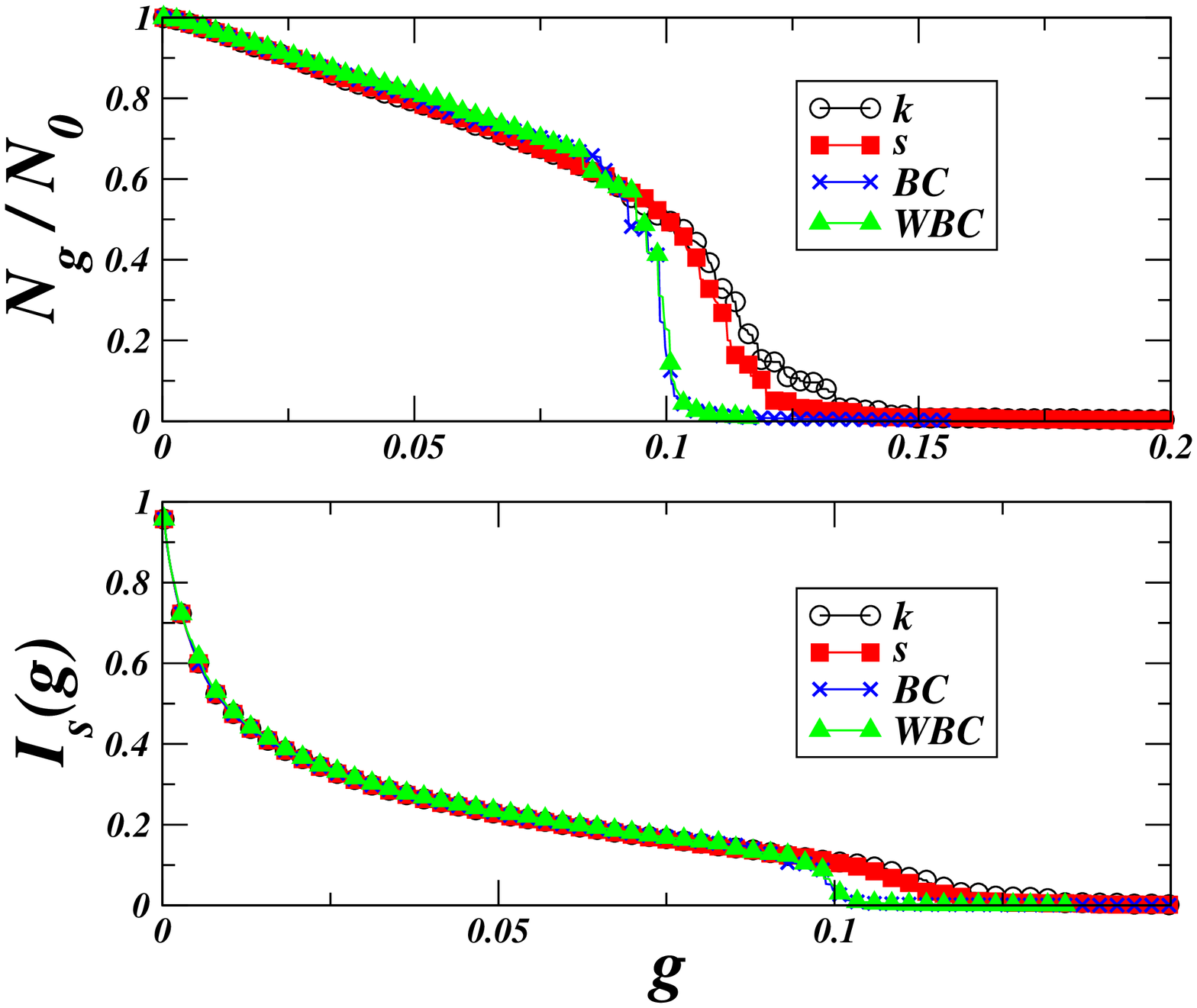} 
\includegraphics[width=7cm,clip=true]{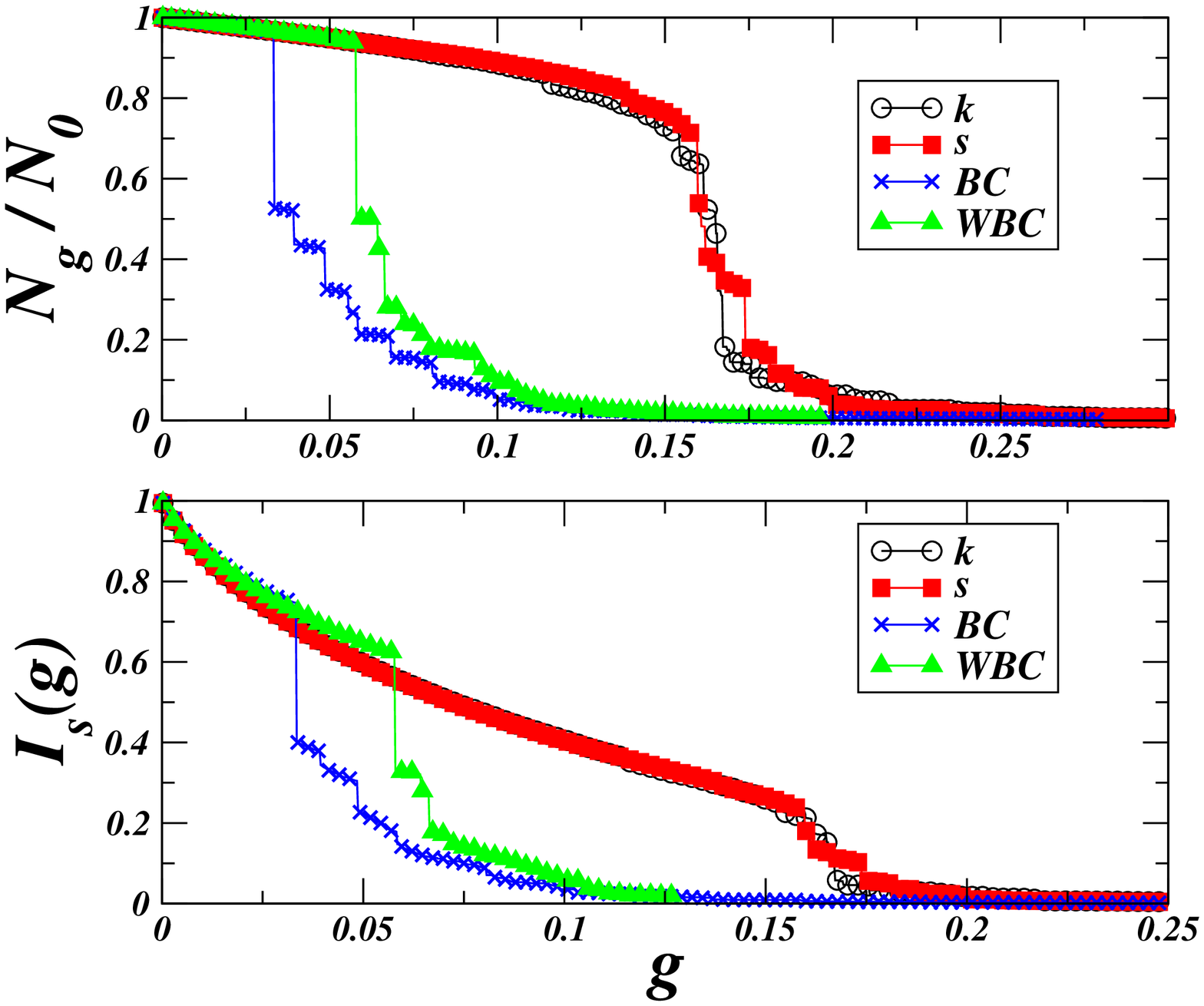}
}
\caption{Vulnerability of a network obtained using the growing model with 
spatial constraints exposed in Section~\ref{CHAP2_4_3} for two values of
the parameter controlling the spatial constraints $\eta=0.5$ (left) and $\eta=0.005$ (right).
The targeted damage is driven removing nodes in order of: 
degree (circles), strength (squares), and topological
(crosses) and weighted (triangles) betweenness. 
The behaviors of the topological ($N_{g}/N_{0}$) and functional ($I_{s}(g)$) integrities 
are shown.}
\label{model_sc}
\end{figure}
Figure~\ref{model_sc} displays the behaviors of the topological and functional integrities under different damage strategies. We have focused only on the main attack strategies, i.e. those driven by removing nodes in order to degree, strength, betweenness and weighted betweenness. In this figure the weighted betweenness has been computed using the weighted distance $\ell_{ij} = 1/w_{ij}$ instead of $\ell_{ij} = d_{ij}/w_{ij}$, but the qualitative differences in the resulting vulnerability properties are negligible. (Note that this is true also for the WAN).  
Comparing the two figures for $\eta=0.5$ (left) and $\eta=0.005$ (right), it is clear that, increasing the spatial constraints, the network becomes more vulnerable to betweenness-driven attacks. This is due to the fact that the system develops a structure composed of many regional networks connected by a limited number of links. The links or the nodes that play as bridges connecting different regional networks are very important for the global structural properties of the system. They are the most central nodes in terms of both topological and weighted betweenness, so that {\em damaging the network removing high-betweenness nodes reveals the fragility of such a regional structure}. This feature is in agreement with the behavior of $N_{g}/N_{0}$ in the real airports network.\\    
In the model, the picture does not change if we look at the functional level, monitoring the decrease of the functional integrity (in this case we have considered $I_{s}(g) = S_{g}/ S_{0}$, the other functional integrity measures show similar behaviors). The two non-local damage strategies are indeed more effective in destroying the network functionality.
This feature is different from what we observe in the real airports network, that shows the same rapid decrease for all the strategies. The difference is possibly due to the fact that, in the model, the relation between topology and traffic is still too strong, while in the real WAN, other variables play an important role in separating the traffic and economic dimension from the purely topological one.\\
In summary, the introduction of spatial constraints in a model of growing weighted networks produces some structural properties that are in agreement with those observed in a real network such as the WAN; but an improvement is required in order to pinpoint the correct mechanisms that are at the origins of those functional properties of real weighted networks that are still not completely understood.

\subsection{Conclusions}\label{CHAP4_2_4}
In this study (see Ref.~\cite{vulnera_dallasta}), we have identified a set of centrality measures 
for technological weighted networks, in which two further ingredients, 
traffic and spatial constraints, are included. 
The main achievements are summarized in the following points:
\begin{itemize}
\item the various definitions of centrality are correlated, but lead to different descriptions of the importance of the nodes;
\item the level of vulnerability of a network depends on the global property we have decided to monitor; in general, the attack strategies that are based on the same property (or similar properties to) that we are monitoring at a global scale are trivially more effective;
\item spatial heterogeneity has to be considered, weighted regional properties are different;  
\item weighted networks are, at a functional level, much more fragile than we would expect looking at a purely topological level.  
\end{itemize}
It is worth noting that, even if the the vulnerability analysis seems to be completely static, 
the previous strategies are based on a recursive re-calculation of
the centrality measures on the network after each damage. 
As noted by Holme et al.~\cite{holme02}, each node removal leads to a
change in the properties of the other nodes. Only the neighboring
nodes of the removed one suffer a change in the degree and strength, but
the structure and number of shortest paths are modified for the whole
graph and therefore the betweennesses of all nodes may potentially be
altered. It is therefore quite natural that, after each node removal,
the choice of the next discarded node should be made according to {\em
recalculated} degree, strengths and betweennesses, and not according
to the original ranking.  Such procedure is somehow akin to a
{\em cascade failure mechanism} in which each failure triggers a
redistribution on the network and changes the next most vulnerable
node. \\
Actually, we have found \cite{vulnera_dallasta} that recalculated rankings do not differ considerably 
from the original ones, even for non-local quantities like the betweenness.
This result has two important consequences. 
First, it points out the validity of considering static analysis, such as that of vulnerability (and indirectly percolation),
in order to study dynamical property of networks, such as spreading and cascade failures or congestions.
The other observation concerns the protection of large scale infrastructures. 
On the one hand, the planning of an effective targeted attack does need only to gather
information on the initial state of the network. On the other hand,
the identification of crucial nodes to protect is an easier task, since it is only weakly 
dependent on the attack sequence.

\newpage
\section{Spreading processes on Weighted Random Graphs}\label{CHAP4_3}

This second part of the chapter is devoted to study spreading phenomena on weighted networks.
According to the static analysis made in the previous section, at a functional level, weighted networks look
much more vulnerable. 
In other words, network's functionality suffers of the disconnection of some high-traffic portions of the system. 
For the same reasons, we could expect to observe the same features in the  behavior of spreading processes on weighted networks.
More precisely, we want to model a process that exploits the traffic to spread throughout the network, the efficiency of the spreading depending on the weights properties.   \\
In Section~\ref{CHAP4_1_2}, we have explained the relation between spreading processes and percolation, stressing that there is 
a correspondence between the existence of a giant component in the percolation process and the possibility of a macroscopic spreading. Of course, the analogy becomes evident if we think to simple examples, like the outburst of  epidemics in a population of individuals, or the diffusion of a virus on the Internet, but also the spreading of information or other physical quantities on a generic network. \\
Nevertheless, the standard percolation cannot encode all the features of such spreading processes, in which the spreading rates depend on the properties of the nodes and the edges, and on the details of the process. 
For instance, in the Internet, the connections between computers are real cables, thus the transmission along them depends on their bandwidth; in the air-transportation network, the edges are weighted with the number of available seats, that is a measure of the their capacity; in social and contact networks as well, the importance of links is related to the type of relationships between the individuals. It is clear that real processes are influenced by these factors.\\
According to this remark, we develop a general theory of spreading on networks, in which nodes and edges are endowed with different spreading properties.
In order to express the ability of an edge $(i, j)$ to transfer some physical quantities (information, energy, diseases, etc), we introduce an {\em edge transition probability} $T_{ij}$, that depends on the properties of the vertices $i$, $j$ and of the edge $(i,j)$ itself. In addition, in real processes the nodes have different ability to spread, thus they should be supplied with a {\em node traversing probability} $q_{i}$.
In this manner, the actual size of the giant component can be affected by the non-optimal or non-homogeneous flow through the nodes and edges, the most general model performing an {\em  inhomogeneous joint site-bond percolation}.
On the other hand, information about simpler processes such as site percolation with inhomogeneous bonds (edges) or bond percolation with inhomogeneous sites (nodes) can be easily obtained assuming, in turn, $q_{i}$ or $T_{i j}$ as uniform.
In these particular cases, we will specify them as node ($q$) or edge ($T$) {\em occupation probability} (in agreement with standard percolation processes).\\
This formalism fits very well the description of spreading phenomena on weighted networks, since we can choose the transition probabilities to be functions of the weights ($T_{ij}=T(w_{ij})$). 
We can also define a {\em weighted giant component} as the part of the giant component that is reached in the spreading process and gives a measure of the functionality of the network. 
For instance, for the spreading of information, it corresponds to the maximal subgraph that can bear and handle a large flow of information, ensuring the global well-functioning of the network. In contrast, if we are dealing with epidemic spreading, the weighted giant component corresponds to the maximal infected region reachable in the pandemic regime.
In both cases, topological giant component is reduced by introducing weights.\\
For the sake of simplicity, the rest of this section is devoted to discuss spreading phenomena 
in weighted random networks, and the possible applications in the description of real processes. 
This is done fixing $q_{i}=q$ and studying a site-percolation problem.
On the other hand, the model of generalized spreading phenomena on networks goes beyond the description in term of weighted networks and can be analyzed in a very general and abstract form.
The general theory \cite{spreading_dallasta}, by means of which the most relevant results have been derived, is reported in the Appendix~\ref{APP4_2}, while a brief introduction to generating functions in graph theory is reported in the Appendix~\ref{APP4_1}.  

\subsection{Generalized Molloy-Reed Criterion in Weighted Random Graphs}\label{CHAP4_3_1}

\begin{figure}[t] 
\vskip 0.3in
\centerline{
\includegraphics[width=7.2cm]{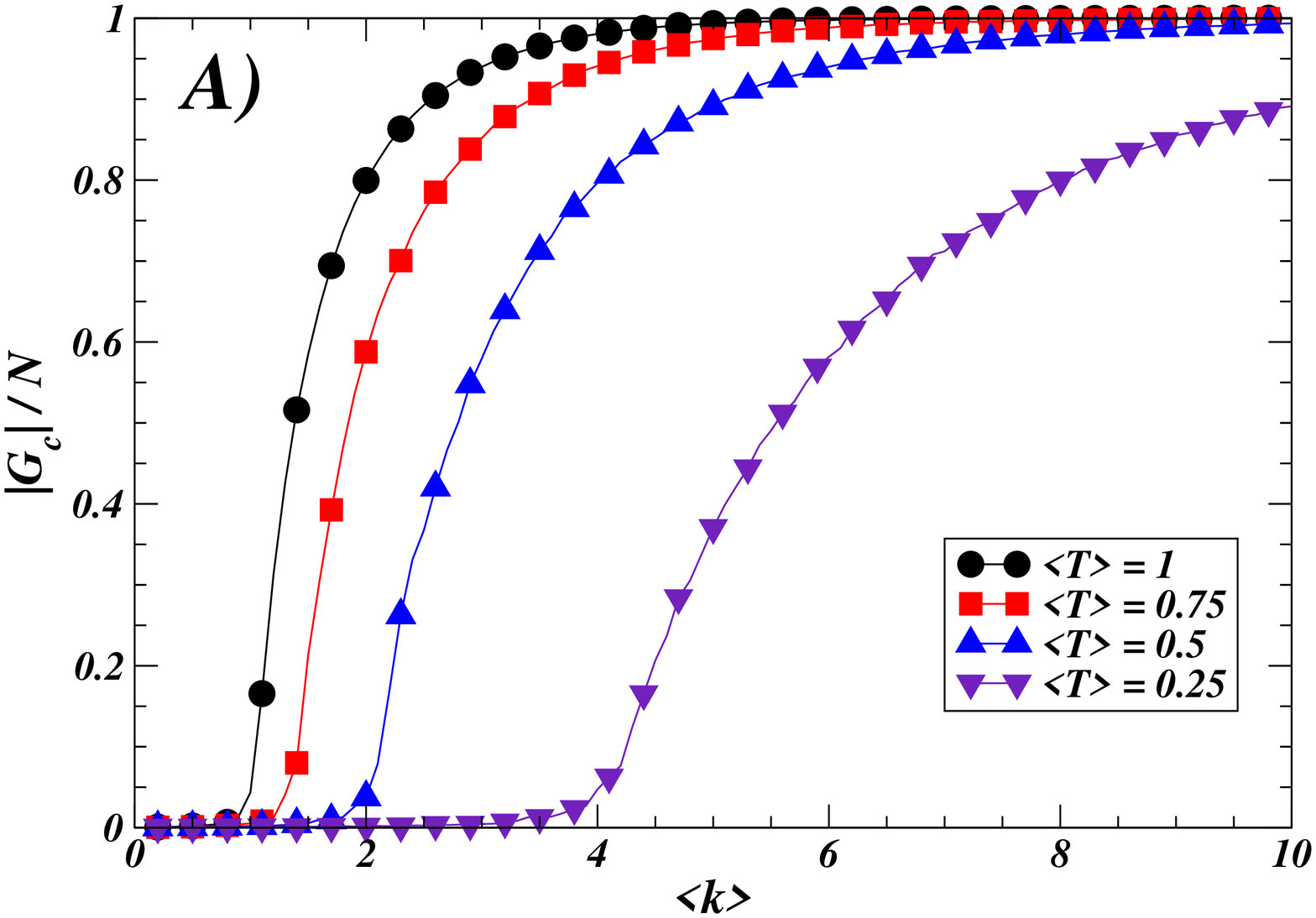}
\includegraphics[width=7.2cm]{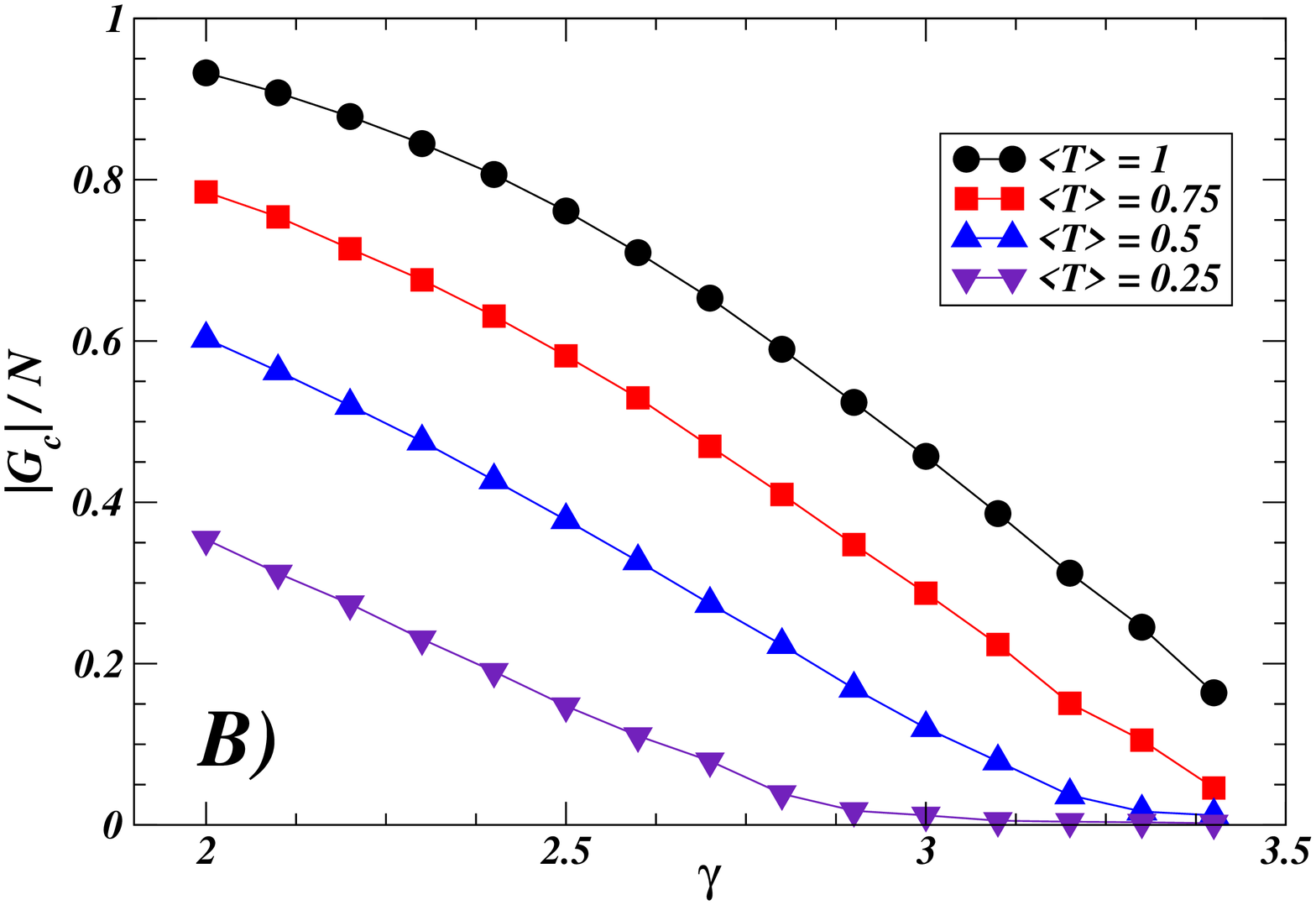}
}
\caption{(A) Relative size of the giant component $\mathcal{G}_{c}$ vs. average degree $\langle k\rangle$ in an  Erd\"os-R\'enyi random graph in which the edge transition probabilities are uniformly random distributed with mean value $\langle T\rangle$. Simulations are performed on a sample of $100$ graphs with $N=10^4$ nodes. The different curves refer to different values of average transition probability: $\langle T \rangle = 1$ (circles), $0.75$ (squares), $0.5$ (up triangles) and $0.25$ (down triangles). 
(B) Relative size of the giant component $\mathcal{G}_{c}$ vs. the exponent $\gamma$ in an power-law graph in which the edge transition probabilities are uniformly random distributed with mean value $\langle T\rangle$.  Simulations are performed on a sample of $100$ graphs with $N=10^4$ nodes, and minimum degree $m=1$. The different curves refer to different values of average transition probability: $\langle T \rangle = 1$ (circles), $0.75$ (squares), $0.5$ (up triangles) and $0.25$ (down triangles).}
\label{bound}
\end{figure}

The idea of a percolation process in which the basic elements (nodes and edges) present inhomogeneous properties goes back to the physical studies of conduction in resistor network models (see Ref.~\cite{kirkpatrick} and references therein). A typical way to introduce disorder on the edges (or on the nodes) is that of assigning to each of them a random number between $0$ and $1$, then removing all edges (nodes) in a selected interval of values. 
Percolation properties are studied as functions of the fraction of remaining edges (nodes) in the strong disorder regime \cite{bruggeman,landauer}. \\
\begin{figure}[t] 
\centerline{
\begin{tabular}{|c|}\hline \\ \includegraphics*[width=0.45\textwidth]{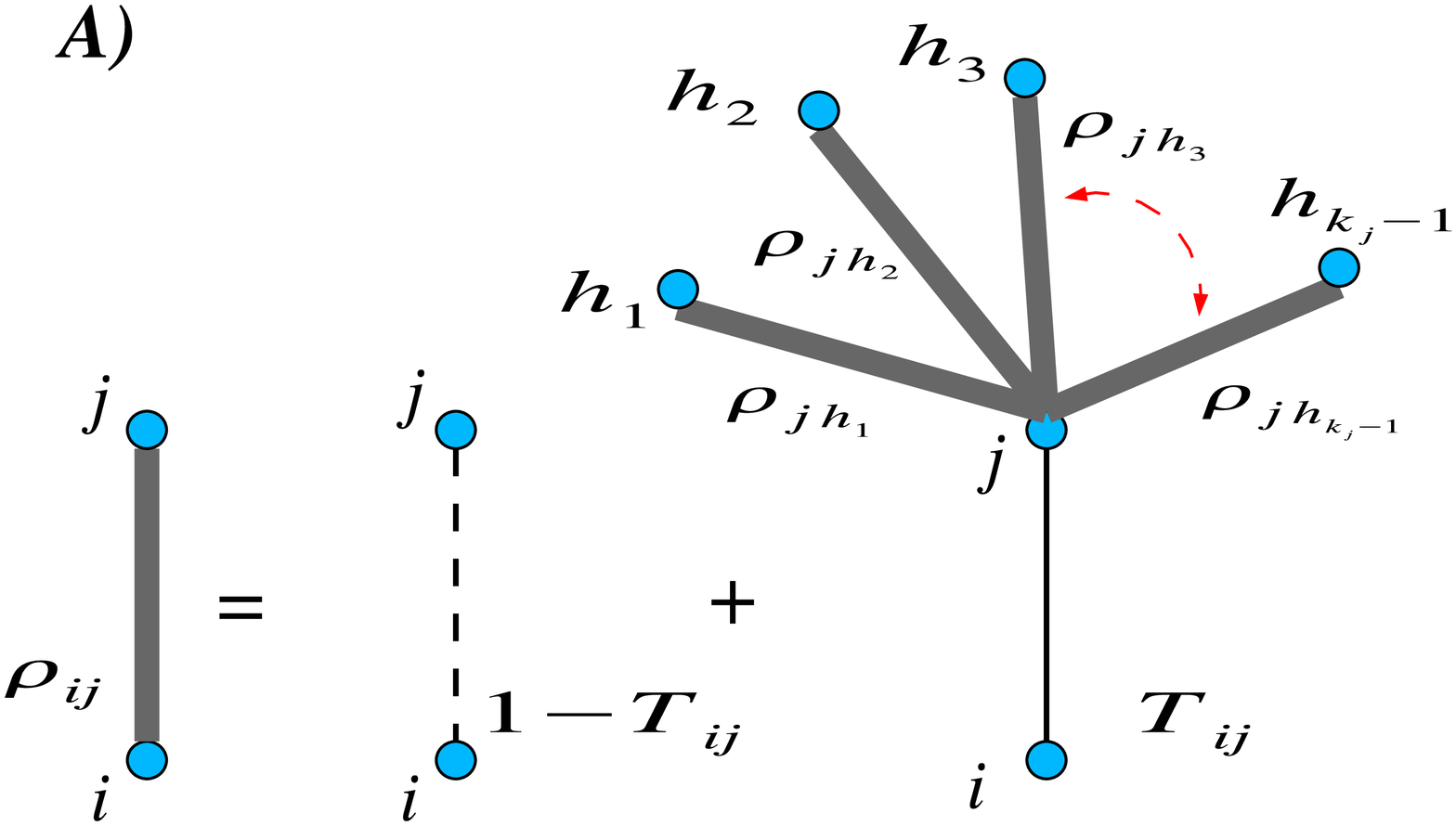}\\ \hline\end{tabular}~~
\begin{tabular}{|c|}\hline \\ \includegraphics*[width=0.45\textwidth]{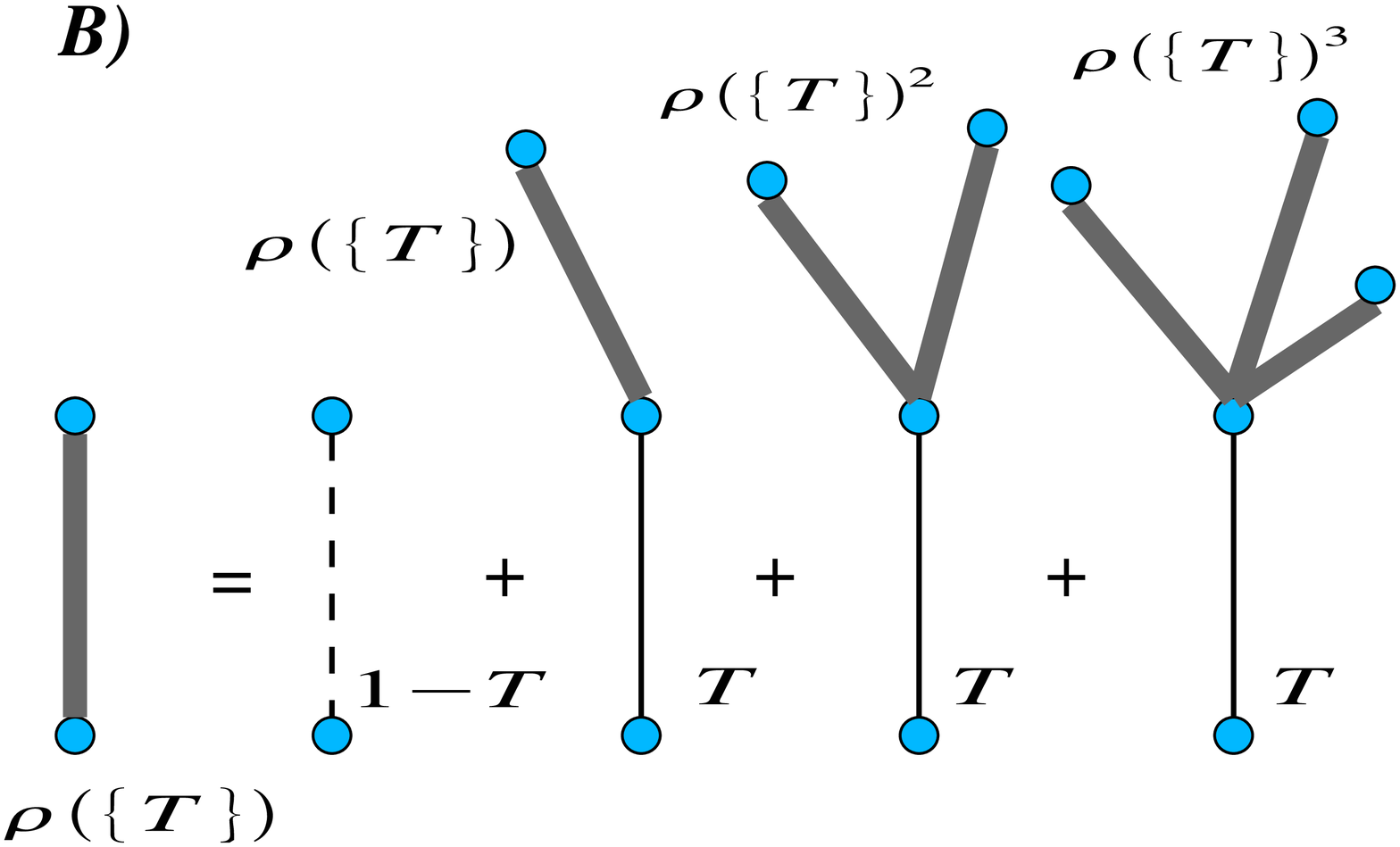}\\ \hline\end{tabular}
}
\caption{(A) Graphical representation of the Eq.~\ref{naive0}. Fat lines indicate the contribution coming from the unknown probability $\rho_{ij}$ that an edge $(i, j)$ does not belong to the infinite cluster. The first term on the right-hand side represents the contribution coming from the probability $1-T_{ij}$ that a flow does not traverse the edge $(i, j)$; the second term accounts for the probability that the edge $(i, j)$ is traversed times the unknown contribution $\rho_{jh}$ of other edges attached to $j$.
(B) Diagrammatic representation of the self-consistent Eq.~\ref{naive1}. Fat lines correspond to the unknown uniform probability $\rho(\{T\})$ that an edge does not lead to the giant component. Full lines mean that edges can be crossed, with probability $T$, while the dashed edge corresponds to the contribution $1-T$ that an edge cannot be crossed. 
}
\label{recursive}
\end{figure}
However, the percolation processes we want to describe is more general than the theory of random resistor networks. 
The latter, indeed, corresponds to the case in which the edge transition probability is a step function (the Heaviside $\theta(T_{i j}-T_{0})$ for a reasonable value $T_0$), so that transmission occurs only along a fraction of the edges (with sufficiently high conductance) and, for those edges, it is optimal. 
In other words, it corresponds to a bond percolation, as already noted in Ref.~\cite{newman1}. 
Here, we are interested in a site percolation process in which, additionally, there is a quenched variable expressing the probability of passing through an edge. This quenched probability rate is a function of the weights along the edges, i.e. 
$T_{ij}=T(w_{ij})$.\\
We use an intuitive derivation of the percolation criterion, that generalizes the (Bethe Approximation) approach introduced by Cohen et al. in Ref.~\cite{cohen2} for uncorrelated sparse random graphs. 
Let us call ${\rho}_{ij}$ the probability that an edge in the network does not lead to a vertex connected via the remaining edges to the giant component (infinite cluster) and $T_{ij}$ the probability that a flow leaving the node $i$ passes trough the edge $(i,j)$, reaching the node $j$. Both ${\rho}_{ij}$ and $T_{ij}$ are defined in the interval $[0,1]$.\\
In general, ${\rho}_{ij}$ depends on the transition probability $T_{ij}$ of the edge $(i,j)$, through a term $1- T_{ij}$, and on the probability that the node $j$, if reached, does not belong to the infinite cluster.
This contribution can be computed as the joint probability that each one of the remaining edges emerging from $j$ does not belong to the giant component.  
Assembling the different contributions, we get the recursive expression (Fig.~\ref{recursive}-A)
\begin{equation}
{\rho}_{ij} = 1-T_{ij} + T_{ij} \prod_{{\begin{array}{c} h\in \mathcal{V}(j)\\ h\neq i \end{array}}} {\rho}_{jh}~,
\label{naive0}
\end{equation}
that contains a product of $k_{j}-1$ terms $\rho_{jh}$ if $k_{j}$ is the degree of the node $j$ and $\mathcal{V}(j)$ is its set of neighbors.
This recursive procedure is not closed, but the introduction of some hypotheses on the set of transition probabilities $\{T\}$ allows a statistical reformulation of Eq.~\ref{naive0}.\\
Let us suppose that the transition probabilities are random variables with a given distribution $p_{T}$, and the average overall probability $\rho$ that a randomly chosen edge does not belong to the giant component depends only on some general properties of that distribution such as the mean, the variance, etc (i.e. $\rho = \rho(\{T\})$). \\
Note that in this section (as well as in Appendices~\ref{APP4_1}-\ref{APP4_2}) we change notation with respect to Chapter~\ref{CHAP3}, in which the degree distribution is indicated as $P(k)$; here, we use instead $p_{k}$ for the degree distribution. \\ 
By picking up an edge at random in a random graph, the probability that it is connected to an edge of degree $k$ is $\frac{k p_{k}}{\langle k \rangle}$, where $p_{k}$ is the degree distribution of the graph. 
We can now write a self-consistent equation for $\rho (\{T\})$ (Fig.~\ref{recursive}-B),
\begin{equation}
\rho (\{T\}) = 1-T + T \sum_{k}\frac{k p_{k}}{\langle k \rangle} {\rho (\{T\})}^{k-1}~,
\label{naive1}
\end{equation}
in which $T$ is a realization of the independent and identically distributed (i.i.d.) random variables $\{T\}$ in the interval $[0,1]$. Averaging both the members of the Eq.~\ref{naive1} on $p_{T}$, we obtain that the probability $\rho$ depends only on the average value $\langle T \rangle$, i.e. $\rho (\{T\})=\rho (\langle T\rangle)$, and the equation becomes
\begin{equation}
\rho (\langle T\rangle) = 1- \langle T\rangle + \langle T \rangle \sum_{k} \frac{k p_{k}}{\langle k \rangle} {\rho (\langle T\rangle)}^{k-1} = I[\rho (\langle T\rangle)]~.
\label{eq_1b}
\end{equation}
Apart from the trivial solution $\rho =1$, another solution $\rho ={\rho}^{*}<1$ exists if and only if $\frac{d I}{d \rho}{\vert}_{\rho =1} \geq 1$. The curve $I[\rho]$, indeed, is positive in $\rho=0$ ($I[0] = 1-\langle T \rangle + \langle T\rangle p_{1} /\langle k\rangle > 0$), therefore $\frac{d I}{d \rho}{\vert}_{\rho =1} \geq 1$ means that it crosses the line $f(\rho)=\rho$ in a point $0 < {\rho}^{*} < 1$.
The condition on the derivative of the r.h.s. of Eq.~\ref{eq_1b} corresponds to
\begin{equation} 
\langle T \rangle \frac{\langle k^{2} \rangle - \langle k \rangle}{\langle k \rangle} \geq 1~.
\label{condgc}
\end{equation}
A generalization of the Molloy-Reed criterion \cite{molloy1} for the existence of a giant component in presence of random weights on the edges immediately follows,
\begin{equation}
\frac{\langle k^2 \rangle}{\langle k \rangle} \geq 1+\frac{1}{\langle T \rangle}~,
\label{mrgc}
\end{equation}
meaning that, when the transition probabilities are i.i.d. random variables, a giant component exists if and only if the inequality is satisfied. The case of uniform transition probabilities is exactly the same, with uniform value $T$ instead of the average value $\langle T \rangle$.
Note that, while in the case of perfect transmission ($T_{ij}=1$) the usual formulation of the Molloy-Reed criterion is recovered \cite{molloy1}, when $\langle T \rangle < 1$ the r.h.s. of Eq.~\ref{mrgc} can grow considerably, affecting the possibility of observing percolation: the smaller is the average transition probability, and the larger are degree fluctuations $\langle k^2 \rangle$ needed to ensure the presence of a giant component.\\
For a random graph with poissonian degree distribution (i.e. $p_{k} = e^{-\langle k \rangle} {\langle k \rangle}^k/k!$), the criterion in Eq.~\ref{mrgc} corresponds exactly to have $\langle k \rangle \geq 1/ \langle T \rangle$.
We show in Fig.~\ref{bound}-A the results of numerical simulations for the giant component's computation on an Erd\"os-R\'enyi random graph with $N = 10^4$ nodes and with random transition probabilities uniformly distributed in $[a,b]$ with $0 \leq a < b \leq 1$. A giant component clearly appears when the mean connectivity $\langle k \rangle$ exceeds the inverse of mean transmissibility value $(b-a)/2$. Considering different distributions (e.g. binomial distributions) for the i.i.d. random variable $T$ does not affect the results.\\
For heterogeneous graphs with broad degree distribution, the inequality in Eq.~\ref{mrgc} is satisfied thanks to the huge fluctuations of the node degrees ensuring the l.h.s. to be larger than $1+ 1/ \langle T \rangle$. In particular, when the graph has power-law degree distribution $p_{k} \sim m k^{-\gamma}$ ($2 < \gamma < 3$, $m$ is the minimum degree), the fluctuations diverge in the limit $N \rightarrow \infty$. In this case, the giant component always exists if $N$ is large enough. If $\gamma > 3$, the second moment is finite, and Eq.~\ref{mrgc} provides the following bound for the average transition probability necessary to have a giant component (computed using the continuum approximation for the degree),
\begin{equation}
{\langle T \rangle} \geq \frac{\gamma-3}{\gamma (m-1) + 3 - 2m}~.
\label{sfth}
\end{equation}
The inequality is always satisfied for $m \geq 1+1/ \langle T\rangle$, while for $m < 1 + 1/ \langle T \rangle$ it is satisfied when $\gamma \leq 3+ \frac{m\langle T\rangle}{1-(m-1)\langle T \rangle}$. For $m=1$, an infinite (uncorrelated) scale-free graph presents a giant component of order $\mathcal{O}(N)$ only when $\gamma \leq 3 + \langle T \rangle \leq 4$. 
At $\gamma = 3$, logarithmic corrections should be taken into account \cite{cohen2}.\\
Actually, real networks are large but finite, and the second moment cannot diverge, thus we expect that the condition 
is not always satisfied. 
This is clearly shown in Fig.~\ref{bound}-B, where we have reported the size of the giant component for power-law random graphs as a function of the exponent $\gamma$ for different values of the average edge transition probability $\langle T \rangle$. The size of the giant component is dramatically affected by low transition probabilities.
%
%
%
%
Now, the argument of Cohen et al. \cite{cohen2} can be used to compute the threshold of site percolation with random disorder on the edges. If $q$ is the (uniform) node occupation probability in an uncorrelated random graph, then a randomly chosen node, whose natural degree in case of $q=1$ is $k'$, will assume a degree $k \leq k'$ with probability 
\begin{equation}
\left(\begin{array}{c} k' \\ k \end{array} \right) {q}^{k} {(1-q)}^{k'-k}~,
\end{equation}
that means that the corresponding degree distribution $p_{k}(q)$ is related to the degree distribution $p_{k'}(q=1)$ by 
\begin{equation}
p_{k}(q) = \sum_{k'=k}^{\infty} p_{k'}(q=1) \left(\begin{array}{c} k' \\ k \end{array} \right) {q}^{k} {(1-q)}^{k'-k}~.
\end{equation}
It follows that ${\langle k \rangle}_{q} = q {\langle k \rangle}_{q=1}$ and ${\langle k^2 \rangle}_{q} = q {\langle k^2 \rangle}_{q=1} +q(1-q){\langle k \rangle}_{q=1}$, that introduced into the expression Eq.~\ref{mrgc} for the Molloy-Reed criterion gives the expression of the threshold value for site percolation,
\begin{equation}
q_{c} = \frac{1}{\langle T \rangle}\frac{\langle k \rangle}{\langle k^2 \rangle - \langle k \rangle}~.     
\label{thres0}
\end{equation}
Eq.~\ref{thres0} shows that decreasing the average transmission capacity of the edges (i.e. the edge transition probability), the value of node occupation probability necessary to ensure percolation increases.\\
The expression of the percolation threshold in Eq.~\ref{thres0} is not very general, since it is correct only
for randomly distributed weights (and thus transition probabilities).
On the contrary, in many real networks (e.g. the WAN) the weights are correlated with the degree, $w_{ij} =w(k_{i},k_{j})$.\\ 
Consequently, a natural generalization is that of considering $T=T_{k_{i} k_{j}}$. Of course, in this case the previous derivation breaks down, since two types of correlations have to be considered: correlations between edge transition probabilities and degree, and degree-degree correlations. 
The correct approach, reported in the Appendix~\ref{APP4_2}, is that of using the generating functions formalism in the framework of Markovian correlated random graphs (i.e. with only degree-degree correlations) \cite{boguna2}.\\
We consider here  the simpler case in which the transition probability depends only on the degree of the departure node (i.e. $T_{ij}=T_{k_{i}}$). The self-consistent equation \ref{naive1} can be straightforwardly generalized to consider the case $T=T_{k}$, obtaining the following expression for the percolation threshold,  
\begin{equation}
q_{c} = \frac{\langle k \rangle}{\langle k^{2} T_{k} \rangle - \langle k T_{k} \rangle}~. 
\label{chap4_eq_2k}
\end{equation}
The striking property is that, {\em for particular forms of the transition probability, a finite value for the percolation threshold can be restored also in infinite scale-free networks}. We will investigate this case and other possible situations in the following section.

\subsection{Examples and Numerical Simulations}\label{CHAP4_3_2}
\label{appsec}

The functional dependence of the transition probabilities $\{T\}$ is strongly related to the details of the system and to the type of spreading process. Therefore, in absence of hints from studies of real data about spreading processes on networks, the present section is devoted to highlight some simple examples of reasonable functional forms for the transition probability, and to illustrate their effects on the percolation condition. We perform our analysis on two main classes of random graphs: homogeneous graphs, in which the connectivity distribution is peaked around a characteristic degree value, and heterogeneous graphs, whose degree distribution is broad, with very large fluctuations in the degree values. 
%
\begin{figure}[t] 
\vskip 0.3in
\centerline{
\includegraphics[width=7.0cm]{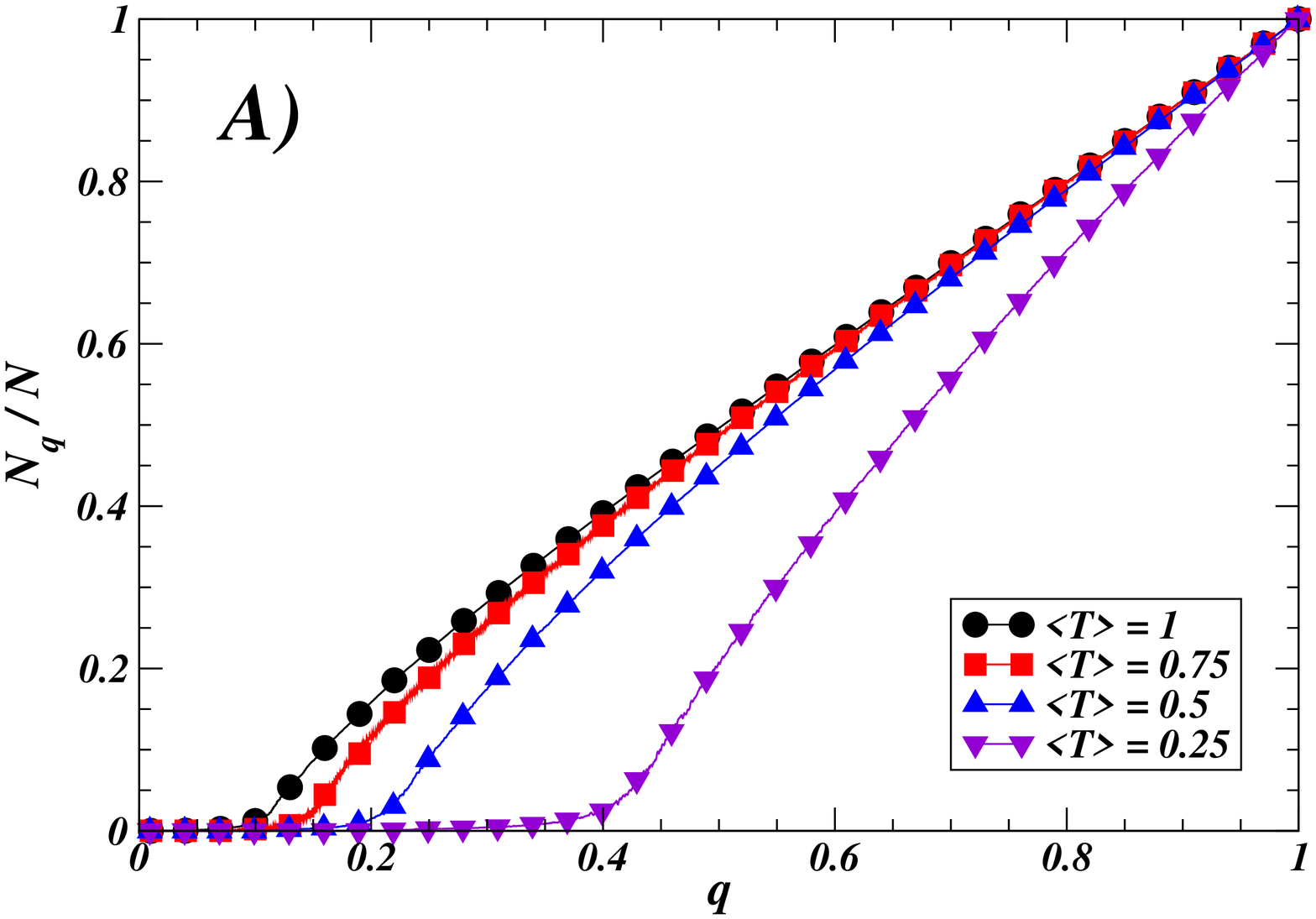}~~
\includegraphics[width=7.0cm]{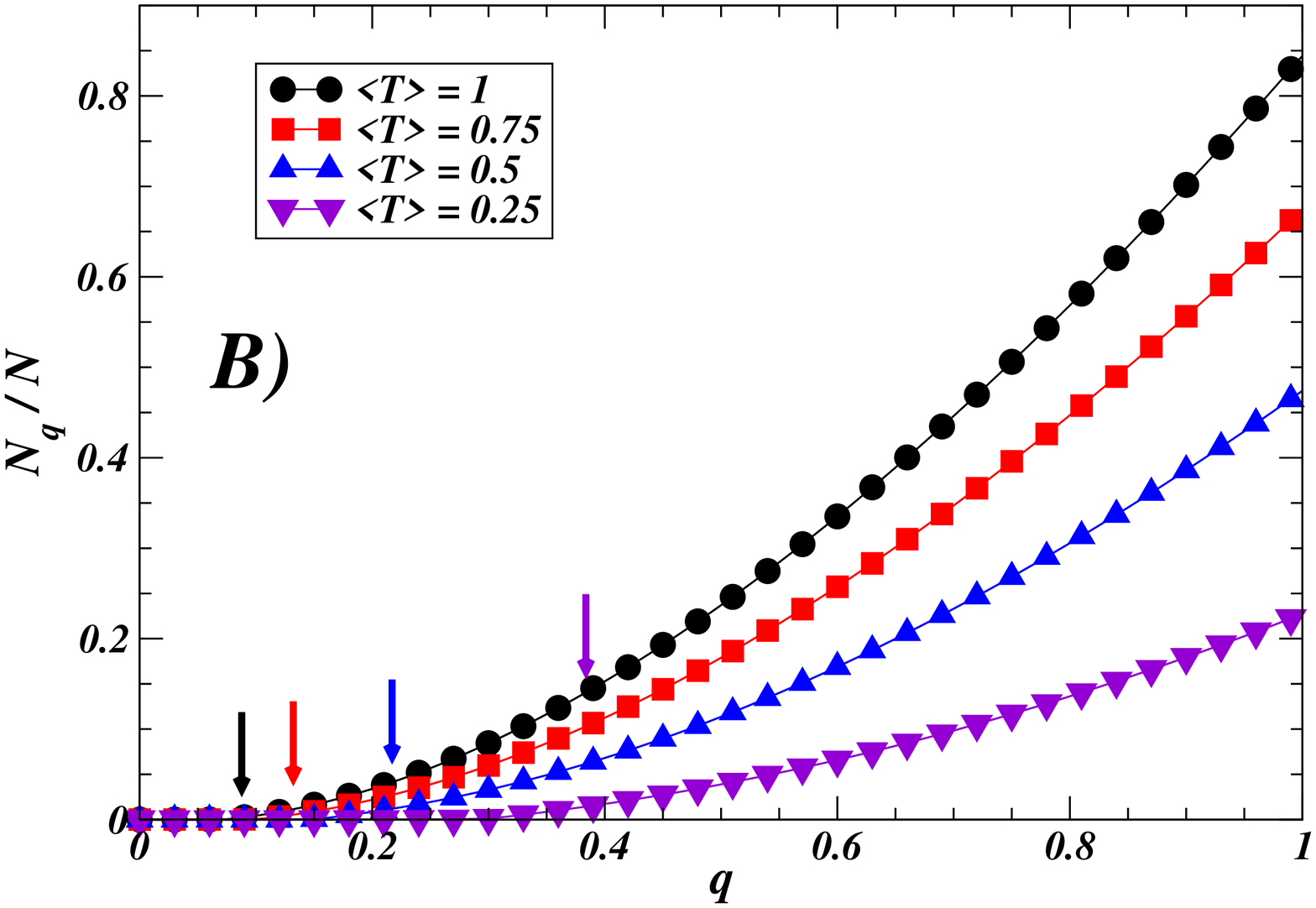}
}
\caption{Size $N_{q}/N$ of the giant component $\mathcal{G}_{g}$ as function of the fraction $q$ of occupied vertices for $\langle T \rangle = 0.25, 0.5, 0.75, 1$ for an Erd\"os-R\'enyi random graph (A) with $N=10^5$ vertices and $\langle k \rangle = 10$ and a power-law random graphs (B) with same size, exponent $\gamma = 2.3$. All curves have been averaged over $100$ realizations. The predicted values of $q_{c} = 0.4, 0.2, 0.133, 0.1$ are well verified in the Erd\"os-R\'enyi graph by the values of $q$ at which $N_{q} \sim \mathcal{O}(\log N)$. The power-law graph gives slightly worse results, but still in agreement with the theoretical values $q_{c} = 0.387, 0.20, 0.19, 0.1$ (that are indicated by arrows on the curves).}
\label{unif}
\end{figure}

{\bf \em Random Transition Probabilities - \quad}
Taking randomly distributed transition probabilities $T_{ij}$ (and weights) can appear unrealistic in technological networks, but it may be interesting to model spreading processes on some very disordered network. 
We have checked the validity of Eq.~\ref{thres0} performing simulations on an Erd\"os-R\'enyi random graph with $\langle k \rangle =10$ and on a power-law random graph with exponent $\gamma = 2.3$. Both graphs have $N=10^5$ nodes and the transition probabilities are assigned randomly between $0$ and $1$.  
Fig.~\ref{unif}-A reports the size of the giant component of the Erd\"os-R\'enyi graph as function of the fraction $q$ of occupied  vertices for $\langle T \rangle = 0.25, 0.5, 0.75, 1$; the predicted values $q_{c} = 0.4, 0.2, 0.133, 0.1$  are well reproduced by simulations. The same measures for a power-law graph are shown in Fig.~\ref{unif}-B. The curves of the simulations are in good agreement with the theoretical values of the percolation threshold $q_{c} \simeq 0.387$ (for $\langle T \rangle = 0.25$), $0.20$ (for $\langle T \rangle = 0.5$), $0.19$ (for $\langle T \rangle = 0.75$), $0.1$ (for $\langle T \rangle = 1.0$). 
\footnote{The estimation of the threshold value, as explicitly shown in Ref.~\cite{spreading_dallasta}, is a very difficult task in power-law graphs, given that the shape of the curves decreases with constant convexity, converging very smoothly to zero.  
Indeed, the probability of nodes with maximum degree is $\mathcal{O}(1/N)$, thus in a finite graph their frequency depends on the size of the system. Since the flows are clearly unbalanced in favor of large degree nodes, finite size effects are more relevant on these types of processes.}

{\bf \em Single-vertex dependent transition probability - \quad}
The assumption of transition probabilities $T=T_{k}$ yields different results depending on the exact functional dependence on the degree of the initial vertex.
%
\begin{figure}[t] 
\centerline{
\includegraphics*[width=0.5\textwidth]{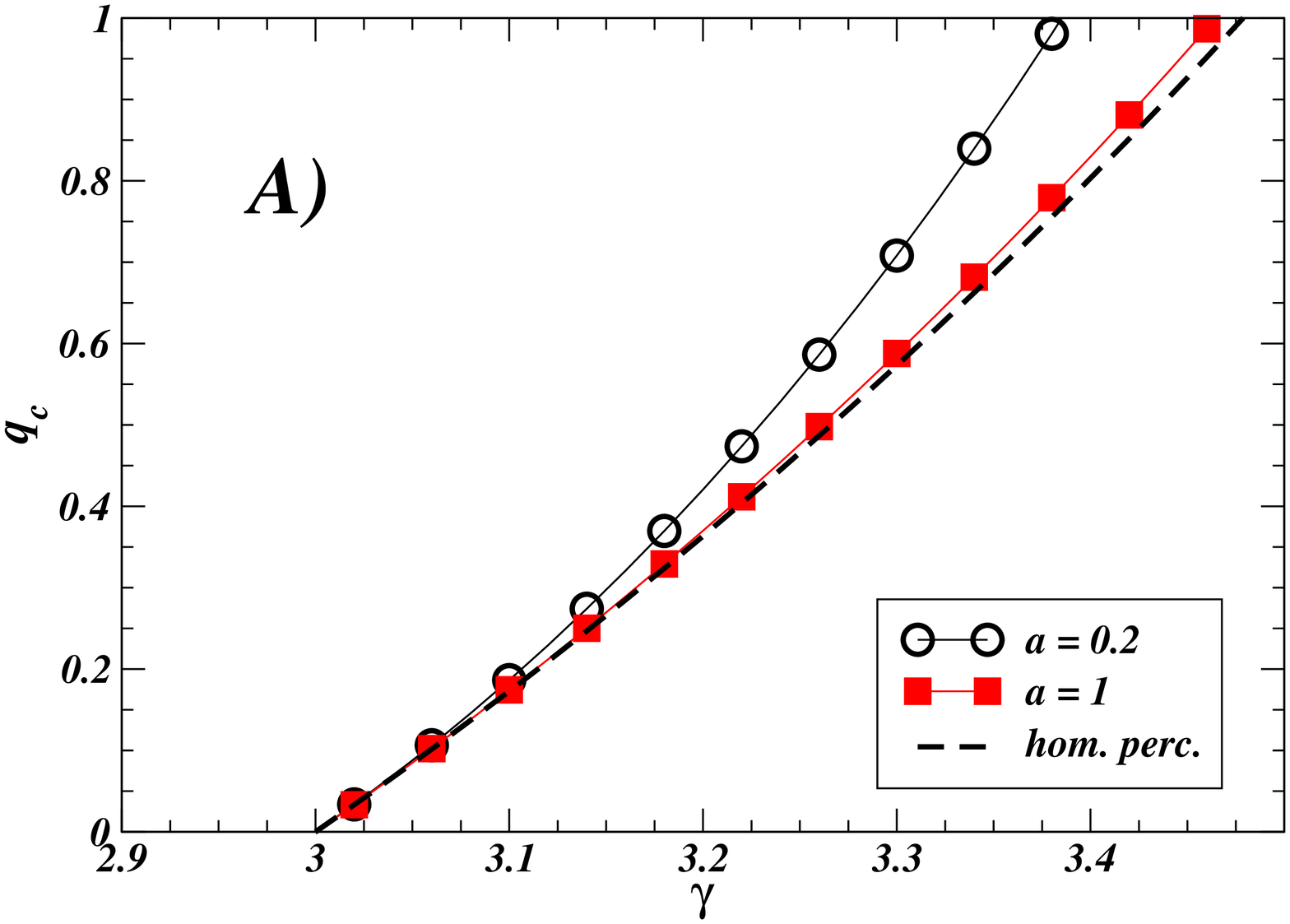}
\includegraphics*[width=0.5\textwidth]{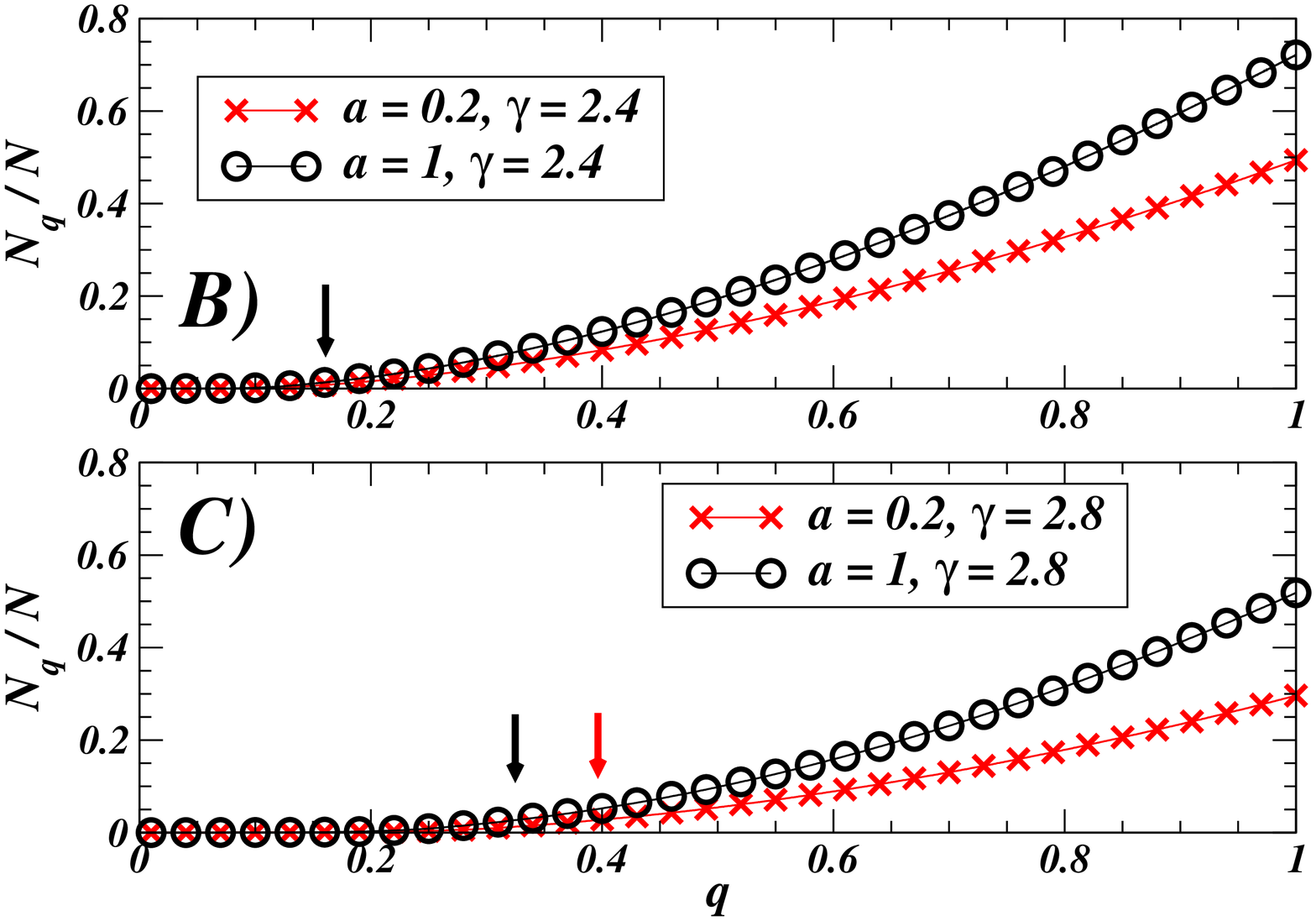}
}
\caption{(A) Critical density $q_{c}$ of nodes ensuring the existence of a giant component $\mathcal{G}_{g}$ as a function of the exponent $\gamma$ for a power-law graph in which the transition probability function is single-vertex depending with the form $T_{k} \sim 1- \exp(-\frac{a k}{\langle k \rangle})$. The figure reports the theoretical curve. The parameter $a$ controls the saturation to $1$ for highly connected nodes: the cases $a=0.2$ (circles) and $a=1$ (squares) are shown. For large values of the parameter $a$, the curves rapidly converge to the limit behavior of homogeneous percolation (limit $a \rightarrow \infty$). For $a <1$, the curves shift from this limit, but in the range $2 < \gamma <3$ diverging fluctuations ensure the  percolation.
(B)-(C) The curves $N_{q}/N$ obtained by numerical simulations on power-law random graphs of size $N=10^5$ and exponent $\gamma = 2.4$ and $\gamma = 2.8$ for the two values of $a = 0.2, 1$. The arrows indicate the positions of the threshold predicted using the theoretical calculation of Ref.~\cite{spreading_dallasta}.
}
\label{figura3}
\end{figure}
%
%
A first example is that of a monotonously increasing function of the degree, that saturates to $1$ at large $k$ values: we consider $T_{i j} = T_{k_{i}} \sim 1-\exp(-{\frac{a k}{\langle k \rangle}})$, that can be applied also to infinite graphs. The parameter $a$ controls the convergence to $1$.
For large scale-free networks, however, the mere presence of a saturation, rather than its rapidity, is sufficient for  
the system to behave as in the standard percolation, showing that graph's heterogeneity ensures a zero percolation threshold.
In general, from Eq.~\ref{chap4_eq_2k} simple calculations lead to the expression of the threshold for site percolation
(see Ref.~\cite{spreading_dallasta}). Its behavior is sketched in Fig.~\ref{figura3}-A as a function
of $\gamma$ and for some values of the parameter $a$. 
It shows that the critical fraction of occupied nodes $q_{c}$ is exactly $0$ for $2 < \gamma < 3$. Also for $\gamma \geq 3$, the behavior is qualitatively the same as for standard percolation on scale-free networks.
The reason is essentially that a transition probability converging to $1$ for highly connected nodes does not affect the network's properties if the network is infinite, because there is always a considerable fraction of nodes with optimal transmission.\\
This condition is not trivially satisfied by finite graphs, and the presence of a cut-off in the degree can have a strong influence on the network's functional robustness.
Two kinds of cut-offs on power-laws have been largely studied in the literature: an abrupt truncation of the degree distribution at a maximum value $\kappa \approx N^{\frac{1}{\gamma-1}}$, or a natural exponential cut-off, i.e. $p_{k} \sim k^{-\gamma} e^{-k/\kappa}$.
Without entering in the details of the analytic and numerical investigation, the main result is that 
the introduction of a cut-off actually reduces the effects of degree fluctuations, producing finite percolation threshold also in the range $2 < \gamma <3$.\\ 
However, this choice of edge inhomogeneity does not seem to enrich the scenario obtained by standard percolation, though the case of slow convergence ($a <1$) produces non negligible effects on finite systems (panels B and C in Fig.~\ref{figura3}). 
The structural properties of homogeneous graphs, on the contrary, are always deeply affected by this type of transition probability, expecially when perfect transmission is reached beyond the peak of the degree distribution ($a < 1$) \cite{spreading_dallasta}.
%
%
\begin{figure}[t] 
\centerline{
\includegraphics*[width=0.5\textwidth]{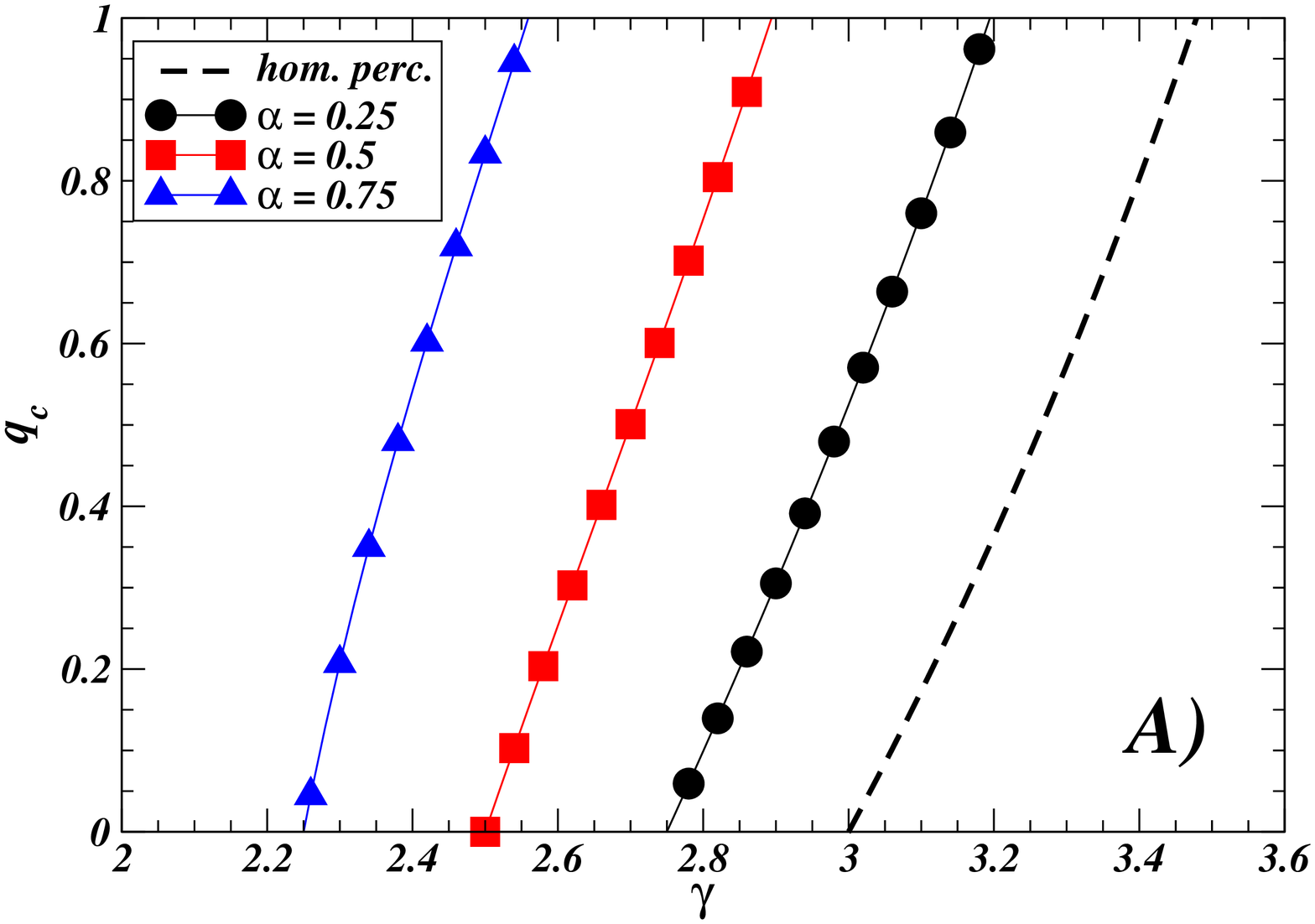}~~
\includegraphics*[width=0.5\textwidth]{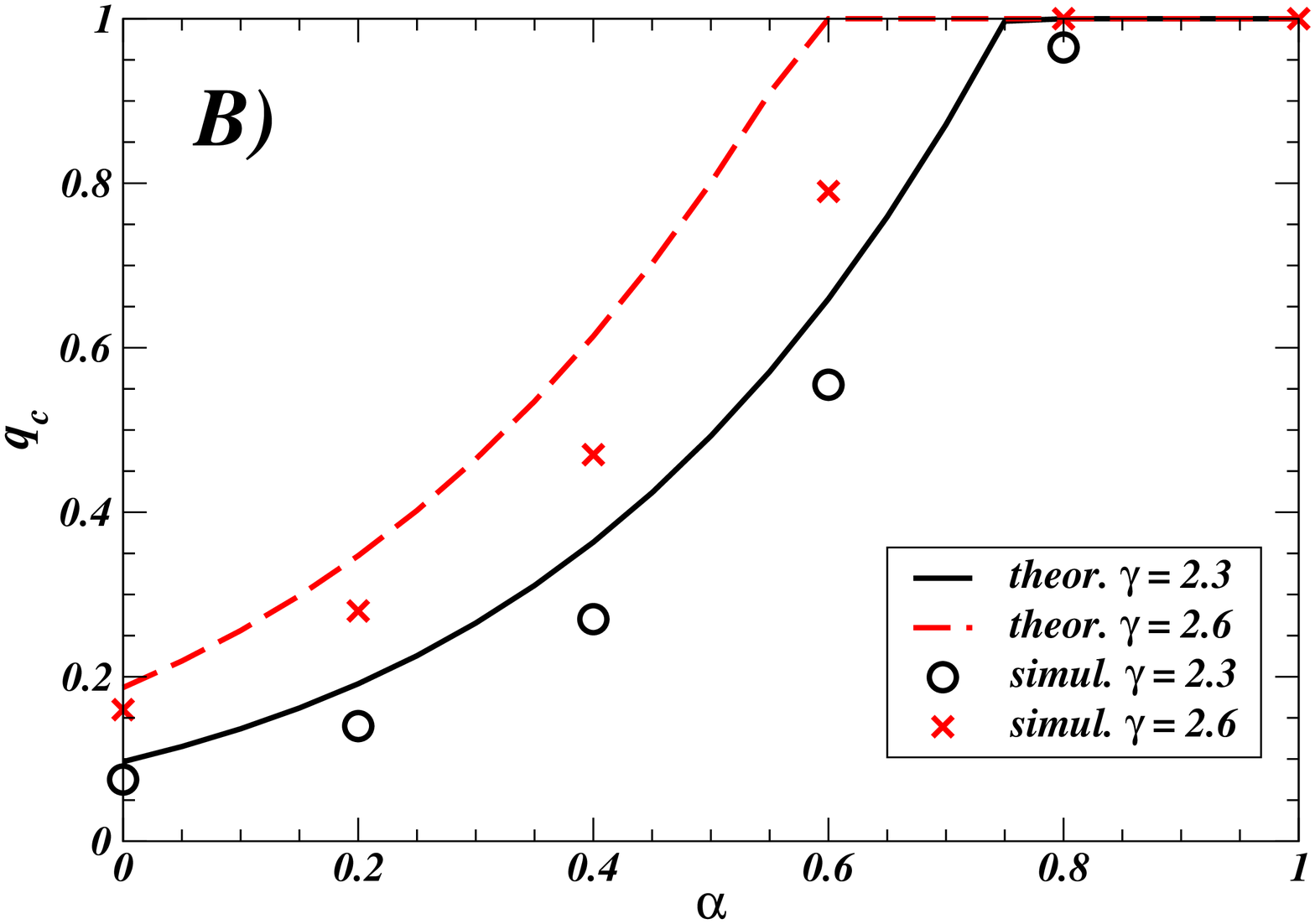}
}
\caption{(A) Fraction $q_{c}$ of occupied nodes required to have a giant component as a function of the exponent $\gamma$ for a power-law graph in which the transition probability is the single-vertex function $T_{k} \sim {k}^{-\alpha}$,  with exponent $0 < \alpha <1$. The curves report the values of $q_{c}$ computed by Eq.~\ref{q_alp}. All curves for $\alpha = 0.25$ (circles), $0.5$ (squares) and $0.75$ (triangles) show that the percolation threshold is  increased compared to the homogeneous result (dashed line), also for an infinite graph. In particular, a finite threshold larger than $0$ appears also in the range $2 < \gamma < 3$.
(B) Threshold value $q_{c}$ of occupied nodes as a function of the parameter $\alpha$ for power-law graphs with cut-off $\kappa = 10^2$ and exponent $\gamma = 2.3$ (full line and circles) and $\gamma = 2.6$ (dashed line and crosses). The symbols (circles and crosses) are results from simulations on random graphs of size $N= 10^5$ (average over $100$ realizations); the lines are the corresponding theoretical predictions.}
\label{figura5}
\end{figure}
%
%
Another wide class of transition probabilities contains those converging to zero with increasing degree. 
At a first glance, it seems a very unphysical condition, but the problem can be inverted asking which is the maximal decay in the degree dependence still ensuring site percolation.  \\
For single-vertex transition probability $T_{k}$, Eq.~\ref{chap4_eq_2k} implies that, independently of the degree distribution, a graph does not admit percolation if $T_{k}$ decays as $T_{1}/{k}^{\alpha}$ with $\alpha \geq 1$. Indeed, by substitution, $q_{c} = \langle k \rangle / [T_{1}(\langle {k}^{2-\alpha} \rangle - \langle {k}^{1-\alpha} \rangle)]$, that is larger than $1$ because $\langle k \rangle >\langle {k}^{2-\alpha} \rangle \geq \langle {k}^{1-\alpha} \rangle$, and $T_{1} <1$ (assuming $\langle k \rangle \geq 1$). This is actually equivalent to the result presented in Ref.~\cite{newman1}.\\
Since $\alpha = 0$ corresponds to standard percolation, we are interested in the intermediate case $0 < \alpha < 1$. 
In scale-free networks the percolation threshold is computed by series summation,
\begin{equation}
q_{c} = \frac{\zeta(\gamma -1)}{\zeta(\gamma+\alpha-2)-\zeta(\gamma+\alpha-1)}~,  
\label{q_alp}
\end{equation}
that has been plotted in Fig.~\ref{figura5}-A as a function of $\gamma$ for different values of $\alpha$ between $0$ and $1$. 
The curves show that a transition probability moderately decreasing with the degree can induce power-law graphs with $2 < \gamma <3$ to present a finite percolation threshold. However, many real networks (Internet, WWW, etc) have a power-law degree distribution with exponent lower than $2.5$, for which the giant component persists even for relatively large $\alpha$ (at least in the approximation of infinite systems). 
Nevertheless, the giant component is reduced in finite systems or in presence of a cut-off on the degree as shown by the results of simulations reported in Fig.~\ref{figura5}-B.
The figure displays the behavior of the threshold value $q_{c}$ as a function of the parameter $\alpha$ for two different exponents $\gamma = 2.3, 2.6$. The lines are the theoretical predictions for (infinite) systems with cut-off $\kappa = 10^2$, the points are numerical results on finite systems with $N=10^5$ and same cut-off.
The threshold values $q_{c}$ for finite sizes are reasonably smaller compared to the theoretical predictions for infinite graphs with the same cut-off on the degree distribution.

\subsection{Discussion and Conclusions}\label{CHAP4_3_3}
This collection of examples is far from being complete, it should rather represent a way to better understand potential applications of calculations and formulae we have presented. 
Among other possible expressions for the transition probability, we mention non-monotonous or peaked degree-dependent transition probabilities, with an optimum of transmissibility for (not necessarily large) characteristic values of the degree. The actual spreading of a virus on the Internet might have such a behavior.
Highly connected nodes have large exchanges of data with their neighbors, for that they should be potentially the nodes with highest transmission capability. On the other hand, there is a common awareness of their importance, thus they are better protected and controlled, so that the effective transition probability for the spreading of viruses from these nodes is strongly reduced.
Very low-degree nodes are less protected but also less exposed to the transmission (their data exchanges are limited). 
Such scenario suggests that also non-monotonous transition probabilities are actually interesting in the study of dynamical processes on networks.\\
In conclusion, this section has been devoted to the definition of a generalized percolation with edge transition probabilities that can be used to describe spreading processes.
In the case of weighted networks, the edge transition probabilities should depend on the weights.
We have obtained two important results: 
\begin{itemize}
\item When weights and degree are not correlated, it is reasonable to assume that also the edge transition probabilities are randomly  distributed (not strictly uniformly random). It follows that when the percolation threshold is present in the topological description, it is increased; but it cannot be restored when it is topologically absent.
\item Real systems present correlations between weights and degree, thus it is reasonable to include correlations also in the expression for the edge transition probability. In this case, it is possible to restore a finite percolation threshold, even when it is topologically absent.
\end{itemize}

\chapter[The Naming Game]{Naming Game: a model of social dynamics on networks}\label{CHAP5}

\section{Introduction}\label{CHAP5_1}
In Chapters~\ref{CHAP3}-\ref{CHAP4}, we have studied the structural and functional properties of complex networks, 
supporting the phenomenological analysis of real infrastructure networks with a numerical and analytical 
investigation of theoretical models.
In such a framework, dynamical processes on networks, like exploration and spreading models, have been principally conceived as tools, by means of which a better characterization of the underlying networks properties can be achieved.\\
We turn now our attention to a more direct investigation of dynamical phenomena on complex networks, 
focusing on the field of social networks, and in particular on models of social interactions between individuals.\\
A theoretical description of social interactions at the individual level is a very complicated task, that is studied by sociologists and goes beyond the purpose of any statistical physics approach.
Though it is not possible to reproduce or predict human decisions and actions, when looking at a global scale, social interactions are much simpler to describe.
In this case, we do not need to understand all details of single individual behaviors, but only few collective aspects and their relation with the other global properties of the population under study.
In particular, the large scale observation of social interactions reveals the existence of striking collective phenomena, i.e. the mechanism, based on individual decision processes, leading to the appearance of an overall homogeneous behavior or some other global property in a population of agents.\\ 
The standard methods of statistical physics are very appropriate to study such collective behaviors, neglecting details and retaining only few general ingredients observed in real social interactions.  
For this reason, physicists have put forward a large number of theoretical models of {\em social dynamics}, borrowing a suite of statistical methods from the theory of interacting particles systems \cite{ligget,oliveira,durlauf,blume}.
The largest part of these models deals with the study of opinion formation and strategic games. For instance, the Ising majority rule has found an unexpected field of application in problems of opinion formation \cite{chen,mobilia,galam1,galam2}. On the same subject, many other models have been proposed, such as the Voter model \cite{ligget,krapivsky1,krapivsky2,krapivsky3,sood,frachebourg}, the Sznajd-Weron model \cite{sznajd}, the Axelrod model for the dissemination of culture \cite{axelrod}, and their variants (e.g. the models proposed by Deffuant et al. \cite{deffuant} and by Krause and Hegselmann \cite{krause}). 
Other types of social dynamics involve diffusion or spreading processes (e.g. the diffusion of knowledge \cite{cowan}, rumors \cite{daley}, innovations \cite{rogers} or ideas \cite{bettencourt}, etc), and the wide field of strategic games (e.g. the minority game \cite{minority_book}, the prisoner's dilemma \cite{pdilemma}, etc).\\
The behavior of these models was usually studied on regular topologies, even though it is clearly inappropriate for representing the topology of social interactions. 
For this reason, for a long time  models of social dynamics have been regarded more as academic exercises than as systems with real potential applications. 
The growing interest for networks science has recently led to a better knowledge of the topological properties of real social groups (Section~\ref{CHAP2_3}), and to the formulation of network models reproducing such properties (Section~\ref{CHAP2_4}). Consequently, many traditional models of social dynamics have been reconsidered, in order to be studied in the new, more appropriate, framework of complex networks.  
 
In this chapter, we report the main results of a series of publications \cite{naming_gameLD,naming_gameSW,naming_gameNET,naming_gameACT,naming_gameOTHER1}, in which we have studied a
recently proposed model of social dynamics, called Naming Game, investigating its dynamical behavior on both 
regular topologies and complex networks.
As we will see in the next section, {\em this model is conceived to grasp the self-organized mechanisms leading to the onset of a communication system in groups of individuals and, more in general, to describe decision processes involving pairwise interactions and negotiation, like those underlying the opinion spreading}.
In addition, with respect to other models of social dynamics, the evolution rules of the Naming Game present new ingredients, whose role will be elucidated along the chapter. They can be summarized in the following points:
\begin{itemize}
\item the introduction of {\em memory} in the individual dynamics;
\item the presence of a feedback interaction with an asymmetric interaction rule;
\item an a priori unlimited number of states (or words, opinions, etc) that is dynamically determined during the system's evolution. 
\end{itemize}
The resulting dynamics is rich of interesting properties, some of them strongly depending on the underlying topological properties of the system. Hence, the main motivation of the work is that of {\em studying the impact of different topological properties on the local and global dynamical patterns generated by the Naming Game and on the process leading to the emergence of collective  phenomena}. 

The present chapter is organized as follows. Section~\ref{CHAP5_2} contains the definition of the model and the analysis of the mean-field case. 
A topology in which agents are allowed to interact with all the others is not realistic; social environments are, on the contrary, more reliably modeled by means of networked structures, that can better account for the disparity of social relations. Thus, in Section~\ref{CHAP5_3}, we provide an exhaustive analysis of the Naming Game dynamics on various network models. We start with low-dimensional lattices and small-world networks, on which some analytical results are available.
Then we turn our attention to general complex networks, looking in particular at the effects of the node heterogeneity. The last part of the Section~\ref{CHAP5_3} is devoted to study real complex networks, whose dynamics is often surprisingly different to those observed on computer generated networks.    
The internal activity of the agents governing the decision process is described in Section~\ref{CHAP5_4}, while the conclusions are exposed in Section~\ref{CHAP5_5}.
We refer to the Appendices~\ref{APP5_1}-\ref{APP5_2} for the analysis of some more technical issues that are omitted in the main text.

\newpage
\section{Naming Game: General features}\label{CHAP5_2}
Social interactions are based on the existence of a communication system among the agents, who are able to understand each other by means of common linguistic patterns or, more generally, by means of a common vocabulary of symbols. \\
Such a communication system is the result of a self-organized process in which individuals select
specific symbols ({\em words}) and associate them to concepts and ideas ({\em objects}).
The emergence of a shared lexicon inside social groups and communities of people is very likely to be driven by simple criteria, like popularity, imitation, negotiation, and agreement.
When a new concept is introduced, people refer to it using several different names or words. These words start spreading among the population, competing one against the other, until the choice of one of them is taken (with a sudden transition or with a long process) and everyone uses the same word (or symbol, etc) \cite{lass,briscoe,hurford}. 
This kind of dynamics has recently become of broad interest after the diffusion of a new generation of web-tools which enable human users to self-organize a system of tags in such a way to ensure a shared classification of information about different arguments (see, for instance, {\texttt del.icio.us} or {\texttt www.flickr.com} and Refs.~\cite{cattuto,huberman}).   
Another application concerns global coordination problems in artificial intelligence, where a group of artificial embodied agents moving in an unknown environment have to exchange informations about the objects they gradually discover.
The emergence of consensus about the objects names allows to establish a communication system.
A practical example of this type of dynamics is provided by the well-known Talking Heads experiment \cite{talking_heads,talking_heads2}, in which embodied software agents develop their vocabulary observing objects through digital cameras, assigning them randomly chosen names and negotiating these names with other agents.\\
In other words, the process leading to the emergence of a communication system in a population of agents (e.g. a social network) is an example of social collective phenomenon with polarization of individual opinions or ideas.
On the base of these observations, a new field of research called {\em Semiotic Dynamics} has been developed \cite{semiotic}, that  investigates by means of simple models {\em how (linguistic) conventions originate, spread and evolve over time in a population of agents endowed with simple internal states and local pairwise negotiation interactions}. \\ 
The fundamental model of Semiotic Dynamics is the so-called Naming Game \cite{firstNG}, in which a population of agents, interacting by pairwise negotiation rules, try to assign a common name to an object. 
Moreover, this model can be studied as an alternative model of opinion formation, since in place of names to be assigned to an object we can think to the competition of different opinions on a given topic.  \\
The next section is devoted to the description of the Naming Game model, while the subsequent one discusses the mean-field case.
%
\begin{figure}[t] 
\centerline{
\includegraphics*[width=10.0cm]{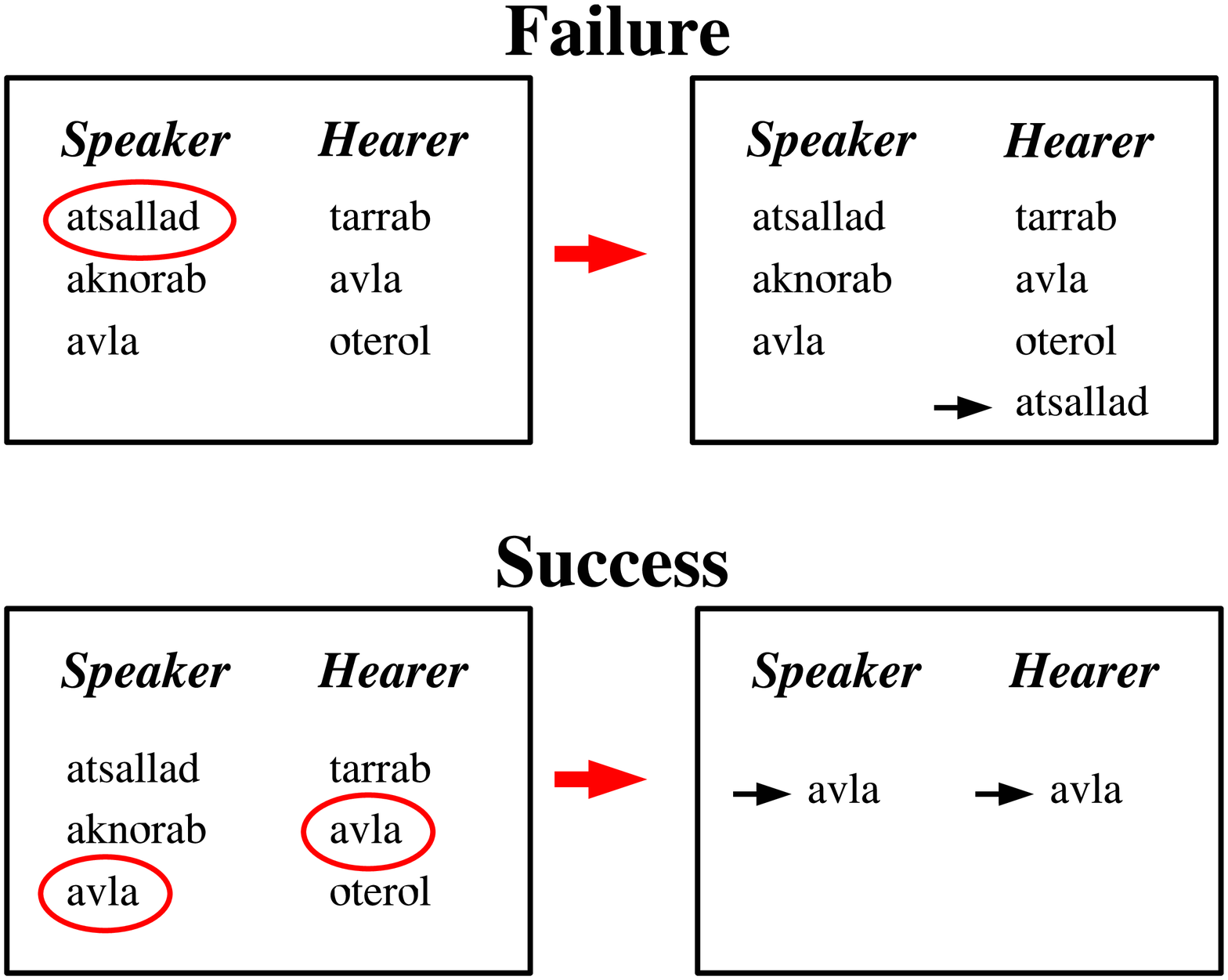}
}
\caption{
Agents interaction rules. Each agent is described by its inventory, i.e. the repertoire of known words.
The speaker picks up at random a name in its inventory and transmits it to the hearer. If the hearer does not know the selected word the interaction is a failure (top), and it adds the new name to its inventory. Otherwise (bottom), the interaction is a success and both agents delete all their words but the winning one. Note that if the speaker has an empty inventory (as it happens at the beginning of the game), it invents a new name and the interaction is a failure.
}
\label{parole}
\end{figure}

\subsection{The Model}\label{CHAP5_2_1}
A minimal model of Naming Game has been put forward by Baronchelli et al. in Ref.~\cite{baronchelli} to reproduce the main features of Semiotic Dynamics and the fundamental results of adaptive coordination observed in the Talking Heads experiment.  The minimal Naming Game model consists of a population of $N$ agents observing a single object, for which they invent names that they try to communicate to one another through pairwise interactions, in order to reach a global agreement. 
The agents are identical and dispose on an internal inventory, in which they can store an a priori unlimited number of names (or opinions). All agents start with empty inventories. 
At each time step, a pair of neighboring agents is chosen randomly, one playing as ``speaker'', 
the other as ``hearer'', and negotiate according to the following rules (see also Fig.~\ref{parole}):
\begin{itemize}
\item the speaker selects randomly one of its words and conveys it to the hearer;
\item if the hearer's inventory contains such a word, the two agents update their inventories in order to keep only the word involved in the interaction ({\em success});
\item if the hearer does not possess the uttered word, the latter is added to those already stored in the hearer's inventory ({\em failure}), i.e. it learns the word.
\end{itemize}
Note that the time unit corresponds to the pairwise interaction, in contrast to usual statistical mechanics models in which it corresponds to $N$ interactions. However, in order to compare the results for the present dynamical rule with well-known results of other models, in many cases, we will use a rescaled time $t/N$.\\ 
Before entering in the detailed description of the dynamics, it is worthy to specify some visible differences of the Naming Game with other commonly studied models of social dynamics and, in particular, of opinion formation \cite{sznajd,deffuant,axelrod,chen,mobilia,ligget}.
\begin{itemize}
\item Each agent can potentially be in an infinite number of possible discrete states (words, names, opinions), and {\em the maximum number of states depends on the dynamical evolution} itself. This is in contrast with traditional models (Voter, Potts, etc) in which the number of states is a fixed external parameter taking finite (and usually small) values.
\item The {\em two-steps decision process} is realistic: an agent can accumulate in its memory different possible names for the object, waiting before reaching a decision. This feature is probably at the origin of the surface tension that emerges from the  dynamics in low-dimensional lattices (see sec.~\ref{CHAP5_3}) and of the non-poissonian behavior at the agents level, a property that will be discussed in Section~\ref{CHAP5_4}.
\item Each dynamical step can be seen as a {\em negotiation} between speaker and hearer, with a certain degree of stochasticity, that is absent in deterministic models such as the Voter model. The stochastic component is however of a different nature compared to that of standard Glauber dynamics used in majority rule models \cite{glauber}, since here it comes from an internal selection criterion, and involves only the speaker, without affecting the (deterministic) decision process of the hearer.         
\end{itemize}
A second important remark concerns the random extraction of the word in the speaker's inventory. Most previously proposed models of semiotic dynamics attempted to give a more detailed representation of the negotiation interaction assigning weights to the words in the inventories. In such models, the word with largest weight is automatically chosen by the speaker and communicated to the hearer. Success and failures are translated into updates of the weights: the weight of a word involved in a successful interaction is increased to the detriment of those of the others (with no deletion of words); a failure leads to the decrease of the weight of the word not understood by the hearer.    
An example of a model including weights dynamics can be found in Ref.~\cite{lenaerts} (and references therein).
For the sake of simplicity the minimal Naming Game avoids the use of weights, that are apparently more realistic, but their presence is not essential for the emergence of a global collective behavior of the system.\\
An important point needs to be stressed: while in the original experiments the embodied agents could observe a set of different objects, in the minimal Naming Game {\em all agents refer to the same single object}. This is actually possible only if we assume that homonymy is excluded, i.e. two distinct objects cannot have the same name. Consequently, in this model, all objects are independent and the general problem reduces to a set of independently evolving systems, each one described by the minimal model. In more realistic situations, however, the occurrence of homonymy cannot be neglected.

\subsection{Mean-field case}\label{CHAP5_2_2}
 
In the original work on the minimal Naming Game model \cite{baronchelli}, Baronchelli et al. have focused on the {\em mean-field} case, in which the agents are placed on the vertices of a complete graph, corresponding to study a population in which all pairwise interactions are allowed. 
By means of numerical simulations, they investigated the overall dynamics of the model, monitoring along the evolution three main global quantities:
\begin{itemize}
\item the total number $N_{w}(t)$ of words in the system at the time $t$ (i.e. the total size of the memory);
\item the number of different words $N_{d}(t)$ in the system at the time $t$;
\item the average success rate $S(t)$ as function of the time, i.e. the probability, computed averaging over many simulation runs, that the chosen agent gets involved in a successful interaction at a given time $t$.
\end{itemize}
%
\begin{figure}[t] 
\centerline{
\includegraphics*[width=10.0cm]{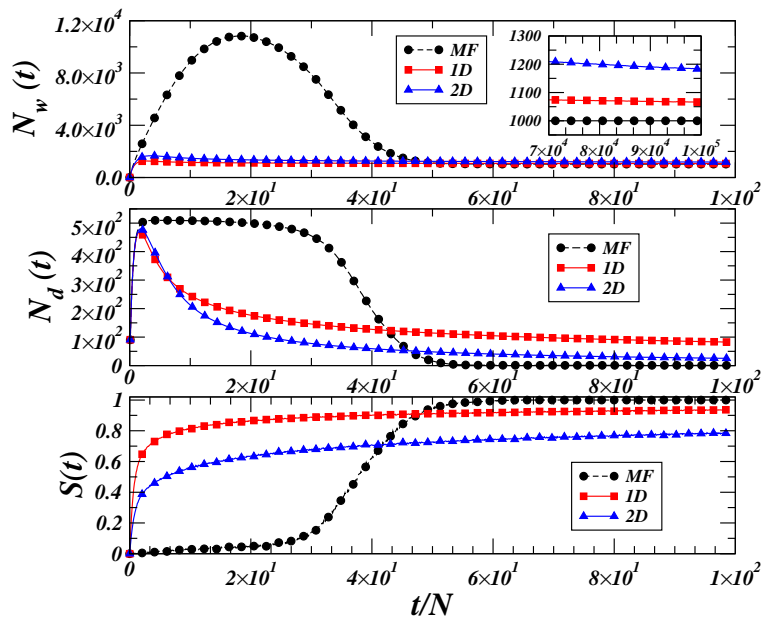}
}
\caption{Evolution of the total number of words $N_{w}$ (top), of the number of different words
$N_{d}$ (center), and of the average success rate $S(t)$ (bottom), for a mean-field system (black circles) and low-dimensional lattices ($1D$, red squares and $2D$, blue triangles) with $N=1024$ agents, averaged over $10^3$ realizations. The inset in the top graph shows the very slow convergence in low-dimensional lattices.}
\label{figMF}
\end{figure}
Note that all these quantities are zero at the beginning of the evolution, when all inventories are empty, while they reach a stable value when the system enters the absorbing state corresponding to a global agreement.
In fact, it is possible to show, using a sort of Liapunov functional, that {\em the mean-field model always converges to the absorbing state (consensus)} \cite{baronchelli}.
The consensus state is  defined by observing $N_{d}=1$ and $N_{w}=N$ (moreover it implies $S=1$).\\
The temporal evolution of the three main quantities is depicted in Fig.~\ref{figMF} (circles). 
At the beginning, many disjoint pairs of agents interact, with empty initial inventories: they invent a large number of different words, that start spreading throughout the system, through failure events. 
Indeed, the number of words decreases only by means of successful interactions. 
In the early stages of the dynamics, the overlap between the inventories is very low, and successful interactions are limited to those pairs which have been chosen at least twice. Since the number of possible partners of an agent is of order $N$, it rarely interacts twice with the same partner, the probability of such an event growing as $t/N^2$. 
Note that this remark is in good agreement with the initial behavior of the success rate $S(t)$ depicted in Fig.~\ref{figMF}. The initial trend of $S(t)$ (black circles) is linear with a slope of order $1/N^{2}$ (whose correct value has been computed in Ref.~\cite{baronchelli}). \\
In this phase of uncorrelated proliferation of words, the number of different words $N_{d}$ invented by the agents grows, rapidly reaching a maximum that scales as $\mathcal{O}(N)$. 
Then $N_{d}$ saturates, displaying a plateau, during which no new word is invented anymore (since every inventory contains at least one word).
The total number $N_{w}$ of words stored in the system has a similar behavior, but it keeps growing after $N_{d}$ has saturated, since the words continue to propagate throughout the system even if no new one is introduced. \\
The peak of $N_{w}$ has been shown to scale as $\mathcal{O}(N^{1.5})$ \cite{baronchelli}, meaning that each agent stores $\mathcal{O}(N^{0.5})$ words. This peak occurs after the system has evolved for a time $t_{max} \sim \mathcal{O}(N^{1.5})$.
In the subsequent dynamics, strong correlations between words and agents develop, driving the system to a rather fast convergence to the absorbing state in a time $t_{conv} \sim \mathcal{O}(N^{1.5})$. \\
The S-shaped curve of the success rate in Fig.~\ref{figMF} summarizes the dynamics: initially, agents hardly understand each others ($S(t)$ is very low); then the inventories start to present significant overlaps, so that $S(t)$ increases until it reaches $1$, and the communication system is completely set in.       
\begin{figure}[t] 
\centerline{
\begin{tabular}{|c|}\hline \\ \includegraphics*[width=8.0cm]{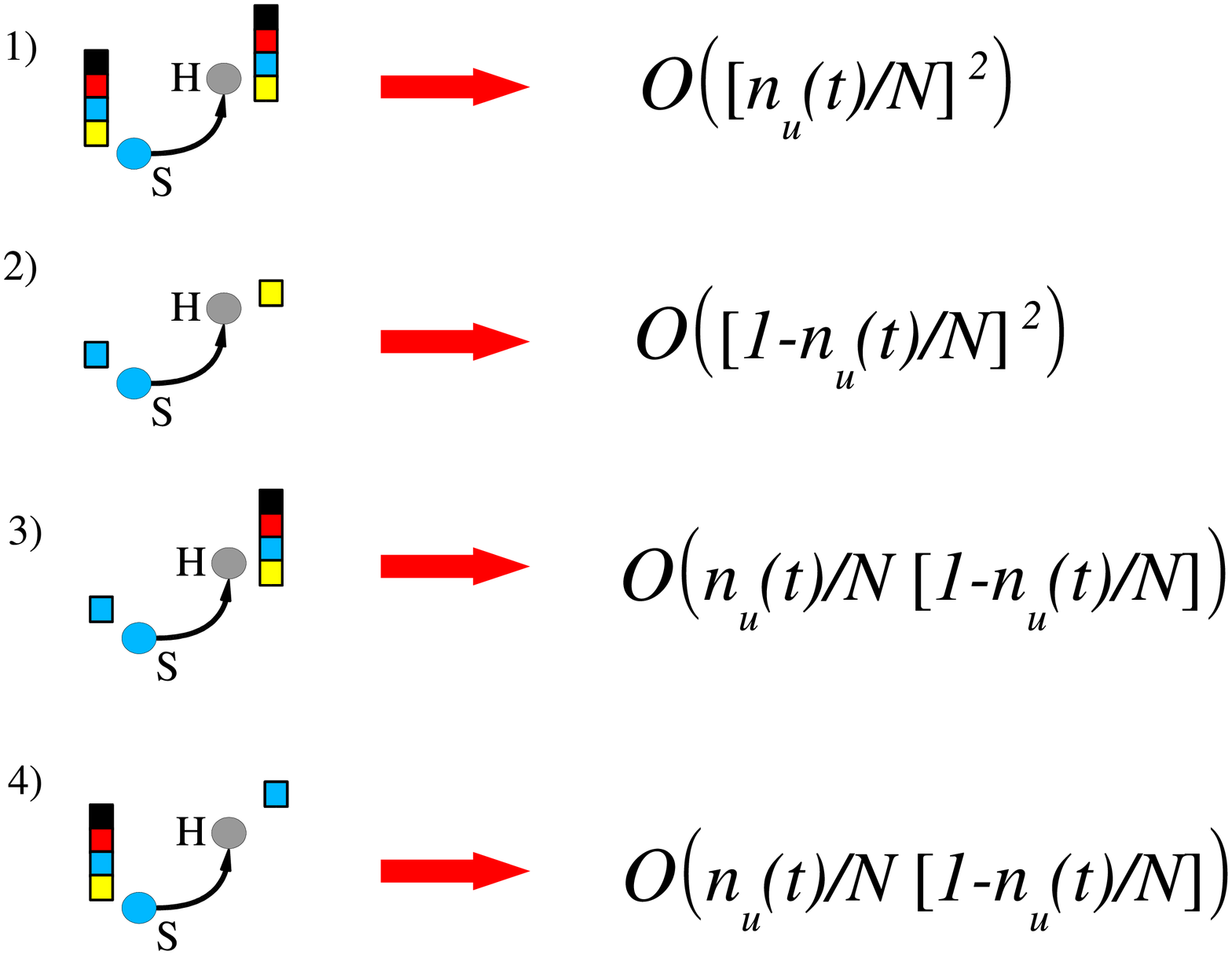}\\ \hline\end{tabular}~~
}
\caption{Graphical representation of the four possible terms contributing to the rate equation for the number of undecided agents and the order of magnitude of the term. Note that, around $t_{max}$, the fraction $n_{u}(t)/N \simeq 1$, and then it decreases to zero; therefore, in principle, all terms play a role during the evolution.}
\label{rateMF}
\end{figure}
Despite the apparent simplicity of the dynamics, it is very difficult to study this model analytically, because the interaction rule is not totally stochastic (the hearer's moves are deterministic) and the update of the inventories after a success is highly non-linear.
Nevertheless, the authors of Ref.~\cite{baronchelli} provided several qualitative theoretical arguments in order to explain the main properties of the population's global behavior.
We consider here only the argument for the scaling behavior of the maximum number of words in the system.\\
From the simulations, we know that the maximum number of words scales as a power of the population size $N$, therefore let us assume that, at the maximum, the number of words per agent scales as $c N^{\alpha}$, for a certain positive value of the exponent $\alpha$, and a constant $c \sim \mathcal{O}(1)$. Note that $\alpha \leq 1$, since the number of different words is at most $\mathcal{O}(N)$.\\  
In order to determine the value of $\alpha$, we use the fact that, at the maximum, the first temporal derivative of the total number of words vanishes, i.e. $\frac{d N_{w}(t)}{d t} \simeq 0$.
On the other hand, the probability that a word chosen by the speaker is present in the inventory of the hearer is approximately $\frac{c N^{\alpha}}{N}$ in absence of correlations. We get the following rate equation \cite{baronchelli},
\begin{equation}\label{scaling_max}
\frac{d N_{w}(t)}{d t} \propto  \left(1-\frac{c N^{\alpha}}{N} \right) - \frac{c N^{\alpha}}{N} 2cN^{\alpha} \enspace. 
\end{equation}
The first term at the r.h.s. of Eq.~\ref{scaling_max} is the gain term (in case of a failure), while the second term represents the loss of $2cN^{\alpha}$ words in case of successful interaction.\\
At the higher order in powers of $N$, i.e. neglecting terms decreasing as $\alpha -1$, the balance condition $\frac{d N_{w}(t)}{d t} \simeq 0$ is satisfied if $\alpha = 1/2$.
This results proves that the maximum number of words in the system scales as $N^{3/2}$. From the same argument we can as well obtain the scaling of the time at which the peak occurs, $t_{max} \sim N^{3/2}$.\\
The study of the convergence time is much more difficult, and no theoretical arguments have been proposed yet (except for some results on the limit of infinite dimensional lattices reported in Section~\ref{CHAP5_3_1}). 
It is however possible to write down a very naive argument for the scaling (with the size $N$) of the convergence time, i.e. $t_{conv} \sim \mathcal{O}(N^{3/2})$.  This argument does not want to be rigorous, but its importance is principally that of showing which kind processes lead the system to a consensus state, justifying the observed behaviors.\\ 
Let us consider the number $n_{u}(t)$ of agents with more than one word in the inventory (we denote them as {\em undecided} agents), around the peak in the number of words $n_{u}\simeq N$, while at the convergence $n_{u}\simeq 0$. 
We can now study the way the system approaches the consensus writing a rate equation for this quantity.\\
According to the actual pairwise interaction rule of the Naming Game, there are many different contributions that should be taken into account (see Fig.~\ref{rateMF}). Moreover, the negotiation process introduces correlations related to the feedback mechanism in case of success (i.e. both the speaker and the hearer update the inventory).
On the other hand, in a mean-field system, the set of words stored in the inventories should be approximately the same for all agents, suggesting to neglect correlations and focus on the behavior of a single node subjected to the average effect of the rest of the system. 
Since the hearer's interaction rule is strictly deterministic, while the speaker chooses randomly in its inventory, it seems more reasonable to consider the (mean-field) average on the hearer's term. \\
\begin{figure}[t] 
\centerline{
\includegraphics*[width=7.5cm]{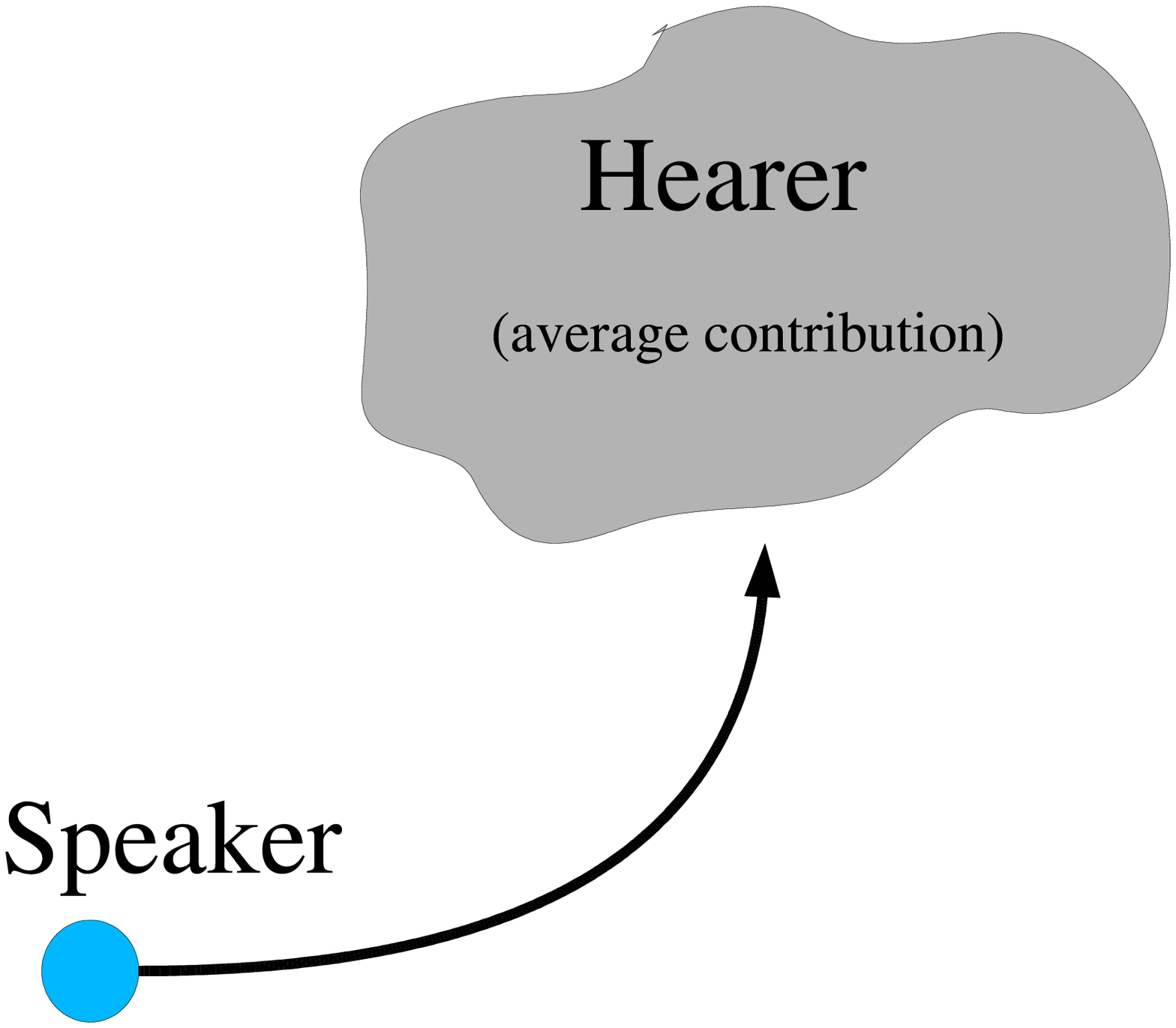}
\includegraphics*[width=7.5cm]{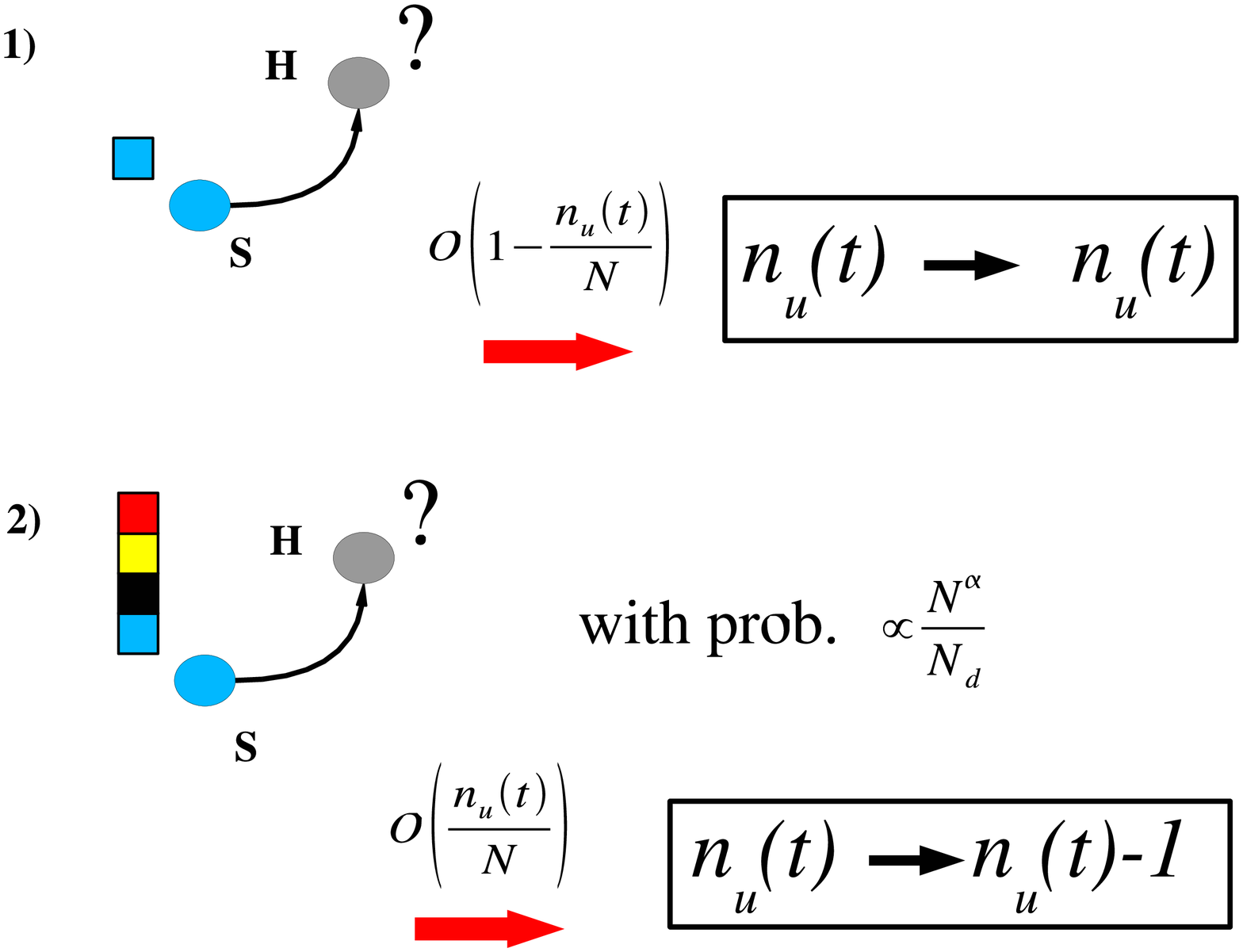}
}
\caption{A naive representation (left) of the mean-field approximation of pairwise interactions: we focus on the speaker's behavior, substituting single hearer's dynamics with a mean-field average quantity. Therefore, only speakers contributions to the rate equation for $n_{u}(t)$ have to be considered, that is provided by only two possible interactions (right). 
We can neglect the first term, in which the speaker has only one word, because it does not change the number of undecided nodes. The second term, that diminishes of one unit $n_{u}(t)$, is the only contribution to Eq.~\ref{rate_undecided} in this rough approximation.}
\label{rateMF2}
\end{figure}
We can now write down an equation for the number of undecided speakers $n_{u}(t)$. 
First, each node is chosen as speaker or hearer with the same probability, thus along the temporal evolution all nodes will play as speakers and the equation can be considered to give a rough but correct approximation of the real dynamics.
Then, the four types of interactions depicted in Fig.~\ref{rateMF} can be grouped in only two terms (see Fig.~\ref{rateMF2}): one in which the speaker is an undecided node, the other in which the speaker has only one word into the inventory. The latter case do not contribute to the rate equation for $n_{u}(t)$, because in case of both success and failure the number of undecided speakers does not change. The only significant contribution is the second term in Fig.~\ref{rateMF2}, in which $n_{u}(t)$ is decreased by $1$ with a fixed probability. \\
The naive rate equation for the number of undecided agents reads  
\begin{equation}\label{rate_undecided}
\frac{d n_{u}(t)}{d t} \propto  -\frac{n_{u}(t)}{N} S(t)\enspace, 
\end{equation}
in which $\frac{n_{u}(t)}{N}$ is the probability of choosing an undecided agent as speaker, and  $S(t)$ is the success rate, i.e. the average probability of having a successfull interaction. 
The functional form of the success rate is not theoretically known, but we see from Fig.~\ref{figMF} that it varies very slowly after the peak until the convergence process sets in, then it shows a super-exponential saturation to $1$, being enhanced by the decrease of the number of redundant words. Using numerical results for $S(t)$ and $n_{u}(t)$ we have checked that the relation in Eq.~\ref{rate_undecided} is in fact approximately satisfied.\\
According to this picture, Eq.~\ref{rate_undecided} sheds light on the type of self-accelerating cascade process leading the system to the absorbing state. Just before the peak of $N_{w}(t)$, the success rate increases only linearly as $t/N^2$, then it slows down before the sudden acceleration leading to the convergence; therefore after the peak $S(t) \sim \mathcal{O}(1/\sqrt{N})$. This result can be obtained also recalling that the average probability of success (i.e. the success rate) is given by the ratio between the average inventory size and the number of different words present in the system. At the beginning of the convergence process, just after the peak, $N_{d}(t)$ is almost constant while the average inventory size is slightly decreasing (starting from $\mathcal{O}(\sqrt{N})$), i.e. $S(t) \sim \mathcal{O}(1/\sqrt{N})$. Then both these quantities start to decrease until they reach $1$ (at the convergence). \\
Solving Eq.~\ref{rate_undecided} in the temporal range just after $t_{max}$, where $S(t) \sim \mathcal{O}(1/\sqrt{N})$ varies very slowly, we realize that the system has already entered the convergence process, that is approximately exponential with characteristic time of order $N^{3/2}$ (justifying the observed scaling). However, this exponential decrease immediately triggers the growth of the success rate, so that the global convergence becomes a self-enhancing process that results into a super-exponential  approach to the absorbing state.

\newpage 
\section{The Role of the Topology}\label{CHAP5_3}
The mean-field case is, from many points of view, rather unrealistic. The agreement process 
described by the Naming Game model is, indeed, an example of social dynamics, that is likely to 
take place on more realistic topological structures like those characterizing social networks.
We expect the topology to have an effect on the dynamics, in particular on the time required to reach the 
absorbing state and on the process of words propagation.
It is thus interesting to study the Naming Game on different topologies, expecially on complex networks, whose structure 
presents properties that are typically observed in social environments.\\
Studies of dynamical processes on complex networks showed that the main properties governing the overall dynamics are the small-world property and the presence of the hubs in heterogenous networks; it is thus important to understand {\em which is the impact of these properties on the dynamics}.
For this reason, we start analyzing, in Section~\ref{CHAP5_3_1}, the behavior of the Naming Game on low-dimensional lattices, which we prove to be governed by coarsening dynamics; then we show that the presence of topological shortcuts (i.e. the smallworld property) is responsible for a crossover from a slowly converging process to a faster one (Section~\ref{CHAP5_3_2}).
Finally, in Section~\ref{CHAP5_3_3}, a survey of the behavior of the Naming Game on complex networks is considered, giving special attention to the role played by the hubs in heterogeneous topologies. \\
The behavior of the Naming Game on different topologies has recently attracted the attention of researchers in the field of A.I., who are interested in knowing {\em which topology ensures the best trade-off between a fast convergence to consensus and an optimization of the required memory per agent}. For this reason our work is expected to be relevant for the application of similar models to the description and/or modeling of learning processes of robots \cite{kirby,kaplan}.   

\subsection{Coarsening dynamics on low-dimensional lattices}\label{CHAP5_3_1}
\begin{figure}[t] 
\centerline{ \includegraphics*[width=10.0cm]{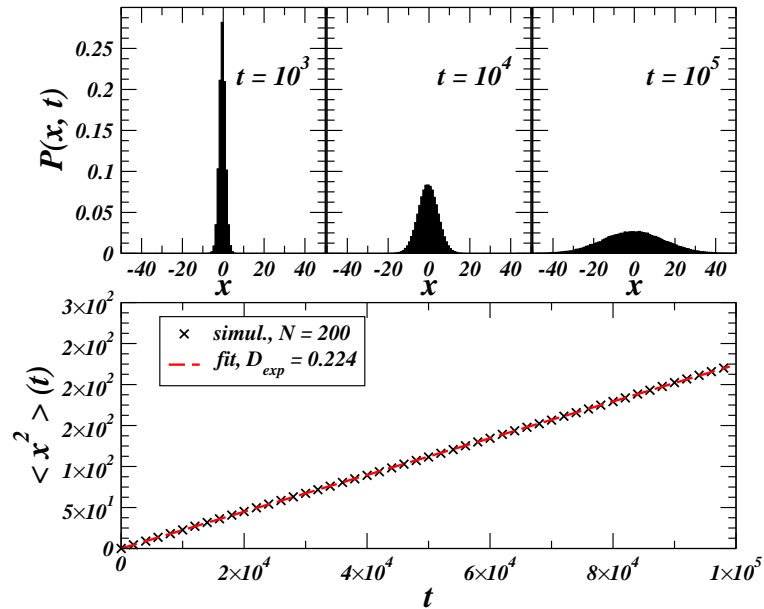} }
\caption{Evolution of the position $x$ of an interface $\cdots
AAABBB\cdots$. (Top) Evolution of the distribution $\mathcal{P}
(x,t)$. (Bottom) Evolution of the mean square displacement, showing a
clear diffusive behavior $\langle x^2\rangle = 2 D_{exp} t/N$ with a
coefficient $D_{exp}\approx 0.224$ in agreement with the theoretical
prediction.  }
\label{gaussian_1D}
\end{figure}

\begin{figure}[t] 
\centerline{
\includegraphics*[width=10.0cm]{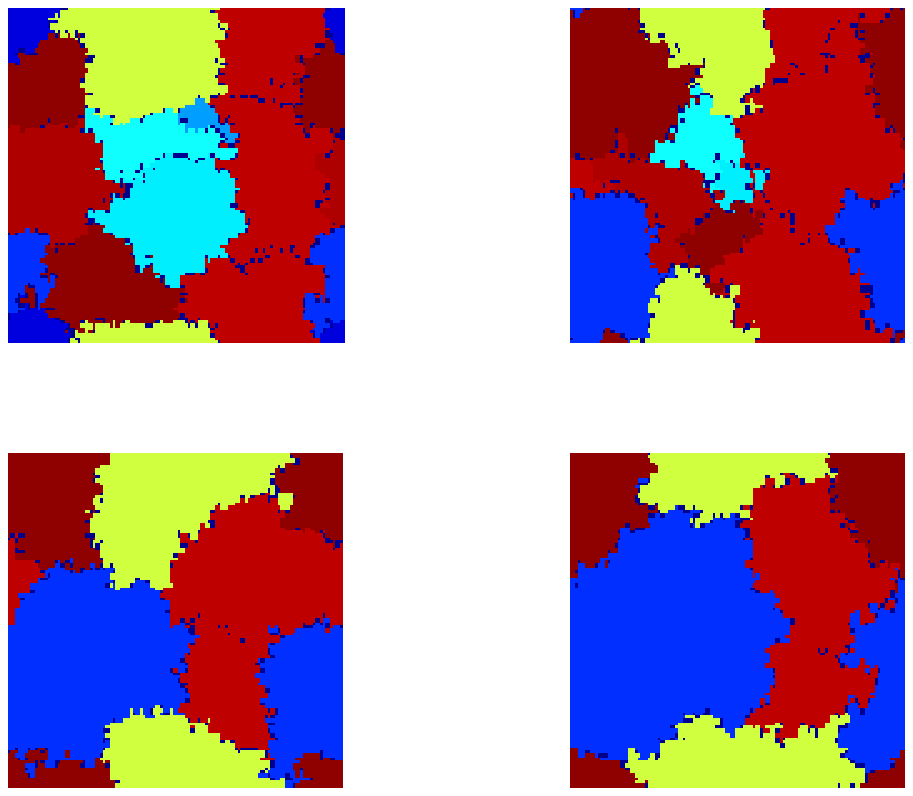}
}
\caption{Four snapshots at different times (lexicographic order from the top-left corner) for a two-dimensional system. 
The form of the clusters (composed of agents with a unique word) clearly show the presence of an effective surface tension, associated with coarsening dynamics. The width of the interfaces (in black) is very small as already seen for the one-dimensional model.
}
\label{evolution_2D}
\end{figure}

\begin{figure}[t] 
\centerline{
\includegraphics*[width=10.0cm]{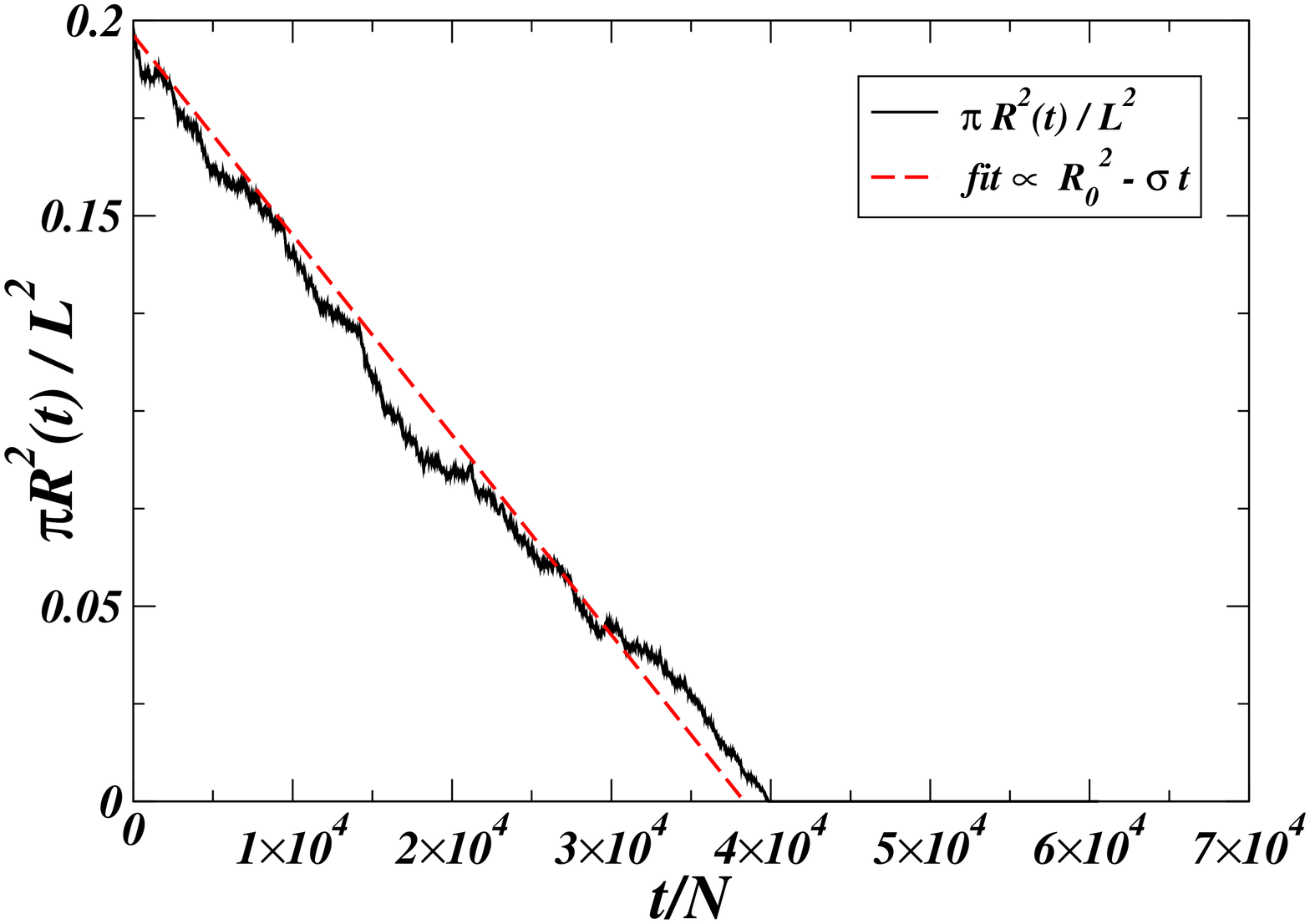}
}
\caption{A measure of the surface tension in the two-dimensional Naming Game. The initial condition is a configuration ($N=L^2=400^2$) with only two clusters, an internal droplet (a bubble) of radius $R_{0}=100$ and a surrounding sea with a different word. Initially all agent already possess one word, thus no new words are created, but the presence of surface tension at the interface between the two clusters provokes the decrease of the bubble's size. We have monitored the normalized area $\pi R^2(t)/L^2$ as a function of the time. According to a coarsening dynamics, the radius $R(t)$ decreases as $\sqrt{t}$. Note that the linear trend is very clear even if the data refer to a single realization. 
}
\label{dornic_2D}
\end{figure}

When the Naming Game is embedded in a regular $d$-dimensional lattice, the agents interact only with their $2d$ neighbors, and the overall dynamical properties turn out to be very different compared to the mean-field case. In particular, the 
time required by the system to reach the global consensus displays a different scaling with the size $N$, and the 
effective size of the inventories is considerably diminished.\\ 
Actually, the existence of different dynamical patterns are clearly visible in Fig.~\ref{figMF}, where we have reported the curves for the total number of words $N_{w}(t)$, the number of different words $N_{d}(t)$ and the success rate $S(t)$ in the cases of mean-field topology (circles),  one-dimensional lattice (squares) and two-dimensional lattice (triangles).  \\
At the early stages of the dynamics, we observe a sharp growth of the success rate, meaning that agents easily find a local agreement with their neighbors; then, the dynamics seem to slow down and the convergence is reached in a much larger time with respect to the mean-field. We know that, in the initial phase, the success rate is equal to the probability that two agents that have already played are chosen again, and is proportional to $t/E$, where $E$ is the number of possible interacting pairs (i.e. the edges). In the mean-field, $E \propto N^2$, while in finite $d$-dimensional systems $E \propto Nd$, explaining for the observed slopes of $S(t)$ (Fig.~\ref{figMF}): this quantity grows $N$ times faster in finite dimensions.
\begin{figure}[t] 
\centerline{
\includegraphics*[width=10.0cm]{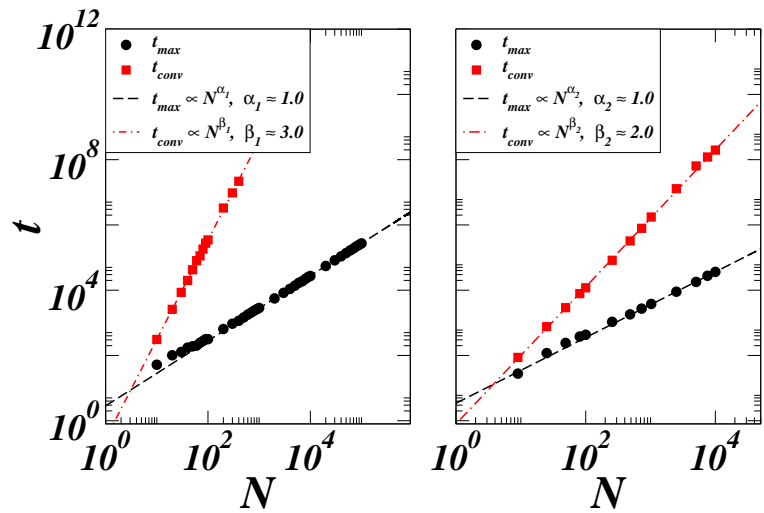} }
\caption{
Scaling of the time at which the number of words is maximal, and
of the time needed to obtain convergence, in $1$ and $2$ dimensions.
}
\label{scaling_LD}
\end{figure}
%
At larger times, the eventual convergence is much slower in finite dimensions than in mean-field.\\
The curves for $N_{w}(t)$ and $N_{d}(t)$ display in all cases a sharp increase at short times, a maximum for a given time $t_{max}$ and then a decay towards the consensus state in which all agents share the same unique word, reached at $t_{conv}$. The short time regime corresponds to the creation of many different words by the agents. 
After a time of order $N$, each agent has played typically once, and therefore $\mathcal{O}(N)$ different words have been invented (typically $N/2$).  
In mean-field, each agent can interact with all the others, so that it can learn many different words; in contrast, 
in finite dimensions words can spread only locally, and each agent has access only to a finite number of different words.
The total memory used scales as $N$, and the time $t_{max}$ to reach the maximum number of words in the system scales as $N^{\alpha_{d}}$, with $\alpha_{1}=\alpha_{2}=1$ (Fig.~\ref{scaling_LD}).
No plateau is observed in the total number of distinct words since coarsening of clusters of agents soon starts to eliminate words.
%
\begin{figure}[t] 
\centerline{
\includegraphics*[width=10.0cm]{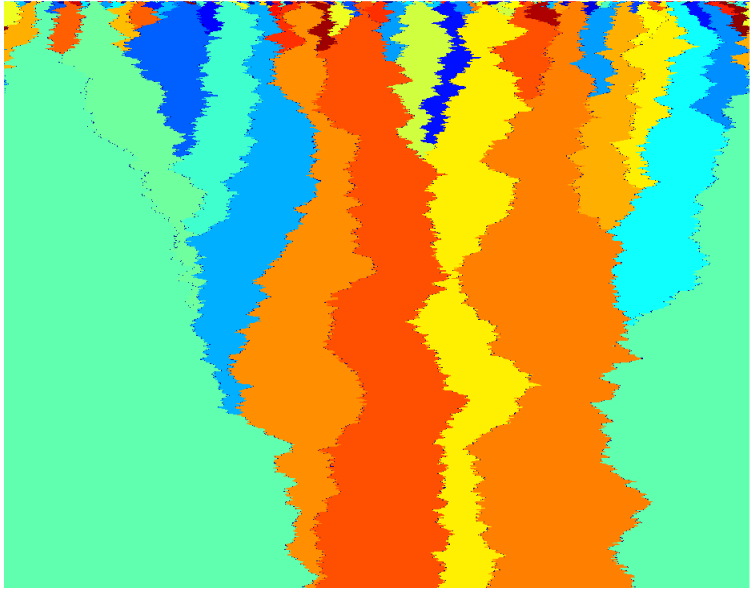}
}
\caption{
Typical evolution of a one-dimensional system ($N=1000$). Black color 
corresponds to interfaces (sites with more than one word). The other 
colors identify different single state clusters. The vertical axis 
represents the time (from top to bottom, $1000 \times N$ sequential steps), the one-dimensional
snapshots are reported on the horizontal axis.
}
\label{evolution_1D}
\end{figure}
Furthermore, the system reaches the consensus in a time $t_{conv}$ that grows as $N^{\beta_{b}}$ with $\beta_{1} \simeq 3$ in one dimension and $\beta_{2} \simeq 2$ in two dimensions (while in the mean-field case $\beta_{MF} \simeq 1.5$).\\
In Fig.~\ref{evolution_1D}, we have represented a typical evolution of agents on a one-dimensional lattice, by displaying one below the others a certain number of (linear) configurations corresponding to successive equally separated temporal steps.
Each agent with a unique word in memory is presented by a colored point, while agents with more than one word in memory are shown in black.
The figure clearly shows the growth of clusters of agents with a unique word by diffusion of interfaces made of agents with more than one word in the inventory. The fact that the interfaces remain small is however not obvious a priori, and requires a deep analytical investigation that is presented in Appendix~\ref{APP5_1}.\\ 
The results exposed in Appendix~\ref{APP5_1}  are confirmed by numerical simulations as illustrated in Fig.~\ref{gaussian_1D}: we have found that the probability $\mathcal{P}(x,t)$ to find an interface in position $x$ at time $t$ is a Gaussian around the initial position, while the mean-square distance reached by the interface at time $t$ follows the diffusion law $\langle x^2 \rangle = 2 D_{exp} t /N$, with experimental diffusion coefficient $D_{exp} \simeq 0.224$.\\
The dynamical evolution of clusters in the Naming Game on a one-dimensional lattice can be described as follows: at short times, pairwise interactions create $\mathcal{O}(N)$ small clusters, divided by $\mathcal{O}(N)$ thin interfaces (see the first lines in Fig.~\ref{evolution_1D}). The interfaces then start diffusing. When two interfaces meet, the cluster situated in between the interfaces disappears, and the two interfaces coalesce. Such a coarsening leads to the well-known growth of the typical size $\xi$ of the clusters as $t^{1/2}$.
The density of interfaces, at which unsuccessful interactions can take place, decays as $1/\sqrt{t}$, so that $1-S(t)$ also decays in the same way.
Moreover, starting from an initial configuration in which agents have no words, a time $N$ is required to reach an average cluster size of order $1$, so that $\xi$ grows as $\sqrt{t/N}$ (as also shown in the Appendix~\ref{APP5_1} by the fact that the diffusion coefficient is $D/N$). This remark explains the time $t_{conv} \sim \mathcal{O}(N^{3})$ needed to reach the global agreement, i.e. $\xi = N$.\\
%
This framework can be extended to the case of higher dimensions, starting from the case $d=2$. 
Fig.~\ref{evolution_2D} shows four different snapshots of the system during the evolution. The interfaces of two-dimensional clusters, although quite rough, are well defined and their width does not grow in time, which points to the existence of an effective surface tension. 
We have substantiated this picture with two types of measures: following Dornic et al. \cite{dornic}, one is the measure of the time dependence in the implosion of a cluster of a given radius; the other is the numerical computation of the equal-time pair correlation function.\\
Let us consider the time dependent behavior of the linear size of a droplet, i.e. a cluster with the form of a bubble in a sea of sites with a different word. As predicted by the theory of coarsening phenomena \cite{bray}, the radius $R(t)$ of the bubble decreases as $\sqrt{R_{0}^2 - \sigma t}$, where $\sigma$ is a coefficient related to the surface tension (see Fig.~\ref{dornic_2D}).
The equal-time pair correlation function $C(r,t)$ in dimension $d=2$ (not shown) can be rescaled using the scaling function $C(r,t) \sim f(r/t^{1/2})$, where $r$ is the linear length,  indicating that the characteristic length scale $\xi$ grows as $\sqrt{t/N}$ (a time $\mathcal{O}(N)$ is needed to initialize the agents to at least one word and therefore to reach a cluster size of order $1$). This result is in agreement with coarsening dynamics for non-conserved fields \cite{bray}. In terms of linear length scale $\xi$, the convergence time  $t_{conv}$ corresponds to the time necessary to reach $\xi = N^{1/d}$, thus we expect $t_{conv} \sim N^{1+2/d}$. This scaling has been verified by numerical simulations in $d=2$ and $d=3$ (not shown).\\ 
Note that the obtained scaling law is the same of coarsening dynamics in non-equilibrium Ising models, but an additional $N$ factor comes from the different timescale used (here, it corresponds to a single interaction instead of $N$ interactions). For the presence of surface tension and the evolution rule with many possible states, the Naming Game model is similar to Potts-like models, but in the Naming Game the number of states (words, opinions, etc) is determined by the dynamics, not fixed a priori. This feature is determinant in the overall behavior of the system, the convergence time depending both on the ordering process and the initial words spreading. \\
In low-dimensional lattices, the time of the peak in the number of words scales as $t_{max} \sim \mathcal{O}(N)$, thus it is dominated by the second part of the dynamics, that is considerably slower. Increasing the dimension, the peak height and time increases, approximately as $N\sqrt{d}$, while the coarsening time decreases as $N^{1+2/d}$. 
For $d=4$, the global consensus is reached in a time that scales as for the mean-field, but the mean-field behavior is dominated by the scaling of the time of the peak ($\mathcal{O}(N^{3/2})$), that is smaller ($\mathcal{O}(N)$) in all finite dimensions. Hence, the general non-equilibrium dynamics of the Naming Game model in finite dimensional lattices is the result of the interplay between two different dynamical regimes (i.e. of creation and elimination of words).   
%

\subsection{Crossover to a fast-converging process in small-world networks}\label{CHAP5_3_2}

\begin{figure}[t] 
\centerline{
\includegraphics*[width=8.0cm]{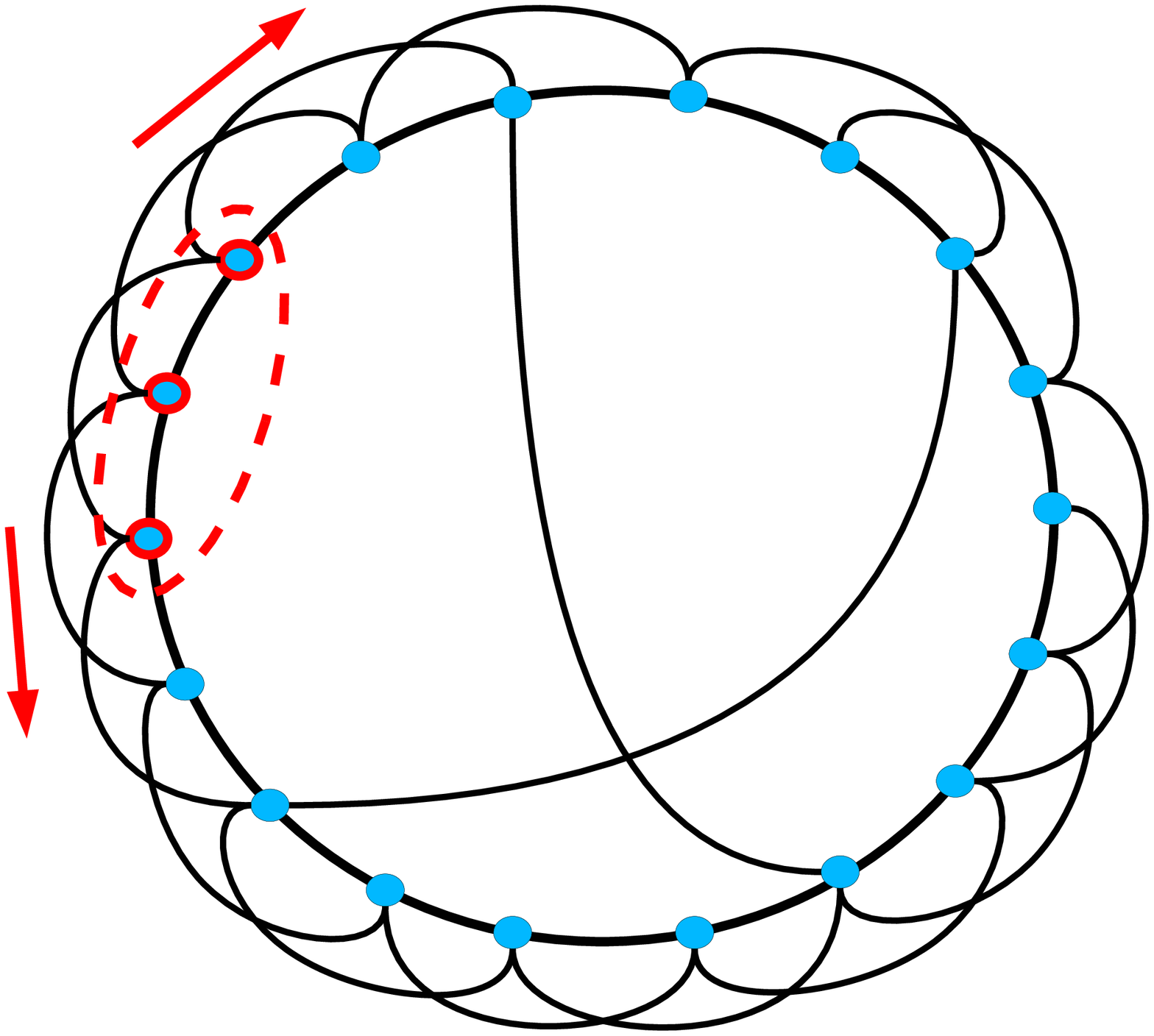}
}
\caption{A naive representation of clusters growth in the small-world model of Watts and Strogatz. A cluster (in red) starts to expand locally by coarsening dynamics like in dimension one. When the size of the cluster is of the order of the average distance between shortcuts, long-range interactions take place. The effect of these long-range interactions is that of boosting up the dynamics.
}
\label{sketch_SW}
\end{figure}

\begin{figure}[t]
\centerline{
\includegraphics*[width=7.5cm]{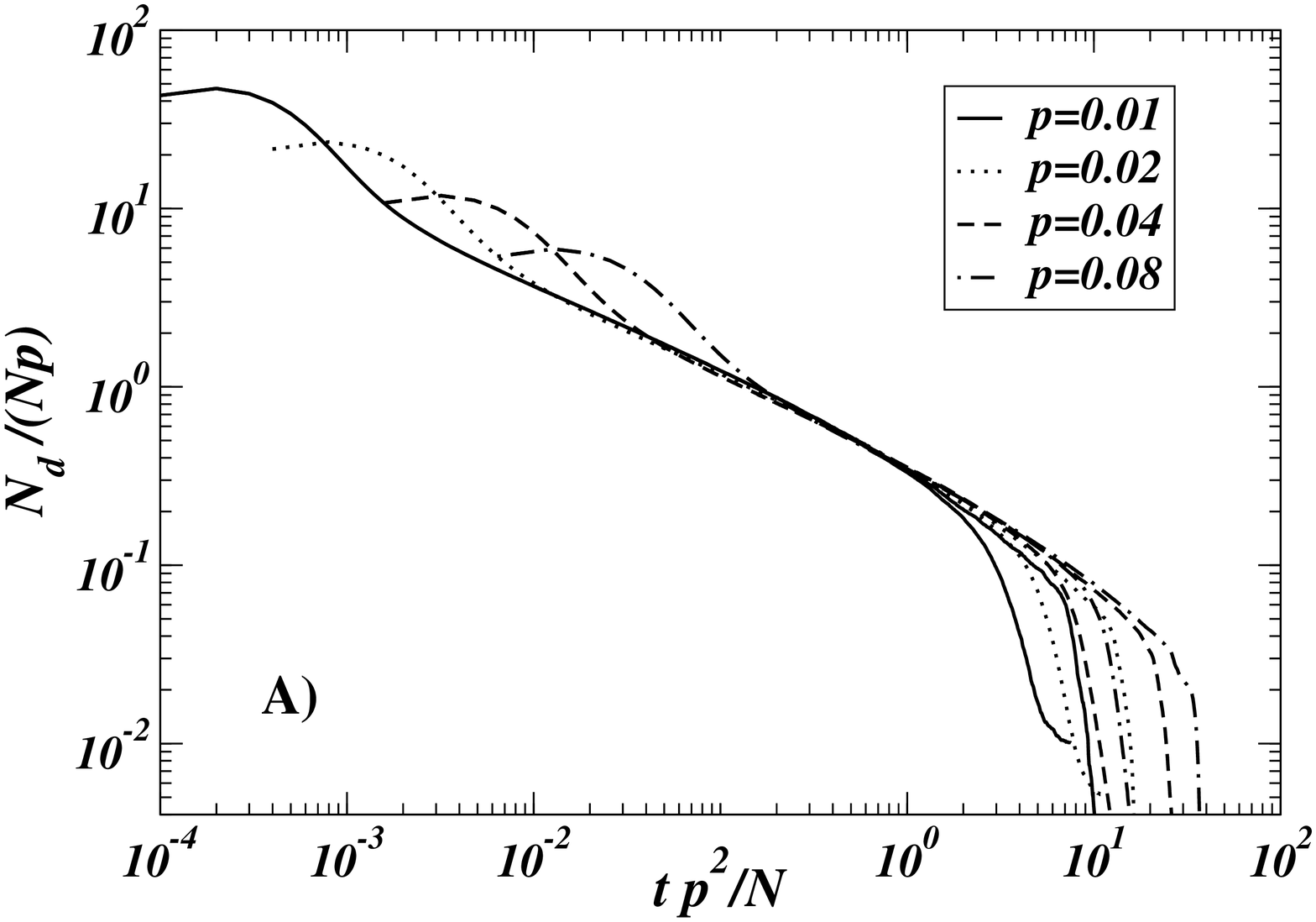}
\includegraphics*[width=7.5cm]{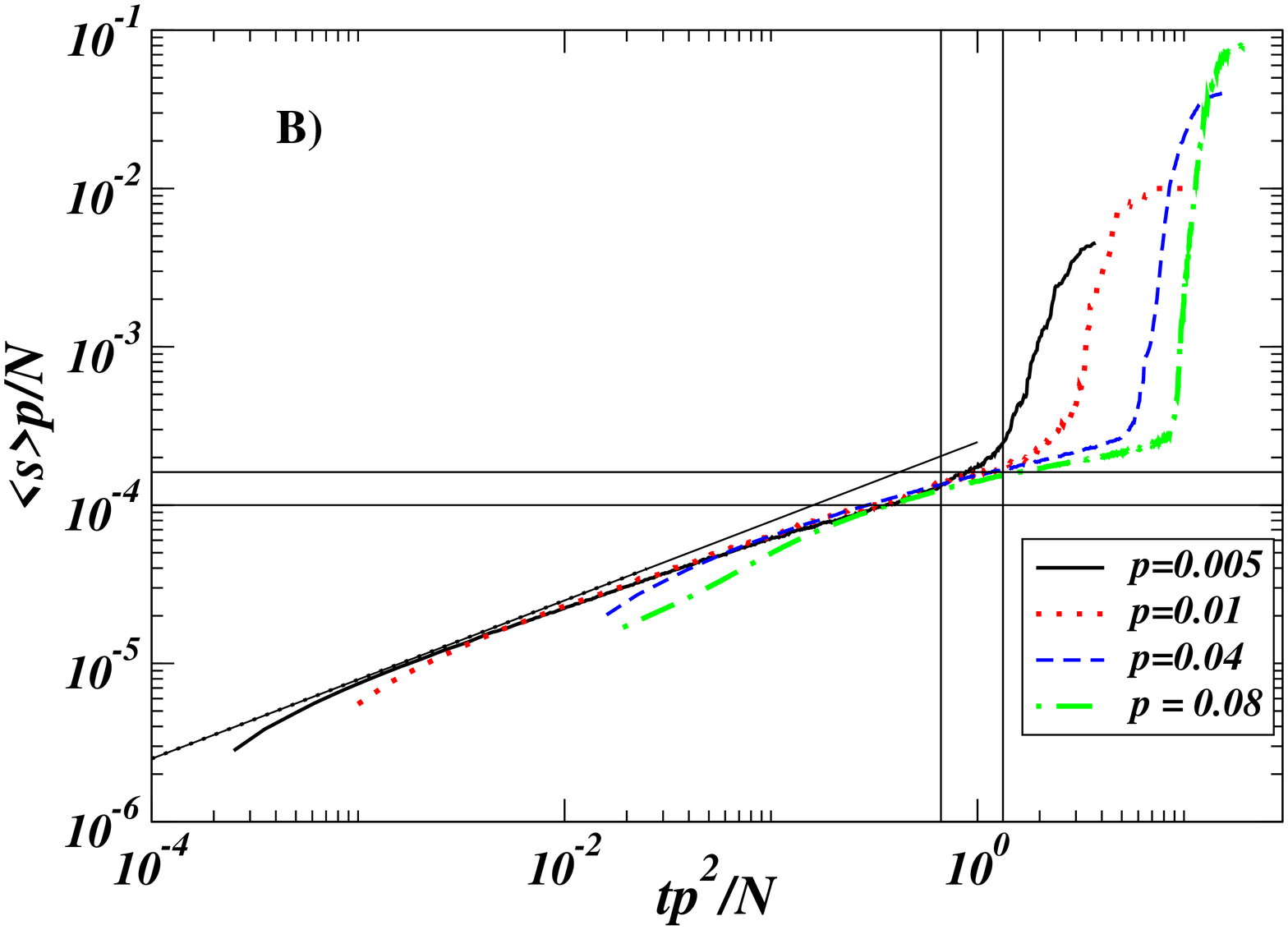}
}
\caption{(A) Number of different words in the system as a function of
time for $\langle k\rangle=8$, $p=10^{-2}, 2. 10^{-2}, 4. 10^{-2}, 8. 10^{-2}$ and
increasing sizes. Data have been rescaled
in order to collapse the curves around the crossover time $N/p^2$. 
Two values of the system size ($N=10^4$
and $N=10^5$) are displayed for each $p$.
(B) Evolution of the cluster size for $N=10^4$ and various values of
$p$. The curves are rescaled around the crossover region.
}
\label{numd_sw}
\end{figure}

The precise knowledge of the dynamical behavior of the Naming Game model on low-dimensional lattices, 
and in particular on the one-dimensional ring, makes possible to understand, by means of simple arguments and numerical simulations, the effect of the small-world property, that is a relevant feature of real complex networks. \\
In the following, indeed, we investigate the effect of introducing long-range connections
which link agents that are far from each other on the regular lattice. In other words, we study the Naming Game on the 
 small-world model proposed by  Watts and Strogatz \cite{watts98}.
The detailed description of the model is reported in Section~\ref{CHAP2_4_1},
however we recall that starting from a quasi-one-dimensional banded network in which 
each node has $2m$ neighbors, the edges are rewired with probability $p$, i.e.
$p$ represents the density of long-range connections introduced in the network. 
For $p=0$ the network retains an essentially one-dimensional topology, while the random network
structure is approached as $p$ goes to $1$. At small but finite $p$
($1/N \ll p \ll 1$), a small-world structure with short distances
between nodes, together with a large clustering, is obtained.\\
When $p=0$, the system is one-dimensional and the dynamics proceeds
by slow coarsening.  
At small $p$, the typical distance between shortcuts is $\mathcal{O}(1/p)$, 
so that the early dynamics is not affected and proceeds as in one-dimensional systems. In
particular, at very short times many new words are invented since the
success rate is small. The maximum number of different words scales as $\mathcal{O}(N)$, as in the other 
cases, while the average used memory per agent
remains finite, since the number of neighbors of each site is
bounded (the degree distribution decreases
exponentially~\cite{barrat_weigt}, see Section~\ref{CHAP2_4_1}).\\
The typical cluster dynamics on a small-world network is graphically represented in Fig.~\ref{sketch_SW}.\footnote{ Following the analysis of Section~\ref{CHAP5_3_1}, we call ``cluster'' a set of neighboring nodes (agents) with the same unique word.}
 %
As long as the typical cluster size is smaller
than $1/p$, the clusters are typically one-dimensional, and the system evolves by means of  
the usual coarsening dynamics. However, as the average cluster size reaches the typical
distance between two shortcuts $\sim 1/p$, a crossover phenomena  
toward an accelerated dynamics takes place. Since the cluster size grows as $\sqrt{t/N}$, 
this corresponds to a crossover time $t_{cross} = {\cal O}(N/p^2)$. For
times much larger than this crossover, one expects that the dynamics
is dominated by the existence of shortcuts, entering  a mean-field
like behavior. The convergence time is thus expected to scale as
$N^{3/2}$ and not as $N^3$. The condition in order for this picture to be possible
is exactly the small-world condition; indeed, the crossover time $N/p^2$ 
has to be much larger than $1$, and much
smaller than the consensus time for the one-dimensional case $N^3$, that together imply $p \gg 1/N$.\\
In summary, {\em the small-world topology allows to
combine advantages from both finite dimensional lattices and
mean-field networks}: on the one hand, only a finite memory per node is
needed, in opposition to the ${\cal O}(N^{1/2})$ in mean-field; on the
other hand the convergence time is expected to be much shorter than in
finite dimensions.
\begin{figure}[t]
\centerline{ 
\includegraphics*[width=8.0cm]{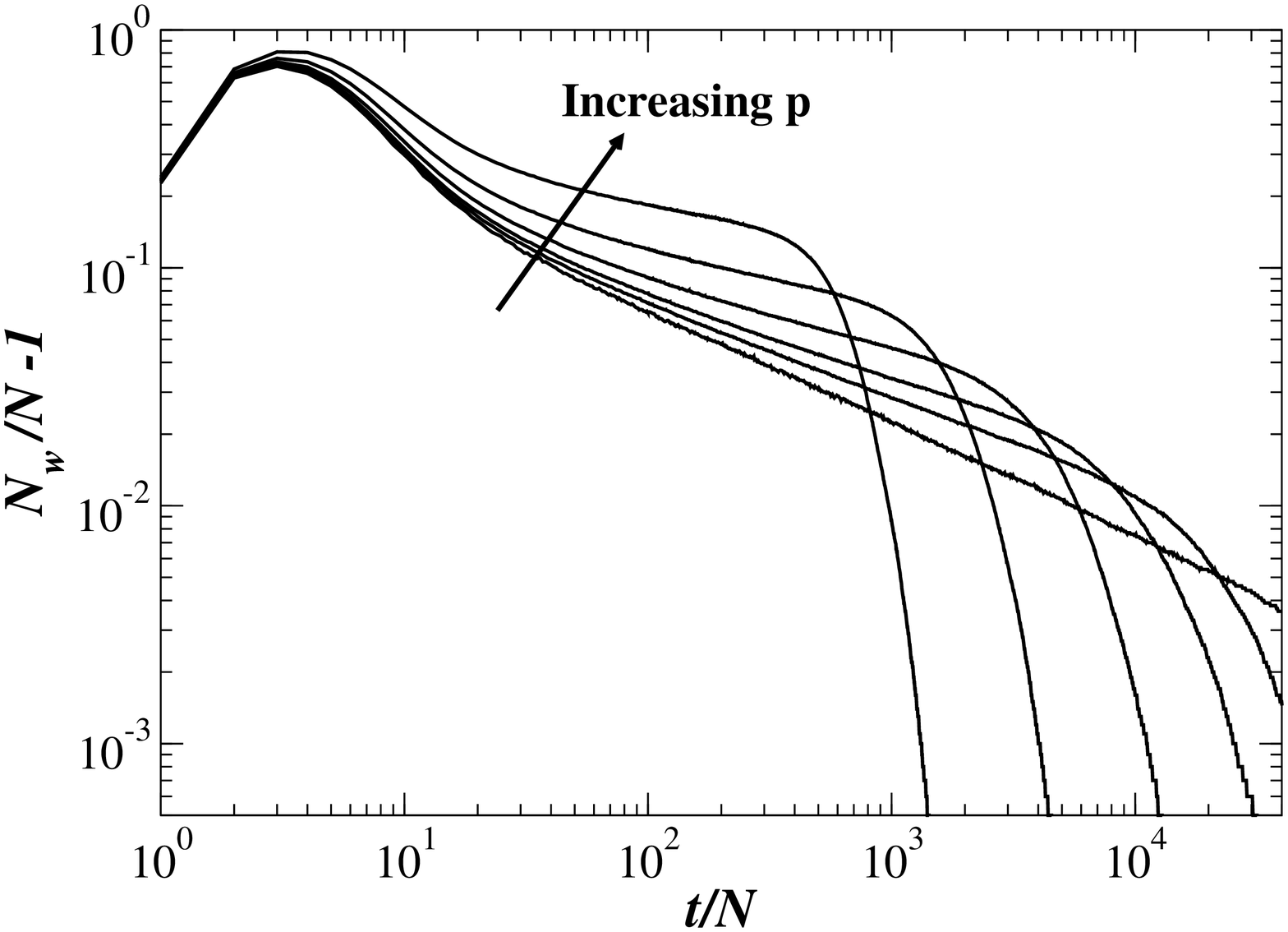}
}
\caption{Average number of words per agent in the system, $N_w/N$
as a function of the rescaled time $t/N$, for small-world networks
with $\langle k \rangle=8$ and $N=10^3$ nodes, for various values of
$p$. The curve for $p=0$ is shown for reference, as well as
$p=5. 10^{-3}$, $p=10^{-2}$, $p=2. 10^{-2}$, $p=4. 10^{-2}$,
$p=8. 10^{-2}$, from bottom to top on the left part of the curves.  
}
\label{numw_sw1}
\end{figure}
The theoretical predictions have been verified monitoring the behavior of the usual global quantities.
Figure~\ref{numw_sw1} displays the evolution of the average number of
words per agent as a function of time, for a small-world network with
average degree $\langle k \rangle=8$, and various values of the
rewiring probability $p$. While $N_w(t)$ in all cases
decays to $N$, after an initial peak whose
height is proportional to $N$, the way in which
this convergence is obtained depends on the parameters. At fixed $N$,
for $p=0$ a power-law behavior $N_w/N - 1 \propto 1/\sqrt{t}$ is
observed due to the one-dimensional coarsening process. As
soon as $p \gg 1/N$ however, we observe deviations getting stronger
as $p$ is increased: the decrease of $N_w$ is first slowed down after
the peak, but leads in the end to an very fast convergence. The effect is more evident
for larger $p$. Moreover, increasing the size of the system the convergence gets slower.\\
\begin{figure}[t]
\centerline{
\includegraphics*[width=7.5cm]{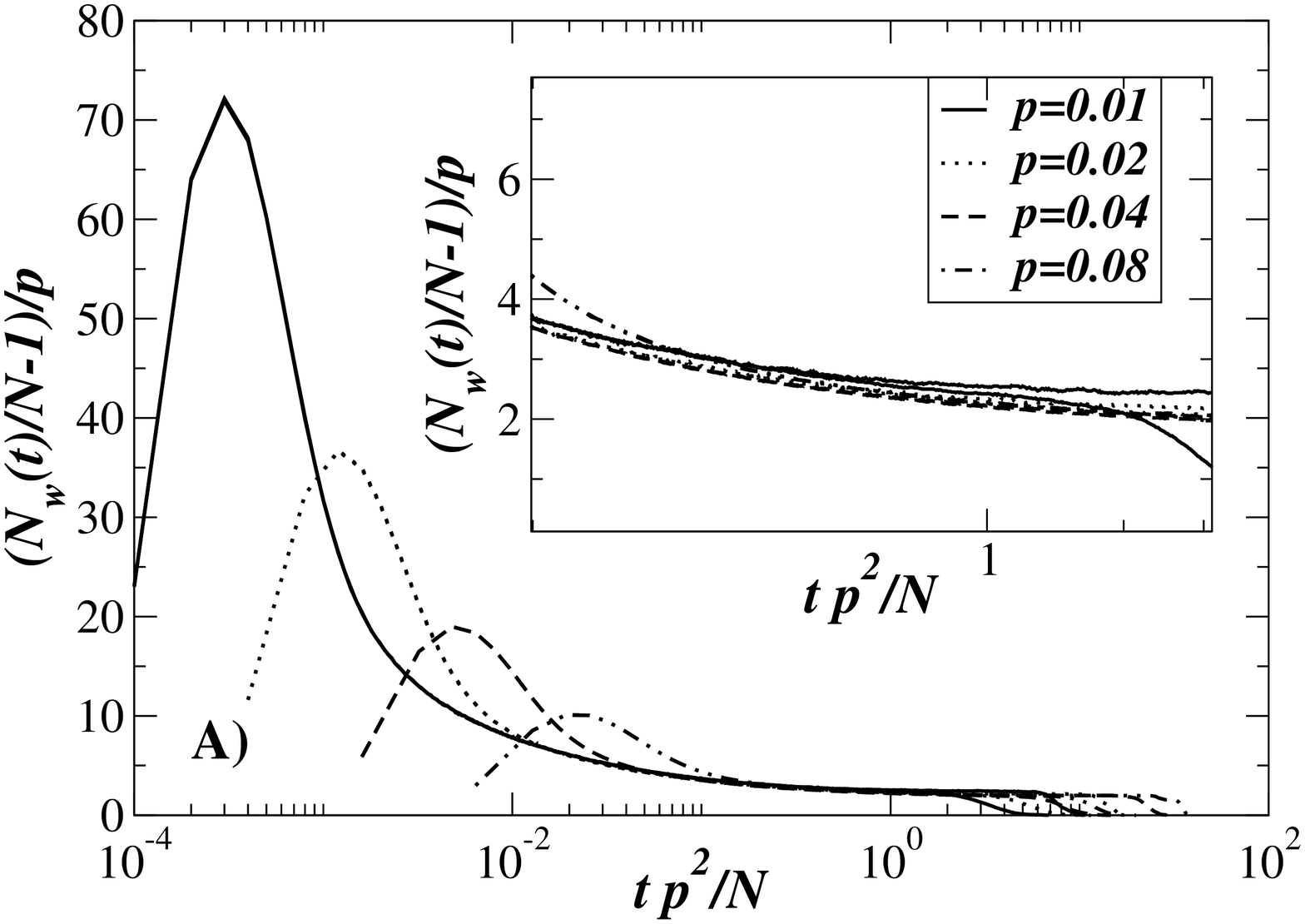}
\includegraphics*[width=7.5cm]{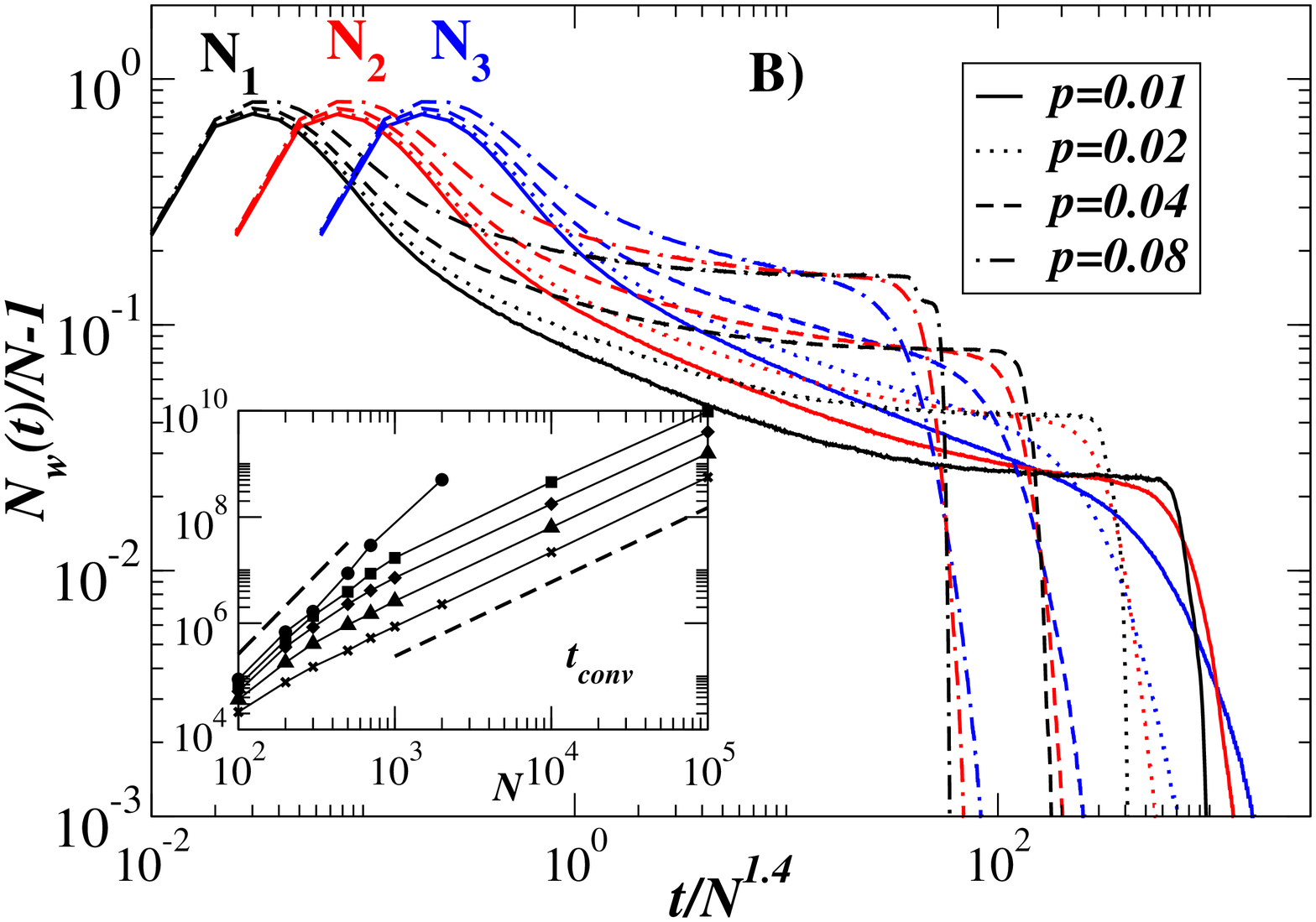}
}
\caption{(A) Rescaled curves of the average number of words per agent
in the system, in order to show the collapse around the crossover time
$N/p^2$. For each value of $p$, two values of the system size
($N=10^4$ and $N=10^5$) are displayed. The curves for different sizes
are perfectly superimposed before the convergence.  
(B) Convergence at
large times, shown by the drop of $N_w/N-1$ to $0$: the time is
rescaled by $N^{1.4}$. For each $p$, three different sizes ($N_1=10^3$ for
the left peak, curves in black, 
$N_2=10^4$ for the middle peak, curves in blue, and $N_3=10^5$, right peak,
curves in red) 
are shown. On the $N^{1.4}$ scale, the
convergence becomes more and more abrupt as $N$ increases. The inset
displays the convergence time as a function of size for $p=0$
(bullets), $p=0.01$ (squares), $p=0.02$ (diamonds), $p=0.04$
(triangles), $p=0.08$ (crosses); the dashed lines are 
proportional to $N^3$ and $N^{1.4}$.}
\label{numw_sw2}
\end{figure}
As previously mentioned, a crossover phenomenon is expected when the
one-dimensional clusters reach sizes of order $1/p$, i.e. at a time of
order $N/p^2$. Since the agents with more than one word in memory 
are localized at the interfaces between clusters, their number is $\mathcal{O}(Np)$.  
The average excess memory per site (with respect to 
global consensus) is thus of order $p$, so that one
expects $N_w/N -1= p {\cal G}(tp^2/N)$. Figure~\ref{numw_sw2}-A
indeed shows that the data of $(N_w/N -1)/p$ 
for various values of $p$ and $N$
collapse when $t p^2/N$ is of order $1$.  On the other hand,
Fig.~\ref{numw_sw2}-B indicates that the convergence towards
consensus is reached on a timescale of order $N^{\beta_{SW}}$, with
$\beta_{SW} \approx 1.4 \pm 0.1$, close to the
mean-field case $N^{3/2}$ and in strong contrast with the $N^3$
behavior of purely one-dimensional systems. Note that the time to converge 
scales as $p^{-1.4\pm.1}$, that is consistent with the fact that for 
$p$ of order $1/N$ one should recover an essentially one-dimensional
behavior with convergence times of order $N^3$.\\
In the small-world regime, the system develops a plateau in the total 
number of words (after the peak and before the convergence), whose duration increases with the size $N$.
However, during this plateau the system evolves
continuously towards consensus by elimination of redundant words, as 
evidenced by the continuous decrease in the number of {\em distinct} words
displayed in Fig.~\ref{numd_sw}-A. 
It shows that curves for various system sizes and 
values of $p$ collapse when correctly
rescaled around the crossover time $N/p^2$.\\
The combination of the results concerning average used memory and
number of distinct words corresponds to a picture in which clusters of
agents sharing a common unique word compete during the time lapse
between the peak and the final consensus. It is thus interesting to
measure how the average cluster size evolves with time and how it
depends on the rewiring probability $p$. Figure~\ref{numd_sw}-B
allows to compare the cluster size $\langle s \rangle$ evolution for
the one-dimensional case and for finite $p$. At $p=0$, a pure
coarsening law $\langle s \rangle \propto \sqrt{t}$ is observed.  As
$p$ increases, deviations are observed when time reaches the crossover
$p^2/N$, at a cluster size $1/p$, as was expected from the intuitive
picture previously developed.  
As expected, the collapse of the curves of $\langle s \rangle p$ vs. $tp^2/N$
takes place for $tp^2/N$ of order $1$.\\
Another interesting remark concerns the slowing down of the curves after the peak and before 
the convergence. This is possibly related to the first interactions between clusters and shortcuts.
When a cluster touches a shortcut, the presence of a branching point slows down the interface movement.
Thus, the clusters are locally more stable, due to the 
presence of an effective 'pinning' of interfaces near a shortcut. This 
effect is reminiscent of what happens for the Ising model on small-world 
networks~\cite{boyer} where, at low temperature, the local field 
transmitted by the shortcuts delays the passage of interfaces. Unlike Ising's zero 
temperature limit, however, the present dynamics only slows down and is
never blocked into disordered configurations. The idea of interfaces pinning at nodes playing as 
branching points is discussed in more detail for tree structures in the next section.

\subsection{Naming Game on complex networks}\label{CHAP5_3_3}

In this section we expose the main results on the dynamics of the minimal Naming Game model on complex networks.
Before entering into the details of the analysis, it is worth noting that the minimal Naming Game model itself, as described in Section~\ref{CHAP5_2_1}, is not well-defined on general networks. 
On regular topologies, the number of neighbors of a node is fixed, thus any possible method used to choose at random a pair of neighboring agents to interact is completely equivalent.  
In a general network, on the contrary, different nodes possess a variable number of neighbors, therefore it is important to take notice of the strategy used to draw the pairs of agents in interaction, since it can produce a different dynamical behavior.
When choosing a pair, one should specify which player is chosen first, the speaker or the hearer. \\
On a generic network with degree distribution $P(k)$, the degree of the first chosen node and of its chosen neighbor are distributed respectively according to $P(k)$ and to $kP(k)/\langle k \rangle$.
The second node will therefore have typically a larger degree, and the asymmetry between speaker and hearer can couple to the asymmetry between a randomly chosen node and its randomly chosen neighbor, leading to different dynamical properties.
This aspect of the dynamical processes evolving on irregular topologies and networks has been first noticed by Suchecki et al. \cite{suchecki} and Castellano et al.~\cite{castellano,castellano1} in the case of the Voter model.   
This is particularly relevant in heterogeneous networks, in which the neighbor of a randomly chosen node is likely to be a hub.\\
\begin{figure}[t] 
\centerline{
\includegraphics*[width=8.0cm]{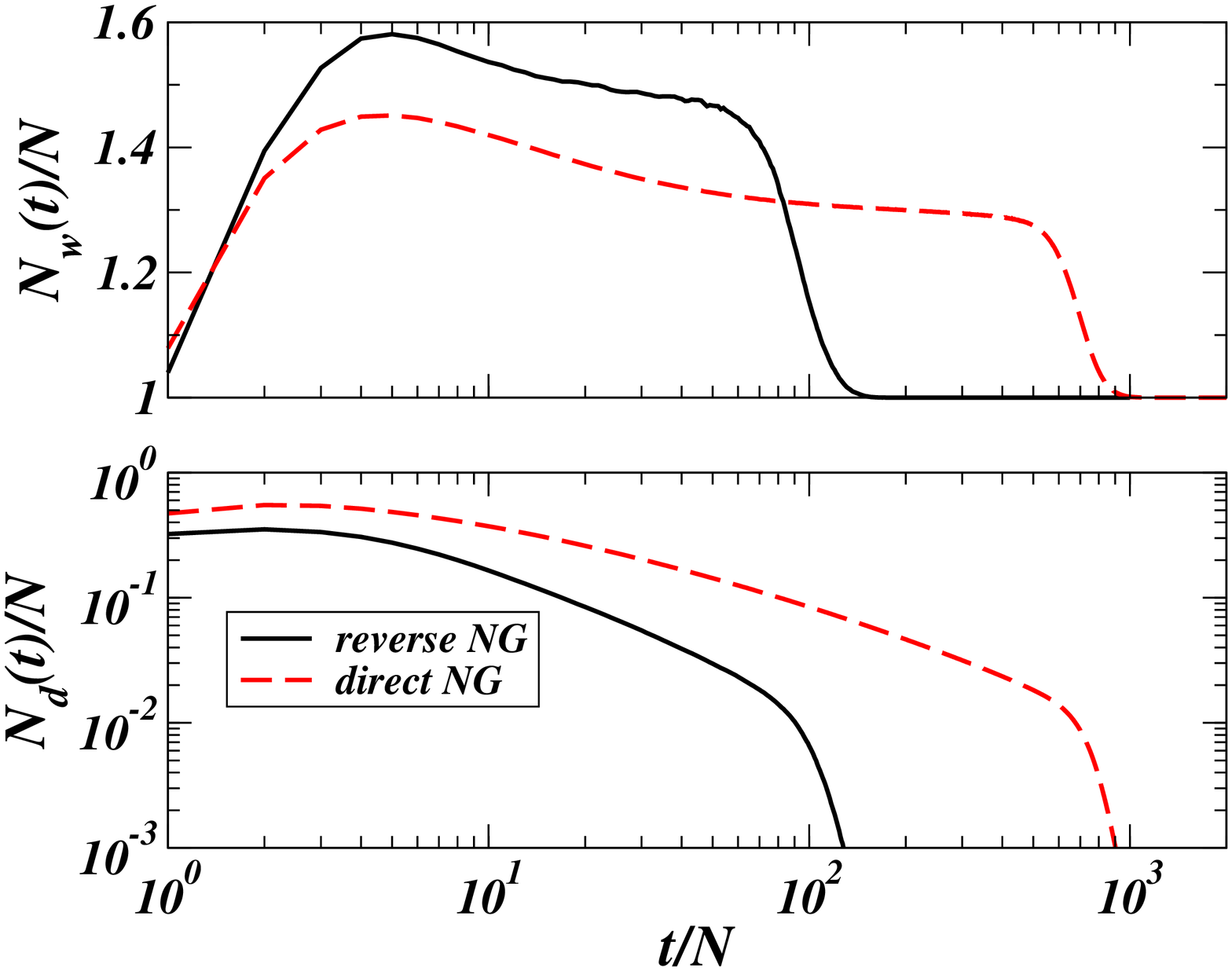}
}
\caption{Total memory $N_{w}$ (top) and number of different words
$N_{d}$ (bottom) vs. rescaled time for two different strategies of
pair selection on a BA network of $N=10^4$ agents, with $\langle k \rangle
=4$.  The reverse NG rule (black full line) converges much faster than
the direct rule (red dashed line). Note that the two strategies
do not lead to the same scaling laws with the system size for the convergence
time (not shown).}
\label{reverseNET}
\end{figure}
This remark on the asymmetry of dynamical processes on networks suggests to define three different pair selection strategies for the Naming Game:
\begin{itemize}   
\item A randomly chosen speaker selects randomly a hearer among its neighbors. This is probably the most natural generalization of the original rule.
We call this strategy {\em direct Naming Game}. In this case, larger degree nodes will preferentially act as hearers.
\item The opposite strategy, called {\em reverse Naming Game}, can also be carried out: we choose the hearer at random and one of its neighbors as speaker. In this case the hubs are preferentially selected as speakers.
\item A {\em neutral} strategy to pick up pairs of nodes is that of considering the extremities of an edge taken uniformly at random. The roles of speaker and hearer are then assigned randomly with equal probability among the two nodes.
\end{itemize}
Figure~\ref{reverseNET} allows to compare the evolution of the direct and the reverse Naming Game for a heterogeneous network, a Barab\'asi-Albert (BA) network with $N=10^4$ agents and $\langle k\rangle = 4$. In the case of the reverse rule, a larger memory is used although the number of different words created is smaller, and a faster convergence is obtained. This corresponds to the fact that the hubs, playing principally as speakers, can spread their words to a larger fraction of the agents, and remain more stable than when playing as hearers, enhancing the possibility of convergence.
Similarly to the case of the Voter model \cite{castellano,castellano1}, the scaling laws of the convergence time for direct and reverse strategies seem to be the same only in some very special cases (power-law degree distribution with exponent $\gamma=3$); however, we do not dispose of an accurate study of the reverse NG properties.
This is due to the fact that from the point of view of a realistic interaction among individuals or computer-based agents, the direct Naming Game, in which the speaker chooses a hearer among its neighbors, seems somehow more natural than the other ones. For this reason we have focused on the direct Naming Game. In future work we will study more in detail the similarities and differences of the three strategies.\\
\begin{figure}[t] 
\centerline{
\includegraphics*[width=10.cm]{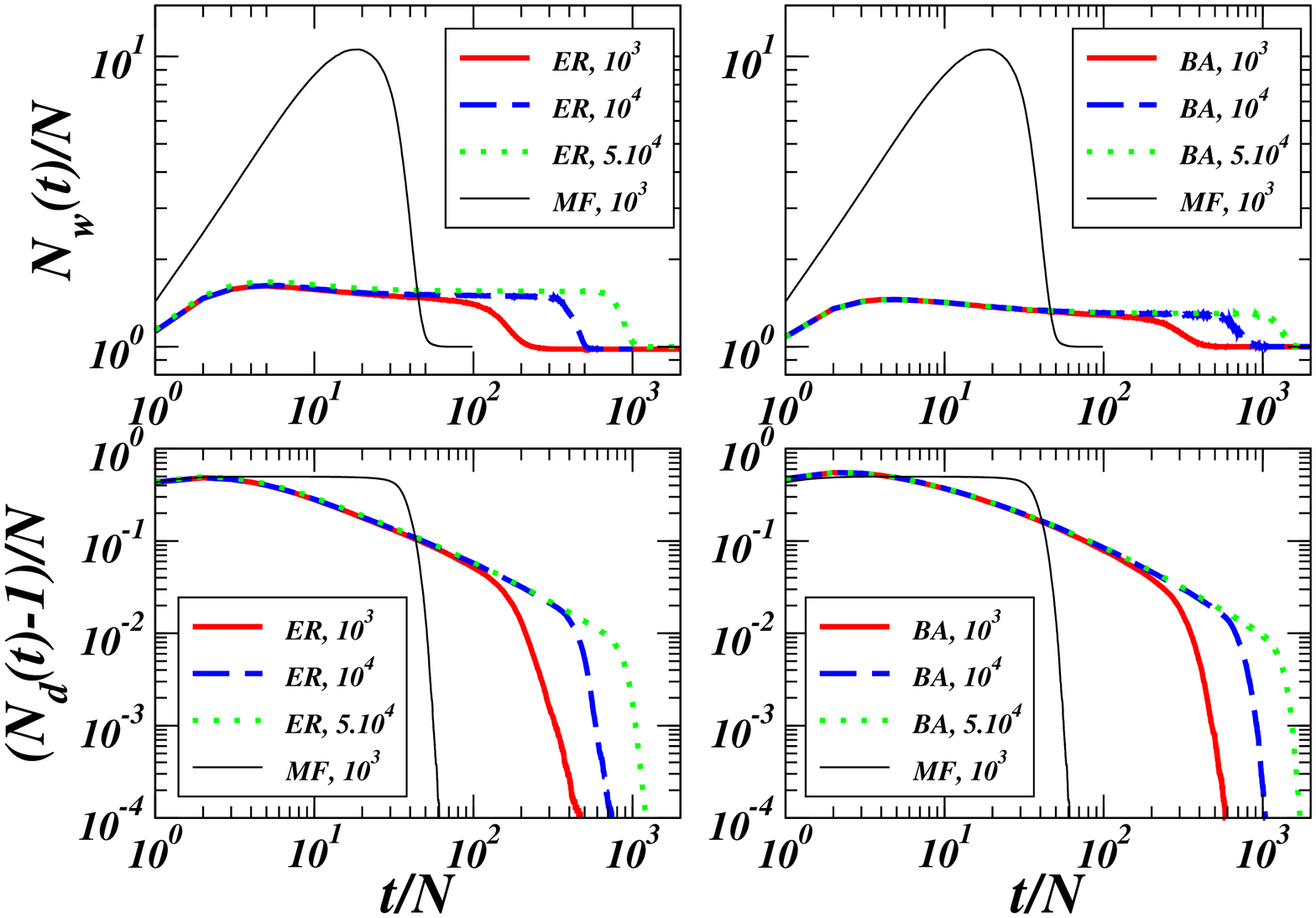}
}
\caption{ER random graph (left) and BA scale-free
network (right) with $\langle k \rangle =4$ and sizes $N=10^3, 10^4,
5.10^4$. Top: evolution of the average memory per agent $N_{w}/N$
versus rescaled time $t/N$.  For increasing sizes a plateau
develops in the re-organization phase preceding the convergence.
The height of the peak and of the plateau collapse in this plot,
showing that the total memory used scales with $N$. Bottom: evolution
of the number of different words $N_{d}$ in the system. $(N_{d}-1)/N$
is plotted in order to emphasize the convergence to the consensus with
$N_d=1$. A steady decrease is observed even if the memory $N_w$ displays a
plateau. The mean-field (MF) case is also shown 
(for $N=10^3$) for comparison. }
\label{generalNET}
\end{figure}

\begin{figure}[t] 
\centerline{
\includegraphics*[width=8.0cm]{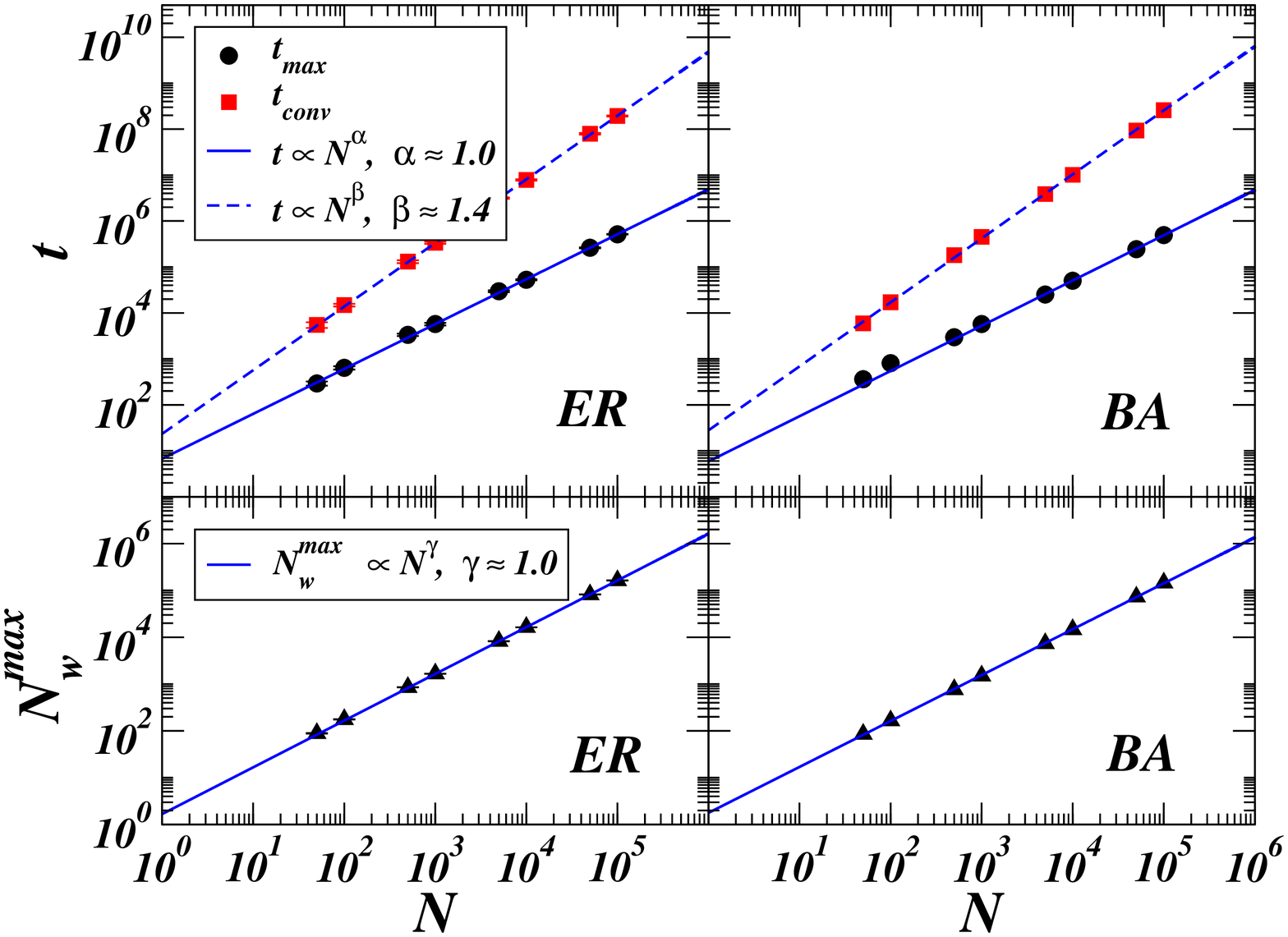}
}
\caption{(Top) Scaling behavior with the system size $N$ for the time
of the memory peak ($t_{max}$) and the convergence time ($t_{conv}$)
for ER random graphs (left) and BA scale-free networks (right) with
average degree $\langle k\rangle=4$.  In both cases, the maximal memory
is needed after a time proportional to the system size, while
the time needed for convergence grows as $N^{\beta}$ with $\beta \simeq 1.4$.
(Bottom) In both networks the necessary memory capacity (i.e. the maximal
value reached by $N_w$) scales linearly with the size of the network.}
\label{scalingNET}
\end{figure}

{\bf \em Global quantities - \quad} 
As already done for the other topologies, we study the global behavior of the system looking at the temporal evolution of three main quantities: the total number $N_{w}(t)$ of words in the system, the number of different words $N_{d}(t)$, and the rate of success $S(t)$. 
In Fig.~\ref{generalNET}, we report the curves of the number of words ($N_{w}(t)$ and $N_{d}(t)$) for homogeneous ER networks (left) and heterogeneous BA networks (right) with $N=10^3, 10^4, 5\cdot 10^4$ nodes and average degree $\langle k \rangle = 4$. The corresponding data for the mean-field case (with $N=10^{3}$) are displayed as well for reference. The curves for the average use of memory $N_{w}(t)$ show a rapid growth at short times, a peak and then a plateau whose length increases as the size of the system is increased (even when time is rescaled by the system size, as in Fig.~\ref{generalNET}). The time and the height of the peak,
and the height of the plateau, are proportional to $N$. A systematic study of the scaling behavior with the size of the system for these quantities is reported in Fig.~\ref{scalingNET}, which shows that the convergence time $t_{conv}$ scales as $N^{\beta}$ with $\beta \simeq 1.4$ for both ER and BA. 
The apparent plateau of $N_{w}$ does however not correspond to a steady state, as revealed by the continuous decrease of the number of different words $N_{d}$ in the system: in this re-organization phase, the system keeps evolving by elimination of words, although the total used memory does not change significantly.\\
%
Note that observed scaling laws for the convergence time is a general robust feature that is not affected by other topological details (average degree, clustering, etc), and more surprisingly it seems to be independent of the particular form of the degree distribution. 
We have indeed checked the value of the exponent $\beta \simeq 1.4 \pm 0.1$ for various $\langle k\rangle$, clustering, and exponents $\gamma$ of the degree distribution $P(k) \sim k^{-\gamma}$ for scale-free networks constructed with the uncorrelated configuration model. 
These parameters have instead an effect on other quantities such as the time and the value of the maximum of memory (that will be analyzed later).\\
\begin{figure}[t] 
\centerline{
\includegraphics*[width=7.5cm]{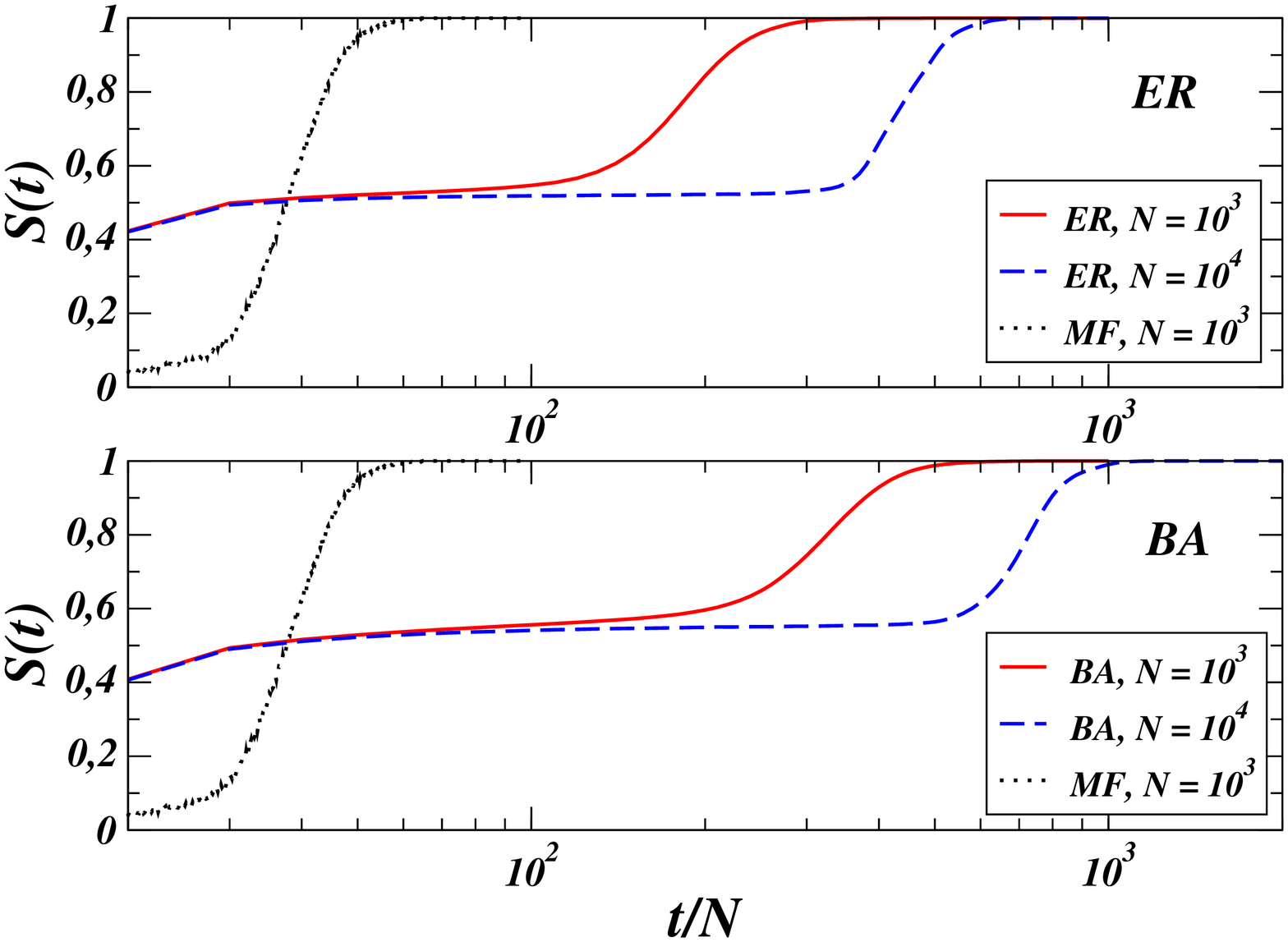}
\includegraphics*[width=7.5cm]{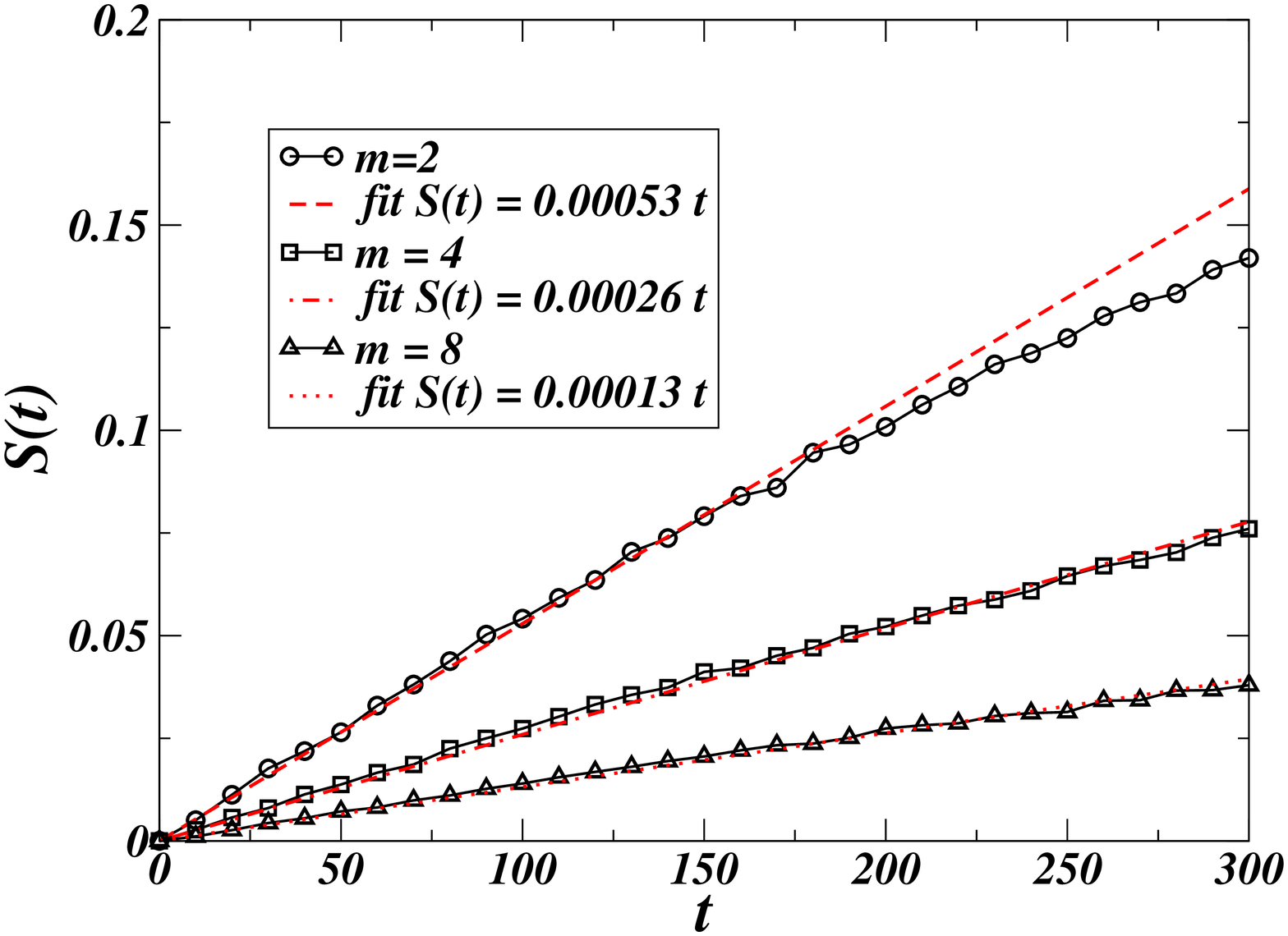}
}
\caption{(Left) 
Temporal evolution of the success rate for ER random graphs
(red continuous line) and BA scale-free networks (blue dashed line) with
$\langle k \rangle =4$ and sizes $N=10^3$ and $10^4$. The dotted black line
refers to the mean-field case ($N=10^3$).
(Right) BA network, $N=10^3$. The short time behavior of the success
rate $S(t)$ is shown for $\langle k \rangle = 4$ (circles),
$\langle k \rangle =8$ (squares), and $\langle k \rangle =16$
(triangles). The curves are linear, with a
slope that is in agreement to the predicted value $2/\langle k\rangle
N$.}
\label{successNET}
\end{figure}
%
The ubiquity of the scaling exponent $\beta \simeq 1.4$ could be related to the fact that all these networks
present the small-world property. In many equilibrium and non-equilibrium statistical physics models defined on general networks, the small-world property is sufficient to ensure a mean-field like behavior of the system.
In the present case, the discrepancy with the mean-field exponent ($\beta_{MF} \simeq 1.5$) may be due to logarithmic corrections that are unlikely to be captured using numerical scaling techniques.\\
Comparing Figures~\ref{generalNET}-\ref{scalingNET} with those for the mean-field (MF)
topology and the regular lattices reported in Sections~\ref{CHAP5_3_1}-\ref{CHAP5_3_2},
some important analogies and differences emerge. 
Thanks to the finite average connectivity, the memory peak scales only linearly with 
the system size $N$, and is reached after a time ${\cal O}(N)$, in contrast with
MF (${\cal O}(N^{1.5})$ for peak height and time) but similarly to the
finite dimensional case. The MF plateau observed in the number of
different words, is replaced here by a slow continuous
decrease of $N_d$ with an almost constant memory used. With respect to
the slow coarsening process observed in finite dimensional lattices on
the other hand, the existence of short paths among the nodes speeds up the convergence
towards the global consensus. Therefore,
complex networks exhibiting small-world properties constitute an
interesting trade-off between mean-field ''temporal efficiency'' 
and regular lattice ''storage optimization''.\\
The success rate $S(t)$ is displayed in Figure~\ref{successNET}-A for ER (top) and
BA (bottom) networks with $N=10^3$ (red full line), and $10^4$ (blue
dashed line) agents and $\langle k \rangle =4$. The success rate for
the mean-field ($N=10^3$) is also reported (black dotted lines). In
both networks the success rate increases linearly at very short times (see also Fig.~\ref{successNET}-B)
then, after a plateau similar to the one observed
for $N_w$, it increases on a fast timescale towards $1$. At short times most
inventories are empty, so that the success rate is equal to the
probability that two agents interact twice, i.e. $t/E$, where
$E=N\langle k\rangle/2$ is the number of possible interacting pairs.
The curves, for BA networks in Fig.~\ref{successNET}-B, give slopes in agreement
with the theoretical prediction $2/\langle k\rangle N$. Compared to the mean-field 
case, in which $E={\cal O}(N^2)$, the initial success rate grows faster.  When $t \sim \mathcal{O}(N)$, no
inventory is empty anymore, words start spreading through
unsuccessful interactions and $S(t)$ displays a bending.\\

\begin{figure}[t]
\centerline{
\includegraphics*[width=7.5cm]{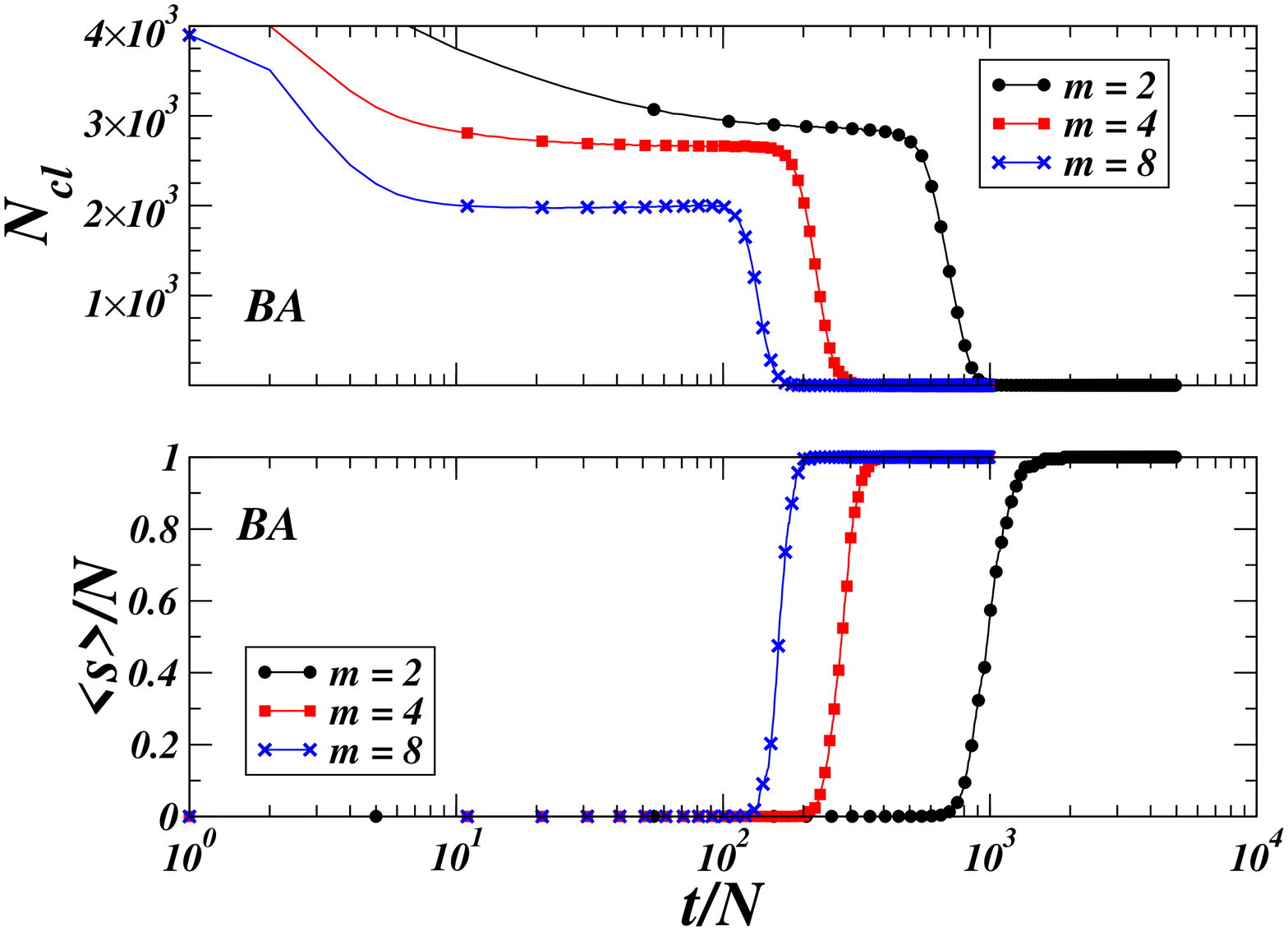}
\includegraphics*[width=7.5cm]{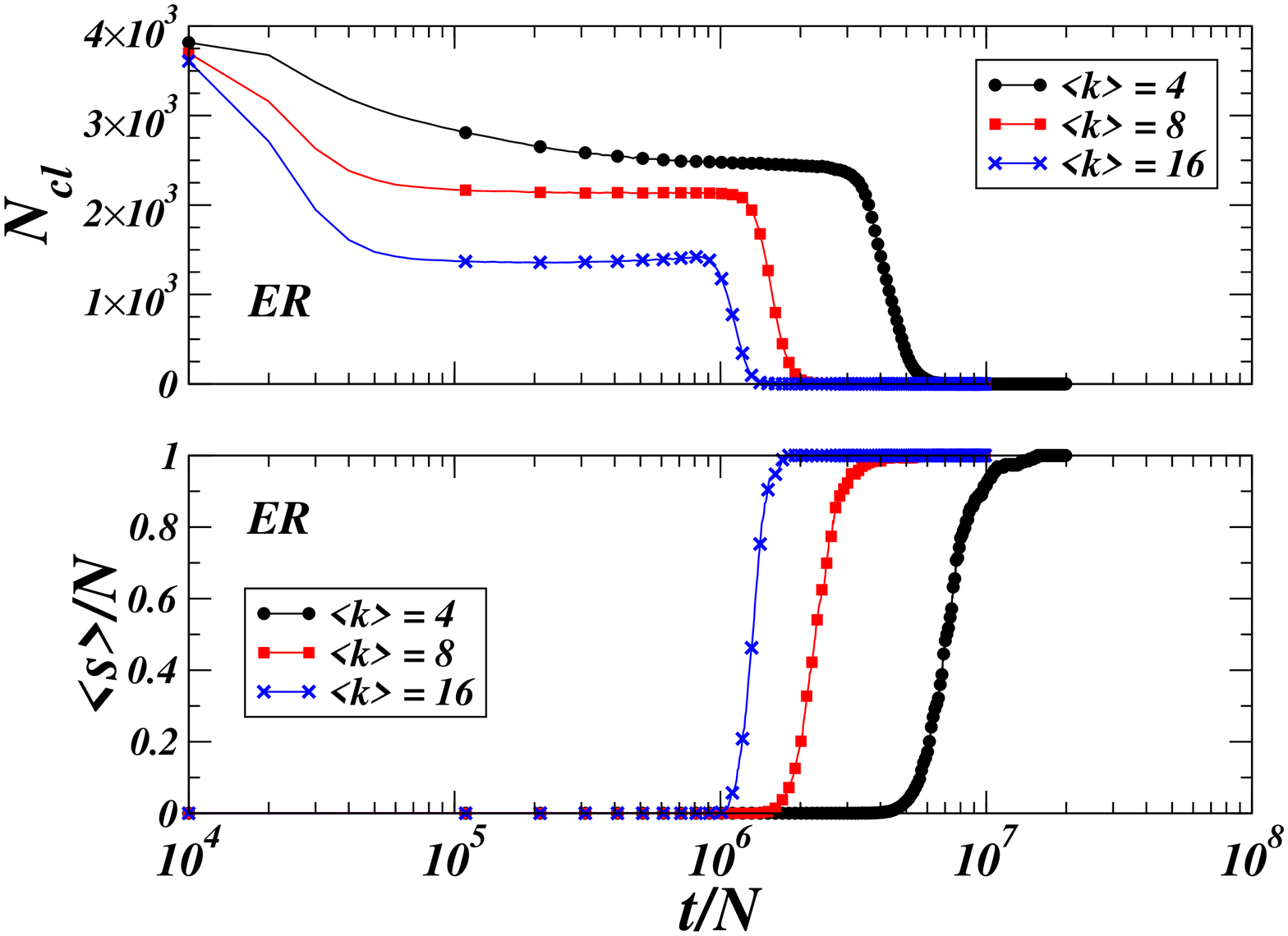}
}
\caption{Number of clusters $N_{cl}$ and normalized average cluster size $\langle s \rangle/N$ vs. time for ER networks (right) and BA networks (left) with $N=10^4$, $\langle k\rangle =4$ (circles), $\langle k\rangle =8$ (squares), $\langle k \rangle =16$ (crosses). }
\label{clustersNET}
\end{figure}

{\bf \em Clusters statistics - \quad}
Without entering the detailed analysis of the behavior of clusters of words, 
for which we refer to Ref.~\cite{naming_gameNET}, it is worthy to spend some words on this aspect of 
the Naming Game dynamics.
We have called ''cluster'' any set of neighboring agents sharing a common unique word.  
In Section~\ref{CHAP5_3_1}, we have shown that, in low-dimensional lattices, 
the dynamics of the Naming Game proceeds by formation of such clusters, 
that grow through a coarsening phenomenon: the average cluster size (resp. the number of clusters)
increases (resp. decreases) algebraically with time. 
As shown instead in Fig.~\ref{clustersNET}, for both models (ER and BA) the
normalized average cluster size remains very close to zero (in fact, of
order $1/N$) during the re-organization phase that follows the peak in the number of words, 
and converges to one with a sudden transition. The same behavior is shown also by the number of clusters $N_{cl}(t)$,
that decreases to one very sharply.\\
The emerging picture is not that of a coarsening or growth of clusters,
but that of a slow process of correlations between inventories, followed by a multiplicative process of 
cluster growth triggered by a sort of symmetry breaking event in the success probability of the words (in favor of the word that will ultimately survive). \\
  
{\bf \em Effect of the degree heterogeneity - \quad}
Global properties of dynamical processes are often affected by the 
heterogeneous character of the network topology~\cite{vespi_book,mendes_book}. 
We have shown however that the global dynamics of the Naming Game is
similar on heterogeneous and homogeneous networks. Nonetheless, a more detailed
analysis reveals that agents with different degrees present
very different activity patterns, whose characterization is necessary
to get additional insights on the Naming Game dynamics~\cite{naming_gameNET,naming_gameACT}.\\
Let us first consider the average success rate $S_{k}(t)$ of nodes
of degree $k$; at the early stages of the dynamics it can be computed
using simple arguments. The probability of choosing twice the edge $(i,j)$ is
\begin{equation}
\frac{t}{N}\left(\frac{1}{k_{i}}+\frac{1}{k_{j}} \right),
\label{prob_twice}   
\end{equation}
i.e. the probability of choosing first $i$ ($1/N$) then $j$
($1/k_{i}$) or viceversa. Neglecting the correlations between $k_{i}$
and $k_{j}$, one can average over all nodes $i$ of fixed $k_{i} = k$,
obtaining
\begin{equation}
S_{k}(t) \simeq \frac{t}{N}\left( \frac{1}{k} +
\left\langle \frac{1}{k} \right\rangle \right)~.
\label{rate_k} 
\end{equation}
Eq.~\ref{rate_k} show that, at the very beginning, the success rate
grows linearly but the effect of the degree heterogeneity is partially
screened by the presence of the constant term $\langle 1/k \rangle$.
The same argument can be used to predict that the success rate should
be essentially degree independent for larger times. $S(t)$ is indeed
always given by two terms, of which only that referring to the node
playing as speaker contains an explicit dependence on $1/k$. \\
%
\begin{figure}[t] 
\centerline{
\includegraphics*[width=8.0cm]{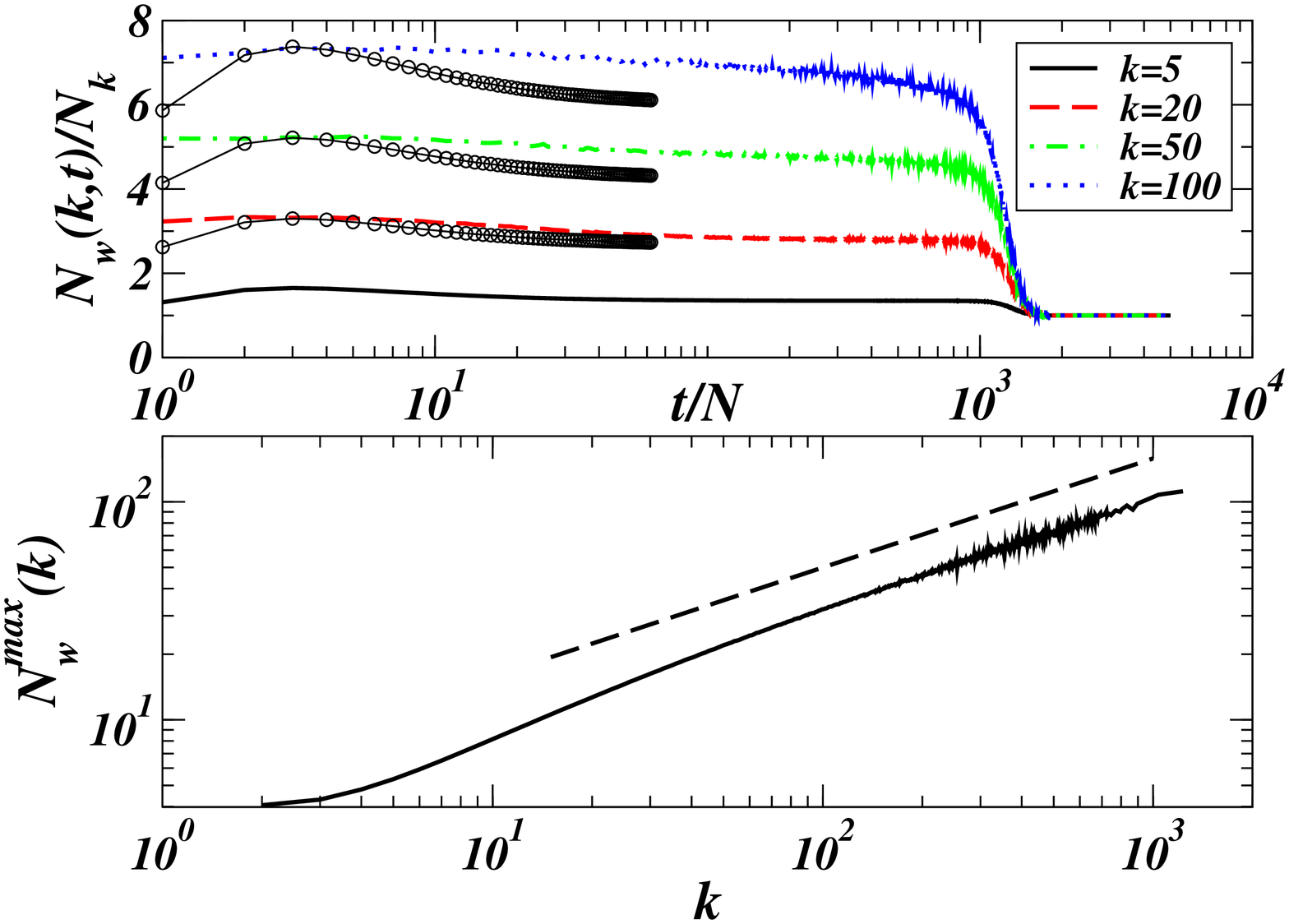}
}
\caption{BA model with $m=2$ (i.e.
$\langle k \rangle=4$), $N=5.10^4$.  (Bottom) Maximum memory used by a
node as a function of its degree. The dashed line is $\propto
\sqrt{k}$.  (Top) Average memory used by nodes of degree $k$, for
various values of $k$. The lines show the total memory 
$N_w(k,t)$ used by nodes of degree $k$ at time $t$, normalized by
the number $N_k$ of nodes of degree $k$.
The circles correspond to the bottom curve
($k_0=5$) rescaled by $\sqrt{k/k_0}$ showing the scaling of the peaks.
Note that the values of $N_w(k,t)/N_k$ are averages over many runs
that wash out fluctuations and therefore correspond to smaller values than
the extremal values observed for $N_w^{max}(k)$.  }
\label{heterogeneityNET}
\end{figure}
Another interesting point concerns the memory peak. In Fig.~\ref{heterogeneityNET} we have computed 
the height of the memory peak reached by different classes of nodes, depending on the degree,
and we have found that it is larger for nodes of larger degree.   
More precisely, the maximal memory used by a node of degree $k$ is proportional to $\sqrt{k}$ (see
bottom panel in Fig.~\ref{heterogeneityNET}).
This is in agreement with what already observed for the mean-field case, in which 
all agents have degree $k=N-1$ and the
maximal value of the total memory $N_w$ scales indeed as
$N\sqrt{k}=N^{3/2}$. Note however that in the general case, the estimation of
the peak of $N_w$ is not as straightforward. This peak is
indeed a convolution of the peaks of the inventory sizes of single
agents, that have distinct activity patterns and may reach their
maximum in memory at different temporal steps.\\
The knowledge of the average maximal memory of a node of degree $k$ is not
sufficient to understand which degree classes play a major role in driving the
dynamics towards the consensus. More insights on this issue can be obtained
observing the behavior of the total number of different words in each degree
class. A detailed analysis is reported in Ref.~\cite{naming_gameNET}, in which we show that 
low degree classes have a larger overall number of different words.
This is due to the fact that during the initial phase, in which words are
invented, low degree nodes are more often chosen as speakers and invent many
different words. The hubs need individually a larger memory, but as classes they retain a
smaller number of different words. Then, words are progressively eliminated
among low-$k$ nodes while the hubs, which act as intermediaries and are in
contact with many agents, still have typically many words in their
inventories. In this sense, the ''super-spreader'' role of the hubs allows a
faster diffusion of words throughout the network and their property of
connecting agents with originally different words helps the system to converge.
The very final phase consists in the late adoption of the consensus by the
lowest degree nodes, in a sort of final cascade from the large to the
small degrees.\\

{\bf \em Effect of the average degree and clustering - \quad}
Social networks are generally sparse graphs, but their structure is
often characterized by high local cohesiveness, that is the result of
a very natural transitive property of social
interactions~\cite{granovetter1}.  The simplest way to take into
account these features on the dynamics of Naming Game is that of
studying the effects of changing the average degree and the clustering
coefficient of the network.
\begin{figure}[t] 
\centerline{
\includegraphics*[width=10.0cm]{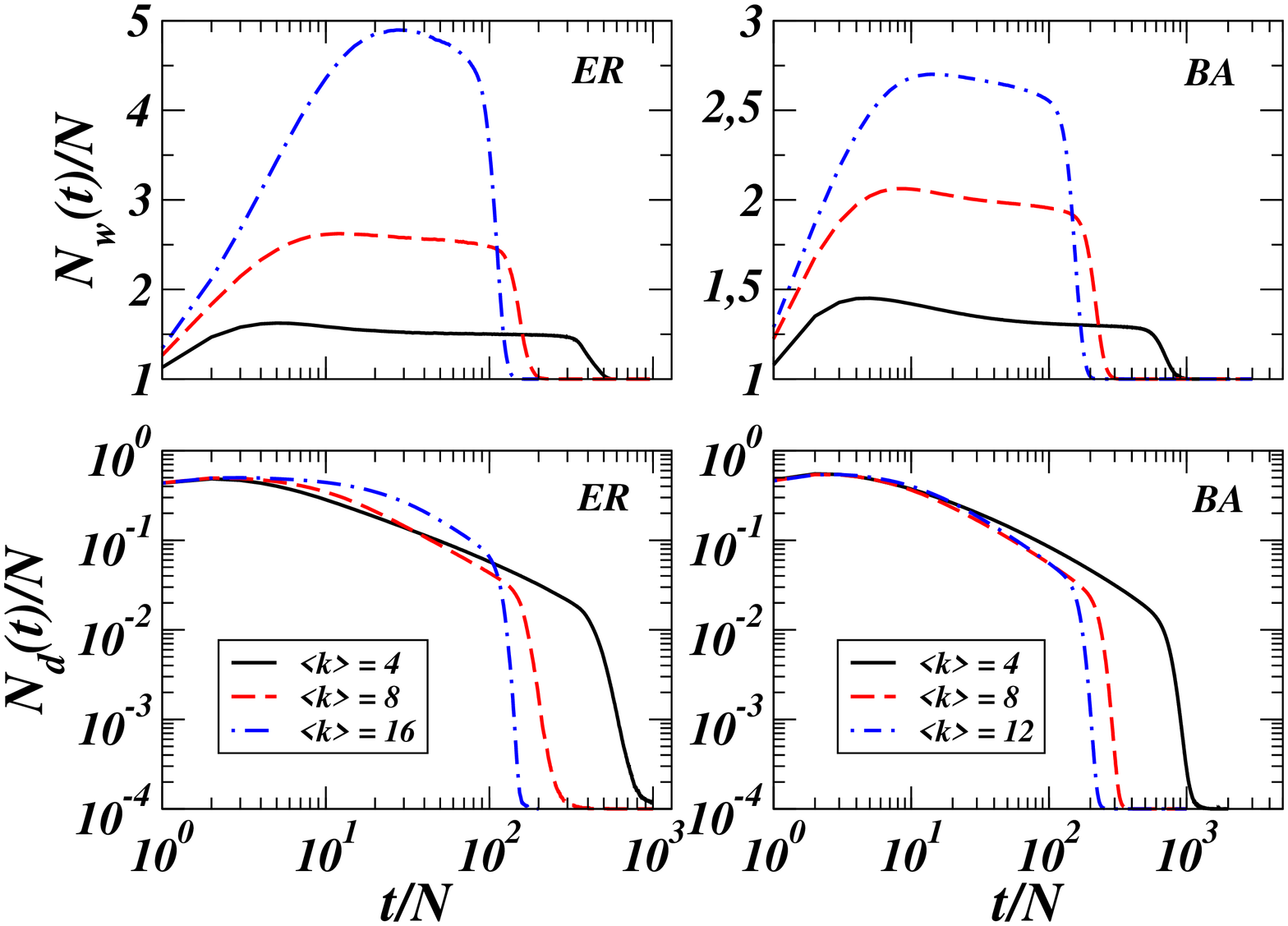}
}
\caption{ER networks (left) and BA networks (right) with $N=10^4$
agents and average degree $\langle k \rangle=4$, $8$, $16$. The increase of
average degree leads to a larger memory used ($N_w$, top)
but a faster convergence. 
The maximum in the number of different words is not affected by
the change in the average degree (bottom). }
\label{keffectNET}
\end{figure}
Fig.~\ref{keffectNET} displays the effects of increasing the average degree 
on the behavior of the main global quantities. In both ER
(left) and BA (right) models, increasing the average degree provokes
an increase in the memory used, while the global convergence time is
decreased. More precisely, the dependence of the height 
$N_{w}^{max}$ and the time $t_{max}$ of the
memory peak as function of $\langle k\rangle$ with fixed population
is approximately power-law, with sub-linear behavior \cite{naming_gameNET}.
This remark suggests that the linear scaling for the memory peak properties 
($N_w^{max} \propto N^\alpha$ and $t_{max}\propto N^\alpha$ with $\alpha=1$) 
are altered by an increase in the average degree (not shown), as expected by the
fact that increasing $\langle k\rangle$ brings the system closer to
the mean-field behavior where the scaling of these quantities is
non-linear ($\alpha_{MF}=1.5$).
It is remarkable that the behavior of the convergence time with $N$
(i.e. a power-law $N^\beta$ with $\beta \approx 1.4$) is instead very robust.
This is possibly due to the fact that, in contrast with the power-law dependence of
the peak, the convergence time depends only logarithmically on the average degree.\\
Note also that the average memory
used by a node of fixed degree is larger for larger average degree
(not shown), therefore such a global argument can also be extended to a
degree based analysis.
The curves of the success rate (not shown) are consistent with the
previous analysis. \\
\begin{figure}[t] 
\centerline{
\includegraphics*[width=10.0cm]{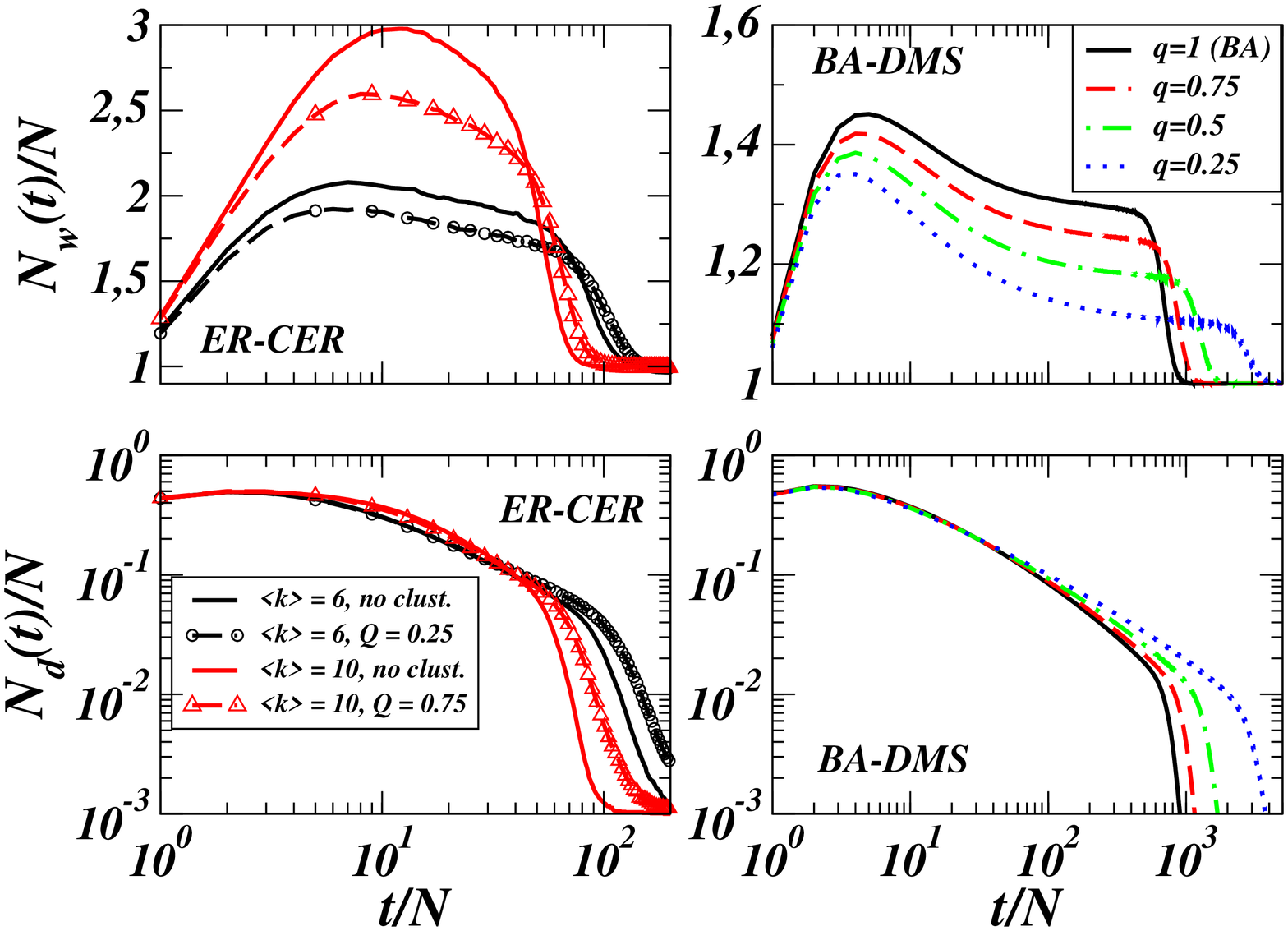}
}
\caption{Effect of clustering on the behavior of the total number of
words $N_{w}(t)$ and of the number of different words $N_{d}(t)$ on
random graphs (left) and scale-free networks (right) with $N=10^4$.
The considered clustered random graphs (CER model, with clustering 
coefficient proportional to $Q$) have been compared to standard ER graphs with equal
average degree ($\langle k\rangle = 6$ and $10$). Scale-free networks
have been generated using the mixed BA-DMS model, in which the clustering 
coefficient is proportional to $1-q$. In both networks higher clustering leads to
smaller memory capacity required but a larger convergence time.}
\label{CeffectNET}
\end{figure}
We are now interested to the effects due to the variation of the clustering coefficient.
First, the clustering is slightly changing when changing the
average degree, but its variation is small enough for the two effects
to be studied separately. Here we use some other mechanisms to enhance
clustering, summarized in the following two models with tunable clustering:
the clustered Erd\"os-R\'enyi (CER) random graphs, and mixed BA-DMS model. 
These networks have been compared to ER and BA networks with the same size and
average degree.
The mixed BA-DMS model is obtained as a generalization of the preferential attachment procedure, 
in the spirit of the Holme-Kim model~\cite{kim}: starting from $m$ connected nodes
(with $m$ even), a new node is added at each time step; with
probability $q$ it is connected to $m$ nodes chosen with the
preferential attachment rule (BA step), and with probability $1-q$ it
is connected to the extremities of $m/2$ edges chosen at random
(DMS-like step). Only the clustering spectrum is different with respect to BA and DMS, it 
can be computed as $c(k)=2(1-q)(k-m)/[k(k-1)]+{\cal O}(1/N)$~\cite{barrat_romu}.
Changing $m$ and $q$ allows to tune the value of the clustering coefficient.\\
Since the ER model also displays a low clustering, we consider
moreover a purposely modified version of this random graph model
(Clustered ER, or CER model) with tunable clustering. Given $N$ nodes,
each pair of nodes is considered with probability $p$; the two nodes
are then linked with probability $1-Q$ while, with probability $Q$, a
third node (which is not already linked with either) is chosen and a
triangle is formed.  The clustering is thus proportional to $Q$ (with
$p\sim {\cal O}(1/N)$ we can neglect the original clustering of the ER
network) while the average degree is approximately given by $\langle k
\rangle \simeq \left[3Q+(1-Q)\right] pN \simeq (2Q+1)pN$
Note that, in order to compare an ER and a CER network with
the same $\langle k \rangle$, we therefore tune $p$ for the
construction of the corresponding CER.\\
Figure~\ref{CeffectNET} shows the effect of increasing the clustering at
fixed average degree and degree distributions: the number of different
words is not changed, but the average memory used is smaller and the
convergence takes more time. Moreover, the memory peak at fixed $k$ is
smaller for larger clustering (not shown): it is more probable for a
node to speak to $2$ neighbors that share common words because they
are themselves connected and have already interacted, so that it is
less probable to learn new words. At fixed average degree, i.e. global number of
links, less connections are available to transmit words from
one side of the network to the other since many links are used in
``local'' triangles. The local cohesiveness is therefore in the
long run an obstacle to the global convergence. This
effect is similar to the observation of an increase in the percolation threshold in clustered networks,
due to the fact that many links are ``wasted'' in redundant local connections~\cite{serrano_clu}. \\
\begin{figure}[t] 
\centerline{
\includegraphics*[width=0.45\textwidth]{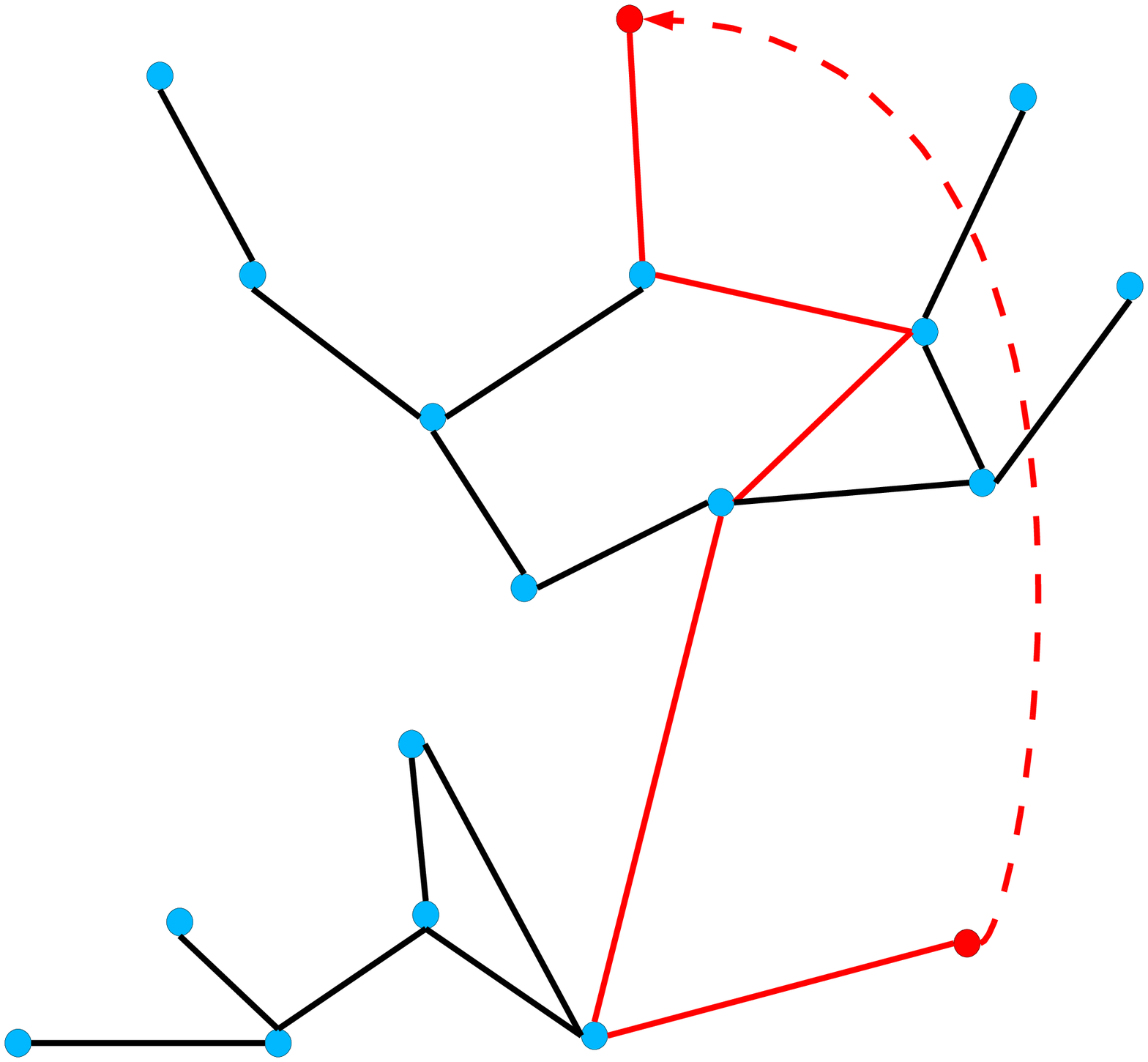}
\includegraphics*[width=0.45\textwidth]{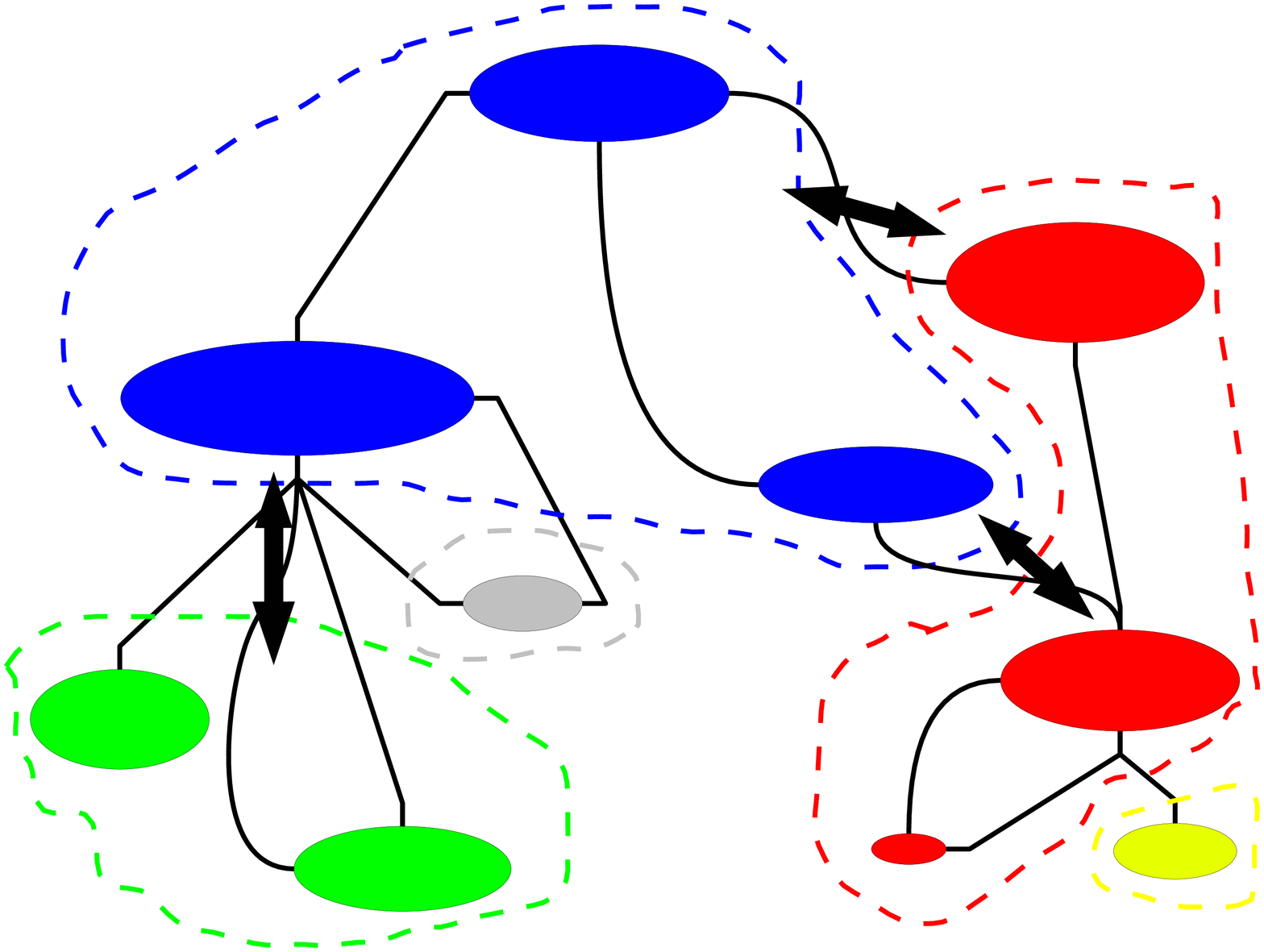}
}
\caption{(A) Sketch of a shortcut connecting two distant regions and reinterpretation as a long loop.
(B) Naive representation of a typical clusters organization in a hierarchical structure.
}
\label{long_loops}
\end{figure}

\begin{figure}[t] 
\centerline{
\includegraphics*[width=10.0cm]{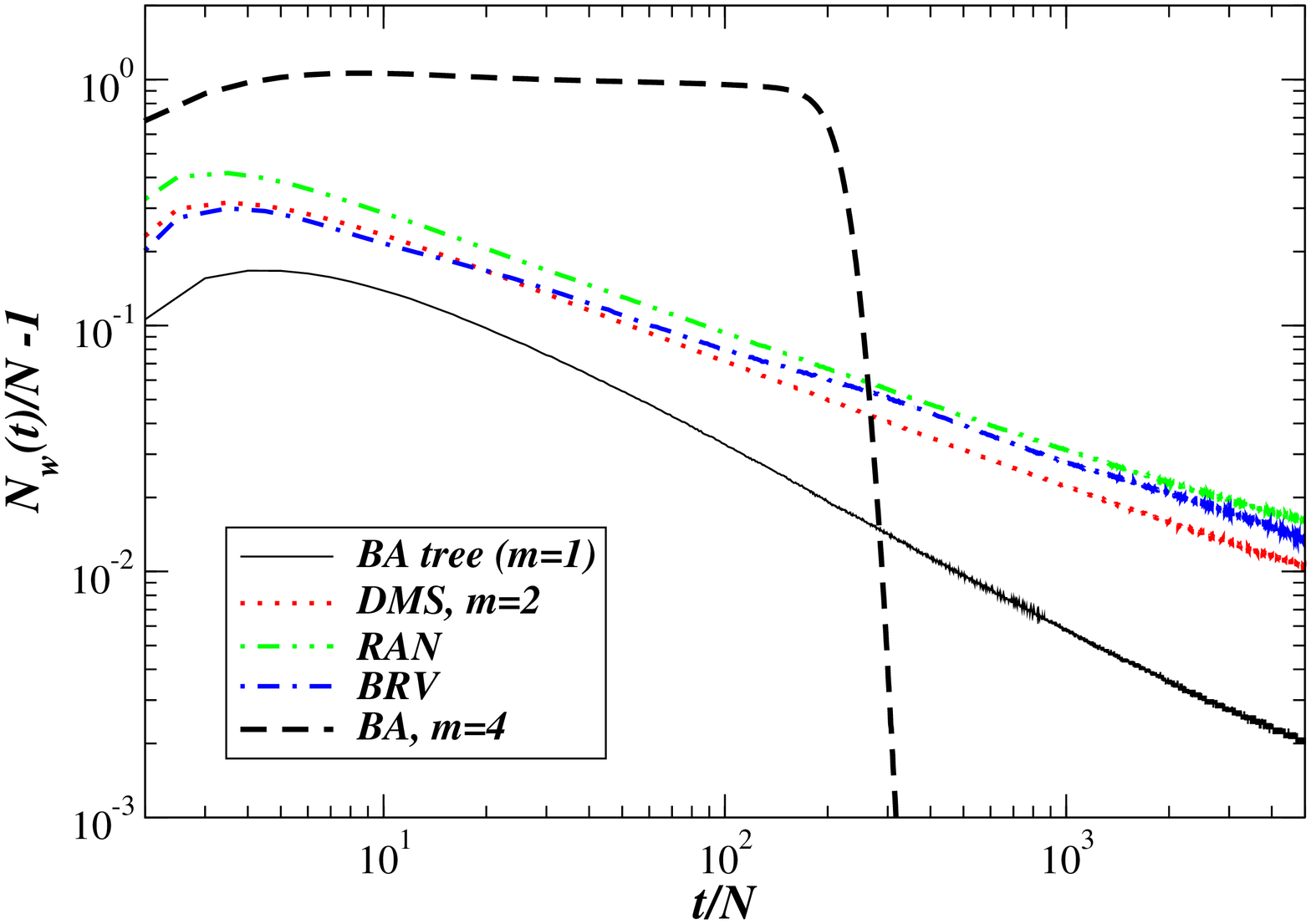}
}
\caption{Power-law decrease in the total number of words $N_{w}(t)$ for the Naming Game on several 
hierarchical networks: the Barab\'asi-Albert scale-free tree (full line), the DMS model with $m=2$ (dotted line), the 
deterministic scale-free network of Barab\'asi, Ravasz and Vicsek (dot-dashed line), the Random Apollonian Network (dot-dot-dashed line). The behavior of hierarchical networks are compared with the mean-field like behavior of a Barab\'asi-Albert network with $m=4$ (dashed line). Note that all hierarchical networks show diffusive coarsening except for the scale-free tree, whose behavior is subdiffusive.}
\label{hierarchyNET}
\end{figure}

{\bf \em Effect of hierarchical structures - \quad}
In the previous sections we have argued that networks with small-world
property have fast (mean-field like) convergence after a
re-organization phase whose duration depends on other properties of
the system. The small-world property holds when the diameter of the network 
grows slowly, i.e. logarithmically or slower, with the size $N$.
However, another requirement is necessary: the small-world must be generated by shortcuts connecting 
regions of the network that are otherwise far away one from the other.
From this point of view, shortcuts correspond to the existence of long loops (see Fig.~\ref{long_loops}-A). 
When shortcuts, and corresponding long loops, are absent the 
topological structure of the network possess an intrinsic metric ordering. In such a situation, regular structures 
like $d$-dimensional lattices admit a real geometric distance, whereas 
disordered topologies are more generally associated to hierarchical structures.\\
On a hierarchical network, each node belongs in fact to a given sub-hierarchical unit
and for going from one node to another node in another sub-unit, it is necessary to follow a hierarchical path.
In the Naming Game, each sub-unit can converge towards a local consensus, which makes the global consensus more difficult to achieve (see Fig.~\ref{long_loops}-B). In other words, the dynamics slows down in the passage between different levels of hierarchy, with a behavior that resembles that observed in models of glassy dynamics with traps \cite{bouchaud} or in ``hierarchical islands models'' of diffusion in turbulent flows \cite{zaslavsky}. Note however that, unlike the Naming Game, in these models there are real energy barriers obstructing the dynamics. 
\begin{figure}[t] 
\centerline{\begin{tabular}{c}
\begin{tabular}{|c|}\hline \\ \includegraphics*[width=7cm]{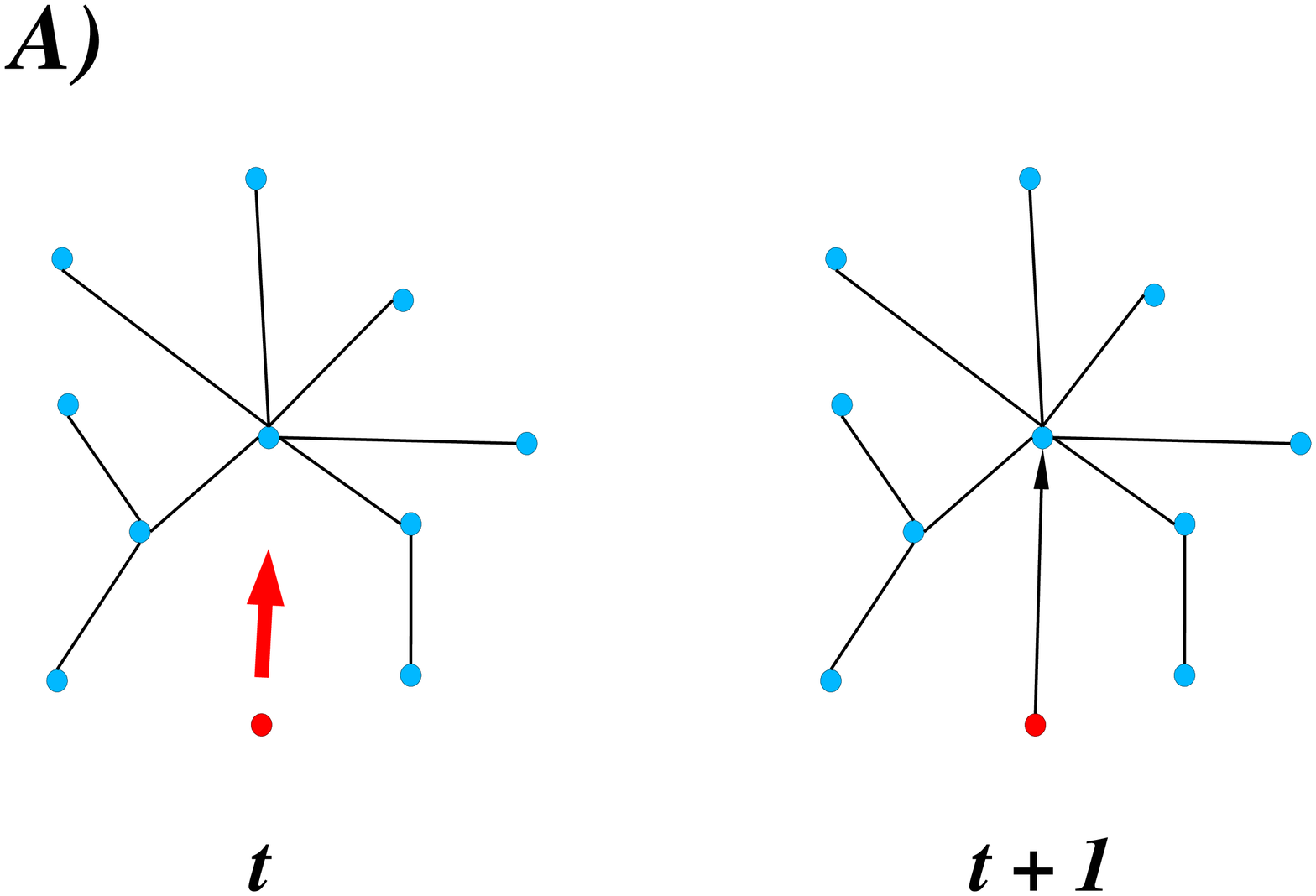}\\ \hline\end{tabular}~~
\begin{tabular}{|c|}\hline \\ \includegraphics*[width=7cm]{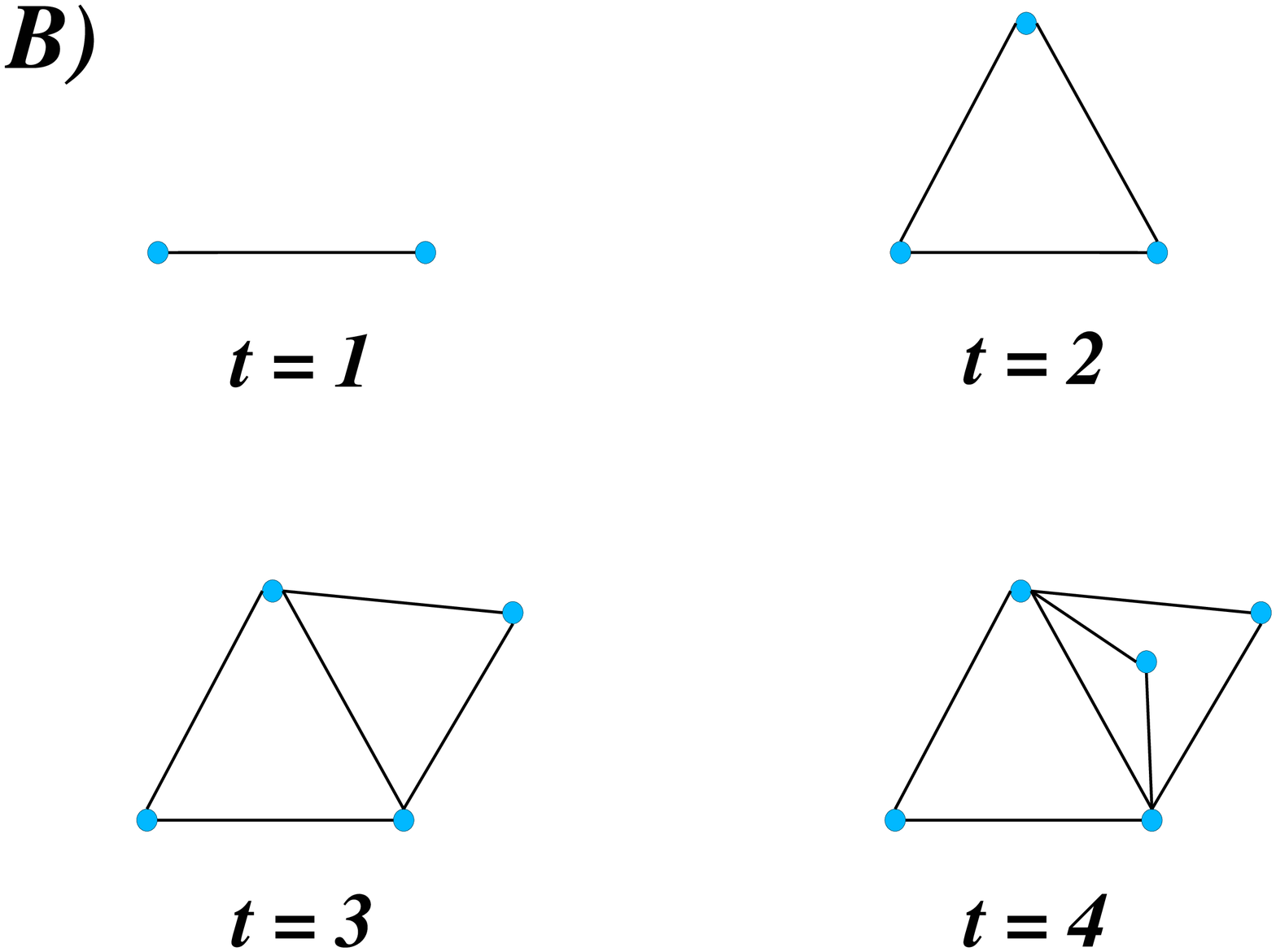}\\ \hline\end{tabular} \\
~\\ 
\begin{tabular}{|c|}\hline \\ \includegraphics*[width=7cm]{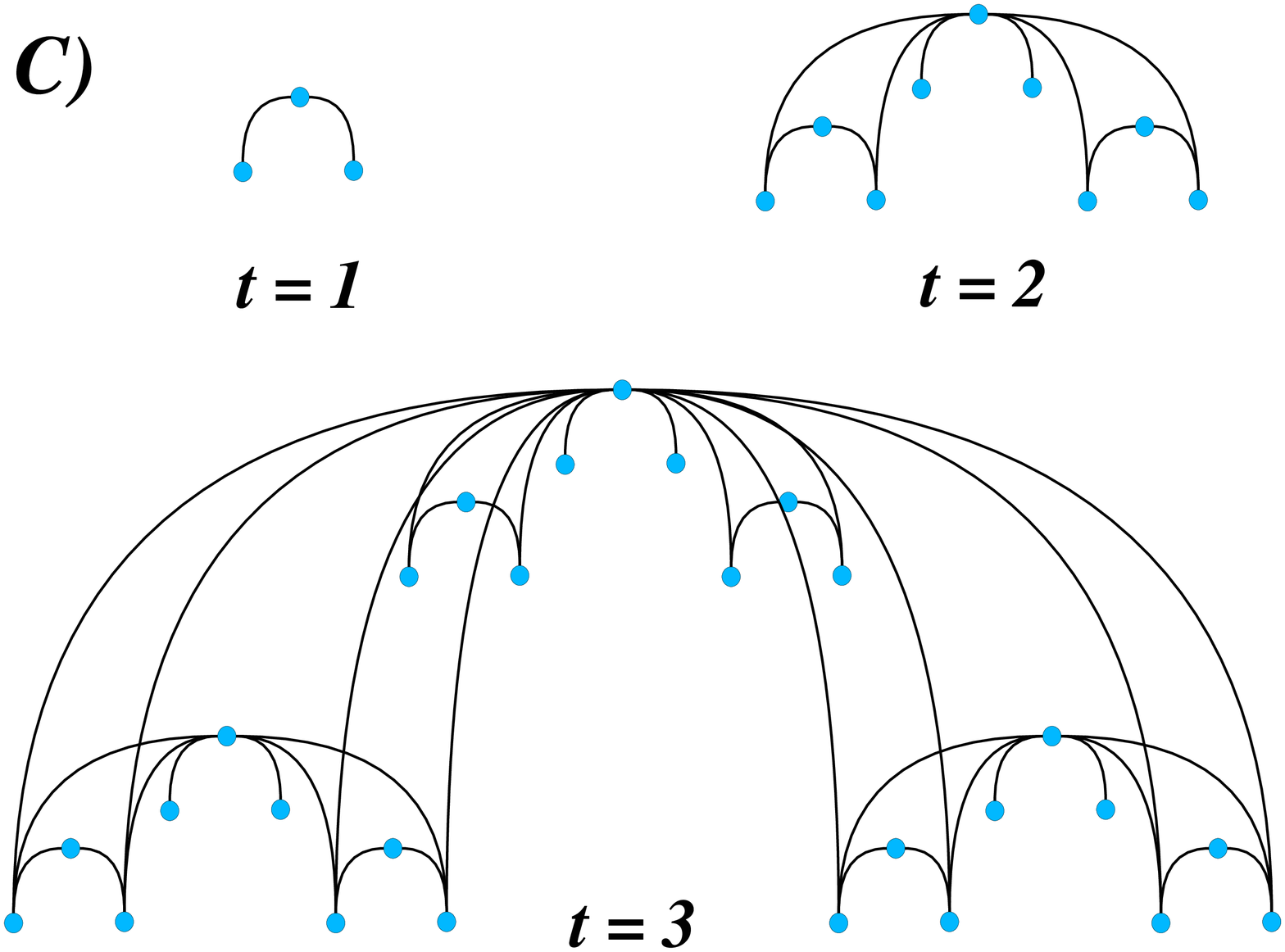}\\ \hline\end{tabular}~~
\begin{tabular}{|c|}\hline \\ \includegraphics*[width=7cm]{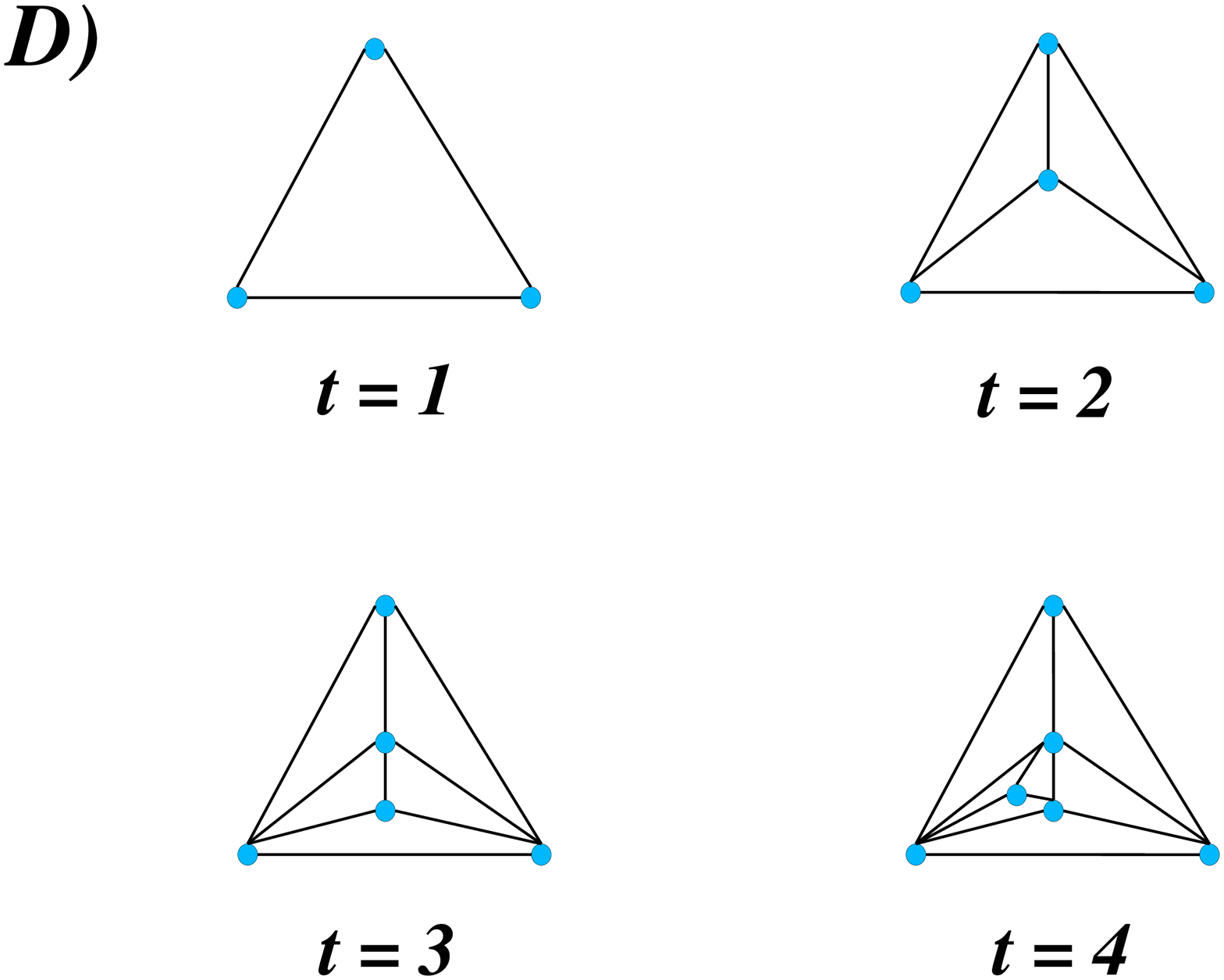}\\ \hline\end{tabular}
\end{tabular}
}
\caption{Sketch of various generation procedures for models of hierarchical networks: 
(A) the Barab\'asi-Albert tree (BA model with $m=1$); (B) DMS model with $m=2$, i.e. only a new triangle enter the system each time; (C) deterministic hierarchical network proposed in Ref.~\cite{ravasz}; (D) random apollonian network \cite{andrade,zhou}.
}
\label{hier_modelsNET}
\end{figure}
\begin{figure}[t] 
\centerline{
\begin{tabular}{|c|}\hline \\ \includegraphics*[width=8.0cm]{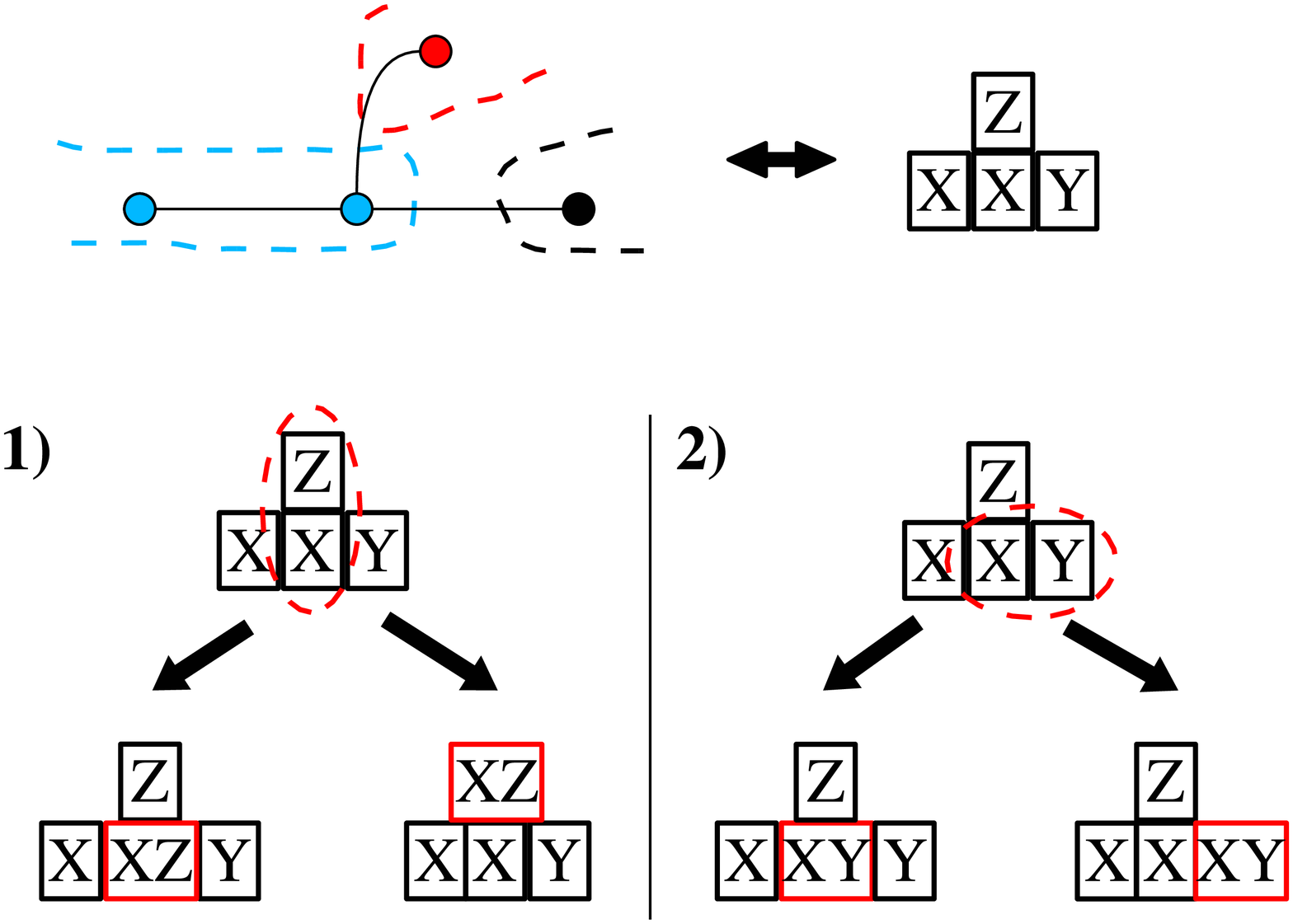}\\ \hline\end{tabular}
}
\caption{A typical branching point at which clusters interfaces may be pinned down. 
The increasing number of possible transitions for larger degree causes the effective transition probability to decrease.
}
\label{branchingNET}
\end{figure}
The results of numerical simulations on network models with a strong hierarchical structure are striking: the Naming Game
converges very slowly, the number of different words decreasing as a power law of the time (in Fig.~\ref{hierarchyNET} we have reported $N_{w}(t)/N-1$ for several hierarchical networks).
The presence of hierarchy in a network is usually hard to quantify, but in some cases, such as those of the networks represented in Fig.~\ref{hier_modelsNET}, it is implicitly introduced in the generating procedure:
\begin{itemize}
\item[A.] Regular or scale-free trees are clearly hierarchical structures (we have checked the behavior of Cayley trees and BA scale-free trees obtained, as sketched in Fig.~\ref{hier_modelsNET}-A, by means of a preferential attachment rule with $m=1$~\cite{BA_model}); 
\item[B.] For the DMS model with $m=2$ \cite{DMS_model}, one adds at each step a new node which is connected to the extremities of a randomly chosen edge, thus the causal structure of the tree introduces a hierarchy (see Fig.~\ref{hier_modelsNET}-B); 
\item[C.] The deterministic scale-free networks are built starting with two nodes connected to a root. At each temporal step $n$, two units (of $3^{n-1}$ nodes) identical to the network formed at the previous step are added, and each of the bottom $2^n$ nodes are connected to the root \cite{ravasz} (Fig.~\ref{hier_modelsNET}-C);
\item[D.] The Random Apollonian Networks (RAN) \cite{andrade,zhou} are embedded in a two-dimensional plane.
One starts with a triangle; a node is added and connected to the three previous nodes; at each step a new node is added in one of the existing triangles (chosen at random) and connected to its three edges, replacing the chosen triangle by three new smaller triangles (Fig.~\ref{hier_modelsNET}-D).  
\end{itemize}
In the particular case of {\em tree structures}, the power-law decay can be justified with 
a more precise qualitative argument.
In general, from the viewpoint of the Naming Game dynamics, a tree is
formed by two ingredients: linear structures on which the interfaces
between clusters diffuse as in one-dimensional systems and branching
points at which the motion of interfaces slows down. 
Following the arguments used in Section~\ref{CHAP5_3_1} and in Appendix~\ref{APP5_1}, 
on a linear structure we can model the motion of the interfaces between
clusters of words as random walks. At branching points, however, the
interfaces can in principle interact with more than one cluster, thus
the effective hopping probability is decreased or, in other terms, the
mean waiting time between two successive steps increases. The average
waiting time can be computed as the inverse of the stationary
probability for the local configurations of the interface (as for a
classic escape-over-a-barrier problem~\cite{chandra}). In principle, such
probabilities can be obtained by solving a truncated Markov
chain for the transition rates for all possible moves of
the interface at the branching point (as we have done for one-dimensional systems in Appendix~\ref{APP5_1}). The
computation is actually very demanding even in simple situations such as that
of an interface going across a node of degree $3$ (we have reported in Fig.~\ref{branchingNET} an example containing some of the transitions one should take into account). 
However, from simple examples, we expect that increasing the degree corresponds 
to a stronger effective pinning of the interfaces and larger waiting times.\\
The above qualitative argument explains the diffusive behavior observed on regular 
trees, such as the Cayley tree; on the other hand, on scale-free trees, such as the BA networks with 
$m=1$, the behavior is even slower, with a clearly subdiffusive exponent.\\
In the theory of random walks, subdiffusion can be associated with an
anomalously long waiting time between successive
walks~\cite{georges,isichenko,west,zaslavsky}. According to our picture, this is exactly what 
happens in a scale-free tree, where the degree of the nodes is broadly distributed. 
The interfaces make random walks but they may be suddenly pinned at some branching 
points with waiting times that depend on the degree of the nodes they try to
by-pass. Consequently, the heterogeneity in the degree induces that of the waiting times and the corresponding
subdiffusive behavior.\\ 
Note that this argument holds only for trees, not for general hierarchical structures, 
whose local dynamics is more complicated preventing us from a detailed analysis. \\
%

{\bf \em Community structures - \quad}
In contrast with other non-equilibrium models, as those based on
zero-temperature Glauber dynamics or the Voter
model~\cite{dornic,boyer,suchecki,castellano2,castellano1}, we do not find any signature of
the occurrence of metastable blocked states in any relevant topology
with quenched disorder.  Even when in several cases the total number of words 
displays a plateau whose length increases with the system size, the number of different words is
continuously decreasing, revealing that the convergence is not triggered by 
fluctuations due to finite-size effects, but it is the result of an
evolving self-organizing process.  Such behavior makes the Naming Game
a robust model of self-coordinated communication in any structured
population of agents.  \\
\begin{figure}[t] 
\centerline{ 
\includegraphics*[width=7.5cm]{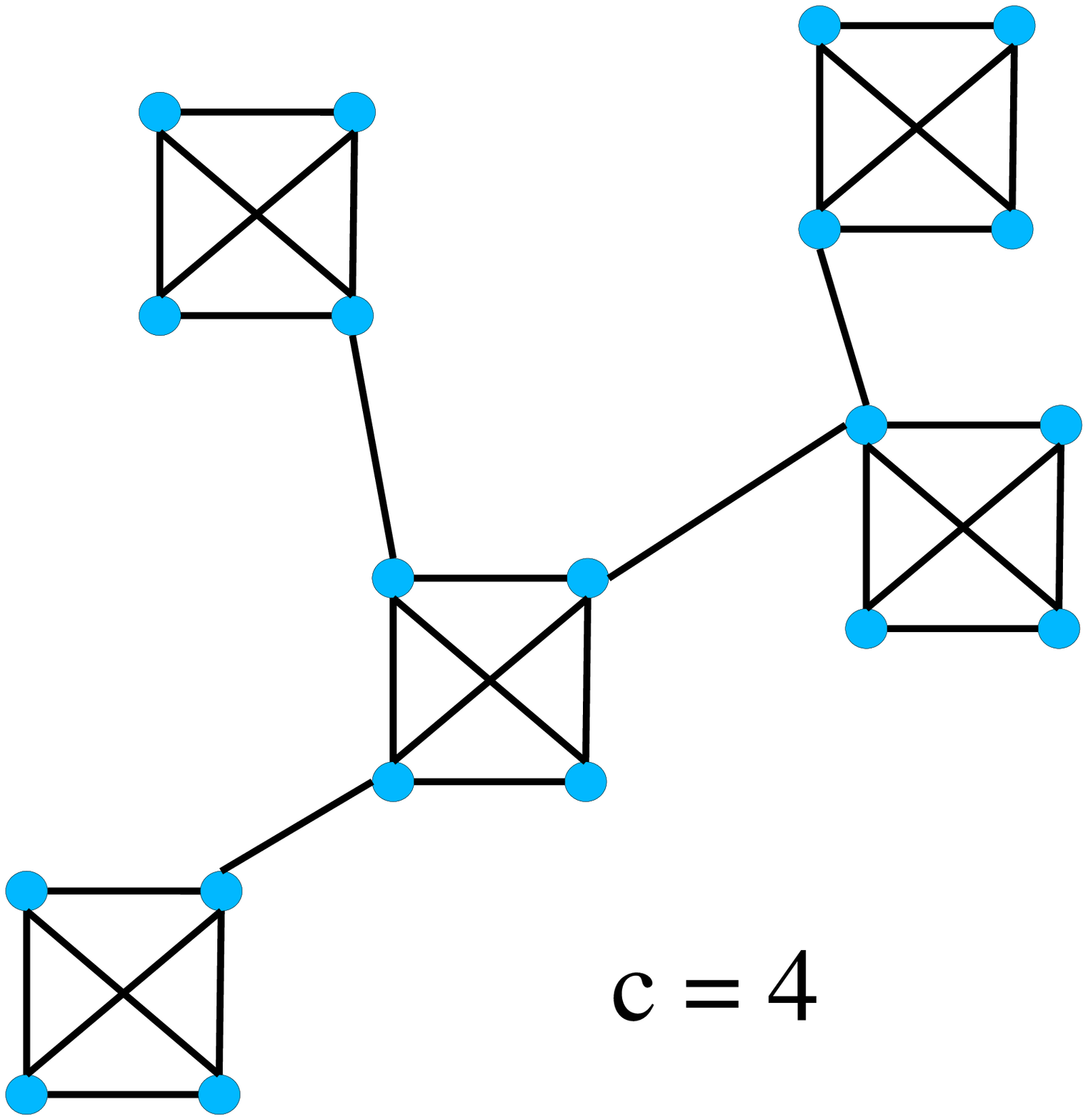}
\includegraphics*[width=7.5cm]{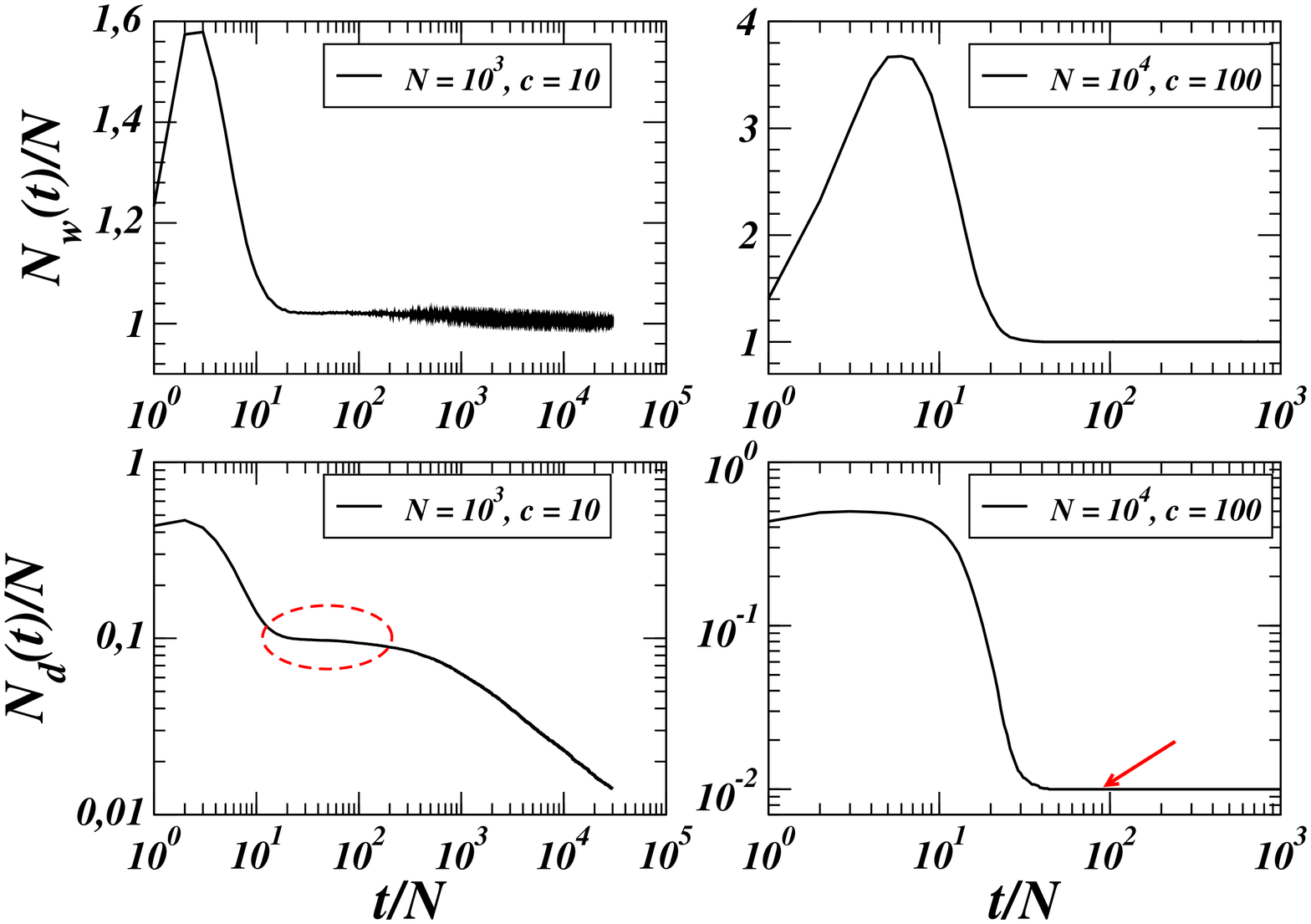}
}
\caption{(Left) A network with strong community structure.
(Right) Metastable states in networks with strong community structure.
Each community is composed of $c$ nodes so that there are $N/c$
communities. }
\label{communityNET}
\end{figure}
Finding topological properties ensuring the existence of metastable states or blocked configurations 
seems to be a non-trivial and intriguing task, that we try to investigate starting from the following main remarks:
\begin{itemize}
\item the model displays slow coarsening dynamics whenever
there is the possibility of cluster formation, i.e. when topological constraints are strong
enough to prevent words propagation;
\item highly clustered regions and cliques of nodes rapidly find a local consensus;
\item at the interfaces between clusters, an effective surface tension is generated; 
\item close to bottlenecks the surface tension may increase, causing the ordering process
to slow down.     
\end{itemize}
According to this analysis, reasonable candidates for observing metastable states are networks with strong community 
structures, i.e. networks composed of a certain number of internally highly connected
groups interconnected by few links working as bridges.
An example of network with strong community structure is represented in Fig.~\ref{communityNET} (left):
fully connected cliques composed of $c=4$ agents are interconnected by single edges.
Figure~\ref{communityNET} (right) reports the behavior of the Naming Game on such a
network, for different clique's size $c$.
From simulations it turns out that, not only the total number of
words, but even the number of different words has a plateau whose
duration increases with the size of the system. The number of
different words in the plateau equals the number of communities, while
the corresponding total number of words per node is about one, {\em proving
the existence of a real metastable state in which the system reaches
a long-lasting multi-vocabulary configuration}. Indeed, each community
reaches internal consensus but the weak connections between communities
are not sufficient for words to propagate from one community to the other.
The chosen network certainly has an extremely strong community structure, 
but preliminary studies on real networks of scientific collaborations give results 
that are in qualitative agreement with our results (i.e. plateaus are observed).\\
More precisely, when a network contains communities of different sizes and the community structure
is not very strong, the corresponding curves $N_{w}(t)$ and $N_{d}(t)$ display a series of plateaus,
with sharp transitions in between.
Several groups have recently put forward methods to distinguish different levels of community structures in real and computer generated networks exploiting dynamical processes evolving on them (see the review articles in Refs.~\cite{community_detection1,community_detection2}). For instance, Diaz-Guilera et al. \cite{diaz-guilera} have
used synchronization properties of non-linear oscillators (deployed on the nodes of a network) in order to determine community structures at different levels of resolution. Communities or groups of highly interconnected nodes are more likely to synchronize, thus looking at the temporal evolution of synchronization properties it is possible to identify  communities at different scales. 
Similar analyses have been carried out by Bornholdt et al. \cite{bornholdt1,bornholdt2} using Potts dynamics. In this case, the process leading to the community detection is the same as for the Naming Game, i.e. a coarsening dynamics of clusters with surface tension at the interfaces. Compared to Potts-based methods, the Naming Game has the relevant property that we do not have to fix the number of states in advance and the strength of the effective surface tension depends on the local topological constraints. Future studies could be addressed to modify the Naming Game model in order to have a more appropriate tool for community detection.

\newpage
\section{Agents activity in heterogeneous populations}\label{CHAP5_4}
The aim of this section, based on the material presented in Ref.~\cite{naming_gameACT}, is that of providing a detailed statistical description of the internal dynamics of single agents, and its relation with the collective behavior of the Naming Game model. \\
The analysis of simulations results points out that the internal dynamics of an agent depends strongly on its degree, highly connected agents being much more active than low-degree nodes.
The existence of different activity patterns is reflected in the shape of the distribution of the number of words stored in the inventory of a node, that turns out to depend on the level of heterogeneity of the network.
In homogeneous networks such distribution is exponential for all agents, while highly-connected agents in heterogeneous networks present a distribution with a clearly gaussian tail (half-normal distribution).\\
From the point of view of the evolution rule, this result shows that {\em the role of the memory is different depending on the connectivity properties of single agents}. The effect on the dynamics are clearer if we consider a closely related quantity, the cumulative distribution of the waiting times (or survival probability) between two consecutive successful interactions, i.e. between two decisions taken by the same agent.  
Indeed, an exponential waiting time distribution is the signature of a poissonian dynamics, while our results point out that the decision process associated with the internal activity of the agents is intrinsically non-poissonian, and it turns out to be poissonian only in the special case of a homogeneous network.
This feature is completely new in non-equilibrium models of social interactions, in which the interaction rules are usually defined in such a way that the agents take decisions at approximately constant rate. \\
Apart from the intrinsic interest for non-poissonian individual dynamics,  our findings are interpreted in order to understand the property of strong convergence towards the absorbing state that the model exhibits in all small-world structures, independently of the degree heterogeneity. 

\subsection{Numerical results on agents activity}\label{CHAP5_4_1}

\begin{figure}[t]
\centerline{
\includegraphics*[width=10.0cm]{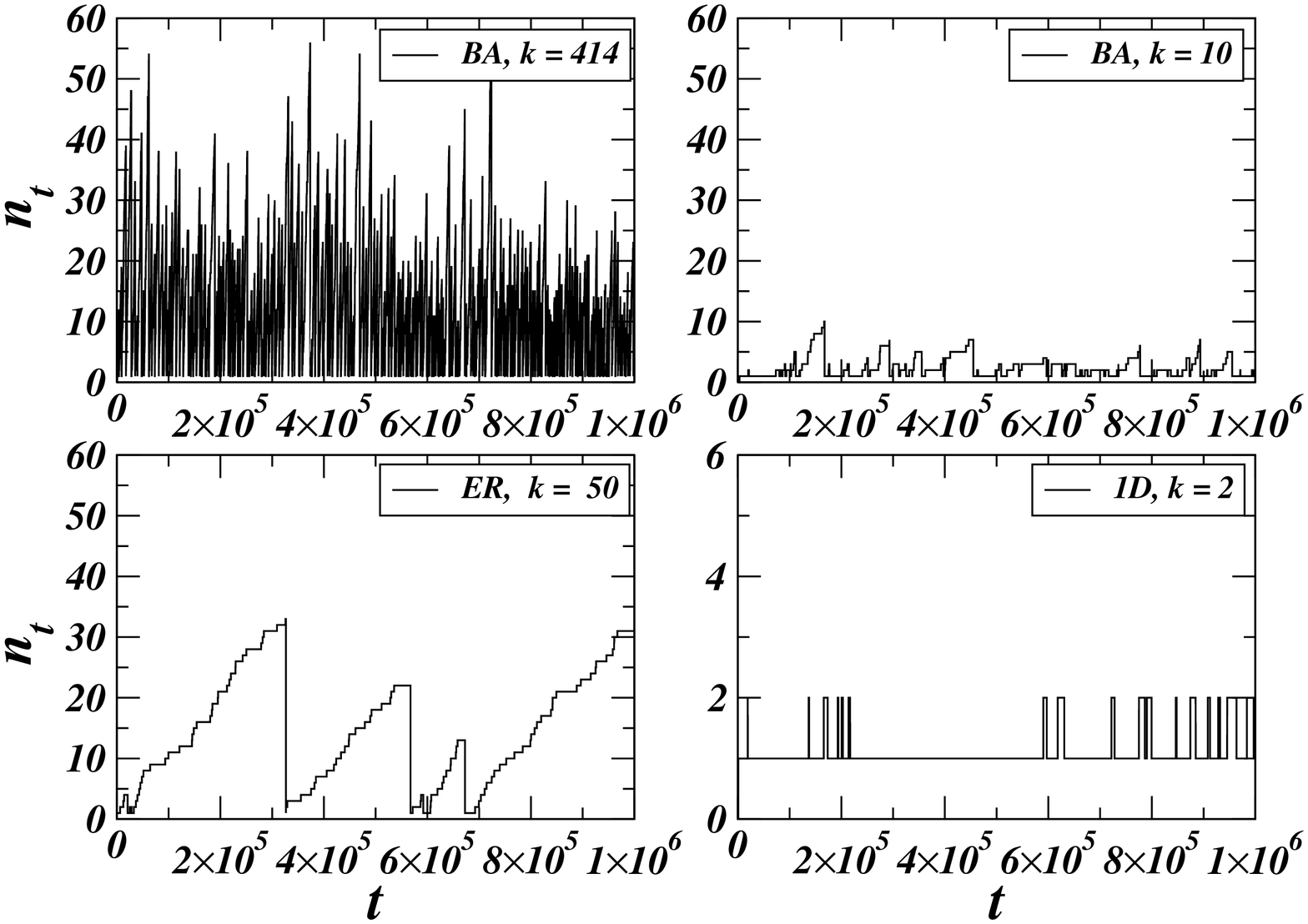}
}
\caption{Examples of temporal series of the number of states at a given node. (Top) Series from a BA network with $N=10^4$ nodes and $\langle k\rangle =10$, for nodes of high degree (e.g. $k=414$) and low degree (e.g. $k=10$). (Bottom) Series for nodes in ER random graph ($N=10^4$, $\langle k \rangle = 50$) and in a one-dimensional ring ($k=2$).}
\label{internal_direct}
\end{figure}

By means of numerical simulations, we characterize the activity patterns of an agent, focusing on the dynamics of its inventory size, i.e. the number of words (opinions, states, etc) of an agent.
In particular, our analysis is conceived for those topologies which present mean-field like dynamics (e.g. complete graph, homogeneous and heterogeneous random graphs, high-dimensional lattices, etc), where we cannot clearly identify a coarsening process leading to the nucleation and growth of clusters containing quiescent agents.
In other topologies, as in low-dimensional lattices, the agents internal activity is biased by the limited number of words locally available (Section~\ref{CHAP5_3_1}). 
An example of the different activity patterns in different topologies is reported in Fig.~\ref{internal_direct}.
In homogeneous networks (e.g. ER random graphs), the nodes have similar topological properties, thus their activity patterns are very similar as well. For heterogeneous networks, instead, highly connected nodes (hubs) play a different role in the dynamics compared to low degree nodes. The hubs are much more active, and their activity is determinant to drive the system to a rapid collective agreement. \\ More precisely, they show opposite behaviors depending on the pairs selection strategy.
As already pointed out in Section~\ref{CHAP5_3_3}, the asymmetry of the NG interaction rules becomes relevant in the case of heterogeneous networks.  
\begin{figure}[t] 
\centerline{
\includegraphics*[width=10.0cm]{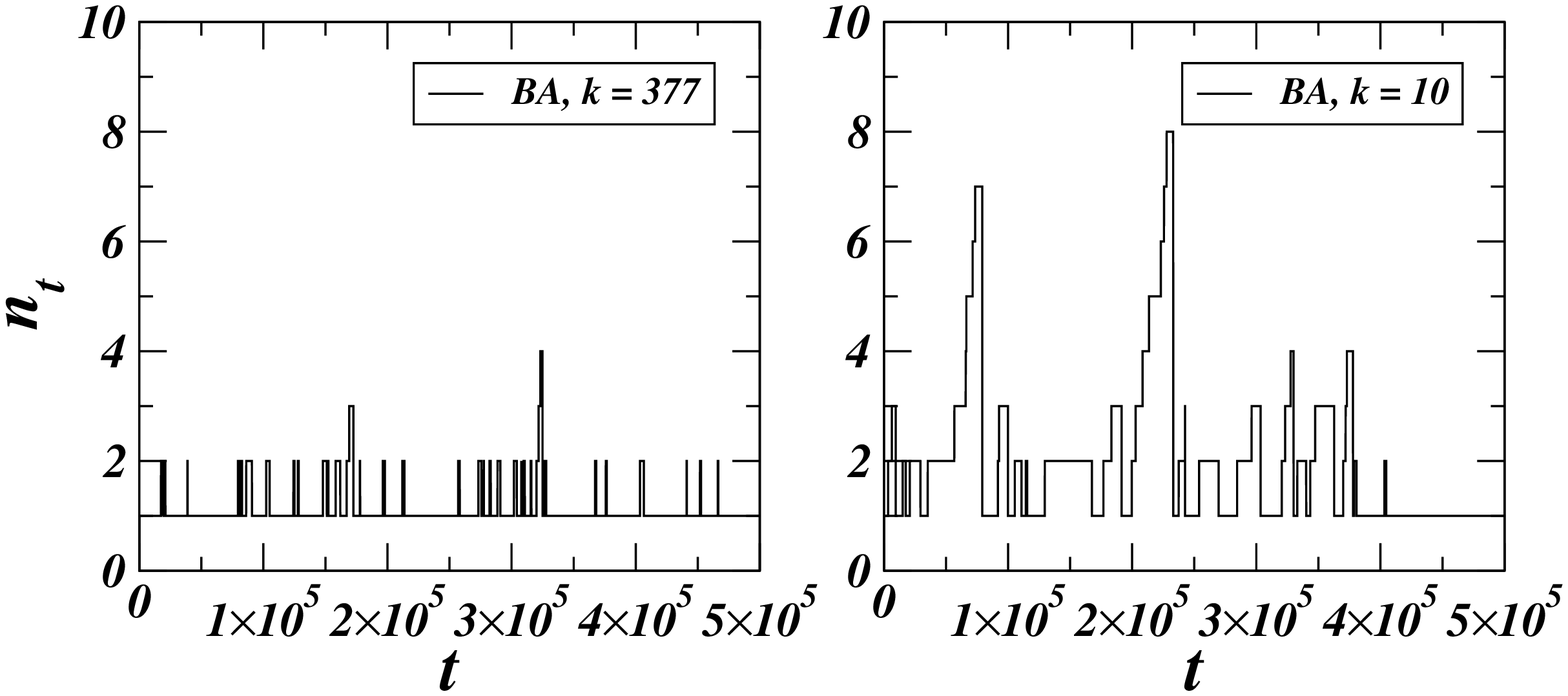}
}
\caption{Example of temporal series for nodes activity in a Reverse Naming Game on a BA network with $\langle k\rangle = 10$. The activity of the hubs (left) is very low (they are preferentially chosen as speakers) while that of low degree nodes (right) is the same as in the direct NG.}
\label{internal_reverse}
\end{figure}
In the direct Naming Game, that most naturally describes realistic speaker-hearer interactions, the speaker is  chosen with probability $p_{k}$ (where $p_k$ is the degree distribution of the network), while the hearer is chosen with probability $q_k = k p_k/\langle k\rangle$. According to this selection criterion, the high-degree nodes are preferentially chosen as hearers. 
Using the opposite strategy, called reverse Naming Game, the hubs are preferentially selected as speakers; whereas
the neutral strategy ensures that the roles of speaker and hearer are assigned with equal probability. 
Figure~\ref{internal_reverse} shows that the reverse strategy produces completely different activity patterns compared to the direct one, with a rather low variability and the absence of high spikes in the number of words. No significant difference between hubs and low-degree nodes is visible. 
The reason is that the inventory size increases because of a failure only if the node is playing as hearer. The speakers never add states to their inventories. Hence, agents that preferentially play as hearers tend to be more unstable, amplifying the number of states in the system. 
Using the direct strategy in heterogeneous networks favors the choice of the hubs as hearers, the degree of the hubs being  orders of magnitude larger than the average degree. 
This is the reason of the large number of states stored in the inventory of the hubs for the direct strategy. 
%
\begin{figure}[t] 
\centerline{
\includegraphics*[width=8.0cm]{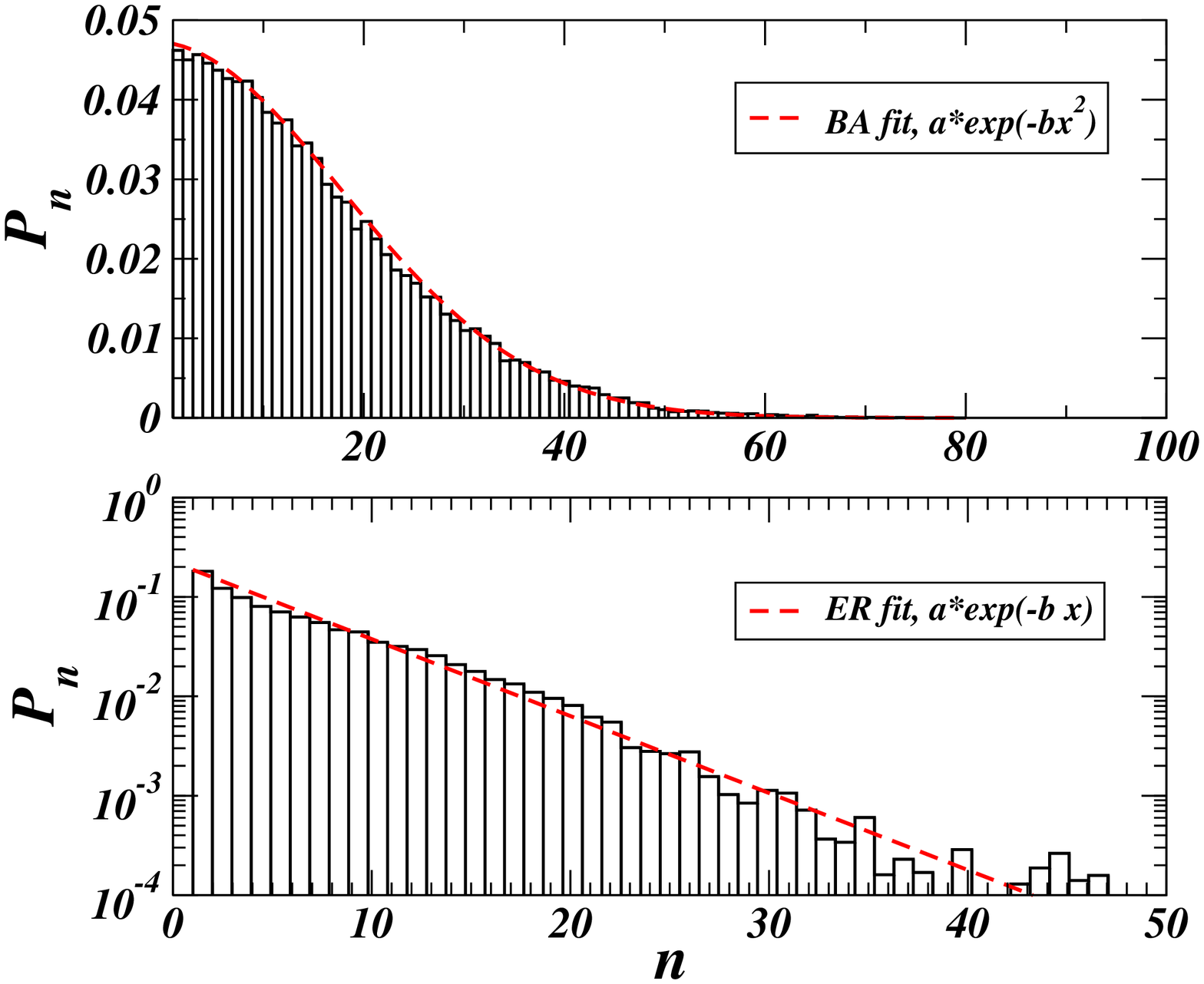}
}
\caption{Distribution of the number of words for a class of high-degree nodes in a BA network (top) with $\langle k \rangle =10$ and for average degree nodes in a $ER$ (bottom) with $\langle k\rangle = 50$. Both networks have $N=10^4$ nodes. The distributions have been computed during the re-organizational phase after the peak in the number of words. 
}
\label{distributionP}
\end{figure}

A first quantity that clearly points out differences in the activity of nodes depending both on their degree and on the  topological structure of the network is the probability distribution $\mathcal{P}_{n}(k|t)$ of the number $n$ of states  stored in the inventory of nodes of degree $k$ at time $t$. This means that the distribution is averaged over the class of nodes of given degree. Actually, as it is shown in Appendix~\ref{APP5_2}, this quantity depends only parametrically on the time $t$. 
Fig.~\ref{distributionP} (top) reports $\mathcal{P}_{n}(k|t)$ for the case of highly connected nodes in a heterogeneous network (the Barab\'asi-Albert network), whereas the same data for nodes of typical degree in a homogeneous network (the Erd\"os-R\'enyi random graph) are displayed in the bottom panel.
In homogeneous networks the shape of the distribution does not actually depend on the degree of the node, since all nodes have degree approximately equal to the average degree $\langle k\rangle$. 
In heterogeneous networks, instead, a deep difference exists between the behavior of low and high degree nodes. 
Low degree nodes have no room to reach high values of $n$, thus their distribution has a very rapid decay (data not shown); for high degree nodes, on the contrary, the distributions extend for more than one decade and their form is much clearer.\\
Apart from the behavior of low degree nodes, it is clear that the functional form of the distribution $\mathcal{P}_{n}(k|t)$ is different in homogeneous and heterogeneous networks. In homogeneous networks the distribution is exponential, while in heterogeneous networks it decays faster, and is well approximated by a half-normal distribution.\\
\begin{figure}[t] 
\centerline{
\includegraphics*[width=8.0cm]{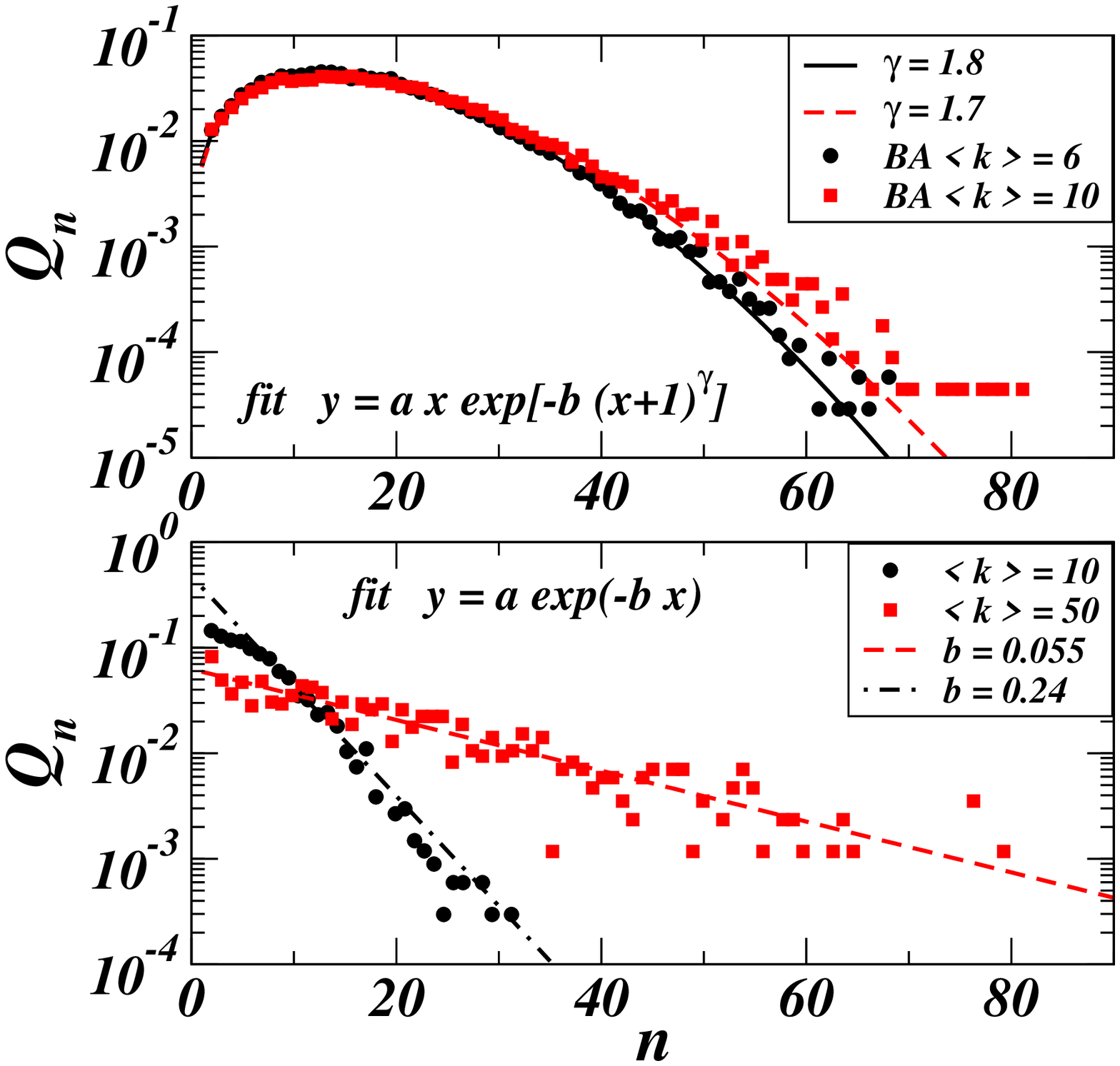}
}
\caption{Probability distribution $\mathcal{Q}_{n}(k|t)$ of the number of states $n$ at which an agent 
gets a success, i.e. the inventory is reset to $1$ state. For highly connected nodes in heterogeneous (top) 
and nodes of typical degree in homogeneous networks (bottom). Same parameters as in Fig.~\ref{distributionP}.
}
\label{distributionQ}
\end{figure}
Another interesting quantity is the probability distribution $\mathcal{Q}_{n}(k|t)$ that an agent of degree $k$ gets a success in an interaction occurring when it has $n$ states into the inventory, i.e. the value at which the inventory is abruptly reset to $1$. This quantity has an exponential shape in homogeneous networks (Fig.~\ref{distributionQ}-bottom) for sufficiently high average degree, and a Weibull-like shape (Fig.~\ref{distributionQ}-top) for the case of heterogeneous networks (high-degree nodes). 
The existing relation between $\mathcal{Q}_{n}(k|t)$ and $\mathcal{P}_{n}(k|t)$ is straightforward: the probability distribution $\mathcal{Q}_{n}(k|t)$ of the level $n$ at which a successful interaction occurs is the product of the probability of having $n$ states and the conditional probability $\mathcal{W}_{k}(n\rightarrow 1|t)$ that an agent (of degree $k$) finds at time $t$ a (temporary) agreement when it has $n$ states; i.e. $\mathcal{Q}_{n}(k|t)$=$\mathcal{W}_{k}(n\rightarrow 1|t)$ $\mathcal{P}_{n}(k|t)$.

\subsection{Theoretical interpretation and future work}\label{CHAP5_4_2}
 
In Appendix~\ref{APP5_2}, we discuss a master equation approach for the jump process associated to the dynamics of single agents, by means of which it is possible to derive the correct expression for the distribution $\mathcal{P}_{n}(k|t)$ and, consequently, that of $\mathcal{Q}_{n}(k|t)$. 
We recover here the same result using a naive argument, based on the concept of survival probability in renewal processes, that is useful to clarify the role of non-poissonian dynamics in relation with other types of non-equilibrium statistical models.\\
A {\em renewal process} \cite{renewal} is a stochastic process characterized by a series of recurrent events that are separated by waiting times $\{\tau_{i}\}$. The waiting times are mutually independent random variables with a common distribution $\mathcal{T}(\tau)$. We call {\em survival probability} $\mathcal{T}_{>}(\tau)$ the probability that the renewal event occurs after a waiting time at least equal to $\tau$.\\
In physics, stochastic processes are usually coarse-grained models for some natural phenomenon, since the observed waiting times statistics is the result of some peculiar properties of the underlying phenomena.
For instance, we observe power-law waiting time distributions in many natural phenomena and in the models used to study or reproduce these phenomena (e.g in solar flames \cite{wheatland,paczuski}, financial markets \cite{scalas,fin_wait}, anomalous transport \cite{zaslavsky}, earthquakes \cite{omori,bak}, etc). Distributions with Weibull (and gaussian) tails are frequent in more general problems of queuing theory \cite{feller} and survival analysis \cite{survival}. 
On the contrary, models of statistical mechanics traditionally used in opinion dynamics show exponentially shaped waiting time distributions (signature of a poissonian dynamics).\\
All these different statistics are based on the factorization property of the corresponding stochastic (renewal) 
processes.  
Let us denote $h(\tau)$ the {\em hazard function} of the process, i.e. the rate of occurrence of the renewal event at a (waiting) time $\tau$. The survival probability, i.e. the probability that the event occurs at time at least $\tau$ satisfies the following recursion equation,
\begin{equation}\label{rec_haz}
\mathcal{T}_{>}(\tau+1)=\mathcal{T}_{>}(\tau)\left[1-h(\tau)\right]~,
\end{equation}
whose general solution is 
\begin{equation}\label{wt_dist}
\mathcal{T}_{>}(\tau) = \frac{\prod_{i=1}^{\tau} \left[ 1- h(i)\right]}{\sum_{\tau=1}^{\infty} \left[\prod_{j=1}^{\tau} \left[ 1- h(j)\right]\right]}. 
\end{equation}
In the particular case of the Naming Game, a renewal event is identified with a successful interaction, bringing back to $1$ the inventory size of the node.  
The pair selection mechanism is purely poissonian, thus an agent interacts with a precise constant rate (that is $p_{k}$ or $q_{k}$ depending on whether it plays as speaker or as hearer). This poissonian external signal can be regarded as a discrete timing for the internal activity of the nodes.
Hence, apart from a time rescaling, the dynamics of the inventory size, described in the previous section and in App.~\ref{APP5_2}, gives a good approximation of the waiting time statistics related to the renewal events. The distributions $\mathcal{P}_{n}(k|t)$ and $\mathcal{Q}_{n}(k|t)$ correspond respectively to $\mathcal{T}_{>}(\tau)$ and $\mathcal{T}(\tau)$. The hazard function has thus the same functional form of the success probability $\mathcal{W}_{k}(n\rightarrow 1|t)$, with the waiting time $\tau$ instead of the number of words $n$. \\
In the Appendix~\ref{APP5_2} (and in Ref.~\cite{naming_gameACT}), we show that the success probability $\mathcal{W}_{k}(n \rightarrow 1|t)$ assumes different forms depending on the underlying topology. In homogeneous networks, it turns out to be almost independent of the number $n$ of words stored in the inventory of the node; in heterogeneous networks it is instead linearly proportional to $n$ (for nodes of sufficiently high degree $k$).
Inserting in Eq.~\ref{wt_dist} hazard functions with these functional forms, we can compute the corresponding expression for $\mathcal{T}_{>}(\tau)$, obtaining an approximate expression also for $\mathcal{P}_{n}(k|t)$.\\
Let us consider a constant hazard function, from Eq.~\ref{wt_dist} the corresponding survival probability distribution is exponential, and consequently also $\mathcal{P}_{n}(k|t)$ decreases exponentially with $n$. On the other hand, when the hazard function grows linearly with the waiting time, with normalization constant $C$, we get $\mathcal{T}_{>}(\tau)\propto \exp(-\frac{\tau^2}{2C})$. 
This simple argument provides an explanation of the gaussian decrease of the distribution $\mathcal{P}_{n}(k|t)$ observed for highly-connected nodes of heterogeneous networks (See Appendix~\ref{APP5_2} for a more rigorous approach.)

A further remark concerns the relation between waiting times and poissonian processes.
In general, the agent-based models studied in statistical physics are spin-like models, in which an individual 
is endowed with a variable, assuming a given set of values, each one corresponding to a different state. 
In such systems, single agent dynamics is intrinsically poissonian.
For instance, in systems evolving by means of Glauber dynamics, spin flips at a site occur independently 
one of the other, i.e. they are poissonian events with (Boltzmann) rate $\lambda \propto e^{-\beta \Delta H}$. 
The corresponding waiting time distribution (and survival probability) is exponential (as $\lambda \exp(-\lambda t)$).
In the present model, on the contrary, {\em the internal activity of an agent is modeled in order to reproduce a sort of  decision process based on information storage, and the waiting time between successive decisions turns out to depend strongly on the underlying topology}.
From the point of view of the global behavior it seems to be important: in the direct strategy, the hubs drive the system to a fast convergence to the absorbing state, as a result of the trade-off between their larger activity and their stronger inclination to reach an agreement (due to their internal memory patterns).\\
The first step toward a better comprehension of the role of non-poissonian dynamics is that of comparing these results and the scaling properties of the convergence time with those for the reverse Naming Game, in which the hubs have exponential waiting time distributions. 
 
Finally, waiting time statistics is also used as a measure of the criticality in the behavior of physical systems, individuals and natural phenomena; in particular, in relation with extreme events \cite{bak}. Waiting time distributions with heavy-tails are signature of the absence of a characteristic scale on which the events occur.
For example, the theory that justifies the observation of power-law distributed waiting times between aftershocks in earthwakes is based on the Omori law \cite{omori}, corresponding to a hazard function inversely proportional to the waiting time.
Actually, a success rate decreasing with the time is necessary to get a broad waiting time distribution.\\
It should be interesting to modify the interaction rules of the Naming Game model (and thus the hazard function)
in order to change the shape of the waiting time distribution and in particular to get a power-law one. 
Such a situation would correspond to a critical decision process, in which agents might store a very large number of words, with an a priori unlimited memory requirement.

\newpage
\section{Conclusions}\label{CHAP5_5}
In the last part of this thesis, we have investigated a model of social dynamics, the Naming Game, that can be considered  an example of a new interesting class of dynamical processes, conceived to describe the onset of  global agreement in a population of individuals by means of pairwise negotiation interactions and memory-driven decision processes.
With respect to well-known models of social dynamics that have been borrowed from statistical mechanics (e.g. majority rules models, Voter model, etc), the Naming Game takes into account more realistic characteristics of social interactions, conserving a sufficient level of simplicity that was a good quality of the former ones. \\
Thanks to this mixture of ingredients, the Naming Game (and possible variants of the model) seems to be more appropriate than previous models for the study of heterogeneous populations of agents, such as social networks, since the dynamics is the result of a strong interplay between topological features and the internal properties of the agents.
We have focused our attention on two aspects: 
\begin{itemize}
\item the {\em dynamical features of the Naming Game on different topologies}; 
\item single {\em agents internal activity} and its relation with the global behavior.
\end{itemize}
In order to understand the behavior of the model on complex networks, we have analyzed the impact of the different topological structures, starting from the rather unrealistic cases of the complete graph (mean-field) and the one-dimensional system. However, they turn out to be precious for the conprehension of more complex topologies. They are, indeed, almost completely analytically solvable, providing two opposite behaviors:
\begin{itemize}
\item the mean-field model is characterized by an initial super-spreading of words throughout the system, whose maximum is reached in a time $t_{max} \sim \mathcal{O}(N^{3/2})$ and corresponds to a state in which each single agent possesses $\mathcal{O}(\sqrt{N})$ words; then, a {\em very fast convergence} (i.e. more than exponential) takes place leading the system to the global consensus in a time $t_{conv} \sim \mathcal{O}(N^{3/2})$;    
\item in the one-dimensional model, agents find immediately a local consensus, many clusters of neighboring agents with a common unique word start to grow, {\em competing in a coarsening process} driven by the diffusion-coalescence process of the interfaces. Consequently, the maximum total number of words, reached very quickly in $\mathcal{O}(N)$ steps, scales as $\mathcal{O}(N)$ as well, but the global agreement requires a time $t_{conv} \sim \mathcal{O}(N^{3})$.
\end{itemize}
The second step towards the comprehension of the Naming Game dynamics has been provided by the study of the Watts-Strogatz model \cite{watts98}. The networks generated by this model are characterized by a tunable parameter (the rewiring probability) that allows to interpolate between a one-dimensional regular lattice and a homogeneous random graph.
For non-zero rewiring probability, the model has the small-world property, i.e. different regions of the network
are connected by shortcuts, so that the average distance between nodes scales logarithmically with the network size.
After an initial phase during which words are created and small local clusters appear, 
the small-world property ensures their propagation out of the local scale, boosting up the spreading process
(contrarily to what happens in low dimensional lattices where words
spreading is purely diffusive).\\
{\em The same acceleration of the dynamics is then observed in many other networks sharing the small-world property},
suggesting that it is sufficient to recover the high temporal efficiency 
observed in the mean-field system. 
For both the homogeneous and heterogeneous network models, we get a scaling law for
the convergence time $t_{conv}$ with the size $N$ of the system of the
type $t_{conv} \sim N^{\beta_{SW}}$, with exponent approximately
$\beta_{SW} \simeq 1.4$.  The discrepancy with the mean-field exponent
($\beta_{MF} \simeq 1.5$) may be due to logarithmic corrections. 
Moreover, small-world networks have higher memory
efficiency than the mean-field model, since the peak in the total
number of words scales only linearly with the size $N$. This is due to
the fact that these networks are sparse (their average
degree $\langle k\rangle$ is small compared to $N$).\\  
Nonetheless, a detailed analysis allows to point out {\em distinct dynamical patterns 
on homogeneous and heterogeneous networks}.  In homogeneous networks
all nodes have a similar neighborhood and therefore similar dynamical
evolution, while in heterogeneous networks classes of nodes with different
degree play different roles in the evolution of the Game. High degree nodes, indeed, 
are more likely chosen as hearers (in the direct Naming Game). 
At the beginning, low degree nodes are much more involved in the
process of word generation than the hubs; local consensus is easily reached
and a large amount of locally stable different words gets in touch with higher
degree nodes. The latter start to accumulate a large number of words in their
inventories, playing as ``super-spreaders'' of names towards less connected
agents and finally driving the convergence.
From this viewpoint, the convergence dynamical pattern of the Naming Game on
heterogeneous complex networks presents some similarities with more studied
epidemic spreading phenomena~\cite{barth_activity}.\\
The shape of the degree distribution and the scaling of the average distance are not the only topological properties that  determine the Naming Game dynamics; for this reason, we have investigated the effects of a number of other quantities.\\ 
On both homogeneous and heterogeneous networks, an {\em increase in the average degree 
induces a larger memory peak and a faster convergence}, while 
the growth of the {\em clustering coefficient has a completely opposite effect}.
This is particularly important in social networks that are usually characterized by a large 
level of cohesiveness.\\
A very striking result concerns the convergence of the Naming Game on networks with well-defined hierarchical organization or community structures. 
{\em On generic hierarchical networks, and particularly on trees, the process leading to the final agreement is very slow, governed by a diffusive or even subdiffusive coarsening}. We have identified the origin of this behavior in the existence of a non negligible surface tension between different hierarchical levels.
{\em A similar behavior is due to the presence of a community structure}: each community finds quickly an internal agreement, but a cluster cannot easily expand outside its own community since the interfaces get pinned on the few bridges interconnecting different communities. 
Therefore, in this case the curve of the number of different words $N_{d}(t)$ is not characterized by a power-law decay, but by a series of plateaus of different size corresponding to different levels of refinement in the community organization of the network. When the duration of a plateau (in $N_{d}(t)$) increases with the size of the network, we say that the system is in a metastable state.        \\
Even in presence of {\em metastable states}, if we wait sufficiently long, the system will converge to the absorbing state, 
meaning that the Naming Game model is a strongly converging dynamical rule.
The origin of this behavior resides in the {\em memory-based decision rule}. We have investigated its {\em implications both at a local and global level}. \\
At a local level we have focused on the internal activity of the agents in different topologies, getting deeper insights 
in the mechanisms governing the decision processes (see Section~\ref{CHAP5_4}).
In particular, we have found that {\em the single agents dynamics is intrinsically non-poissonian}, resulting in a stronger tendency to take a decision (and then to converge) for the high-degree nodes. This attitude balances the instability due to the fact that high-degree nodes, being in contact with many different words, are more exposed to perturbations in the dynamics. \\
{\em The role of the memory in the global behavior of the system is that of generating an effective surface tension}
(see Section~\ref{CHAP5_3_1}), that is responsible for the coarsening of clusters. 
In fact, the surface tension associated with the coarsening process of clusters of agents with a unique word is strong enough to ensure the convergence of the Naming Game in any dimension, but sufficiently weak to prevent the system to block in metastable states. From this point of view, the Naming Game is similar to a low-temperature Potts model, but without the typical bulk noise due to an externally imposed temperature-like parameter.\\
Note that this form of {\em surface tension} does not have an energetic microscopic nature, as for the Ising model, but it {\em is due to the introduction of temporal correlations in the decision process} (i.e. the introduction of memory) similar to what happens in diffusion-limited aggregation when a mechanism of noise-reduction is taken into account. 
Some preliminary results on the comparison between clusters dynamics in models with memory (like the Naming Game) and  other lattice spin models seem to corroborate this picture, that will be developed in a future work.

In summary, as other models of opinion formation, {\em the Naming Game shows a non-equilibrium dynamical evolution from a disordered state to a state of global agreement}. However, with respect to most opinion models, in which the agents may accept or refuse to conform to the opinion of someone else, {\em the Naming Game gives more importance to the bilateral negotiation process between pairs of agents}. 
For this reason, the Naming Game should be regarded as a model for the emergence of a globally accepted linguistic convention or, in other terms, the establishment of a self-organized communication system; but it can be also reasonably used to describe opinion formation and other social polarization phenomena.   
The main novelty resides certainly in the introduction of pairwise interactions endowed with memory and feedback, that make the Naming Game phenomenology closer to that observed in real systems, in particular when we consider the system embedded in a complex network topology. 

\chapter{General Conclusions and Outlook}
\label{CHAP6}

In this thesis, I have studied both numerically and analytically some structural and dynamical aspects of complex networks.
From the point of view of a theoretical physicist, the problems that phenomenological observations on complex networks are raising are undoubtedly exciting, because of the possibility of getting striking results applying rather simple statistical physics approaches.   
As a consequence of the small-world property, indeed, the general behavior of complex networks and of the dynamical processes taking place on them can be fairly described using mean-field arguments.
On the other hand, when mean-field like methods are not applicable, or the picture we obtain from them is not satisfactory,
in most of the cases the only possible approach is that of  numerical simulations. \\
This picture corresponds in general to the approach of research followed in this thesis. 
For example, in the case of the exploration of networks, we have provided a mean-field analysis that gives very  good results at a qualitative level, but for any quantitative characterization of networks sampling the use of numerical simulations cannot be avoided.  
In the interdisciplinary field of complex networks, however, the quantitative aspect is very important, since the original problems are usually closely related with applications and theoretical results have to be compared with the abundance of phenomenological data.
Thus, future works will be addressed to improve separately these two approaches. 
First, from the numerical side, it is worthy to extend the investigations of this thesis to more realistic models of Internet mapping, in order to verify the reliability of our results when the condition of shortest path probes is relaxed.
Preliminary results have been obtained following two different approaches: one corresponds to use a model in which local path inflations are introduced (i.e. distortions of the shortest path that should reproduce the effect of traffic and policies \cite{algo_crovella}); the other considers \texttt{traceroute} probing by means of weighted shortest paths, in which the weights are randomly distributed on the edges (we have only studied low disorder regimes, but also the strong disorder limit should provide interesting results \cite{havlin_strong}).     
In both cases, the quantitative estimation of the relevant topological quantities seems not to change considerably from the results here exposed. \\
From an analytical point of view, I am currently interested in understanding the origins of the biases introduced by tree-like explorations in relations with the causal structure of the spanning trees generated in the sampling or more generally the framework of hidden variable models. It has been shown \cite{bialas,jamming} that scale-free topological properties emerge naturally when the networks are endowed with a causal structure; according to this picture, sampling biases could be the natural result of the systematic introduction of causality in the network's topology. \\
Of course, theoretical improvements of the mean-field approach are possible, for instance taking into account correlations in the expressions for the node and edge discovery probability.\\
The most promising possible application of the work exposed in Chapter~\ref{CHAP3} is however the definition of statistical estimators able to correct the biases due to the sampling process. We have successfully introduced estimators for the number of nodes in the Internet, but the present work is aimed to define unbiased estimators also for other quantities (e.g. the number of edges). \\
A similar twofold approach holds for the subjects discussed in Chapter~\ref{CHAP4}.
In order to achieve a more satisfactory comprehension of which mechanisms are responsible of the non-trivial structural and  functional organization of real complex networks, such as the world-wide airports network, we need first 
to refine the analysis of the real data, selecting those quantities better pinpointing the functional and economic dimension of the system; on the other hand, we have to improve the current models of weighted growing networks to produce more  realistic effects.\\
For the inhomogeneous spreading on complex networks, whose theory is treated in Chapter~\ref{CHAP4} and expecially in Appendix~\ref{APP4_2}, the situation is the opposite one: we have a good understanding of the theoretical description of the process, but we do not dispose of phenomenological data (e.g. those regarding the rates of infection of a real virus on the Internet if the heterogeneity of the nodes as well as their functional properties are taken into account).
It should be interesting to retrieve real data on the level of functional inhomogeneity in infrastructure networks and verify if the corresponding spreading properties can be analyzed within the theoretical framework here provided. \\
The Naming Game is a rather recent topic of research, thus many aspects of the dynamics of the model are not completely clear. One of the most interesting phenomena displayed by this model is the presence of an effective surface tension, that is comparable with that of a low temperature Potts model, even if the pairwise evolution rule looks more similar to that of the Voter model \cite{krapivsky1}, in which the surface tension is absent. We have reason to think that the surface tension is a consequence of the presence of memory in local rule, therefore a future work will be addressed to show, using a simplified model, that the presence of memory in the nodes is sufficient to produce a coarsening dynamics in analogy with some techniques of noise reduction studied in the problems of surface growth \cite{meakin}.
Moreover, my personal opinion is that a further simplified model of Naming Game, conceived in such a way to retain all its relevant properties, could allow to study analytically the dynamics on the whole class of mean-field like models, maybe elucidating the relation between the small-world property (and other topological properties) and the scaling of the convergence time.  
It should be also interesting to add some external source of noise to the system, maybe coupled with the internal activity of the agents, in order to study if a phase transition toward a non-trivial state in which agents do not find a consensus emerges. 

In conclusion, the study of dynamical phenomena on complex networks is a fascinating subject of research that is expected to become more and more popular in the next years, because of the large amount of data that is still not available on real dynamical processes and for the possibility of theoretical modeling by means of known statistical physics approach and for the considerable number of issues in which the analytical and numerical analysis of simple models can be successfully applied.

\addcontentsline{toc}{chapter}{Acknowledgements}

\chapter*{Remerciements} 
 
Cette th\`ese s'est d\'eroul\'ee de novembre 2003 \`a juin 2006 au Laboratoire de Physique Th\'eorique d'Orsay.
Je remercie les directeurs du labo Dominique Schiff, puis Henk Hilhorst de m'y avoir tr\`es bien accueilli et permis de 
r\'ealiser ce travail de th\`ese dans les meilleures conditions.\\
\vspace{0.3cm}

Je tiens tout d'abord \`a remercier Alain Barrat, pour avoir accept\'e de diriger ma th\`ese.
Sa disponibilit\'e, sa patience et la confiance qu'il a toujours montr\'ee \`a mon \'egard ont \'et\'e
tr\`es importantes pour moi. 
Je tiens \'egalement \`a remercier Alessandro Vespignani, avec lequel j'ai eu l'honneur et le plaisir de travailler durant toute ma th\'ese. 
Je leur suis profond\'ement reconnaissant.\\
\vspace{0.3cm}

Je remercie aussi chaleureusement tous les membres du groupe de Physique Statistique du LPT avec lesquels j'ai partag\'e les joies et les peines, ou plus simplement la vie quotidienne durant ces trois ann\'ees. En particulier,  `Nacho' Alvarez-Hamelin, Vivien Lecomte, Aur\'elien Gautreau, Fr\'ed\'eric van Wijland et `les italiens' de mon bureau, Andrea Puglisi et Paolo Visco, qui m'ont evit\'e le mal du pays. Leur sympathie et leur amiti\'e a \'et\'e tr\`es importantes pour moi.\\ 
Je souhaite aussi remercier tout le personnel du labo et le directeur de l'Ecole Doctorale Yves Charon; leurs comp\'etences et leur disponibilit\'e m'ont rendu la vie  plus facile  \`a Orsay.  \\
\vspace{0.3cm}

Je remercie Messieurs Olivier Martin, R\'emi Monasson, et Alessandro Vespignani d'avoir bien voulu participer au jury de th\`ese et Messieurs Romualdo Pastor-Satorras et Cl\'ement Sire d'avoir accept\'e la lourde t\^ache d'en \^etre les rapporteurs. \\
\vspace{0.3cm}

Je ne pourrai jammais oublier tous ceux qui ont travaill\'e avec moi, en particulier Marc Barth\'elemy, Vittorio Loreto et mon grand ami, `malheureusement int\'eriste', Andrea Baronchelli. \\
\vspace{0.3cm}

Finalement, je voudrais t\'emoigner ma profonde gratitude \`a mes parents et \`a ma soeur Chiara, pour leur soutien et l'affection qu'ils me donnent. Je salue \'egalement du fond du coeur tous mes amis avec qui j'ai partag\'e les meilleurs moments et les difficult\'es, et en particulier ma copine Deborah. Je ne trouve pas suffisamment de mots pour te remercier (surtout en francais!); ta presence a \'et\'e pour moi la source de motivation et de s\'er\'enit\'e la plus grande.

\appendix
\chapter{Generating Functions in Percolation Problems}
\label{APP4_1}

In this appendix, we provide a brief introduction to the formalism of generating functions in the study of percolation on random graphs. We consider infinite graphs without isolated vertices, self-edge or multiple edges. 
The generating function for the degree distribution $p_k$ of a randomly chosen vertex is
\begin{equation}
G_{0}(x)=\sum_{k=1}^{\infty}p_{k}x^{k}~,
\end{equation}
with $G_{0}(1)=1$, and $\langle k \rangle =\sum_{k}k p_{k} = G_{0}'(1)$. 
Similarly, the generating function for the probability that a randomly chosen edge leads to a vertex of given degree is
\begin{equation}
\frac{\sum_{k}k p_{k} x^{k-1}}{\sum_{k}k p_{k}} = G_{1}(x)= \frac{G_{0}'(x)}{G_{0}'(1)}~.
\end{equation}
A useful property is that the probability distribution and its moments can be computed by simple derivative of the corresponding generating function. 

If we call $q_{k}$ the probability that a vertex of degree $k$ is occupied (or node traversing probability if regarded as a spreading phenomena), the probability that, choosing randomly a vertex, we pick up an occupied vertex of degree $k$ is the product of the probabilities of two independent events, i.e. $p_{k}q_{k}$. Repeating the same operation with the edges, we need the probability that the randomly chosen edge is attached to an occupied vertex of degree $k$. This event happens with probability $k p_{k} q_{k} /\langle k \rangle$.  
Hence, we define the generating functions for both these probabilities that are very important in the site percolation,
\begin{subequations}
\begin{align}
 F_{0}(x; \{q\})& =\sum_{k=1}^{\infty}p_{k} q_{k} x^{k}~,\label{f0}\\
 F_{1}(x; \{q\})& =\frac{\sum_{k=1}^{\infty} k p_{k} q_{k} x^{k-1}}{\sum_{k}k p_{k}} = \frac{F_{0}'(x)}{\langle k\rangle}~.\label{f1}
 \end{align}
\end{subequations}
The function $F_{0}(x; \{q\})$ is the generating function of the probability that a vertex of a given degree exists and is occupied, while $F_{1}(x; \{q\})$ is the generating function for the probability of reaching a vertex of a given degree starting by a randomly chosen edge and that it is occupied.

The solution of the site percolation problem is the set of values $\{q_k\}$ for which an infinite cluster (giant component) exists. 
In order to compute the probability that a randomly chosen vertex belongs to the giant component, we start by computing the probability $P_{s}$ that a randomly chosen vertex belongs to a connected cluster of a certain size $s$. 
The use of generating functions allows to do it simultaneously for all the possible sizes. Then, the mean cluster size is obtained as the first derivative of the generating function of $P_{s}$. 
Finally, the condition for the divergence of the mean cluster size gives the condition for the existence of the giant component as a function of the parameters of the system, that are the degree distribution and the node occupation probability.
%
%
\begin{figure}[thb] 
\centerline{
\begin{tabular}{|c|}\hline \\ \includegraphics*[angle=-90,width=0.45\textwidth]{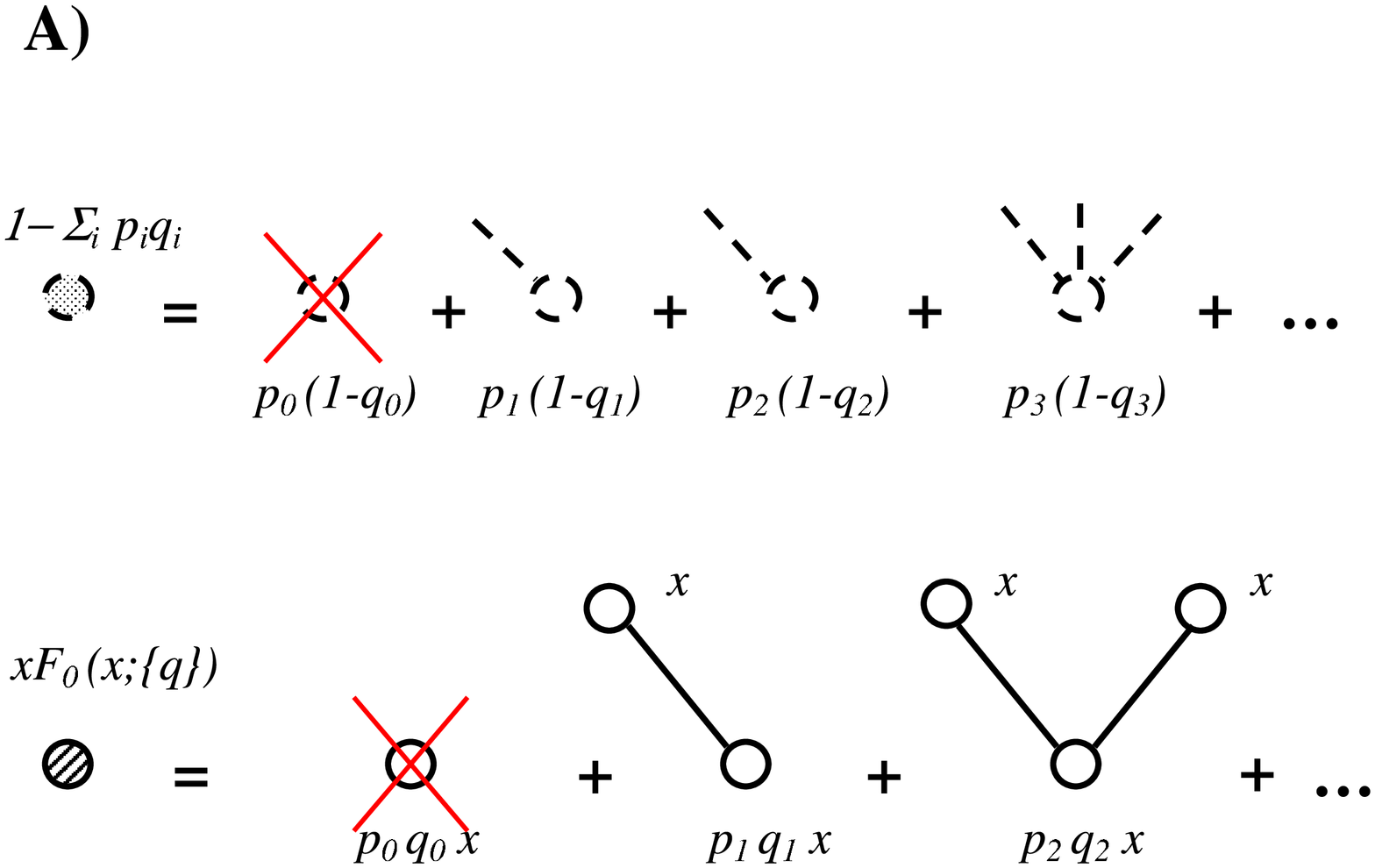}\\ \hline\end{tabular}~~
\begin{tabular}{|c|}\hline \\ \includegraphics*[angle=-90,width=0.45\textwidth]{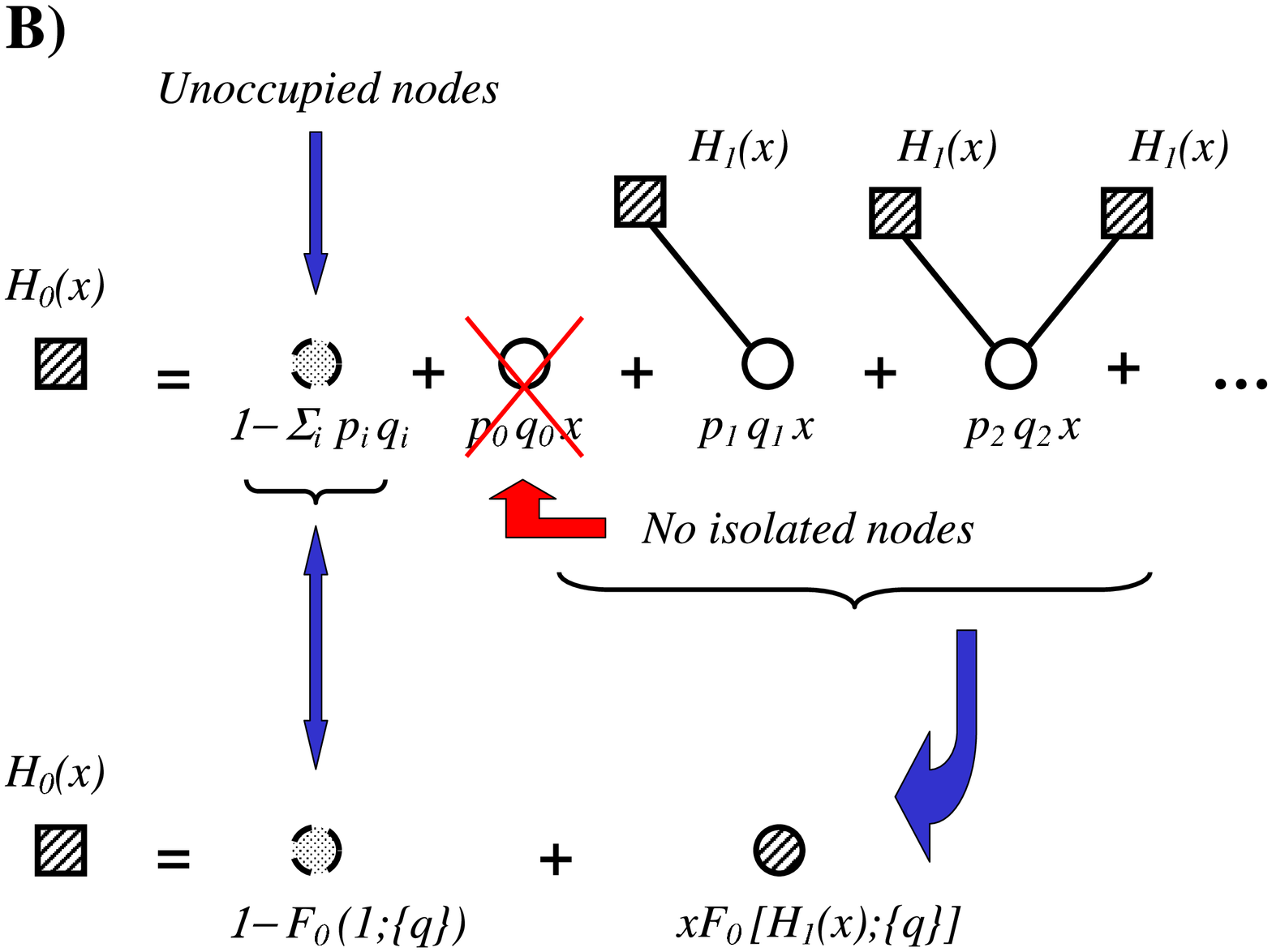}\\ \hline\end{tabular}
}
\caption{(A) A full dotted bullet with dashed contour line corresponds to the probability that a vertex is unoccupied. This is given by a particular series of diagrams, in which we sum the contributions of unoccupied vertices of all possible degrees.  
A striped bullet with full contour line represents the generating function $F_{0}(x;\{q\})$, whose diagrammatical expansion
contains all possible combinations of occupied vertices reachable by an occupied vertex with a certain degree. The $x$ accounts for the occupation of a vertex. 
The contributions of the isolated vertices ($p_{0}q_{0}$) have been deleted in agreement with our convention of considering only graphs with non isolated vertices.  
(B) Diagrammatical representation of the generating function $H_{0}(x)$. The first terms of the infinite series correspond to the summation of the Eqs.~\ref{series} as presented in Eq.~\ref{lungaH0}. These contributions can be expressed in a compact form using Eq.~\ref{h00}, that contains the  generating functions $F_{0}(x;\{q\})$ and $H_{1}(x)$.}
\label{diag1}
\end{figure}
%
%

Firstly, we consider the probability $P_{s}$ that a randomly chosen vertex in the network belongs to a connected cluster of a certain size $s$. We call $H_{0} (x; \{q\})$ its generating function,
\begin{equation}
H_{0}(x; \{q\}) = \sum_{s=0}^{\infty} P_{s} x^{s}~,
\end{equation}
in which we have conventionally grouped in the term for $s=0$ the probability $1-\sum_{k}q_{k}p_{k}$ that a vertex is not occupied.
Similarly, let $\hat{P_{s}}$ be the probability that a randomly chosen edge leads to a cluster of a given size $s$, and $H_{1}(x; \{q\})$  its generating function. 
  
Since we choose the starting vertex at random, each possible degree $k$ gives a different contribution to each possible cluster size probability $P_{s}$, meaning that each term $P_{s} x^{s}$ is itself given by an infinite sum of terms labeled by the degree $k$. 
For instance, the first terms of $H_{0}(x; \{q\})$ are:
\begin{subequations}\label{series}
\begin{align}
 P_{0}~~& = 1 - \sum_{k} p_{k} q_{k}~,\label{ser1}\\
 P_{1}x~& = \sum_{l=1}^{\infty} {p_{l} q_{l} x ~\hat{P_{0}}}^{l}~,\label{ser2}\\
 P_{2}{x}^{2}& = \sum_{l=1}^{\infty} {l p_{l} q_{l} x ~(\hat{P_{1}} x)}~{\hat{P_{0}}}^{l-1}~,\label{ser3}\\
 P_{3}{x}^{3}& = \sum_{l=1}^{\infty} {l p_{l} q_{l} x ~(\hat{P_{2}} {x}^{2})~\hat{P_{0}}^{l-1}} + \sum_{l\geq 2} p_{l} q_{l} x ~{\left(\begin{array}{c} l\\ 2\end{array}\right)}~{(\hat{P_{1}} x)}^{2}~\hat{P_{0}}^{l-2}~,\label{ser4}\\
 P_{4}{x}^{4}& = \sum_{l=1}^{\infty} l p_{l} q_{l} x ~(\hat{P_{3}} {x}^{3})~{\hat{P_{0}}}^{l-1} + \sum_{l\geq 2} p_{l} q_{l} x ~2 {\left(\begin{array}{c} l \\ 2 \end{array} \right)}~{(\hat{P_{2}} x^2)}~(\hat{P_{1}} x)~{\hat{P_{0}}}^{l-2}\\ \quad & \quad\quad +\sum_{l\geq 3} p_{l} q_{l} x ~{\left(\begin{array}{c} l \\ 3 \end{array} \right)}~{(\hat{P_{1}} x)}^3~{\hat{P_{0}}}^{l-3}~,\label{ser5}\\
\nonumber ~& \vdots
\end{align}
\end{subequations}
Now, summing these terms and grouping similar contributions we get
\begin{equation}\label{lungaH0}
\begin{split} H_{0}(x; \{q\})& = 1 - \sum_{k} p_{k} q_{k} + p_{1} q_{1} x ~ \left[ \hat{P_{0}} + \hat{P_{1}} x + \hat{P_{2}} {x}^{2} + \dots \right] \\
&\quad\quad + p_{2} q_{2} x ~ \left[ {\hat{P_{0}}}^{2} + 2 \hat{P_{0}} ~\hat{P_{1}} x + {(\hat{P_{1}} x)}^{2} + 2 \hat{P_{0}} ~\hat{P_{2}} x^{2} \right. \\ 
&\left. \quad\quad\quad + 2 \hat{P_{1}} x ~ \hat{P_{2}} {x}^{2} + {(\hat{P_{2}} {x}^{2})}^{2} + \dots \right] + \dots \\
&= 1 - \sum_{k} p_{k} q_{k}  + p_{1} q_{1} x \left[ \hat{P_{0}} + \hat{P_{1}} x +  \hat{P_{2}} {x}^{2} + \dots \right]\\ 
&\quad\quad+ p_{2} q_{2} x ~{\left[ \hat{P_{0}} + \hat{P_{1}} x +  \hat{P_{2}} {x}^{2} + \dots \right]}^{2} + \dots~\\
&= 1 - \sum_{k} p_{k} q_{k}  + x p_{1} q_{1} H_{1}(x; \{q\}) + x p_{2} q_{2} {\left[ H_{1}(x; \{q\})\right]}^{2} + \dots~.
\end{split}
\end{equation}
Note that no term contains $p_{0}q_{0}$ according with the convention of considering only non isolated nodes.
A compact form for this expression is written using Eq.~\ref{f0},
\begin{eqnarray}
H_{0}(x; \{q\}) = 1-F_{0}(1;\{q\}) + x F_{0}(H_{1}(x; \{q\}); \{q\})~.
\label{h00}
\end{eqnarray}
The structure of Eq.~\ref{h00} can be represented diagrammatically as shown in Fig.~\ref{diag1}, associating the variable $x$ to each ``bare'' vertex and a variable $H(x; \{q\})$ to each ``dressed'' vertex, while the function $F_{0}$  weights the contributions over all possible degrees.  

Moreover, the generating function $H_{1}(x; \{q\})$ satisfies a similar self-consistent equation,
\begin{eqnarray}
H_{1}(x; \{q\}) = 1-F_{1}(1;\{q\}) + x F_{1}(H_{1}(x; \{q\}); \{q\})~,
\label{h01}
\end{eqnarray}
that is obtained following completely similar arguments starting from picking up an edge at random. 

Taking the first derivative of $H_{0}$ in Eq.~\ref{h00} with respect to $x$ computed in $x=1$, we get the mean cluster size $\langle s \rangle$. Then, imposing the divergence of the latter allows to find the condition for the existence of a giant component, that corresponds to the Molloy-Reed criterion as presented in Ref.~\cite{callaway}. Finally, considering a uniform occupation probability $q_{k}=q$ (or uniform node traversing probability), the expression for the site percolation threshold $q_{c}$ can be computed.

\chapter{A General Percolation Theory for Spreading Processes}
\label{APP4_2}

It is possible to develop a general theory for {\em inhomogeneous joint site-bond percolation} exploiting 
the method of generating functions \cite{callaway,dorogoR,wilf}, that is briefly introduced in Appendix~\ref{APP4_2}. 
The theory holds for Markovian networks, i.e. random graphs with two-point degree correlations \cite{boguna2}, 
therefore the edge transition probabilities are at least dependent on the degree of the two extremities. 
Since real spreading rates may depend on many other features, we account for these properties introducing multi-type nodes, and assigning to each edge transition probability  a pair of additional labels indicating the types of the nodes at the extremities.

The main result of this appendix is a general version of the Molloy-Reed criterion for the existence of a giant component in the case of joint site-bond percolation and, consequently, the expressions of the critical threshold for the two separate cases of site percolation and bond percolation.

\section{Markovian Networks with Multi-Type Nodes}\label{APP4_2_1}

As already mentioned in Chapter~\ref{CHAP2}, vertex-vertex correlations are usually expressed using the {\em degree conditional probability} $p(k'|k)$, i.e. the probability that a vertex of degree $k$ is connected to a vertex of degree $k'$. This has led to the definition of a class of correlated networks, called {\em Markovian} random networks \cite{boguna2} and defined only by their degree distribution $p_{k}$ and by the degree conditional probability $p(k'|k)$. 
The function $p(k' |k)$ satisfies the normalization constraint and a detailed balance condition (see Chapter~\ref{CHAP2} for details). \\
In this context, the edge transition probability $T_{ij}$ must depend on the degrees $k_{i}$ and $k_{j}$ of the extremities.
Note that, while the analysis of the standard site (bond) percolation is based on the relation between the degree distribution $p_{k}$ and the degree-dependent node occupation probability $q_{k}$, in the inhomogeneous joint site-bond percolation the relation is between the pair of distributions $\{p_{k}, p(k'|k)\}$ and the pair of probability functions $\{q_{k}, T_{k k'}\}$.

The multi-type classification of the nodes consists in dividing the nodes of a graph into $n$ classes, each one specified by a particular quality or ``type''. The meaning is generic, it can be a distinctive feature of the node (e.g. the gender or the age in a population of individuals) or it can be associated to some quantity that have been measured on the network (e.g. the strength or the betweenness of the node). 
Then, we consider the degree distribution $p_{k}^{(h)}$ of nodes of class $h=1, \dots, n$, conventionally normalized on the relative set of nodes, i.e. $\sum_{k} p_{k}^{(h)} = 1$. This condition ensures the normalization to $1$ for the generating functions.   
Inside the classes there are no restrictions on the transition probabilities and they might be very different. \\
Summarizing, {\em our approach considers a Markovian correlated graph with multi-type vertices, in which each vertex is given an occupation probability depending on its degree and type, and each edge is endowed with a transition probability depending on the degrees and the types of the extremities}. 

The fundamental brick for the construction of generating functions in correlated graphs is the rooted edge composed of a starting vertex $i$ and the pending edge $(i, j)$ connecting it to a second vertex $j$, without explicitly considering this second extremity. For this reason we will always average on the degree of the second extremity of the edge.  
Let us consider a vertex $i$ chosen at random, it will be characterized by a class $h$ and by a degree $k_{i}$. In principle, the $k_{i}$ edges departing from that node are connected to $k_{i}$ other nodes belonging to different classes. Actually, only $m_{i}$ of them are really reached by a flow  because of the presence of transition probabilities on the edges (we call such transmitting edges {\em open}). Therefore, they identify a partition of $\{m_{i}^{(1)}, m_{i}^{(2)}, \dots, m_{i}^{(n)}\}$, with $\sum_{l} m_{i}^{(l)} = m_{i} \leq k_{i}$, in which $m_{i}^{(l)}$ is the number of these neighbouring nodes belonging to the class $l$ and linked to $i$ by an open edge.\\
Suppose that $m_{i}^{(l)}$ of the $k_{i}$ edges emerging from a node of class $h$ and degree $k_{i}$ are successfully connected to nodes of a same class $l$ and (possibly different) degrees $k_{j}$. The average probability that an edge among them allows the flow to pass is $\sum_{k_{j}} T^{(h \rightarrow l)}_{k_{i} k_{j}} p^{(h\rightarrow l)}(k_{j}|k_{i})$, where $p^{(h\rightarrow l)}(k_{j}|k_{i})$ is the degree conditional probability between vertices of states $h$ and $l$ and $T^{(h\rightarrow l)}_{k_{i} k_{j}}$ is the transition probability along an edge from a node of degree $k_{i}$ in the class $h$ to a node of degree $k_{j}$ in the class $l$.
The origin of this term is trivial: the probability to pass along the edge is the product of two independent events, i.e the edge exists and it is open; then, being interested in rooted edges, we have to average over all possible degrees $k_{j}$. 
The probability that there are $m_{i}^{(l)}$ of these edges produces a term ${[\sum_{k_{l}} T^{(h\rightarrow l)}_{k_{i} k_{j}} p^{(h\rightarrow l)}(k_{j}|k_{i})]}^{m_{i}^{(l)}}$. 
Positive events give $n$ contributions of this kind, while the $k_{i}-m_{i}$ negative events contribute to a single term ${\left[ 1- \sum_{l=1}^{n} \sum_{k_{j_{l}}} T^{(h\rightarrow l)}_{k_{i} k_{j_{l}}} p^{(h\rightarrow l)}(k_{j_{l}}|k_{i}) \right]}^{k_{i}-m_{i}}$, that is the probability that $k_{i} - m_{i}$ edges do not admit the flow's passage whichever class they belong to.
Computing the probability of the whole event associated with the {\em partition} $\{m^{(0)}_{i}=k_{i}-m_{i}, m^{(1)}_{i}, m^{(2)}_{i}, \dots, m^{(n)}_{i}\}$ of the neighbours of the node with degree $k_{i}$, we get the multinomial distribution 
\begin{equation}
\begin{split}
P^{(h)}(k_{i},\{m^{(l)}_{i}\}) &= {k}_{i}! \frac{1}{m_{i}^{(0)}!} {\left[ 1- \sum_{j_{l}=1}^{n} \sum_{k_{j_{l}}} T^{(h\rightarrow l)}_{k_{i} k_{j_{l}}} p^{(h\rightarrow l)}(k_{j_{l}}|k_{i}) \right]}^{m^{(0)}_{i}}\\ 
\quad & \quad \times \prod_{l=1}^{n} \frac{1}{m^{(l)}_{i}!} {\left[ \sum_{k_{j_{l}}} T^{(h\rightarrow l)}_{k_{i} k_{j_{l}}} p^{(h\rightarrow l)}(k_{j_{l}}|k_{i}) \right]}^{m^{(l)}_{i}}. 
\end{split}
\label{multi1}
\end{equation}
A simpler version of this multinomial distribution appears in Ref.~\cite{newman4}.\\
The following step consists in using the expression of the multinomial distribution to obtain the generating function of the probability that a physical quantity spreading from a vertex of class $h$ successfully flows through $\{m^{(l)}\}$ of its edges that point to vertices in the class $\{l\}$ ($l=1,2,\dots,n$). Summing over all possible values of $k_{i}$ and over all possible partitions of $k_{i}$ in $n+1$ values $\{m^{(l)}\}$, we obtain the generating function 
\begin{equation}
F_{0}^{(h)}(x_{1},x_{2}, \dots, x_{n};\{q, T\}) = \sum_{k_{i}=1}^{\infty} p_{k_{i}}^{(h)} q_{k_{i}}^{(h)} \sum_{\{m^{(l)}_{i}\}} \delta(k_{i}, \sum_{l=0}^{n} m^{(l)}_{i}) P^{(h)}(k_{i}, \{m^{(l)}_{i}\}) \prod_{l=1}^{n} x_{l}^{m^{(l)}_{i}}, 
\label{multi2}
\end{equation}
in which $q_{k_{i}}^{(h)}$ is the occupation (traversing) probability of a vertex belonging to the class $h$ with degree $k_{i}$, $\delta(\cdot,\cdot)$ is a Kronecker's symbol and the $x_{1}, \dots, x_{n}$ variables represent the average contributions of the rooted edges of the different classes.
Introducing Eq.~\ref{multi1} in Eq.~\ref{multi2}, the sum over the partitions $\{m^{(l)}_{i}\}$ corresponds to the extended form of a multinomial term, providing the following expression for the generating function  
\begin{equation}\label{multi3}
\begin{split}
F_{0}^{(h)}(x_{1},x_{2}, \dots, x_{n};\{q, T\}) & = F_{0}^{(h)}(\mathbf{x};\{q, T\}) \\
& = \sum_{k_{i}=1}^{\infty} p_{k_{i}}^{(h)} q_{k_{i}}^{(h)} {\left[ 1+ \sum_{l=1}^{n} (x_{l}-1) \sum_{k_{j_{l}}} T^{(h\rightarrow l)}_{k_{i} k_{j_{l}}} p^{(h\rightarrow l)}(k_{j_{l}}|k_{i}) \right]}^{k_{i}}.
\end{split}
\end{equation}
With a completely similar argument, we compute the generating function $F_{1}^{(h)}(\mathbf{x};\{q,T\})$ of the probability that a randomly chosen edge leads to a vertex of class $h$ from which the
spread toward its neighbours successfully flows through $\{m^{l}\}$ edges pointing to nodes of class $\{l\}$ ($l=1,2,\dots,n$). 
Hence, observing that now the number of emerging edges available to the spreading process reduces to $k_{i}-1$ and that the probability to reach the starting vertex (from an edge pointing to a generic vertex of class $h$) is $\frac{k_{i} p_{k_{i}}^{(h)}}{\sum_{k} k p_{k}^{(h)}}$, the generating function $F_{1}^{(h)}(\mathbf{x};\{q, T\})$ reads
\begin{equation}\label{multi4}
\begin{split}
F_{1}^{(h)}(x_{1},x_{2}, \dots, x_{n};\{q, T\}) & = F_{1}^{(h)}(\mathbf{x};\{q, T\}) \\
&= \sum_{k_{i}=1}^{\infty} \frac{k_{i} p_{k_{i}}^{(h)}}{\sum_{k} k p_{k}^{(h)}} q_{k_{i}}^{(h)} {\left[ 1+ \sum_{l=1}^{n} (x_{l}-1) \sum_{k_{j_{l}}} T^{(h\rightarrow l)}_{k_{i} k_{j_{l}}} p^{(h\rightarrow l)}(k_{j_{l}}|k_{i}) \right]}^{k_{i}-1}. 
\end{split}
\end{equation}

As recalled in the Appendix~\ref{APP4_1}, the two generating functions are useful in the computation of a system of self-consistent equations (similar to those in Eqs.~\ref{h00}-\ref{h01}) from which the expression of the average cluster size $\langle s \rangle$ should be derived. The main difference concerns the form of the generating functions $F_{0}^{(h)}(\mathbf{x};\{q, T\})$ and $F_{1}^{(h)}(\mathbf{x};\{q, T\})$, that are partitioned in classes (of nodes in different states) and contain the contributions of the transition probabilities. Firstly, we consider the probability $P_{s}^{(h)}$ that a randomly chosen edge leads to a vertex of class $h$ belonging to a connected component of a given size $s$. Its generating function $H_{1}^{(h)}(x;\{q, T\}) = \sum_{s} \hat{P}_{s}^{(h)} x^{s}$ satisfies the self-consistent equation
\begin{equation}
H_{1}^{(h)}(x; \{q, T\}) = 1-F_{1}^{(h)}(\mathbf{1};\{q, T\}) + x~ F_{1}^{(h)}[H_{1}^{(1)}(x;\{q, T\}), \dots, H_{1}^{(n)}(x;\{q, T\});\{q, T\}]~,
\label{multi5}
\end{equation}
where the presence of $H_{1}^{(h)}(x;\{q, T\})$ for all $h=1, \dots, n$ on the r.h.s. means that the constraint on the value of $h$ is required only on the starting node, not on the others reachable from it. Moreover, $x$ refers to the cluster distribution and does not need any label. \\
The first term in the r.h.s. of Eq.~\ref{multi5} is due to the probability that the node of class $h$ to which a chosen edge leads is not occupied, therefore it should not depend on the transmissibility of any outgoing edge. As required, the term $1-F_{1}^{(h)}(\mathbf{1};\{q, T\})$ computed in $x=1$ does not depend on $\{T\}$. This corresponds exactly to the term $\hat{P}_{0}^{(h)}$ in the cluster expansion. The second term of Eq.~\ref{multi5} refers to the contribution of an occupied vertex. Let us suppose that its degree is $k_{i}$ and consider one of its outgoing edges leading to a vertex in the class $l$: its contribution is given by the probability $1-\sum_{k_{j_{l}}} T^{(h\rightarrow l)}_{k_{i} k_{j_{l}}} p^{(h\rightarrow l)}(k_{j_{l}}|k_{i})$ that the flow does not reach the second extremity of the edge, and by the probability that it passes ($\sum_{k_{j}} T^{(h \rightarrow l)}_{k_{i} k_{j}} p^{(h\rightarrow l)}(k_{j}|k_{i})$). The latter has to be multiplied by the vertex function associated to the probability that this second vertex (of class $l$) belongs to a cluster of a given size. 
This probability is generated by the function $H^{(l)}_{1}(x; \{q, T\})$ that is the correct vertex term for this contribution.
Then, all these quantities have to be averaged over the set of degrees $k_{i}$, with weights $p^{(h)}_{k_{i}}$ and $q^{(h)}_{k_{i}}$, easily recovering the second term on the r.h.s.. Since the spirit of the derivation is completely the same, we refer again to Fig.~\ref{diag1} for a diagrammatical representation of the Eq.~\ref{multi5} (details are different). \\
The other equation, for the generating function $H_{0}^{(h)}(x; \{q, T\})$ of the probability that a randomly chosen vertex of class $h$ belongs to a cluster of fixed size $s$, reads
\begin{equation}
H_{0}^{(h)}(x; \{q, T\}) = 1-F_{0}^{(h)}(\mathbf{1}; \{q, T\}) + x~ F_{0}^{(h)}[H_{1}^{(1)}(x; \{q, T\}), \dots, H_{1}^{(n)}(x; \{q,T\}); \{q, T\}]~,
\label{multi6}
\end{equation}
Note that, by definition, both $H_{0}^{(h)}(x; \{q, T\})$ and $H_{1}^{(h)}(x; \{q, T\})$ are $1$ in $x=1$ for all $h$.

Now, taking the derivative of Eq.~\ref{multi6} with respect to $x$ in $x=1$, we obtain the average number of vertices reachable starting from a vertex in the class $h$,
\begin{equation}\label{multi7}
\begin{split} \langle s_{h} \rangle & = \frac{d H_{0}^{(h)}}{d x}{\bigg \vert}_{x=1}\\  
&= F_{0}^{(h)}(\mathbf{H}_{1}(1)=\mathbf{1}; \{q, T\}) + \sum_{l} \frac{\partial F_{0}^{(h)}}{\partial x_{l}}{\bigg \vert}_{\mathbf{x}=\mathbf{1}} {H_{1}^{(l)}}'(1; \{q, T\})~.
\end{split}
\end{equation}
The second term in the r.h.s. contains linear contributions from other classes of vertices, therefore Eq.~\ref{multi7} can be written in a matrix form (in the $n\times n$ product space generated by pairs of classes) as
\begin{equation}
\langle \mathbf{s} \rangle = \nabla_{x} \mathbf{H}_{0}(x; \{q, T\}) {\bigg \vert}_{x=1} = \mathbf{F}_{0}[\mathbf{1}; \{q, T\}] + \nabla_{x} \mathbf{F}_{0}[\mathbf{x}; \{q, T\}]{\bigg \vert}_{\mathbf{x}=\mathbf{1}} \cdot \nabla_{x} \mathbf{H}_{1}(x;\{q, T\}){\bigg \vert}_{x=1}~,
\label{multi8}
\end{equation}  
with $\mathbf{s} = (s_{1}, s_{2}, \dots, s_{n})$.
Taking the derivative of Eq.~\ref{multi5} with respect to $x$ in $x=1$, we obtain an implicit expression for ${H_{1}^{(h)}}'(1; \{q, T\})$ 
\begin{equation}
{H_{1}^{(h)}}'(1; \{q, T\}) = F_{1}^{(h)}[\mathbf{1}; \{q, T\}] + \sum_{l} \frac{\partial}{\partial x_{l}} F_{1}^{(h)}[1; \{q, T\}] {H_{1}^{(l)}}'(1; \{q, T\})~, 
\label{multi9}
\end{equation}
and putting together the contributions in a matrix formulation, we get
\begin{equation}
{\nabla_{x} \mathbf{H}_{1}(x; \{q, T\})}{\bigg \vert}_{x=1}  = \mathbf{F}_{1}[\mathbf{1}; \{q, T\}] + \nabla_{x} \mathbf{F}_{1}[\mathbf{x}; \{q, T\}]{\bigg \vert}_{\mathbf{x}=\mathbf{1}} \cdot ~\nabla_{x} \mathbf{H}_{1}(x; \{q, T\}){\bigg \vert}_{x=1}~. 
\label{multi10}
\end{equation}
Explicitly, Eq.~\ref{multi10} becomes  
\begin{equation}
{\nabla_{x} \mathbf{H}_{1}(x; \{q, T\})}{\bigg \vert}_{x=1} =  {\left[ I - \nabla_{x} \mathbf{F}_{1}[\mathbf{x}=\mathbf{1}; \{q, T\}] \right]}^{-1} \cdot \mathbf{F}_{1}[\mathbf{1}; \{q, T\}] = {\left[ I - \mathcal{F} \right]}^{-1} \cdot  \mathbf{F}_{1}[\mathbf{1}; \{q, T\}]~,
\label{multi11}
\end{equation}
with $\mathcal{F} = \nabla_{x} \mathbf{F}_{1}[\mathbf{x}=\mathbf{1}; \{q, T\}]$. Introducing the expression in Eq.~\ref{multi8},
we obtain 
\begin{equation}
\langle \mathbf{s} \rangle = \mathbf{F}_{0}[\mathbf{1}; \{q, T\}] + \nabla_{x} \mathbf{F}_{0}[\mathbf{x}=\mathbf{1}; \{q, T\}] \cdot {\left[ I - \mathcal{F} \right]}^{-1} \cdot  \mathbf{F}_{1}[\mathbf{1}; \{q, T\}]~.
\label{multi12}
\end{equation}
The condition for a giant component to emerge consists in the divergence of the mean cluster size, i.e. of at least one of the components of $\langle \mathbf{s} \rangle$. 
It corresponds to the following {\em generalized Molloy-Reed criterion} for inhomogeneous joint site-bond percolation in multi-type Markovian correlated random graphs: 
\begin{equation}
\det \left[I - \mathcal{F} \right] \leq 0~,
\label{pre_multi13}
\end{equation}
where $\mathcal{F} = \nabla_{x} \mathbf{F}_{1}[\mathbf{x}=\mathbf{1}; \{q, T\}]$ (whose elements are of the type 
$\frac{\partial}{\partial x_{l}} F_{1}^{(h)}[1;\{q, T\}]$).

It is evident that such a general result strongly depends on which kind of node partition we are considering.
Two examples of partitions are particularly relevant: i) a single class collecting all the nodes and ii) a degree-based classification of the nodes.

\section{Degree-based Multi-type Solution}\label{APP4_2_2}
In this situation the types correspond to the different degrees, thus each class gathers all vertices with a given degree and there is in principle an infinite number $n$ of types (in finite networks there are as many types as the number of different degrees). Using the relation $p_{k_{i}}^{(k_{i})}=1$ and $\frac{k_{i} p_{k_{i}}^{(k_{i})}}{\sum_{k} k p_{k}^{(k_{i})}} =1$, the Eqs.~\ref{multi3}-\ref{multi4} become 
\begin{subequations}\label{multi145}
\begin{align}
F_{0}^{(k_{i})}(x_{1}, \dots, x_{n}; \{q, T\})& = q_{k_{i}} {\left[ 1+ \sum_{k_{j}} (x_{k_{j}} -1) T_{k_{i} k_{j}} p(k_{j}|k_{i})\right]}^{k_{i}},\label{multi14}\\
F_{1}^{(k_{i})}(x_{1}, \dots, x_{n}; \{q, T\})& = q_{k_{i}} {\left[ 1+ \sum_{k_{j}} (x_{k_{j}} -1) T_{k_{i} k_{j}} p(k_{j}|k_{i})\right]}^{k_{i}-1}.
\label{multi15}
\end{align}
\end{subequations}
The self-consistent system of equations for the generating functions of clusters size probability looks like in Eqs.~\ref{multi5}-\ref{multi6}, and the condition for the existence of a giant component is still the divergence of the mean cluster size, but now the elements of the matrix $\mathcal{F}$ are
\begin{equation}
\mathcal{F}_{ij} = {(\nabla_{x} \mathbf{F}_{1} [\mathbf{x=1}; \{q, T\}])}_{ij} = (k_{i}-1) q_{k_{i}} T_{k_{i} k_{j}} p(k_{j}|k_{i})~.
\label{multi16}
\end{equation}
The generalized Molloy-Reed criterion becomes
\begin{equation}
\det \left[ (k_{i}-1) q_{k_{i}} T_{k_{i} k_{j}} p(k_{j}|k_{i}) -\delta_{ij} \right] \geq 0~.
\label{multi17}
\end{equation}
This expression corresponds to the criterion of Ref.~\cite{vazquez} for the existence of percolation in correlated random graphs (apart from a matrix transposition), but with the difference that in this case we are dealing with inhomogeneous joint site-bond percolation, then both the degree-dependent node traversing probability and edge transition probability appear in the expression. The condition that percolation threshold is related to the largest eigenvalue (see Ref.~\cite{vazquez}) is recovered if we assume that all nodes have equal traversing probability $q_{k_{i}} = q = const$. 
In other words, we are switching from a joint site-bond percolation to a simple site percolation with an occupation probability $q$. If $q \neq 0$ (otherwise the percolation condition cannot be satisfied), we can write the condition as 
\begin{equation}
 q \det \left[ (k_{i}-1) T_{k_{i} k_{j}} p(k_{j}|k_{i}) - \Lambda \delta_{ij} \right] \geq 0~,
\label{multi17bis}
\end{equation}
with $\Lambda = 1/q$.
Since in $q=0$ the determinant is negative, the smallest positive value of $q$ ensuring Eq.~\ref{multi17bis} to be satisfied corresponds to the largest eigenvalue $\Lambda_{max}$ of the matrix $(k_{i}-1) T_{k_{i} k_{j}} p(k_{j}|k_{i})$. 
It follows that the critical value of site occupation probability is $q = 1/\Lambda_{max}$. In the case $T_{k_{i} k_{j}}=1$, the condition gives exactly the results obtained by V\'azquez et al.~\cite{vazquez}.  

However, the complete knowledge of the correlation matrix $p(k_{j}|k_{i})$ is very unlikely for real networks, and the analytical solution of Eq.~\ref{multi17} can be problematic also for simple artificial networks.

\section{Local Homogeneity Approximation}\label{APP4_2_3}
 
When all nodes belong to a unique class, the single-state network is recovered and the analytical treatment becomes easier. 
We recover indeed a sort of local homogeneity approximation for the probability that an edge emerging from a vertex of degree $k_{i}$ is traversed by the flow. 
When $n=1$, the terms on the r.h.s. of Eqs.~\ref{multi3}-\ref{multi4} reduces to the average over the contributions of all the second extremities of the edges emerging from a node, obtaining a degree-dependent {\em effective average transmission coefficient}  $\tau_{k_{i}} = \sum_{k_{j}} T_{k_{i} k_{j}} p(k_{j}|k_{i})$ ($0 \leq \tau_{k_{i}} \leq 1$). 
The advantage of this approximation is that of providing analytically solvable equations for the percolation condition. Moreover, even if the sum on $k_{j}$ introduces an approximation, the effective term $\tau_{k_{i}}$  does not neglect the contributions due to the edge transition probabilities, weighting them with the correct degree conditional probability as required for Markovian networks. If these contributions are similar the approximation is very good, while when the edge transition probabilities or the nodes correlations are highly heterogeneous the local effective medium approximation breaks down, and we have to use the general approach. \\

{\bf \em Derivation of the Molloy-Reed Criterion - \quad}
According to this approximation, the generating function $F_{0}(x; \{q, T\})$ of the probability that a spreading process emerging from an occupied node flows through exactly $m$ nodes (whatever their degrees are) is written as  
\begin{equation}
F_{0}(x; \{q, T\}) = \sum_{k_{i}=1}^{\infty} p_{k_{i}} q_{k_{i}} {\left[1+ (x-1) \sum_{k_{j}=1}^{\infty} T_{k_{i} k_{j}} p(k_{j}|k_{i}) \right]}^{k_{i}}~. 
\label{ff0}
\end{equation}
This expression can be alternatively derived using simple arguments (see Ref.~\cite{spreading_dallasta}). 
The value $F_{0}(1;\{q,T\})$ assumed by this function in $x=1$ is the average occupation probability $\langle q \rangle = \sum_{k} p_{k} q_{k}$, that is consistent with the fact that summing over the contributions of all the possible amounts of emerging edges traversed by the flow means that we are simply considering the number of starting nodes, i.e. the average number of occupied nodes. 
The first derivative with respect to $x$ computed in $x=1$
\begin{equation}
\frac{\partial}{\partial x}F_{0}(x; \{q, T\}){\bigg \vert}_{x=1} = \sum_{k_{i}=1}^{\infty} k_{i} p_{k_{i}} q_{k_{i}} \sum_{k_{j}=1}^{\infty} T_{k_{i} k_{j}} p(k_{j}| k_{i})~,
\end{equation}
is the average number of open edges emerging from an occupied vertex.\\
Using similar arguments, we get the generating function $F_{1}(x; \{q, T\})$ of the probability that the flow spreading from a vertex, reached as an extremity of an edge picked up at random, passes through a given number of the remaining edges,
\begin{equation}
F_{1}(x; \{q, T\}) = \sum_{k_{i}=1}^{\infty} \frac{k_{i}  p_{k_{i}}}{\sum_{k} k p_{k}} q_{k_{i}} {\left[1 + (x-1)\sum_{k_{j}=1}^{\infty} T_{k_{i} k_{j}} p(k_{j}| k_{i}) \right]}^{k_{i}-1}~. 
\label{ff1}
\end{equation}

The next step consists in considering the self-consistent equations,
\begin{subequations}
\begin{align}
H_{0}(x; \{q, T\})& = 1 - F_{0}(1; \{q, T\}) + x F_{0}\left[ H_{1}(x; \{q, T\}); \{q, T\} \right]~, 
\label{apph0}\\
\nonumber ~\\
H_{1}(x; \{q, T\})& = 1 - F_{1}(1; \{q, T\}) + x F_{1}\left[H_{1}(x; \{q, T\}); \{q, T\}\right]~. 
\label{apph1}
\end{align}
\end{subequations} 
Now the two quantity $H_{0}(x; \{q, T\})$ (and $H_{1}(x; \{q, T\})$) is scalar, and represents the generating function of the probability that a randomly chosen vertex (and, respectively, vertex reached by a randomly chosen edge) belongs to a cluster of given size.
Since using scalar quantities allows to explain some passages, we derive also the expression for 
the mean cluster size $\langle s \rangle$, 
\begin{equation}
\langle s \rangle = \frac{d}{d x} H_{0} (x; \{q, T\}) {\bigg \vert}_{x=1} = H_{0}' (1; \{q, T\})~,  
\end{equation}
where $x=1$ implies the average over all possible degrees $k_{i}$.
From Eq.~\ref{apph0}-\ref{apph1}, using the relation $H_{1}(1; \{q, T\}) =1$, we get 
\begin{equation}
\langle s \rangle = F_{0}(1; \{q, T\}) + F_{0}'(1; \{q, T\}) H_{1}'(1; \{q, T\})~,
\label{apps1}
\end{equation}
where the expression for $H_{1}'(1; \{q, T\})$ comes directly from deriving Eq.~\ref{apph1} with respect to $x$ in $x=1$,
\begin{equation}
H_{1}'(1; \{q, T\}) = \frac{F_{1}(1; \{q, T\})}{1 - F_{1}'(1; \{q, T\})}~.
\end{equation}
Inserted into Eq.~\ref{apps1} it leads to the well-known expression (\cite{callaway})
\begin{equation}
\langle s \rangle = F_{0}(1; \{q, T\}) + \frac{F_{0}'(1; \{q, T\})}{1 - F_{1}'(1; \{q, T\})}~;
\label{apps2}
\end{equation}
hence, the giant component exists if and only if $F_{1}'(1; \{q, T\}) \geq 1$.
Explicitly, it means that, in the local effective medium approximation, the generalized Molloy-Reed criterion for the inhomogeneous joint site-bond percolation on Markovian random graphs takes the following form, 
\begin{equation}
\sum_{k_{i} = 1}^{\infty} p_{k_{i}} k_{i} \left[(k_{i} -1) q_{k_{i}} \sum_{k_{j} = 1}^{\infty} T_{k_{i} k_{j}}  p(k_{j}| k_{i}) - 1 \right] \geq 0~.
\label{mrcgg}
\end{equation}
The threshold for the site percolation with inhomogeneous edges occurs when the equality holds. Inserting uniform occupation probability $q_{k} \equiv q$, its critical value is  
\begin{equation}
q_{c} = \frac{\langle k \rangle}{\sum_{k_{i}=1}^{\infty} k_{i} (k_{i}-1) p_{k_{i}} \sum_{k_{j} = 1}^{\infty} T_{k_{i} k_{j}} p(k_{j}| k_{i})}~.
\label{thresh1}
\end{equation}
\\

{\bf \em Transition probability factorization - \quad}
Note that, all the expression for the site-percolation threshold reported in Section~\ref{CHAP4_3_1} have been derived starting from Eq.~\ref{mrcgg}. Henceforth, we will rapidly derive them, making explicit assumptions on the form of the edge transition probabilities.
The most natural assumption consists of their factorization in two single-vertex contributions, i.e. 
\begin{equation}
T_{k_{i} k_{j}} = \Theta_{i}(k_{i}) \Theta_{f}(k_{j})~, 
\label{teta}
\end{equation}
where subscripts $i$ and $f$ indicate an initial and a final term respectively, in order to stress the fact that first and second vertices of an edge can give different contributions in the inhomogeneous percolation process.  \\
Inserting Eq.~\ref{teta} in Eq.~\ref{mrcgg}, the condition for the existence of a giant component becomes
\begin{equation}
\sum_{k_{i}, k_{j} = 1}^{\infty} p_{k_{i}} k_{i} \left[ (k_{i} -1) \Theta_{i}(k_{i}) \Theta_{f}(k_{j}) q_{k_{i}} - 1 \right] p(k_{j}| k_{i}) \geq 0~. 
\label{tetateta}
\end{equation}

In the case of uncorrelated graphs, the conditional probability also factorizes in $p(k'|k) = \frac{k' p_{k'}}{\langle k \rangle}$, and Eq.~\ref{tetateta} gets simpler, leading to an interesting expression for the site percolation threshold, 
\begin{equation}
q_{c} = \frac{{\langle k \rangle}^{2}}{[ \langle k^{2} \Theta_{i}(k) \rangle - \langle k \Theta_{i}(k) \rangle ] \langle k \Theta_{f}(k) \rangle }~.
\label{tetasoil}
\end{equation}
 
As an alternative, it seems interesting to study situations in which the transition probability is a function of only
one of the two extremities of an edge. 
When it depends only on final nodes means that $\Theta_{i}(k) = const. = 1$ and $T_{k_{i} k_{j}} = \Theta_{f}(k_{j})$ and the  site percolation threshold takes the form 
\begin{equation}
q_{c} = \frac{{\langle k \rangle}^{2}}{[ \langle k^{2} \rangle - \langle k \rangle ] \langle k \Theta_{f}(k) \rangle}
= q_{c}^{hom}\frac{\langle k \rangle}{\langle k \Theta_{f}(k) \rangle}~,
\label{tetasoil2}
\end{equation}
where $q_{c}^{hom}$ is the value of the critical occupation probability for the correspondent standard homogeneous site percolation.
The opposite situation, $\Theta_{f}(k) = const. = 1$ and $\Theta_{i}(k)= T_{k}$, leads to the following form for the Molloy-Reed criterion
 \begin{equation}
\sum_{k=1}^{\infty} k \left[(k - 1)q_{k}T_{k} - 1 \right] p_{k}  \geq 0~. 
\label{eq_1k}
\end{equation} 
The inhomogeneous site percolation threshold $q_{c}$ follows directly imposing uniform occupation probability $q_{k}=q$,
\begin{equation}
q_{c} = \frac{\langle k \rangle}{\langle k^{2} T_{k} \rangle - \langle k T_{k} \rangle}~. 
\label{eq_2k}
\end{equation}
From Eq.~\ref{eq_1k}, also the threshold's value for the inhomogeneous bond percolation is immediately recovered. 
If we suppose uniform transition probabilities $T_{k} =T$, indeed, we can explicit $T$ in Eq.~\ref{eq_1k} as a function of the set of $\{q_{k}\}$ and compute the value of the threshold as
\begin{equation}
T_{c} = \frac{\langle k \rangle}{\langle k^{2} q_{k} \rangle - \langle k q_{k} \rangle}~, 
\label{eq_3k}
\end{equation}
that represents the case in which percolation on the edges is affected by refractory nodes. 

The last case we consider is that of uniform transition probabilities $T_{k_i k_j} = T$, with $0 < T \leq 1$.
In the limit $T = 1$, the original Molloy-Reed criterion is recovered, otherwise, introducing $T_{k_i k_j} = T$ in  Eq.~\ref{thresh1}, we find the same expression for the threshold of site percolation with noisy edges (as in Eq.~\ref{thres0} with $\langle T \rangle = T$)
\begin{equation}
q_{c} = \frac{1}{(\langle k^{2}\rangle/\langle k \rangle - 1) T}~. 
\label{eq_1}
\end{equation} 

\chapter[Naming Game solution in $d=1$]{Solution of the Naming Game on one-dimensional lattices}
\label{APP5_1}

\begin{figure}[thb] 
\centerline{
\includegraphics*[width=10.0cm]{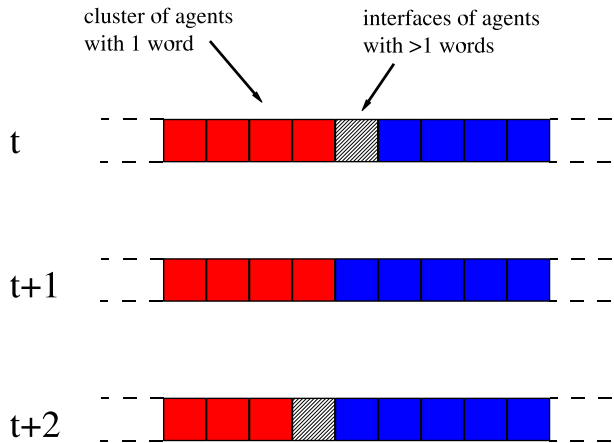}
 }
\caption{Sketch of the evolution of an interface in a one-dimensional Naming Game.
At the interface between two clusters (red and blue sites), one or more sites can assume 
more than one word.
}
\label{fig_interface}
\end{figure}

\begin{figure}[thb] 
\centerline{
\includegraphics*[width=10.0cm,angle=-90]{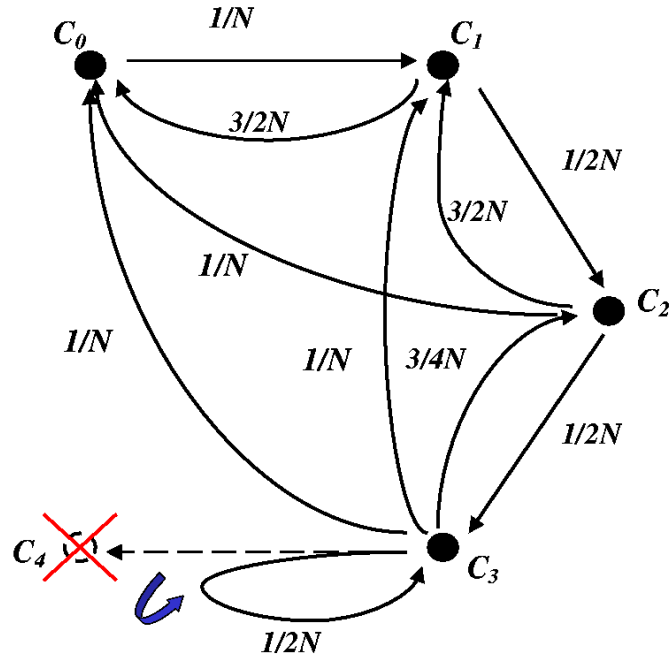}
 }
\caption{Truncated Markov process associated with interface width dynamics -
  schematic evolution of a $C_0$ interface $\cdots AAABBB\cdots$, cut at
  the maximal width $m=3$.}
\label{fig_markov}
\end{figure}

\begin{figure}[thb] 
\centerline{
\includegraphics*[width=10.0cm]{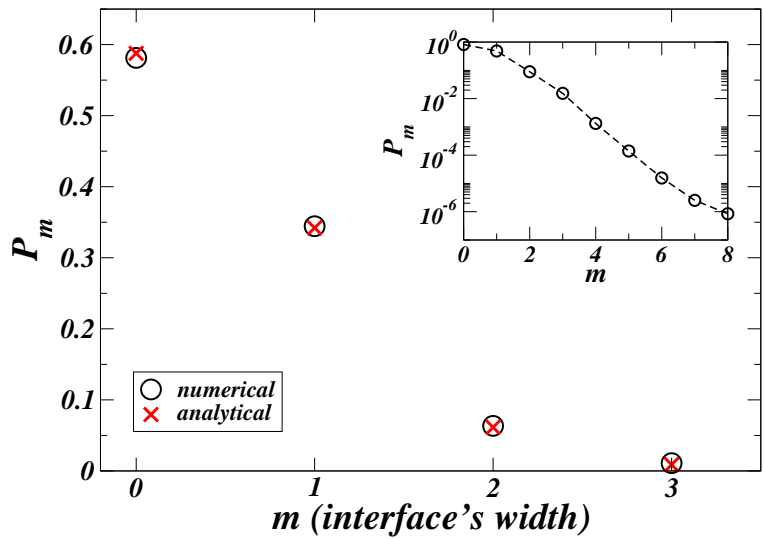} 
 }
\caption{
Numerical values of the probability of the interface's width (circles) and corresponding theoretical predictions (crosses) obtained as solution of the truncated Markov chain. The two sets of values are in excellent agreement.}
\label{fig_statC}
\end{figure}

In Section~\ref{CHAP5_3_1}, we have discussed the behavior of the Naming Game model on low-dimensional 
regular lattices by means of numerical simulations.
Local consensus is easy to reach, thus clusters composed of adjacent agents with the same unique word start to grow, 
developing a sort of effective surface tension at the interfaces.
This effective surface tension, whose origins are partially investigated in Section~\ref{CHAP5_3_1} 
is responsible for the slow coarsening dynamics leading the system to the consensus state.  \\
In one-dimensional lattices, it is possible to solve almost exactly the dynamics mapping the Naming Game on a
problem of diffusion and coalescence of interfaces. 
Let us consider a single interface between two linear clusters of agents: 
in each cluster, all agents share the same unique word, say $A$ in the
left-hand cluster and $B$ in the other. The interface is a string of
length $m$ composed of sites in which both states $A$ and $B$ are
present. We call $C_{m}$ this state ${(A+B)}^m$. A $C_{0}$ corresponds
to two directly neighboring clusters ($\cdots AAABBB\cdots$), while
$C_{m}$ means that the interface is composed by $m$ sites in the state
$C=A+B$ ($\cdots AAAC\cdots CBBB\cdots$).  Note that, in the actual
dynamics, two clusters of states $A$ and $B$ can be separated by a
more complex interface. For instance a $C_{m}$ interface can break down into two (or more)
smaller sets of $C$-states spaced out by $A$ or $B$ clusters, causing
the number of interfaces to grow.  
Numerical investigation shows that such configurations are however 
eliminated in the early times of the dynamics.
A sketch representing the evolution of the interface between two different clusters of 
words is reported in Fig.~\ref{fig_interface}. \\
Since the local dynamics is much faster than the global one, the probability that two neighboring 
clusters are separated by an interface of a given width is well approximated by 
the stationary solution of a Markov process governing the transitions between interfaces of different width.
The Markov chain is easily generated by means of an iterative method. \\
Let us consider a one-dimensional line composed of $N$ sites, initially divided into
two clusters of $A$ and $B$, with an interface of zero width. 
The probability to select for interaction exactly the link associated with this $C_{0}$ 
interface is $1/N$. 
Moreover, the outcome of such an interaction is the appearance of a $C_{1}$ interface, 
since one of the two agents (that chosen as hearer) learns the word ($A$ or $B$) uttered by the other. 
Thus, there is a probability $p_{0,1}=1/N$ that a $C_{0}$ interface becomes a $C_{1}$
interface in a single time step, otherwise it stays in $C_{0}$. \\
We can now compute all possible evolutions, and relative probabilities, for a $C_{1}$ interface.
From $C_{1}$ the interface can evolve into a $C_{0}$ or a $C_{2}$ interface
with probabilities $p_{1,0} = \frac{3}{2 N}$ and $p_{1,2} = \frac{1}{2
N}$ respectively.  This procedure is easily extended to higher values
of $m$, even if the number of the possible cases grows considerably and the enumeration becomes harder. 
In principle there is no limit to the growth of the interface's width; in practice, we can safely truncate 
this enumeration at $m \leq 3$, as suggested by numerical experiments in which we have counted the number of times an interface of a given width appears during the evolution of one-dimensional models.  \\
In this approximation, the problem corresponds to determine the stationary probabilities of the truncated Markov chain 
reported in Fig.~\ref{fig_markov} and defined by the transition matrix
\begin{equation}\label{markov_matrix}
\mathcal{M} =\left(
\begin{array}{cccc} 
\frac{N-1}{N} & \frac{1}{N} & 0 & 0 \\
\frac{3}{2 N} & \frac{N-2}{N} & \frac{1}{2 N} & 0 \\
\frac{1}{N} & \frac{3}{2 N} & \frac{N-3}{N} & \frac{1}{2 N} \\
\frac{1}{N}  & \frac{1}{N} & \frac{3}{2 N} & \frac{N-4}{N}+\frac{1}{2 N} \\  
\end{array} \right),  
\end{equation}
in which the basis is $\{C_{0}, C_{1}, C_{2}, C_{3}\}$ and the
contribution $\frac{1}{2 N}$ from $C_{3}$ to $C_{4}$ has been
neglected (see Fig.~\ref{fig_markov}).  \\
%
The stationary probability vector ${\bf{P}} = \{P_{0}, P_{1}, P_{2}, P_{3}\}$ is computed by imposing
${\bf{P}} (t+1) - {\bf{P}} (t) = 0$, i.e. $(\mathcal{M}^{T} - I)
{\bf{P}} = 0$. 
This condition gives 
\begin{equation}
P_{0} = \frac{133}{227} \approx 0.586, \quad P_{1}= \frac{78}{227} \approx 0.344, \quad P_{2} = \frac{14}{227}\approx 0.062, \quad P_{3} = \frac{2}{227} \approx 0.0088 \enspace. 
\end{equation}
Direct numerical simulations of the evolution of a line $\cdots AAABBB\cdots$ (of the type reported in Fig.~\ref{fig_interface}) 
yields $P_{0} \simeq 0.581$, $P_1=0.344$, $P_2=0.063$, $P_3=0.01$,
that are plotted in Fig.~\ref{fig_statC} together with the theoretical prediction, confirming 
that our approximation is correct. \\
Having shown that the width of the interfaces remains small, 
we assume that they are punctual objects localized around their central 
position $x$. For instance, in the previously mentioned example with $A$ and $B$
clusters, we denote by $x_{l}$ the position of the
right-most site of cluster $A$ and by $x_{r}$ the position of the
left-most site of cluster $B$: the position of the interface is $x
=\frac{x_{l}+x_{r}}{2}$. An interaction involving sites at the 
interface, i.e. an interface transition $C_{m} \rightarrow C_{m'}$,
corresponds to a set of possible movements for the central position
$x$. 
Our purpose is that of writing down a master equation for the movement of 
the interfaces.\\
The transition rates $W(x \rightarrow x\pm\delta)$ of an interface (its center) 
from a position $x$ to a position $x\pm\delta$ are obtained by enumeration of all
possible cases. Using the above approximation only three symmetric
contributions are present, corresponding to displacements of $\pm 1/2$, $\pm 1$ and $\pm 3/2$.
These transition probabilities receive contributions from interfaces with different width.
For instance, the Markov matrix in Eq.~\ref{markov_matrix} says that the $C_{1}$ interface $AACBB$
evolves into the following configurations
\begin{equation}\label{ex_displacement}
AACBB \Longrightarrow \left\{ \begin{array}{ccc} 
  AAABB & \frac{3}{2N} & x \rightarrow x+\frac{1}{2}~,\\ 
  AABBB & \frac{3}{2N} & x \rightarrow x-\frac{1}{2}~,\\ 
  ACCBB & \frac{1}{2N} & x \rightarrow x-\frac{1}{2}~,\\ 
  AACCB & \frac{1}{2N} & x \rightarrow x+\frac{1}{2}~, \\
  AACBB & \textnormal{otherwise} & x \rightarrow x~. \end{array} \right.
\end{equation}
in which we have put also the corresponding displacement of the center's position $x$.\\
Computing all possible transitions and grouping them according to the displacement $\delta$, we obtain
\begin{eqnarray} 
 W(x \rightarrow x\pm\frac{1}{2}) =& \frac{1}{2N} P_{0} + \frac{1}{N} P_{1} + \frac{1}{N} P_{2} +
\frac{1}{2N} P_{3}~, \\
 W(x \rightarrow x\pm 1) =& \frac{1}{2N} P_{2} + \frac{1}{2N} P_{3}~, \\
 W(x \rightarrow x\pm\frac{3}{2}) =& \frac{1}{2N} P_{3}~.  
\end{eqnarray}
Using the expressions for the stationary probability $P_{0},
\dots, P_{3}$, we finally get 
\begin{equation}\label{trans_rate}
W(x \rightarrow x\pm\frac{1}{2}) = \frac{319}{454N}, \quad W(x \rightarrow x\pm 1) = \frac{8}{227N}, \quad
W(x \rightarrow x\pm\frac{3}{2}) = \frac{1}{227N}.
\end{equation}

The knowledge of these transition probabilities allows us to write the
master equation for the probability $\mathcal{P}(x,t)$ to find the
interface in position $x$ at time $t$.
In the limit of continuous time and space, expanding the single terms 
\begin{eqnarray}
 \mathcal{P}(x,t+1) - \mathcal{P}(x,t) \approx \delta t \frac{\partial \mathcal{P}}{\partial t}(x,t),\\
 \mathcal{P}(x+ \delta x, t) \approx \mathcal{P}(x,t)
+ \delta x \frac{\partial \mathcal{P}}{\partial x}(x,t) +
\frac{{(\delta x)}^2}{2} \frac{{\partial}^2 \mathcal{P}}{{\partial
x}^2}(x,t)~,
\end{eqnarray} 
the master equation reduces to a standard diffusion equation 
\begin{equation}\label{diffusion_eq}
\frac{\partial \mathcal{P} (x,t)}{\partial t} = \frac{D}{N} \frac{{\partial}^2 \mathcal{P} (x,t)}{{\partial x}^2} \enspace,
\end{equation}
where $D = 401/1816 \simeq 0.221$ is the theoretical value of the diffusion coefficient (in the
appropriate dimensional units ${(\delta x)}^2 / \delta t$).\\
These results are confirmed by numerical simulations as illustrated in
the Section~\ref{CHAP5_3_1} by Fig.~\ref{gaussian_1D}, 
where the numerical probability $\mathcal{P}(x,t)$ is reported. 
The latter is a Gaussian centered around the initial position, with diffusion 
coefficient $D_{exp}\simeq 0.224 \approx D$. \\
The diffusion of interfaces and the consequent coarsening dynamics governing clusters growth
are the causes of the slow convergence of one-dimensional Naming Game model, that reaches the 
consensus absorbing state in $t_{conv} \sim \mathcal{O}(N^{3})$ steps ($\mathcal{O}(N^{2})$ if 
we rescale the time step by $N$ as in statistical mechanics models).

\chapter{Master equation approach to agents internal dynamics}
\label{APP5_2}

In this appendix, we discuss a more rigorous derivation of the probability distribution $\mathcal{P}_{n}(k|t)$ that an agent of degree $k$ has $n$ words stored in its inventory (at time $t$) (see also Ref.~\cite{naming_gameACT}).
Formally, the master equation associated to the jump process observed in the numerical simulations for the number of words (opinions, etc.) $n$ in the inventory of a fixed agent of degree $k$ can be written as 
\begin{eqnarray}
 \mathcal{P}_{n}(k,t+1)-\mathcal{P}_{n}(k,t)=& ~\mathcal{W}_{k}(n-1 \rightarrow n|t)\mathcal{P}_{n-1}(k,t)- \mathcal{W}_{k}(n\rightarrow n+1|t)\mathcal{P}_{n}(k,t)\\
\nonumber &~-\mathcal{W}_{k}(n\rightarrow 1|t)\mathcal{P}_{n}(k,t) ~~~~~ N_{d}(t) \geq n > 1\\ 
 \nonumber \mathcal{P}_{1}(k,t+1)-\mathcal{P}_{1}(k,t)=& ~\sum_{j=2}^{N_{d}(t)}\mathcal{W}_{k}(j \rightarrow 1|t)\mathcal{P}_{j}(k,t)- \mathcal{W}_{k}(1\rightarrow 2|t)\mathcal{P}_{1}(k,t)~, \label{rand}
\end{eqnarray} 
where $N_{d}(t)$ is the maximum number of different words at a time $t$ and $ \mathcal{P}_{n}(k,t+1)$ depends a priori explicitly on the time. \\
%
\begin{figure}[thb] 
\centerline{
\includegraphics*[width=10.0cm]{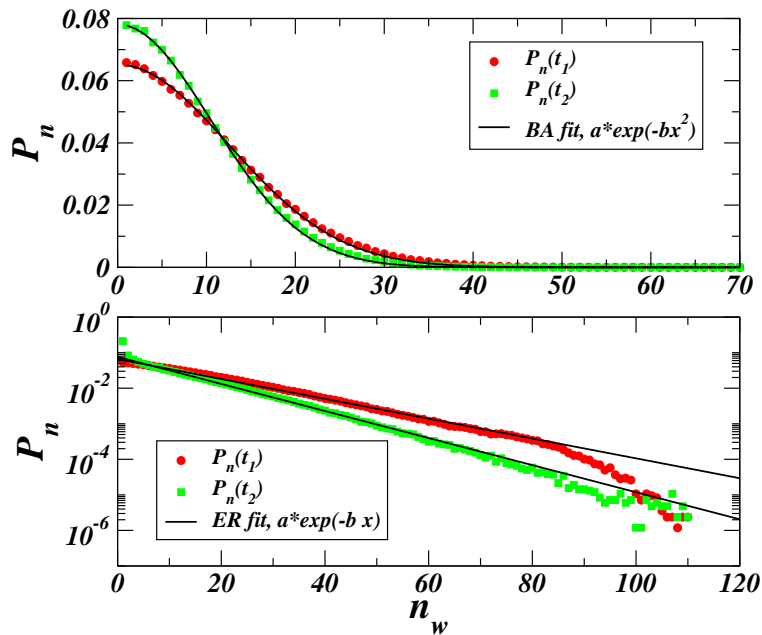}
}
\caption{Parametric dependence on time of the distribution of the number of words:
the time has the effect of deforming the shape of the distribution, but does not change the functional form.
(Top) BA graph with $N=10^4$ nodes with $\langle k\rangle =10$. Only the set of nodes with large $k$ (hubs) is monitored. Histograms come from measurements at different times $t_{1}$ and $t_{2}$ with $t_{2}-t_{1}= 5\cdot 10^{5}$ time steps. (Bottom) ER graph of $N=10^{4}$ nodes and $\langle k\rangle =10$. Measures refer to a set of high-degree nodes with $t_{2}-t_{1}= 5\cdot 10^{5}$ time steps.
}
\label{two_times}
\end{figure}
In order to get an expression for the transition rates, we call $C_{t}(k)$ the number of different words that are accessible to a node (of degree $k$) at time $t$, i.e. that are present in the neighborhood of the node. 
This number is not known, so that we consider it as a parameter to be estimated a posteriori.
Nevertheless, the mean-field type of dynamics that characterizes all small-world topologies, ensures that the quantity $C_{t}(k)$ depends mildly on $k$, so that in the following analysis we will often consider it as function only of the time $C_{t}$. In small-world topologies, indeed, there is an initial spreading of words throughout the network that destroys local correlations. The case of low-dimensional lattices is different since states can spread only locally, causing strong correlations between the inventories. We define also the inverse quantity, $\rho_{t}(k)=1/C_{t}(k)$, that will be used in the following formulae for notation's convenience.\\
The probability of a successful negotiation is $n \rho_{t}(k)$, where $n$ is the number of words in the inventory
of an hearer of degree $k$. Considering both the probabilities of the agent playing as hearer and speaker, the transition rate $\mathcal{W}_{k}(n\rightarrow 1|t)$ reads
\begin{equation}
\mathcal{W}_{k}(n\rightarrow 1|t) \simeq p_{k} \langle \rho_{t} \rangle \langle n \rangle + q_{k} \rho_{t}(k) n~.\label{pwin}
\end{equation}
Analogously, using the failure probability $1-n\rho_{t}(k)$,  
\begin{equation}
\mathcal{W}_{k}(n\rightarrow n+1|t) \simeq p_{k} (1 - \langle \rho_{t} \rangle \langle n \rangle) + q_{k} (1-\rho_{t}(k) n)~.
\label{ploose}
\end{equation} 
where the average terms $\langle \rho_{t} \rangle$ and $\langle n \rangle$ comes from an uncorrelated mean-field hypothesis 
for the neighboring sites of the node playing as speaker.\\
An important remark concerns the temporal evolution of the master equation. 
From Eqs.~\ref{pwin}-\ref{ploose}, we note that the timescale of the internal dynamics is fast, of order $1/p_{k} \propto N$. On the other hand, the only time-dependent parameter is $\rho_{t}(k)$ (or $C_{t}(k)$), that depends on the global behavior of the system, whose dynamics is slower. 
Unfortunately, we do not know the exact behavior of this global quantity, thus the only possible approach consists of 
considering it as an external modulation to the internal dynamics. 
According to this argument, we study the stationary condition of Eq.~\ref{rand}, and compute the solution $\mathcal{P}_{n}(k|t)$, in which we assume a parametric dependence of the distribution on the time, governing its relaxation dynamics.  
We have verified this assumption measuring the distribution at different times. As displayed in Fig.~\ref{two_times}, 
the solution depends on the time, but the dependence has the only effect of deforming the shape during the evolution, without changing the functional form.\\
Plugging the expressions of the transition rates into the stationary master equation, we get 
the following recursion relation,
\begin{equation}
\mathcal{P}_{n}(k|t) = \frac{q_{k} \left[ 1- \rho_{t}(k)(n-1)\right]}{q_{k}\left[ 1- n \rho_{t}(k)\right] + q_{k}\rho_{t}(k) n + p_{k} \langle \rho_{t}\rangle \langle n\rangle} \mathcal{P}_{n-1}(k|t)~.\label{rec_0} 
\end{equation}
Then, introducing $q_{k} = k p_{k}/\langle k\rangle$ $=$ $b(k)p_{k}$ and replacing $\rho_{t}(k)(n-1)$ with $\rho_{t}(k) n$ (that is true for $n \gg 1$), Eq.~\ref{rec_0} can be rewritten as
\begin{equation}
\mathcal{P}_{n}(k|t) = \frac{b(k) [1-\rho_{t}(k) n]}{b(k)+\langle \rho_{t}\rangle \langle n\rangle} \mathcal{P}_{n-1}(k|t)~. 
\end{equation}
Since $n \rho_{t}(k) \ll 1$, we can write $1-\rho_{t}(k)n \simeq e^{-\rho_{t}(k)n}$, thus solving the recurrence relation,
\begin{equation}
\mathcal{P}_{n}(k|t) \simeq {\left[\frac{b(k)}{b(k)+\langle \rho_{t} \rangle \langle n\rangle} \right]}^{n-1} e^{-\rho_{t}(k)\frac{(n+2)(n-1)}{2}} \mathcal{P}_{1}(k|t)~.
\end{equation}
The normalization relation gives the constant $\mathcal{P}_{1}(k|t)$. 
The controlling parameter of the curve is $s_{t}(k)= b(k)/(b(k)+\langle \rho_{t} \rangle \langle n\rangle)$, that allows to tune the decay of the distribution between an exponential and a gaussian-like tail. \\
In the next two paragraphs, we focus on these two cases separately, showing that the two different results derive from  different forms of the transition rates, that can be also inferred from simulations.\\
%
\begin{figure} 
\centerline{
\includegraphics*[width=10.0cm]{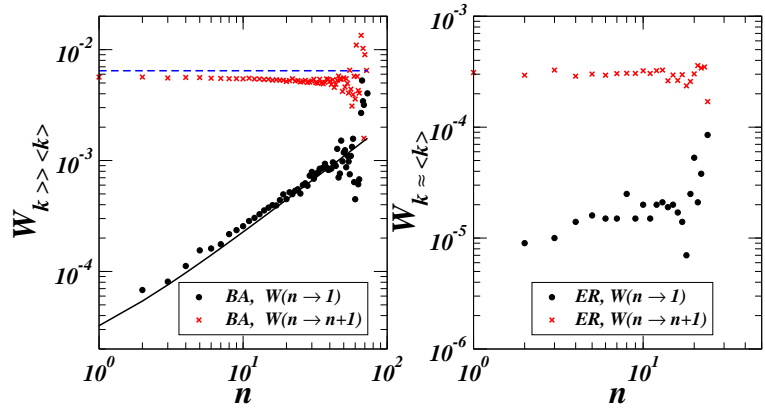}
}
\caption{Probability of $\mathcal{W}_{k}(n\rightarrow 1|t)$ and $\mathcal{W}_{k}(n\rightarrow n+1|t)$ for BA and ER models.
Both with $N=5000$ nodes and $k\simeq 200$ (for a BA with $\langle k \rangle =10$) and $k\simeq 70$ (for a ER with $\langle k \rangle =50$).
}
\label{pwinloose}
\end{figure}

{\bf \em The case of homogeneous networks - \quad}
Exploiting the remark coming from simulations, in homogeneous networks, we can safely put $q_{k} \simeq b p_{k}$, with $b \simeq \mathcal{O}(1)$ and $\rho_{t}(k) \simeq \langle \rho_{t} \rangle$ (i.e. $C_{t}(k) \simeq C_{t}$), since the nodes are almost equivalent.
The number of states as well is almost the same for every node that implies $n \simeq \langle n \rangle$. 
The approximated expressions of the success and failure probabilities are 
\begin{eqnarray}
\mathcal{W}_{k}(n\rightarrow 1|t)  \approx p_{k} \langle n \rangle (1+b) /C_{t} \approx 2 p_{k} \langle n \rangle/C_{t}\\
\mathcal{W}_{k}(n\rightarrow n+1|t) \approx p_{k} (1- \langle \rho_{t} \rangle \langle n \rangle) (1+b) \approx 2 p_{k}.\label{trans_hom}
\end{eqnarray}
We have verified numerically the validity of these approximated expressions for the quantities $\mathcal{W}_{k}(n\rightarrow n+1|t)$ and $\mathcal{W}_{k}(n\rightarrow 1|t)$ (just the term for nodes playing as hearers, since the term for speakers is almost constant). The data reported in Fig.~\ref{pwinloose} (right) show that for an ER model with $N=5 \cdot 10^3$ nodes and $\langle k \rangle =50$, both quantities are almost constant with respect to $n$.\\
The equilibrium condition for the master equation becomes
\begin{eqnarray}\nonumber
0 =& ~ 2 p_{k} \mathcal{P}_{n-1}(k|t)- 2 p_{k}\mathcal{P}_{n}(k|t)
- p_{k} \langle n \rangle \frac{2}{C_{t}} \mathcal{P}_{n}(k|t) ~~~~~ n > 1\\ 
0 =& ~ \sum_{j=2}^{\infty} \frac{2}{C_{t}} \langle n\rangle  p_{k} \mathcal{P}_{j}(k|t)- 2 p_{k} \mathcal{P}_{1}(k|t)~. \label{rand_hom}
\end{eqnarray} 
We can neglect the dependence on $k$, since $p_{k}$ is canceled, in agreement with the approximation done for the homogeneity of the network. The solution by recursion is very simple,
\begin{equation}
\mathcal{P}_{n}(t) \approx (1-\theta_{t}) {\theta_{t}}^{n-1}~, ~~~~~~~~~ \theta_{t} = \frac{1}{1+\frac{\langle n \rangle}{C_{t}}}~.
\label{statd_ER}
\end{equation}
Using the expansion of logarithm $\log(1-\epsilon_{t}) \simeq -\epsilon_{t}$, with $\epsilon_{t} = 1-\theta_{t} \simeq \langle n\rangle /C_{t}$, the previous formula gives the following exponential decay for the distribution of the number of words,
\begin{equation}
\mathcal{P}_{n}(t) \simeq \frac{\langle n\rangle}{C_{t}} e^{-\frac{\langle n\rangle n}{C_{t}}}~. 
\label{distrER}
\end{equation}
The exponential decay is in agreement with the numerical data, providing a value $\simeq 0.16$ for $\langle n \rangle/C_{t}$. \\

{\bf \em High-degree nodes in heterogeneous networks - \quad}
The other important case is that of the hubs in heterogeneous networks. In a direct NG a hub is preferentially chosen as hearer, by a factor $k/\langle k \rangle \gg 1$, then we can neglect the first terms in both success and failure probability. We consider the following approximated expressions
\begin{eqnarray}
\mathcal{W}_{k}(n\rightarrow 1|t) \simeq  q_{k} \rho_{t}(k) n~,\\
\mathcal{W}_{k}(n\rightarrow n+1|t) \simeq  q_{k} (1- \rho_{t}(k) n) \simeq q_{k}~, \label{phubsBA}
\end{eqnarray}
in which the last approximation is justified by the fact that the number of words accessible to a hub is large, thus $\rho_{t}(k)$ is expected to be small.  
For heterogeneous networks, the numerical data in Fig.~\ref{pwinloose} (left) for $\mathcal{W}_{k}(n\rightarrow 1|t)$ show
a clear linear growth of the quantity with $n$, in agreement with Eq.~\ref{pwin}, while the almost constant behavior of  
$\mathcal{W}_{k}(n\rightarrow n+1|t)$ with $n$ can be fitted with an expression of the form Eq.~\ref{ploose} only for very small values of the product $\rho_{t}(k)n$. 
On the other hand, Fig.~\ref{pwinloose} (bottom) points out that in the case of homogeneous networks both the quantities are almost constant with respect to $n$.
We prove now that these different behaviors of the transition rates are responsible of the different shape of the probability distribution $\mathcal{P}_{n}(k|t)$. \\
We can easily compute the quasi-stationary distribution $\{\mathcal{P}_{n}(k|t)\}$, from the first equation in Eq.~\ref{rand}
\begin{equation}
0= q_{k} \mathcal{P}_{n-1}(k|t) - q_{k} \mathcal{P}_{n}(k|t) - (q_{k} + q_{k} \frac{n}{C_{t}(k)} )\mathcal{P}_{n}(k|t)~,
\end{equation}
and we find recursively
\begin{eqnarray}
\mathcal{P}_{n}(k|t) &= \frac{C_{t}(k)}{C_{t}(k)+n} \mathcal{P}_{n-1}(k|t) = \frac{{C_{t}(k)}^2}{(C_{t}(k)+n)(C_{t}(k)+n-1)} \mathcal{P}_{n-2}(k|t) \\
\quad & \approx \frac{{C_{t}(k)}^{n-1}\Gamma(C_{t}(k)+2)}{\Gamma(C_{t}(k)+n+1)} \mathcal{P}_{1}(k|t)~.
\end{eqnarray}
Now, from the closure relation $\sum_{n=1}^{\infty}\mathcal{P}_{n}(k|t) = 1$ we get the expression of $\mathcal{P}_{1}(k|t)$, and the final form for  $\mathcal{P}_{n}(k|t)$ becomes
\begin{equation}
\mathcal{P}_{n}(k|t) = \frac{{C_{t}(k)}^{n-1}}{\Gamma(C_{t}(k)+n+1)} {C_{t}(k)}^{C_{t}(k)+1} e^{-C_{t}(k)}\left[ \frac{\Gamma(C_{t}(k)+1)}{\gamma(C_{t}(k)+1,C_{t}(k))}\right],
\label{statd_BA}
\end{equation}
where $\gamma(a,x)$ is the lower incomplete Gamma function. 
The functional form of the quasi-stationary distribution is complicated, but exploiting some approximations for Gamma functions, it is possible to write it in a simpler form (see Ref.~\cite{naming_gameACT} for details),   
\begin{equation}
\mathcal{P}_{n}(k|t) \simeq \sqrt{\frac{2}{\pi C_{t}(k)}} e^{\frac{-{(n+1)}^{2}}{2 C_{t}(k)}}~.
\label{distrBA} 
\end{equation}
This expresssion, corresponding to a Half-Normal distribution, is in very good agreement with the result of simulations.
It is remarkable that fitting numerical results in Fig.~\ref{distributionP} (top) with this expression gives a consistent value for $C_{t}(k)$. Indeed, from simulations the constant prefactor is $\simeq 0.0046$ and $C_{t}(k) \simeq 357$, that plugged into $\sqrt{\frac{2}{\pi C_{t}(k)}}$ gives the very close value $0.0042$.

\newpage
\pagestyle{fancy}

\backmatter

\end{document}